\documentclass[aps,prb,showpacs,superscriptaddress,twocolumn,longbibliography]{revtex4-1}

\usepackage{amsfonts}
\usepackage{subfigure}
\usepackage{amsmath}
\usepackage{txfonts}
\usepackage{amssymb}
\usepackage{amsbsy} 
\usepackage{epsfig}
\usepackage{graphicx}
\usepackage{epstopdf}

\def\be{\begin{equation}} \def\ee{\end{equation}}
\def\bea{\begin{eqnarray}} \def\eea{\end{eqnarray}}

\def\nn{\nonumber}

\def\bk{{\bf k}}
\def\br{{\bf r}}

\def\be{{\bf e}}
\def\bd{{\bf d}}
\def\bA{{\bf A}}

\newcommand{\bra}[1]{\langle#1|}
\newcommand{\ket}[1]{|#1\rangle}

\def\ra{\rangle}

\def\rw{\rightarrow}

\begin{document}

\title{Topological invariants of Floquet systems: General formulation, special properties, and Floquet topological defects}

\author{Shunyu Yao}
 \affiliation{ Institute for
Advanced Study, Tsinghua University, Beijing,  100084, China }

\author{Zhongbo Yan}
\affiliation{ Institute for
Advanced Study, Tsinghua University, Beijing,  100084, China }

\author{Zhong Wang} \altaffiliation{ wangzhongemail@gmail.com }
\affiliation{ Institute for
Advanced Study, Tsinghua University, Beijing,  100084, China }

\affiliation{Collaborative Innovation Center of Quantum Matter, Beijing, 100871, China }


\begin{abstract}

Periodically driven (Floquet) systems have been under active theoretical and experimental investigations. This paper aims at a systematic study in the following aspects of Floquet systems:
(i) A systematic formulation of topological invariants of Floquet systems based on the cooperation of topology and symmetries. Topological invariants are constructed for the ten symmetry classes in all spatial dimensions, for both homogeneous Floquet systems (Floquet topological insulators and superconductors) and Floquet topological defects. Meanwhile, useful representative Dirac Hamiltonians for all the symmetry classes are obtained and studied. (ii) A general theory of Floquet topological defects, based on the proposed topological invariants. (iii) Models and proposals of Floquet topological defects in low dimensions. Among other defect modes, we investigate Floquet Majorana zero modes and Majorana Pi modes in vortices of topologically trivial superconductors under a periodic drive. In addition, we clarified several notable issues about Floquet topological invariants. Among other issues, we prove the equivalence between the effective-Hamiltonian-based band topological invariants and the frequency-domain band topological invariants.

\end{abstract}

\maketitle


\section{Introduction}

Many phases and phase transitions of condensed matters can be understood by the unifying concepts of local order parameters and broken symmetries. Nevertheless, the discovery of quantum Hall (QH) effects demonstrated convincingly that this paradigm is incomplete\cite{klitzing1980,laughlin1981,laughlin1983}. The quantized Hall conductance in the integer QH effects is proportional to the Thouless-Kohmoto-Nightingale-den Nijs (TKNN) number\cite{thouless1982} (a Chern number), which, as a topological invariant, depends only on the global topology of Bloch wave functions in the entire Brillouin zone. The recent discoveries of topological insulators and topological superconductors\cite{hasan2010,qi2011,Chiu2016rmp,Bansil2016, bernevig2013topological,shen2013topological} have renewed the interests in topological matters, which are now among the central concepts of condensed matter physics. In the noninteracting limit, the interplay of topology and symmetry gives rise to the tenfold way classifications of topological phases\cite{kitaev2009,ryu2010,schnyder2008,stone2010symmetries,Chiu2016rmp}.

The most salient and ubiquitous feature of topological phases is the existence of robustly gapless boundary states, which are immune to disorders. Among the well known examples are the chiral edge states of the quantum Hall insulators, the helical liquids\cite{wu2006helical,xu2006} at the edge of two-dimensional (2d) time-reversal-invariant topological insulators, the surface Dirac cone of 3d topological insulators\cite{fu2007b}, and the half-integer spin at the end of integer-spin Haldane chain\cite{haldane1983continuum,chen2012symmetry}. From a more general perspective, the boundary of a material is a topological defect sandwiched between the material and vacuum, and the gapless boundary modes are examples of topological defect modes.  There are many other types of topological defects, and various potential applications of topological materials rely on these defects; for instance, as a point defect, a vortex core of a 2d chiral topological superconductor carries a Majorana zero mode (MZM)\cite{read2000,volovik1999fermion}, whose braiding obeys non-Abelian statistics\cite{moore1991nonabelions,wen1991a,ivanov2001,nayak1996,sarma2005}, which is potentially important for Majorana-based topological quantum computation\cite{kitaev2001unpaired,nayak2008}. Topological defects have remarkably regular patterns, and a systematic tenfold way classification of topological defects in all spatial dimensions have been put forward by Teo and Kane\cite{teo2010}.

The topological invariants are usually material constants, with rather limited tunability for a given sample. Recently, periodic driving has been explored as a promising approach to create and engineer topological materials with high tunability\cite{Oka2009,lindner2011floquet,Kitagawa2011,Inoue2010,
Gu2011,Kitagawa2010a,Kitagawa2010b,Jiang2011,cayssol2013floquet}, potentially offering a new fruitful platform for topological phenomena. In solid-state systems, a laser beam provides a periodic driving by its time-dependent electromagnetic potential $\bA(t)$. Among many other interesting proposals, it has been predicted that monochromatic light can drive graphene-like Dirac bands to Floquet Chern bands\cite{Oka2009,Usaj2014,Perez2014graphene,Perez2015,Dahlhaus2015}, trivial insulators and semimetals to Floquet topological insulators\cite{lindner2011floquet,Kundu2014,Torres2014, Lago2015,Klinovaja2016,Dutreix2016,nodal-link}, and nodal lines to Weyl points \cite{Yan2016tunable,Chan2016nodal,
Narayan2016nodal,Taguchi2016nodal} or multi-Weyl points\cite{Ezawa2017Photoinduced,Yan2017Floquet}. Experimentally, Floquet-Bloch bands have been observed at the surface of topological insulators\cite{wang2013observation,
mahmood2016selective,Fregoso2013}. In cold atom systems, periodic driving can be implemented by shaking the optical lattice\cite{Eckardt2017,jotzu2014experimental,parker2013direct,Hauke2012, Zheng2014,Jimenez2015,flaschner2016experimental,Mei2014,Lellouch2017}, which has enabled the experimental realization of the Haldane model\cite{jotzu2014experimental}. Photonic and acoustic materials are also platforms of Floquet topological materials\cite{rechtsman2013photonic,peng2016experimental,maczewsky2017observation, Mukherjee2016}. Recently, periodically driven topological\footnote{Other interesting aspects of driving-induced physics, such as light-induced superconductivity\cite{fausti2011light,mitrano2016possible}, will not be discussed here.} systems have attracted widespread attentions\cite{rudner2013anomalous,Carpentier2015,Lindner2013, Dahlhaus2011,Gomez2013,Goldman2014,Goldman2015,
Zhou2011Optical,Delplace2013,wang2014floquet,Dehghani2014,Seetharam2015, Hubener2016,Wang2016Network,Chan2016hall,Fleury2016,Morimoto2017, Xu2017space,bukov2015universal,nathan2015topological,Calvo2015, Fulga2016,morimoto2016topological,Budich2017,Else2016,Roy2016periodic,Yao2017, Martin2017,Roman2017strained-graphene}.

In addition to providing a controllable tool for engineering  topological phases, periodic driving can also create fundamentally new topological states without static counterparts\cite{Kitagawa2010b,rudner2013anomalous,Leykam2016,Mukherjee2016, titum2016,Kundu2017Quantized}. For instance, robust chiral modes can appear at the edge of a 2d driven system even though all the Chern numbers of the bulk bands are zero\cite{Kitagawa2010b,rudner2013anomalous}, which suggests topological classifications and topological invariants beyond the static systems\cite{rudner2013anomalous}.

Topological invariants are central tools in the study of topological matters. The value of a topological invariant unambiguously tells the topological class to which a system belongs. For Floquet systems, although topological invariants  have been constructed for a few symmetry classes and spatial dimensions\cite{rudner2013anomalous,Carpentier2015,Asboth2014,Fruchart2016}, a complete list (in the sense of the tenfold way classification) of topological invariants has so far been lacking. Recently, a periodic table of Floquet topological insulators has been obtained\cite{Roy2016periodic} via the K-theory, which nevertheless does not rely on topological invariants. The first purpose of our present paper is to put forward a complete and unified formulation of topological invariants of Floquet systems. The symmetries constrain the forms and possible values of topological invariants, leading to a systematic treatment for all the tenfold-way symmetry classes in all spatial dimensions. We also address and clarify several subtle points of topological invariants of Floquet systems.  One of them is the equivalence between the effective-Hamiltonian-based band topological invariants and the frequency-domain band topological invariants (see Appendix \ref{sec:proof}).

The second purpose of this paper is to put forward a general theory of topological defects in Floquet systems. In addition to the intrinsic theoretical interest, Floquet topological defects may  potentially offer highly tunable devices for applications. In solid-state systems, Floquet topological defects may be created optically, which can be controlled with high speed. There have been a few scattered studies of topological defects in Floquet systems\cite{Katan2013modulated,Lovey2016,bi2016}; for instance, it has been shown that a light beam with a vortex-like phase modulation can generate a Floquet zero mode in a 2d system\cite{Katan2013modulated}, and a spatially modulated driving can create a Floquet line defect hosting chiral modes in 3d Dirac semimetals\cite{bi2016}, even though the static system is defect-free. However, a systematic study of Floquet topological defects is lacking. Our general theory fills this gap.

This general theory of Floquet topological defects is based on
our unified formulation of topological invariants, which are defined not only for Floquet topological insulators and superconductors with translational symmetry, but also for Floquet topological defects.
The topological invariants are formulated in terms of the time evolution operator defined on certain parameter space (to be explained in details in the following sections). We provide a systematic classification of Floquet topological defects in all spatial dimensions, and a complete list of topological invariants for these defects. The dimensions of defects, the dimensions of space in which the defects live, and the symmetries of the system,  jointly impose constraints on the forms and possible values of the topological invariants.  We prove that the defect topological invariants reduce to the Teo-Kane topological invariants\cite{teo2010} in the static limit. The bulk topological invariants of homogeneous Floquet systems are obtained as special cases (the $D=0$ cases, see below) of our formulation. We also study representative Dirac Hamiltonians for general spatial dimensions, which are useful in model construction of Floquet topological insulators and Floquet topological defects.

The third purpose of this paper is to study a number of interesting examples of Floquet topological defects, some of which may have potential applications. In particular, we show that Majorana Pi modes (MPMs), which are Floquet versions of the MZMs, can be created inside vortices of driven topologically trivial superconductors, which host no MZM in the static case. We apply our topological invariants to study genuine Floquet topological defects without static counterparts.

The rest of this paper is organized as follows. We first introduce the basic concepts of Floquet systems and topological defects, and briefly explain our scheme, followed by a discussion on symmetries in Floquet systems. We then put forward the explicit constructions of topological invariants and discuss their numerical implementation (including simplified algorithms). Finally, exploiting these topological invariants at hand, we study a number of low-dimensional examples of Floquet topological defects and discuss their physical significance. The Floquet topological invariants are numerically evaluated. The technical details of calculation are provided in the appendices. The paper is written in a self-contained manner, so that it can also be read by beginners as an introduction to both Floquet systems and topological defects.

\section{ The scheme of constructing topological invariants }\label{sec:outline}

In this section, we will introduce a few basic concepts of Floquet systems and topological defects, which are indispensable for understanding the rest parts of this paper. We will also briefly introduce the scheme of constructing topological invariants.

In a periodically driven or Floquet system, the Hamiltonian is time-periodic by definition, namely,
\bea \hat{H}(t)=\hat{H}(t+\tau), \eea with a period $\tau$ and angular frequency $\omega=2\pi/\tau$. If the system has translational symmetry, then the Bloch wavevector $\bk$ is a good quantum number\footnote{in this paper, we take the unit that $\hbar=1$, therefore, the wavevector is equivalent to the crystal momentum (or Bloch momentum).}, and we may take the time-dependent Bloch Hamiltonian $H(\bk,t)$ as a starting point of investigation. In this paper, we would like to formulate topological invariants not only for homogeneous systems (i.e., systems with translational symmetry), but also for topological defects; therefore, we will focus on the general problem of topological defects, and treat homogeneous systems as their special cases, to which the general formulation is also applicable.

\begin{figure}
\subfigure{\includegraphics[width=3.8cm, height=3.8cm]{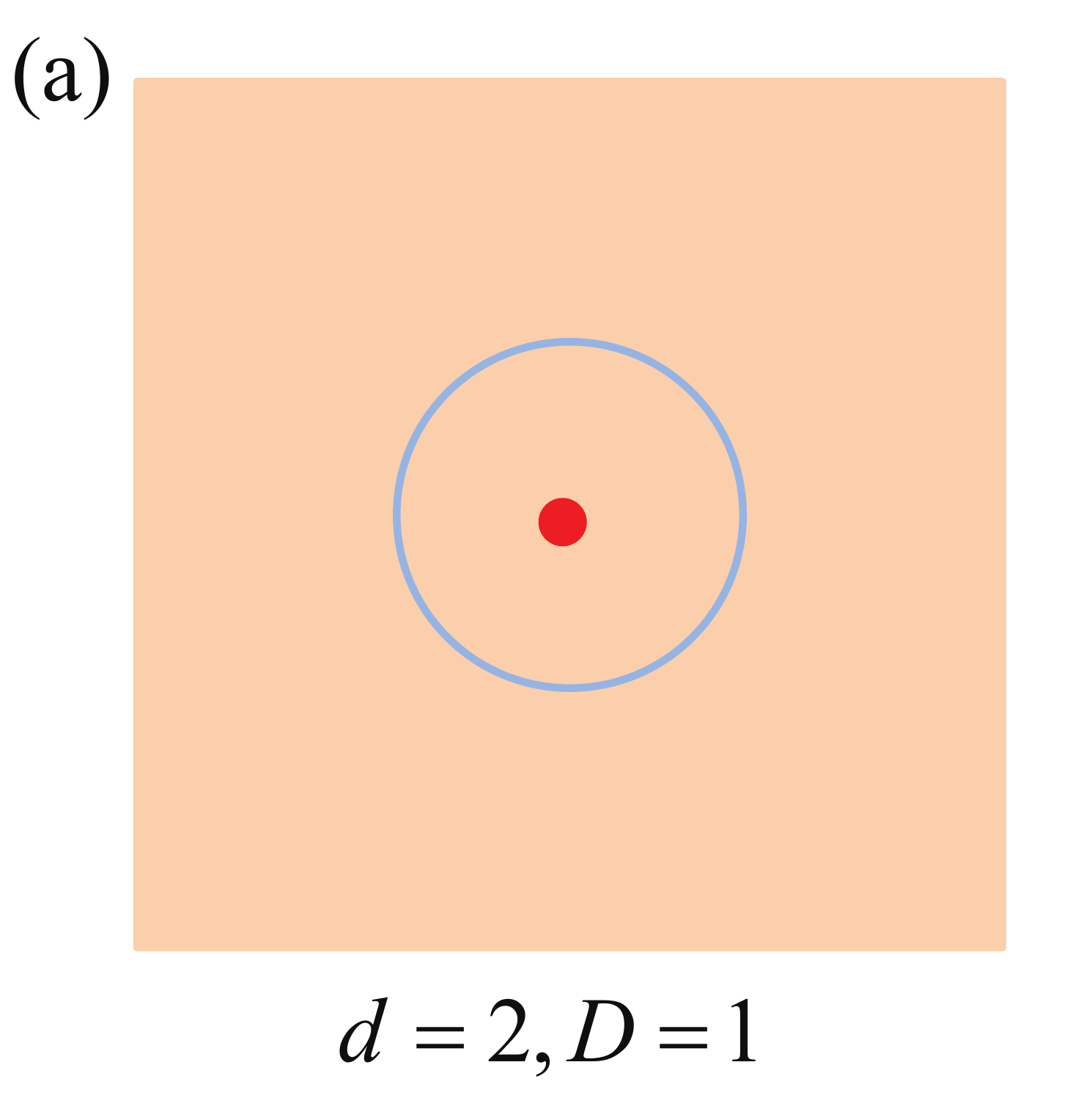}}
\subfigure{\includegraphics[width=3.8cm, height=3.8cm]{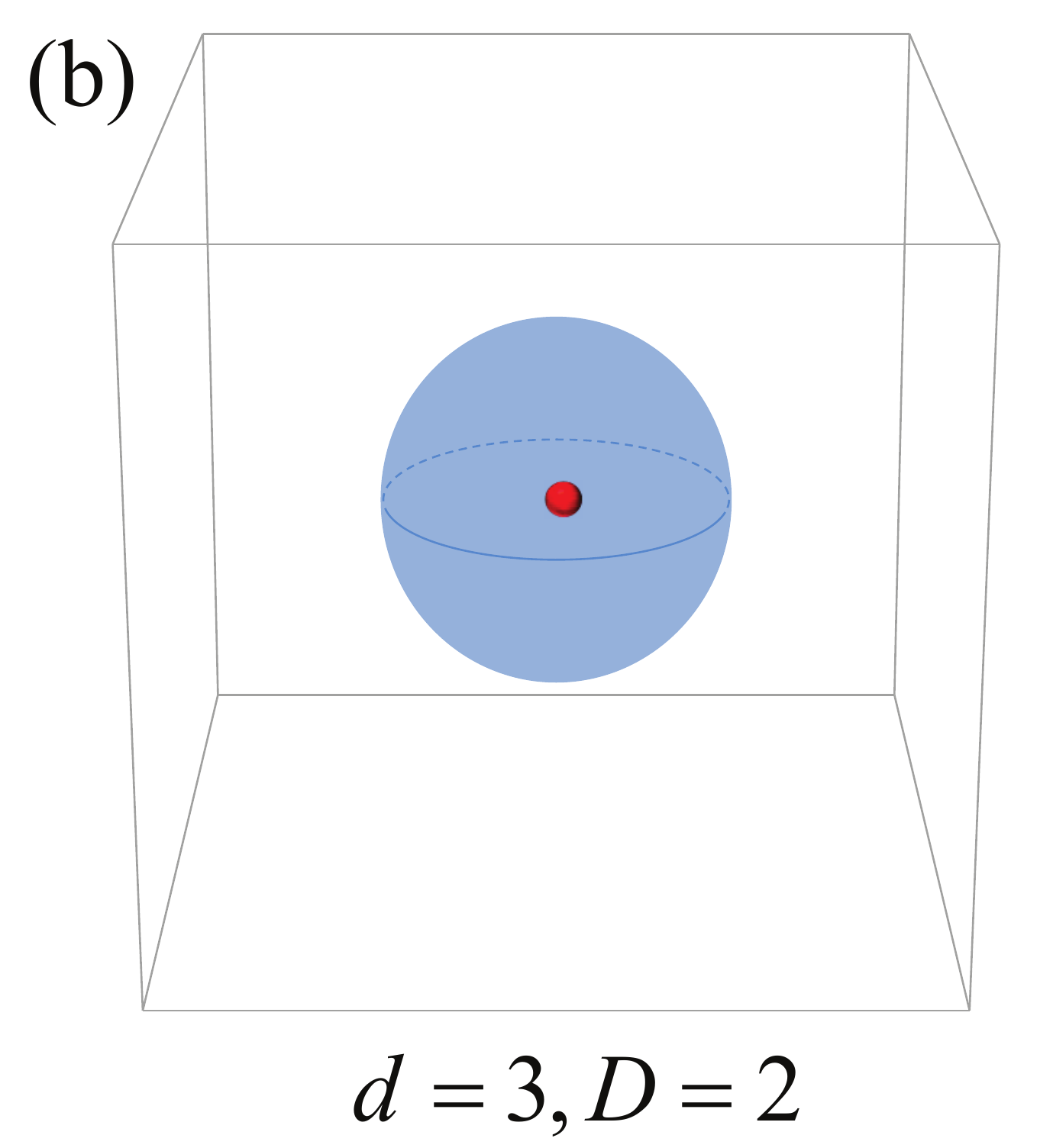}}
\subfigure{\includegraphics[width=3.8cm, height=3.8cm]{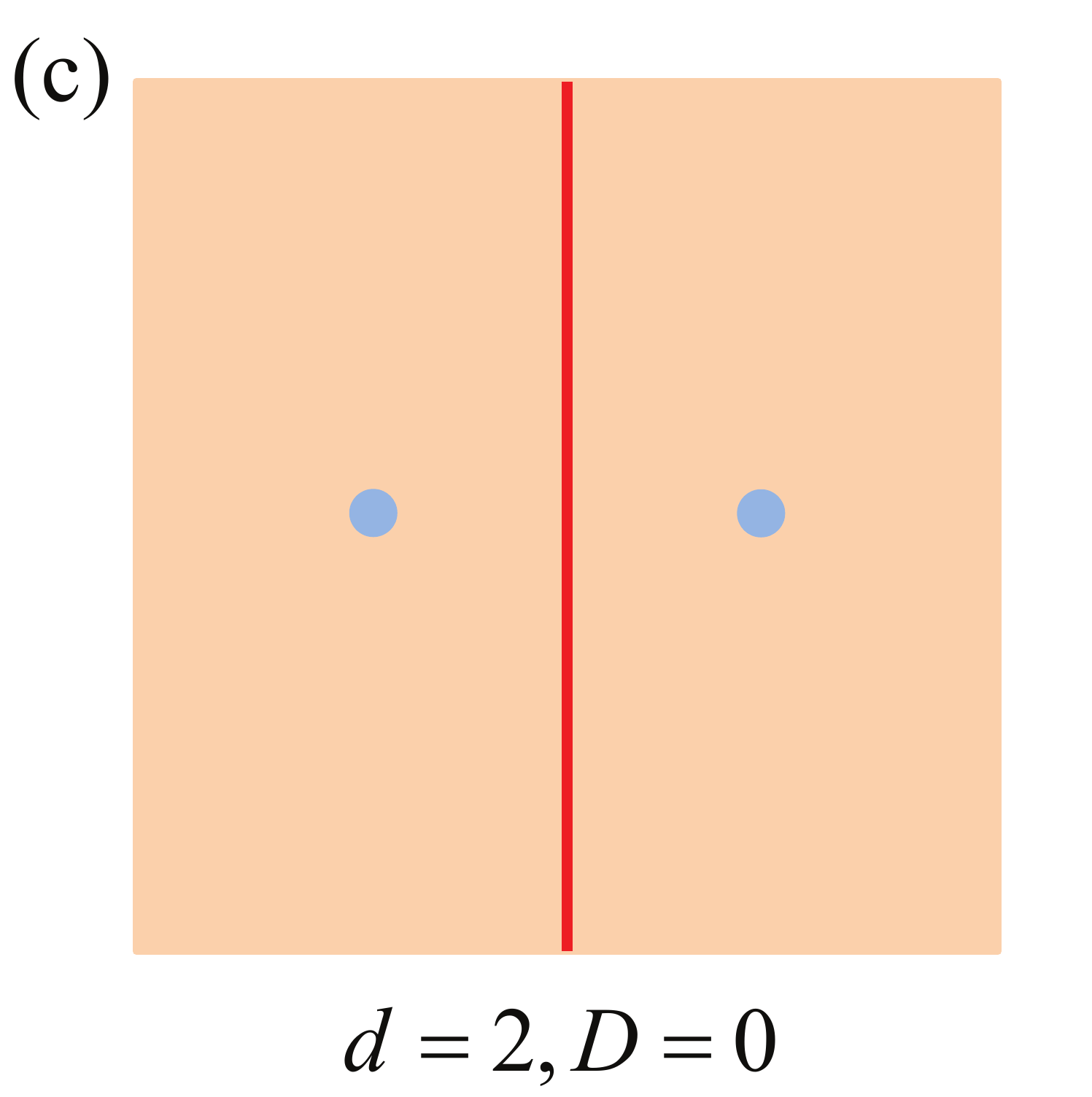}}
\subfigure{\includegraphics[width=3.8cm, height=3.8cm]{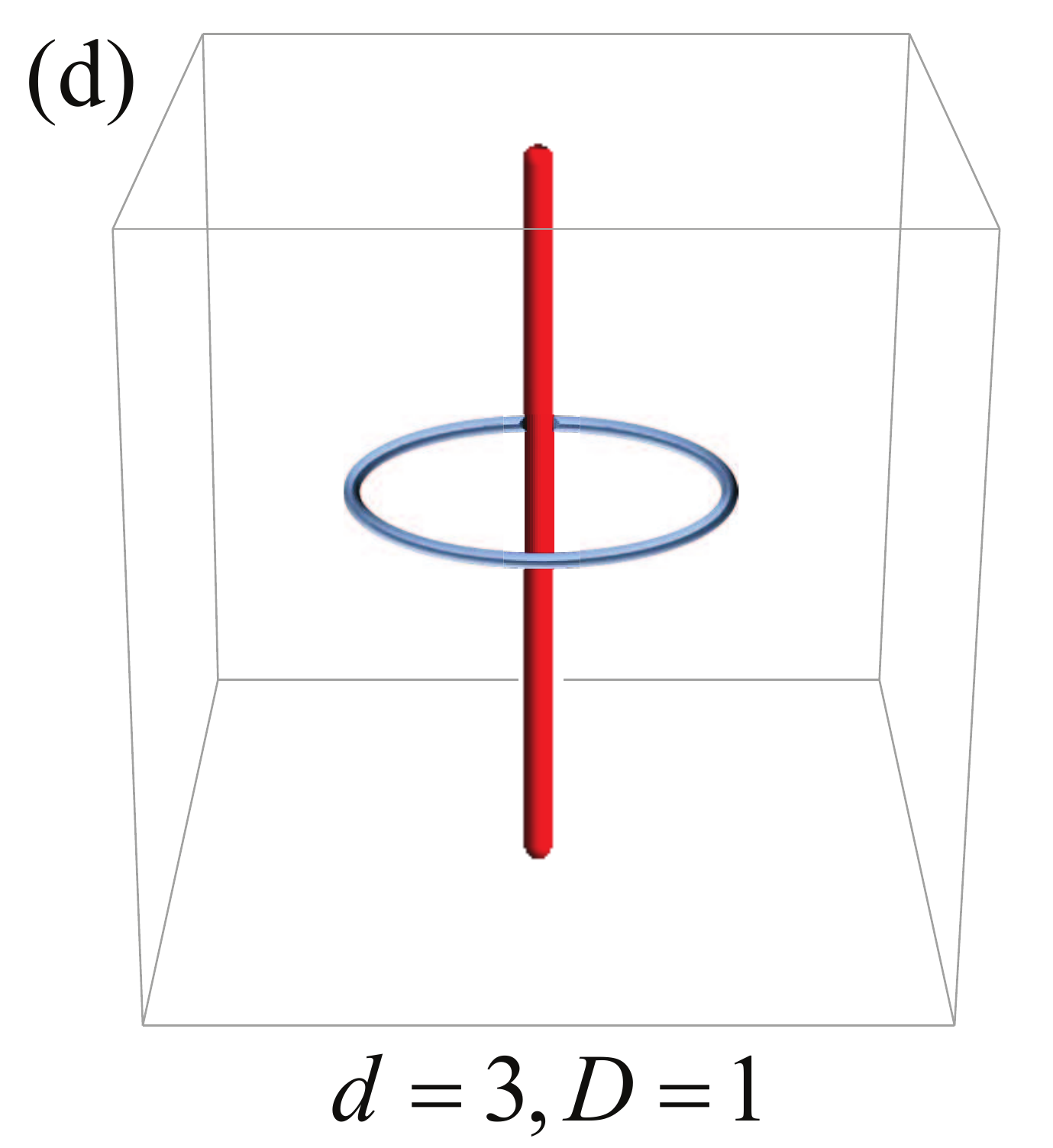}}
\caption{ Illustrations of defects. Here, $d$ stands for the spatial dimension, and $D$ stands for the dimension of the surrounding surface (shown in blue) of defect.  (a) Point defect in two dimensional space. (b) Point defect in three-dimensional space. (c) Line defect in two-dimensional space. (d) Line defect in three-dimensional space. The red points and lines stand for the defect, and the blue points, circles, and sphere denote the $D$-dimensional surrounding surface. }\label{sketch}
\end{figure}

In the presence of a topological defect (several examples of defects are shown in Fig.\ref{sketch}), the translational symmetry is broken, and the wavevector $\bk$ is not a good quantum number. Fortunately, the robust topological properties of a defect can be fully determined by the information far away from the defect. For instance, the Burgers vector of a dislocation can be read from a large contour around the dislocation, which is independent of the details in the vicinity of defect. In the region sufficiently far away from a defect, translational symmetry is asymptotically restored, and the description in terms of the wavevector $\bk$ and the time-dependent Bloch Hamiltonian becomes valid. To describe the topology of a defect, we can take a sufficiently large surface surrounding the defect\cite{teo2010,Teo2010Majorana,teo2017topological}, and seek topological classification and topological invariant based on information on this surface. Let $d, d_{\rm def}, D$ stands for the dimension of the entire space in which the defect lives, the dimension of defect, and the dimension of the surrounding surface, respectively. They automatically satisfy $d_{\rm def}+D+1=d$ [see Fig.\ref{sketch} for a few examples]. We shall take the surrounding surface to be a $D$-dimensional sphere $S^D$. We remark that $S^{D=0}$ consists of two points ($\{+1,-1\}$), as shown in Fig.\ref{sketch}(c). The defect topological invariants for the $D=0$ cases are simply the difference between two bulk topological invariants of homogeneous systems (on the $+1$ side and $-1$ side, respectively). Thus, the bulk topological invariants are essentially the special cases ($D=0$) of defect topological invariants.

On the surrounding sphere $S^D$, the Bloch Hamiltonian is a slowly varying function of the position $\br$.
Thus, we have a time-dependent Bloch Hamiltonian $H(\bk,\br,t)$  defined on the $(d+D+1)$-dimensional parameter space with coordinates $(\bk,\br,t)$, where $\br$ stands for the position on $S^D$. It satisfies the periodicity \bea
H(\bk,\br,t)=H(\bk,\br ,t+\tau).
\eea For the static topological defects, one may take the Bloch  Hamiltonian $H(\bk,\br)$ (without $t$ dependence) as the starting point to define topological invariants\cite{teo2010}. For Floquet topological defects, we will adopt a refined version of the time evolution operator as the generator of topological invariants. The time-evolution-operator-based approach was pioneered by Rudner \emph{et al}\cite{rudner2013anomalous} and has proved useful in a few cases\cite{rudner2013anomalous,Carpentier2015,Fruchart2016} of bulk Floquet topological insulators.

When $t_a>t_b$, the time evolution operator from $t_b$ to $t_a$ is defined as:
\begin{equation}
\label{}
\begin{aligned}
U(\bk,\br;t_a,t_b)=\mathcal{T}\exp[-i\int_{t_b}^{t_a}dt'H(\bk,\br,t')],
\end{aligned}
\end{equation} where $\mathcal{T}$ stands for the time ordering;
when $t_a<t_b$, we define
\begin{equation}
\label{}
\begin{aligned}
U(\bk,\br;t_a,t_b)=U^{-1}(\bk,\br;t_b,t_a).
\end{aligned}
\end{equation}
With this definition, $U(\bk,\br;t_a,t_b)U(\bk,\br;t_b,t_a)=I$ is satisfied.
In most parts of this paper, we will fix $t_b=0$, and take the more concise notation:
\begin{equation}
\label{}
\begin{aligned}
U(\bk,\br,t)\equiv  U(\bk,\br;t_a=t,t_b=0).
\end{aligned}
\end{equation}
One can check that it satisfies the differential equation
$i\partial_t U(\bk,\br,t) = H(\bk,\br,t)U(\bk,\br,t)$. For Floquet systems, a useful quantity is the full-period time evolution operator $U(\bk,\br,\tau)$, which we can expand as \bea
U(\bk,\br,\tau)=\sum_{n=1}^N\lambda_n(\bk,\br)\ket{\psi_n(\bk,\br)}\bra{\psi_n(\bk,\br)},
\eea where $N$ is the rank of $U$, namely, the number of bands. It is customary to define an \emph{effective Hamiltonian} $H^{\text{eff}} (\bk,\br)=(i/\tau)\ln (U(\bk,\br,\tau))$, whose eigenvalues are known as the \emph{quasienergies}. In this paper, the quasienergy will be denoted as $\epsilon$. We also define a dimensionless quasienergy $\varepsilon=\epsilon\tau$, which will be used extensively in this paper. Since the effective Hamiltonian involves a logarithm, the branch cut has to be carefully defined.
A rigorous and unambiguous definition of the effective Hamiltonian is given as  \begin{equation}
\label{Heff}
\begin{aligned}
H^{\text{eff}}_{\varepsilon}(\bk,\br) =\frac{i}{\tau}\sum\limits_{n}\ln_{-\varepsilon}(\lambda_n)\ket{\psi_n(\bk,\br)}\bra{\psi_n(\bk,\br)},
\end{aligned}
\end{equation}
or more compactly,
\begin{equation}
\label{}
\begin{aligned}
H^{\text{eff}}_{\varepsilon}(\bk,\br)=\frac{i}{\tau}\ln_{-\varepsilon}(U(\bk,\br,\tau)).
\end{aligned}
\end{equation}
The subscript $-\varepsilon$ has been introduced to specify the branch cut.
In this paper, $\ln_{\alpha}e^{i\phi}$ stands for the logarithm with the branch cut located at $\exp(i\alpha)$, namely, we take \bea \ln_{\alpha}e^{i\phi}=i\phi \quad\text{for}\quad \alpha-2\pi<\phi<\alpha. \eea  It follows from this definition that, when $\alpha-2\pi<\phi<\alpha$, we have $\ln_{\alpha}e^{i(\phi+2\pi l)}=\ln_{\alpha}e^{i\phi}=i\phi$ for any integer $l$. It also follows that
\bea \ln_{-\varepsilon}e^{i\phi}=i\phi \quad\text{for}\quad -\varepsilon-2\pi<\phi<-\varepsilon, \eea which has been adopted in Eq.(\ref{Heff}). Apparently, the effective Hamiltonian $H^{\rm eff}_\varepsilon(\bk,\br)$ is a Hermitian matrix, and we have \bea U(\bk,\br,\tau)=\exp[-i\tau H^{\rm eff}_\varepsilon(\bk,\br)].\eea The branch cut in $H^{\rm eff}_{\varepsilon}(\bk,\br)$ will be an essential ingredient in the construction of topological invariants for Floquet systems. We mention in advance that, to properly define topological invariants, $\varepsilon$ must be in the bulk (dimensionless) quasienergy gap, namely, $\lambda_n(\bk,\br)\neq e^{-i\varepsilon}$ must be satisfied for all $\bk,\br$ and $n$; otherwise,  due to the branch cut, $H^{\rm eff}_{\varepsilon}(\bk,\br)$ is not a smooth function of $(\bk,\br)$.

The effective Hamiltonian does not, however, contain complete information for topological invariants. The effective Hamiltonian captures only the stroboscopic evolution at integer multiples of $\tau$, losing key information of the evolution within each period. As such, the effective Hamiltonian should only play an auxiliary role in constructing topological invariants. Let us define the periodized time evolution operator\cite{rudner2013anomalous,Carpentier2015}:
\begin{equation}
\label{}
\begin{aligned}
U_{\varepsilon}(\bk,\br,t)=U(\bk,\br,t)\exp[iH^{\text{eff}}_{\varepsilon}(\bk,\br)t],
\end{aligned}
\end{equation}
which satisfies \bea U_{\varepsilon}(\bk,\br,t)=U_{\varepsilon}(\bk,\br,t+\tau). \label{U-periodic} \eea
To see this periodic property, we notice that $U(\bk,\br,t+\tau)=U(\bk,\br,t)U(\bk,\br,\tau) =U(\bk,\br,t)\exp[-iH^{\text{eff}}_{\varepsilon}(\bk,\br)\tau]$.
Eq.(\ref{U-periodic}) is a crucial property. In fact, one cannot define topological invariants directly in terms of $U(\bk,\br,t)$, which is generally not time-periodic: $U (\bk,\br,t)\neq U (\bk,\br,t+\tau)$. Due to the periodicity, $U_{\varepsilon}(\bk,\br,t)$ is defined essentially on the compact parameter space $T^{d+1}\times S^D$, where $T^{d+1}$ stands for the $(d+1)$-dimensional torus. Here, $T^{d+1}=T^d\times T^1$, $T^d$ being the Brillouin zone, and $T^1\equiv S^1$ being the circle with length $\tau$ along the $t$ direction. We can define all the topological invariants as certain mathematically natural winding numbers on the $(\bk,\br,t)$ or $(\bk,\br)$ parameter space in the cases of integer ($\mathbb{Z}$ or $2\mathbb{Z}$) topological invariants, or on certain extended parameter spaces in the cases of $\mathbb{Z}_2$ topological invariants (to be introduced in the following sections).  As we will show in the following sections, the tenfold way symmetries, cooperating with topology, impose powerful constraints on the forms and values of the topological invariants. The combination $\delta=d-D$ enters as a key number in a natural manner. Notably, the $\mathbb{Z}_2$ topological invariants take the forms of Wess-Zumino-Witten (WZW) terms, and their definitions crucially rely on the symmetries. Furthermore, the topological properties of the unitary groups, as manifested in their homotopy groups, dictates that each $\mathbb{Z}$ topological invariant has two and only two $\mathbb{Z}_2$
descendants.

Compared to the static cases\cite{Chiu2016rmp,teo2010}, the Floquet topological invariants to be formulated will take quite different forms. This is understandable because they should be able to describe various ``anomalous topological modes''\cite{rudner2013anomalous}, which are intrinsic to Floquet systems and have no static counterpart. The definitions [such as Eq.(\ref{class-A}) and Eq.(\ref{DC-Z2})] and properties of the topological invariants will be given in the following sections. All the topological invariants are given in explicit and plain formulas rather than more formal languages such as the K-theory, so that they can be used directly in analytical and numerical calculations.

\section{Symmetries of Floquet systems}\label{sec:symmetries}

Since symmetries play important roles, let us first introduce them in this section as a preparation for topological invariants. It should be mentioned that some of the symmetry identities to be presented below, with the spatial variable $\br$ removed, have been discussed in Refs.\cite{Fruchart2016,Roy2016periodic}. We will focus on the symmetries in the tenfold-way classifications\cite{kitaev2009,ryu2010,stone2010symmetries,Chiu2016rmp}. In this framework, there is the time-reversal symmetry (TRS), the particle-hole symmetry (PHS), which is sometimes called the charge conjugation symmetry, and the sublattice symmetry, which is also called the ``chiral symmetry''(CS).

The particle-hole or charge conjugation symmetry is defined as
\begin{equation} \label{}
\Xi H(\bk,\br,t)\Xi^{-1}=-H(-\bk,\br,t),
\end{equation} where $\Xi=C \mathcal{K}$, $C$ is a unitary matrix, and $\mathcal{K}$ is the complex conjugation operator. Equivalently, the PHS can be written as
\begin{equation} \label{particlem}
C^{-1} H(\bk,\br,t)C =-H^*(-\bk,\br,t).
\end{equation} Note that the symmetry operation does not change the spatial coordinate $\br$.

The time-reversal symmetry takes the form of
\begin{equation} \label{}
\Theta H(\bk,\br,t)\Theta^{-1}=H(-\bk,\br,-t),
\end{equation}  where $\Theta=T\mathcal{K}$, $T$ is the unitary TRS matrix. It can be written equivalently as
\begin{equation} \label{timem}
T^{-1} H(\bk,\br,t)T =H^*(-\bk,\br,-t),
\end{equation} Note that the time $t$ is reversed under the time-reversal operation.

The chiral symmetry is defined by
\begin{equation} \label{chiralm}
S^{-1} H(\bk,\br,t)S =-H(\bk,\br,-t).
\end{equation} There is no complex conjugation for the chiral symmetry.

There are three possibilities\cite{Chiu2016rmp,ryu2010} for the TRS: TRS with $T^*T=1$, TRS with $T^*T=-1$, or no TRS; similarly, PHS has three possibilities: $C^*C= 1$, $C^*C=-1$, or no PHS. Therefore, there are $3\times 3=9$ possibilities coming from the TRS and PHS. The product of TRS and PHS is a CS, which cannot be freely assigned when the TRS and PHS are specified. This is true for $8$ of the $9$ cases. The only exception is the case that both PHS and TRS are absent. In this case, the CS can be present or absent, yielding two choices. Therefore, there are $(3\times 3-1)+2=10$ symmetry classes\cite{Chiu2016rmp,ryu2010} (``tenfold way''). Eight of them contain one or two anti-unitary symmetries (PHS or TRS), and the other two do not. They are called real classes and complex classes, respectively. In the context of random matrices, these symmetry classes are known as the Altland-Zirnbauer\cite{Altland1997} (AZ) symmetry classes.

One may wonder whether there are other possibilities. For instance, what happens if there are two TRS operations, denoted by $T_1^{-1} H(\bk,\br,t)T_1 =H^*(-\bk,\br,-t)$ and $T_2^{-1} H(\bk,\br,t)T_2 =H^*(-\bk,\br,-t)$? In this case, we have $T_2 T_1^{-1} H(\bk,\br,t)T_1 T_2^{-1}=H(\bk,\br,t)$, therefore, $H(\bk,\br,t)$ commutes with $T_2 T_1^{-1}$, thus $H(\bk,\br,t)$ can be written in block-diagonal form, $T_2 T_1^{-1}$ being a constant in each block. Within each block, $T_2$ is determined by $T_1$; only one of them is independent.

From the symmetries of the time-dependent Hamiltonian, we can derive symmetry properties of the time evolution operator, and more importantly, of the periodized time evolution operator  $U_{\varepsilon}(\bk,\br,t)$.  To derive them,
we divide $[0,t]$ into $N$ small intervals, each of which has length $\Delta t=t/N$ (we take $t>0$ for concreteness; the $t<0$ case can be done similarly), and then expand the time evolution operator as a continued product
\begin{equation} \label{divide}
\begin{aligned}
U(\bk,\br,t)=&[1-i\Delta tH(\bk,\br,t)][1-i\Delta tH(\bk,\br,t-\Delta t)]\cdots\\&\cdots[1-i\Delta tH(\bk,\br,2\Delta t)][1-i\Delta tH(\bk,\br,\Delta t)],
\end{aligned}
\end{equation} which is accurate in the $\Delta t\rw 0$ limit.
Using this expansion, we can derive the actions of symmetry operators on the time evolution operator. Leaving technical details to Appendix \ref{symmu}, we summarize the main results as follows. For the PHS, we have
\begin{equation} \label{particleu}
C^{-1} U(\bk,\br,t)C =U^*(-\bk,\br,t);
\end{equation} for the TRS, we have
\begin{equation} \label{timeu}
T^{-1} U(\bk,\br,t)T =U^*(-\bk,\br,-t);
\end{equation}
and finally, for the CS, we have
\begin{equation} \label{chiralu}
S^{-1} U(\bk,\br,t)S =U(\bk,\br,-t).
\end{equation} The topological invariant will be defined in terms of the periodized time evolution operator $U_\varepsilon$, whose symmetry properties should be addressed.
To this end, let us first study the symmetry operations on the effective Hamiltonian $H^{\text{eff}}_{\varepsilon}(\bk,\br)=\frac{i}{\tau}\ln_{-\varepsilon}(U(\bk,\br,\tau))$. Again, we summarize the main results here, leaving details to  Appendix \ref{symmh}. They read
\begin{equation}
\label{particleh}
\begin{aligned}
C^{-1} H^{\text{eff}}_{\varepsilon}(\bk,\br)C =-H^{\text{eff}*}_{-\varepsilon}(-\bk,\br)+\frac{2\pi}{\tau},
\end{aligned}
\end{equation}
\begin{equation}
\label{timeh}
\begin{aligned}
T^{-1} H^{\text{eff}}_{\varepsilon}(\bk,\br)T =H^{\text{eff}*}_{\varepsilon}(-\bk,\br),
\end{aligned}
\end{equation}
\begin{equation}
\label{chiralh}
\begin{aligned}
S^{-1} H^{\text{eff}}_{\varepsilon}(\bk,\br)S =-H^{\text{eff}}_{-\varepsilon}(\bk,\br)+\frac{2\pi}{\tau}.
\end{aligned}
\end{equation}
With these preparations, we can finally obtain the symmetry properties of the peroidized time evolution operator, which are listed as
\begin{equation}
\label{particle}
\begin{aligned}
C^{-1} U_{\varepsilon}(\bk,\br,t) C =U^*_{-\varepsilon}(-\bk,\br,t)\exp(i\frac{2\pi t}{\tau}),
\end{aligned}
\end{equation}
\begin{equation}
\label{time}
\begin{aligned}
T^{-1} U_{\varepsilon}(\bk,\br,t)T =U_{\varepsilon}^*(-\bk,\br,-t),
\end{aligned}
\end{equation}
\begin{equation}
\label{chiral}
\begin{aligned}
S^{-1} U_{\varepsilon}(\bk,\br,t) S =U_{-\varepsilon}(\bk,\br,-t)\exp(i\frac{2\pi t}{\tau}).
\end{aligned}
\end{equation}
The details of calculations are provided in
Appendix \ref{symmuu}.

The symmetry properties of the periodized time evolution operator will be most useful in the study of topological invariants.

\section{The periodic table of Floquet topological defects}

\begin{table*}[!htp]
\caption{ The periodic table of Floquet topological defects. TRS with $\Theta^2=\pm 1$ (or $T^*T=\pm 1$) is shown compactly as ``$\pm 1$'', and the absence of TRS is shown as ``0''.  The same notation is taken for the PHS. For the CS, ``$1$'' and ``$0$'' stands for its presence and absence, respectively.  The integer $n$ is the number of quasienergy gaps.  }
\centering
\begin{tabular*}{15cm}{@{\extracolsep{\fill}}c cc c c c| c c c c ccccc }
\hline  \hline
 \multicolumn{5}{c}{Symmetry} &&&  \multicolumn{8}{c}{$\delta=d-D$}       \\

     s & AZ &     $T$ &     $C$ &     $S$ &&&     $0$ &     $1$ &     $2$ &     $3$ &     $4$ &     $5$ &     $6$ &     $7$ \\
\hline
      0&   A&     $0$ &     $0$ &     $0$ & &&    $\mathbb{Z}^n$ &     $0$ &     $\mathbb{Z}^n$ &     $0$ &     $\mathbb{Z}^n$ &     $0$ &     $\mathbb{Z}^n$ &     $0$ \\

     1&    AIII&     $0$ &     $0$ &     $1$&& &     $0$ &    $\mathbb{Z}^2$ &     $0$ &     $\mathbb{Z}^2$ &     $0$ &     $\mathbb{Z}^2$ &     $0$ &     $\mathbb{Z}^2$  \\
\hline
      0&   AI &     $+1$ &     $0$ &     $0$ &  &&   $\mathbb{Z}^n$ &     $0$ &     $0$ &     $0$ &     $2\mathbb{Z}^n$ &     $0$ &     $\mathbb{Z}_2^n$ &     $\mathbb{Z}_2^n$\\

      1&   BDI &     $+1$ &     $+1$ &     $1$  & &&    $\mathbb{Z}_2^2$&     $\mathbb{Z}^2$ &     $0$ &     $0$ &     $0$ &     $2\mathbb{Z}^2$ &     $0$ &     $\mathbb{Z}_2^2$  \\

      2&   D &     $0$ &     $+1$ &     $0$ &&&     $\mathbb{Z}_2^2$&     $\mathbb{Z}_2^2$&     $\mathbb{Z}^2$ &     $0$ &     $0$ &     $0$ &     $2\mathbb{Z}^2$ &     $0$  \\

     3&    DIII &     $-1$ &     $+1$ &     $1$  & &&    $0$ &     $\mathbb{Z}_2^2$&     $\mathbb{Z}_2^2$&     $\mathbb{Z}^2$ &     $0$ &     $0$ &     $0$ &     $2\mathbb{Z}^2$\\

      4&   AII &     $-1$ &     $0$ &     $0$ &  &&   $2\mathbb{Z}^n$&     $0$ &     $\mathbb{Z}_2^n$&     $\mathbb{Z}_2^n$&     $\mathbb{Z}^n$ &     $0$ &     $0$ &     $0$  \\

     5&    CII &     $-1$ &     $-1$ &     $1$ & &&    $0$ &     $2\mathbb{Z}^2$&     $0$ &     $\mathbb{Z}_2^2$&     $\mathbb{Z}_2^2$&     $\mathbb{Z}^2$ &     $0$ &     $0$  \\

      6&   C &     $0$ &     $-1$ &     $0$  &  &&   $0$ &     $0$ &     $2\mathbb{Z}^2$&     $0$ &     $\mathbb{Z}_2^2$&     $\mathbb{Z}_2^2$&     $\mathbb{Z}^2$ &     $0$  \\

      7&   CI &     $+1$ &     $-1$ &     $1$ &   &&  $0$ &     $0$ &     $0$ &     $2\mathbb{Z}^2$&     $0$ &     $\mathbb{Z}_2^2$&     $\mathbb{Z}_2^2$&     $\mathbb{Z}^2$  \\
\hline \hline
\end{tabular*}\label{table}
\end{table*}

In static systems, the classification of topological insulators shows a highly regular pattern, which is summarized in Kitaev's periodic table\cite{kitaev2009}. Remarkably, topological defects also display a regular pattern in a periodic table\cite{teo2010}.  It is notable that the topological classification depends only on $\delta=d-D$, thus the shift $(d,D)\rw(d+1,D+1)$ does not alter the classification\cite{teo2010}.

Before giving derivations, we present the periodic table of Floquet topological defects in Table \ref{table}. For Floquet topological insulators and superconductors with translational symmetry, which are special cases of our formulation, we only need to take $\delta=d$  (i.e., $D=0$) in the table. As shown in the table,
the topological classification of Floquet defects shares the feature of static systems that $d$ and $D$ enter as $\delta=d-D$. This feature will be explained in a natural way in the formulation of topological invariants.

For Floquet topological insulators with translational symmetry, a periodic table has been obtained in an interesting recent work\cite{Roy2016periodic}. It is worthwhile to compare it with ours. First, the periodic table of Ref.\cite{Roy2016periodic} is obtained in an economical way via the K-theory, which does not rely on topological invariants. As such, Ref.\cite{Roy2016periodic} does not provide explicit topological invariants. In the present work, the focuses are topological invariants, which are constructed and then taken as the main tools, and the periodic table is obtained as one of the consequences of topological invariants. Second, the periodic table in Ref.\cite{Roy2016periodic} is that of Floquet topological insulators with translational symmetry, while Table \ref{table} here is more general in that it also includes Floquet topological defects. The periodic table of Floquet topological insulators is a special case of Table \ref{table} (the $D=0$ case).

Compared to the periodic table of static topological defects\cite{teo2010}, a few notable differences should be mentioned. In static systems, the integer topological invariant of the system comes from a summation over all the valence bands (or occupied bands), whose energies are below the Fermi level. For instance, the topological invariant of a two-dimensional homogeneous insulator in class A is the sum of the Chern numbers of all the valence bands. In Floquet systems, the concept of valence band in general is problematic because the quasienergy is periodically defined modulo the driving frequency, and consequently, we cannot unambiguously say that a certain band has a higher or lower energy than another one. The Floquet topological invariants are attached to quasienergy gaps (Admittedly, Floquet band topological invariants can also be defined, however, as we will explain later, they are not as informative as the gap topological invariants). Suppose that there are $n$ quasienergy gaps to be preserved, each of which enjoys an integer topological invariant, we will have the $\mathbb{Z}^n$ classification (Here, $\mathbb{Z}^n\equiv \underbrace{\mathbb{Z}\times\mathbb{Z}\cdots\times\mathbb{Z}}_n$). This is the origin of ``$\mathbb{Z}^n$'' in Table \ref{table}.

In the presence of particle-hole symmetry or chiral symmetry (or both), we have ``$\mathbb{Z}^2\equiv\mathbb{Z}\times\mathbb{Z}$'' or ``$\mathbb{Z}_2^2\equiv\mathbb{Z}_2\times\mathbb{Z}_2$'' classification in Table \ref{table}, in contrast to ``$\mathbb{Z}$'' or ``$\mathbb{Z}_2$'' of static systems with the same symmetries. The reason is that, in Floquet systems, there are two special dimensionless quasienergies satisfying $\varepsilon= -\varepsilon$ (mod $2\pi$), namely, $\varepsilon=0$ or $\pi$ (mod $2\pi$). These two quasienergies are both analogous to the zero-energy point of static systems. If we would like to preserve the quasienergy gaps open at both $\varepsilon=0$ and $\pi$, then there is a topological invariant for each one of these two gaps, and the classification is $\mathbb{Z}^2$ or $\mathbb{Z}_2^2$. On the other hand, if one is concerned only with one of these two dimensionless quasienergies ($0$ or $\pi$), ignoring the gapped/gapless nature of the other one, then the classification is $\mathbb{Z}$ or $\mathbb{Z}_2$.

More than deriving the periodic table of Floquet topological defects, we would like to obtain a complete list of topological invariants, which contains more information than the periodic table, and is directly applicable to analytic and numerical calculations. The periodic table will be obtained entirely as a byproduct of topological invariants, which will be explained in the following sections.

\section{Topological invariants for complex classes}

In this section, we introduce topological invariants of the two complex classes, class A and class AIII, for both homogeneous Floquet systems (i.e., Floquet systems with translational symmetry) and Floquet topological defects.

Before moving on, let us add a general remark about the physical meaning of the topological invariants. This remark applies to all the symmetry classes, including the eight real classes to be studied in the next section. The bulk-boundary or bulk-defect correspondence, which has been widely tested in static systems, is expected to hold in Floquet systems as well. It suggests that the value of an appropriate topological invariant is equal to the net number of boundary modes or defect modes. In the cases of integer ($\mathbb{Z}$ or $2\mathbb{Z}$) topological invariants, each boundary or defect mode has a ``chirality''\footnote{This ``chirality'' has no direct relation to the ``chiral symmetry''.}(e.g., for one-dimensional modes, the chirality is simply the direction of mode propagation), and the net number of modes is the difference between the number of positive-chirality modes and that of negative-chirality modes.  The only subtle point is that, for the chiral classes (AIII, BDI, DIII, CII, CI), the net number of mode in the $\varepsilon=\pi$ gap is equal to the value of topological invariant with a minus sign (We will discuss this point in detail  in Sec.\ref{AIII-point}). In the cases of $\mathbb{Z}_2$ topological invariants, there is no concept of chirality of the boundary or defect modes, and the topological invariant just gives the number of modes modulo two.

\subsection{Topological invariants for class A}\label{sec:class-A}

When $\delta=d-D$ is an even integer (thus, $d+D$ is also an even integer), we can define a winding number
\begin{equation}
\begin{aligned}
&W(U_\varepsilon(\bk,\br,t))=K_{d+D+1}\int_{T^d\times S^D\times S^1} d^dk d^Dr dt\\
&\times\text{Tr}[\epsilon^{\alpha_1\alpha_2\cdots\alpha_{d+D+1}} (U_{\varepsilon}^{-1}\partial_{\alpha_{1}}U_{\varepsilon}) \cdots(U_{\varepsilon}^{-1}\partial_{\alpha_{d+D+1}}U_{\varepsilon})],
\end{aligned} \label{class-A}
\end{equation}
where $\alpha_1,\alpha_2,\cdots,\alpha_{d+D+1}$ run through all the coordinates $(\bk,\br,t)$ of the parameter space, and $\epsilon^{\alpha_1\alpha_2\cdots\alpha_{d+D+1}}$ stands for the Levi-Civita symbol (the sign of permutation). The integration range of time $t$ can be taken as any interval of length $\tau$; in this paper, we take it as  $[-\tau/2,\tau/2]$. This choice of integration range facilitates the discussion of symmetries later on. The periodic property $U_\varepsilon(\bk,\br,t)=U_\varepsilon(\bk,\br,t+\tau)$ tells us that this interval can be regard as a circle $S^1$.
The coefficient
\begin{equation} \label{}
K_{d+D+1}=\frac{(-1)^{\frac{d+D}{2}}(\frac{d+D}{2})!}{(d+D+1)!}\left(\frac{i}{2\pi}\right)^{\frac{d+D}{2}+1}
\end{equation} ensures that the topological invariant $W(U_\varepsilon)$ is quantized as integers\cite{bott1978some,witten1983,wang2010b,ryu2010,wang2012a}.
The $i^{\frac{d+D}{2}+1}$ factor guarantees the reality of the winding number density $w(U_\varepsilon)=K_{d+D+1} \text{Tr}[\epsilon^{\alpha_1\alpha_2\cdots\alpha_{d+D+1}} (U_{\varepsilon}^{-1}\partial_{\alpha_{1}}U_{\varepsilon}) \cdots(U_{\varepsilon}^{-1}\partial_{\alpha_{d+D+1}}U_{\varepsilon})]$, namely,
\begin{equation}
\label{windingreal}
\begin{aligned}
w^*(U_\varepsilon)=w(U_\varepsilon),
\end{aligned}
\end{equation}
which is proved in Appendix.\ref{realwind}.

As illustrated in Fig.\ref{sketch}(c), when $D=0$, the surrounding sphere $S^D$ consists of just two signed points ($\{+1,-1\}$), therefore, the integral on $S^D$ becomes a two-point summation: $\int_{T^d\times S^D\times S^1} d^dk d^Dr dt=\int_{T^d\times\{+1\}\times S^1} d^dkdt - \int_{T^d\times\{-1\}\times S^1} d^dk dt$, and the defect topological invariant in Eq.(\ref{class-A}) is the difference between two winding numbers:
\begin{equation}
\begin{aligned}
W(U_\varepsilon(\bk,\br,t))= W(U_\varepsilon(\bk,+1,t)) -W(U_\varepsilon(\bk,-1,t)),
\end{aligned}
\end{equation}
where $W(U_\varepsilon(\bk,\pm 1,t))$ is the bulk topological invariant at the $\pm 1$ side of the defect:
\begin{equation}
\begin{aligned}
&W(U_\varepsilon(\bk, \pm 1, t))=K_{d+1}\int_{T^d \times S^1} d^dk dt\\
&\times\text{Tr}[\epsilon^{\alpha_1\alpha_2\cdots\alpha_{d+1}} (U_{\varepsilon}^{-1}\partial_{\alpha_{1}}U_{\varepsilon}) \cdots(U_{\varepsilon}^{-1}\partial_{\alpha_{d+1}}U_{\varepsilon})].
\end{aligned} \label{class-A-bulk}
\end{equation} Apparently, the bulk topological invariant is defined in the $(\bk,t)$ space. The bulk topological invariant of 2d Floquet systems in class A, corresponding to the $(d,D)=(2,0)$ case here, was first studied by Rudner \emph{et al} in Ref.\cite{rudner2013anomalous}. Generalizations of Ref.\cite{rudner2013anomalous} to higher-dimensional homogenous Floquet systems can be found in Ref.\cite{Fruchart2016}, which correspond to the $(d>2,D=0)$ cases.  When $D\neq 0$, Eq.(\ref{class-A}) generalizes the bulk topological invariants of class A to Floquet topological defects in the same symmetry class. One of the applications of Eq.(\ref{class-A}) in low dimensions is to determine the number of Floquet chiral modes along a line defect in 3d space, namely, $(d,D)=(3,1)$, for which a concrete lattice model will be put forward in Sec. \ref{sec:line-class-A}.

Eq.(\ref{class-A}) seems to be the only natural topological invariant that one can write down using the periodized time evolution operator $U_\varepsilon$ for class A. Due to its dependence on $\varepsilon$, it is naturally taken as the topological invariant for the quasienergy gap of $\varepsilon$. This topological invariant can be defined only when $d+D$ is an even integer (equivalently, $\delta=d-D$ is an even integer). When $\delta$ is odd, Eq.(\ref{class-A}) is zero by definition, as can be proved using the invariance of the trace of a matrix under cyclic permutations. This is consistent with the topological fact that the stable homotopy groups\cite{nakahara2003,dubrovin1985,wang2010b} (``stable'' here means that $N$ is sufficiently large) of the unitary groups have the following periodicity
\bea
\pi_p(U(N))=\left\{\begin{array}{cc} \mathbb{Z},& p=\text{odd \,integer}, \\0,&p=\text{even\,integer}.\end{array}\right. \label{homotopy} \eea  In this sense, the topology of $U(N)$ groups completely determines the topological classifications and topological invariants of class A.  We should note that, because we are concerned about strong topological invariants, ignoring the weak ones\cite{fu2007b}, the homotopy group $\pi_{d+D+1}(U(N))$ is able to capture the relevant topological classes of the mappings from $T^{d+1}\times S^D$ to $U(N)$. Thus, we can simply take the integer-valued winding number as the definition of homotopy class.

Eq.(\ref{class-A}) can be defined for any value of $\varepsilon$ in the quasienergy gap. If $n$ quasienergy gaps are maintained, there are $n$ integer topological invariants, one for each gap. Thus, the topological invariant in Eq.(\ref{class-A}) leads to the first row of Table \ref{table}.

To gain more confidence in Eq.(\ref{class-A}), we should check that this topological invariant reduces, in the static limit,  to the previously known topological invariant of static defects. Since all static Bloch Hamiltonian can be smoothly deformed to flat-band ones, we consider a general static flat-band Bloch Hamiltonian
\begin{equation} \label{flat}
H_0(\bk,\br)=-E_0P(\bk,\br)+E_0[1-P(\bk,\br)],
\end{equation}
where $P$ is the occupied-band projection operator satisfying $P^2(\bk,\br)=P(\bk,\br)$, and $-E_0$ is the occupied-band energy ($E_0>0$). The static Hamiltonian can be regarded as a time-periodic Hamiltonian with an arbitrary periodicity $\tau$ or frequency $\omega$, the driving term being infinitesimal.  By a straightforward calculation (see Appendix \ref{statich}), we can prove that, for sufficiently large $\omega$ (for this flat-band case,  $\omega>E_0$ suffices), the time-independent limit of winding number is \bea
\label{static1}
W(U_{\varepsilon=0})=C_{(d+D)/2}(P(\bk,\br)), \eea
where
\bea
C_{(d+D)/2}(P)&=&\tilde{K}_{d+D}\int_{T^d\times S^D} d^dk d^Dr\nn \\
&&\times\text{Tr}[\epsilon^{\alpha_1\alpha_2\cdots \alpha_{d+D}}P\partial_{\alpha_1}P\cdots\partial_{\alpha_{d+D}}P],
\eea
whose numerical coefficient $\tilde{K}_{d+D}$ is
\begin{equation}
\label{static2}
\begin{aligned}
\tilde{K}_{d+D}&=i\omega (d+D+1)\frac{2\pi}{\omega}\frac{(D+d)!}{(\frac{d+D}{2})!(\frac{d+D}{2})!}(-1)^{\frac{d+D}{2}}K_{d+D+1}\\
&=-\left(\frac{i}{2\pi}\right)^{\frac{d+D}{2}}\frac{1}{(\frac{d+D}{2})!}.
\end{aligned}
\end{equation} The expression of $C_{(d+D)/2}(P)$ is exactly the $((d+D)/2)$-th Chern number of the occupied bands in the $(\bk,\br)$ parameter space.

\begin{figure}
\includegraphics[width=9.0cm, height=5.0cm]{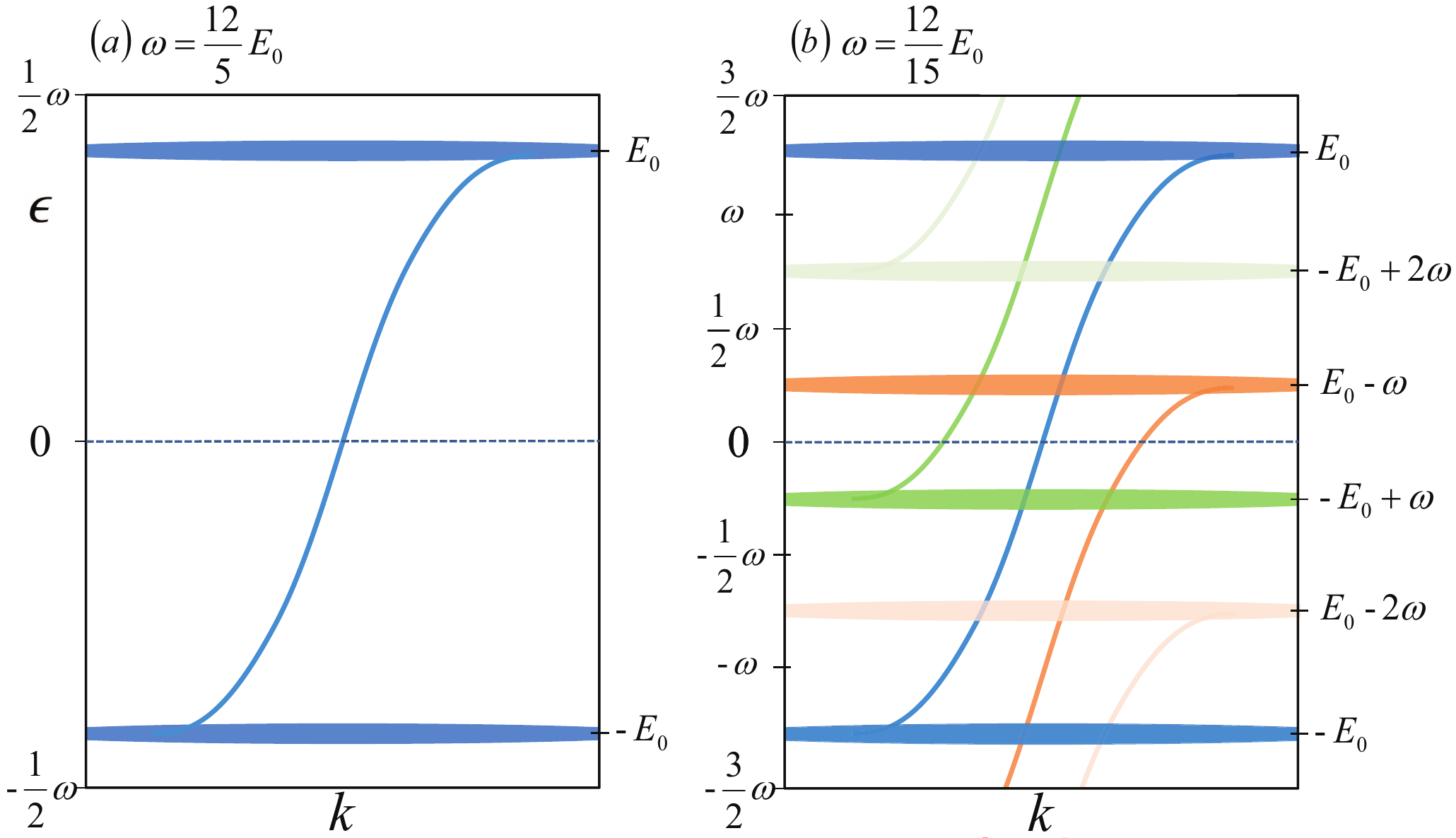}
\caption{ The quasienergy bands of a static two-dimensional Chern insulator with a semi-infinite geometry under an infinitesimal driving with frequency $\omega$. The wave vector $k$ is parallel to the edge. (a): The case $E_0/\omega<1$.  (b) The case $1<E_0/\omega<2$. The number of Floquet chiral modes at $\epsilon=0$ is $1$ and $3$ for (a) and (b), respectively, which is consistent with the value of Floquet topological invariant attached to the $\epsilon=0$ gap ($2\lfloor E_0/\omega\rfloor +1$, see text). In (b), three repeated frequency zones are shown for a better illustration, though the $[-\omega/2,\omega/2]$ zone contains complete information of the quasienergy bands.   }  \label{fig-static}
\end{figure}

For a general frequency, we find that the winding number reduces to the Chern number with an integer coefficient: \bea W(U_{\varepsilon=0})=(2\lfloor E_0/\omega\rfloor +1) C_{(d+D)/2}(P(\bk,\br)), \label{static-small-freq} \eea where $\lfloor E_0/\omega\rfloor$ is the floor function, which stands for the greatest integer smaller than $E_0/\omega$ (e.g. $\lfloor 1.25\rfloor=1$). The derivation of Eq.(\ref{static-small-freq}) is given in Appendix \ref{static-small-f}, in which the branch cut of logarithm plays a crucial role.  Eq.(\ref{static-small-freq}) may look somewhat unexpected at first sight, nevertheless, it is quite intuitive. As a simplest example, let us consider the $(d,D)=(2,0)$ case, namely, the chiral edge states of a static Chern insulator. Suppose that the Chern insulator has two flat bands whose Chern numbers are $\pm 1$, respectively. When $\omega>E_0$, the quasienergy dispersion of the chiral edge states only crosses $\omega=0$ [Fig.\ref{fig-static}(a)]; in contrast, when $1<E_0/\omega<2$ (with $\lfloor E_0/\omega\rfloor=1$), the quasienergy dispersion crosses $0$ and $\pm\omega$ [see Fig.\ref{fig-static}(b)]. Since $\pm\omega$ should be identified as $0$ in the Floquet theory, we have three $\epsilon=0$ points with the same chirality (right-moving). This is consistent with the coefficient $2\lfloor E_0/\omega\rfloor +1=3$ in Eq.(\ref{static-small-freq}). Put simply,  folding the static energy bands into the quasienergy bands increases the number of chiral modes crossing the zero energy [Fig.\ref{fig-static}(b)].

The Floquet topological invariant in Eq.(\ref{class-A}) is attached to a quasienergy gap $\varepsilon$. This is the primary topological invariant for topological defects of class A (the special case $D=0$ gives a bulk topological invariant for homogeneous systems). In addition to this quasienergy gap topological invariant, we can also define Floquet band topological invariants. In fact, for two quasienergies $\varepsilon$ and $\varepsilon'$ satisfying $0\leq\varepsilon<\varepsilon'<2\pi$,  we can prove that \bea H^{\rm eff}_{\varepsilon'}-H^{\rm eff}_{\varepsilon}=\omega P_{\varepsilon,\varepsilon'}, \label{HH'} \eea in which $P_{\varepsilon,\varepsilon'}=\sum_{\varepsilon<\varepsilon_n<\varepsilon'} \ket{\psi_n(\bk,\br)}\bra{\psi_n(\bk,\br)}$ is the projection operator of the Floquet bands with quasienergy $\varepsilon_n \in [\varepsilon,\varepsilon']$, or equivalently, $\arg(\lambda_n^{-1})\in[\varepsilon,\varepsilon']$. To prove Eq.(\ref{HH'}), we notice that when $\varepsilon<\varepsilon_n<\varepsilon'$, we have $\ln_{-\varepsilon} e^{-i\varepsilon_n}=-i\varepsilon_n$ and $\ln_{-\varepsilon'} e^{-i\varepsilon_n}=-i\varepsilon_n-2\pi i$ (see the definition of branch cut), thus we have $(i/\tau)[\ln_{-\varepsilon'}(\lambda_n)-\ln_{-\varepsilon}(\lambda_n)]=2\pi/\tau \equiv\omega$; when $\varepsilon_n<\varepsilon$ or $\varepsilon_n>\varepsilon'$, we have $(i/\tau)[\ln_{-\varepsilon'}(\lambda_n)-\ln_{-\varepsilon}(\lambda_n)]=0$ by similar calculations, thus Eq.(\ref{HH'}) is proved.

For the Floquet bands in $[\varepsilon,\varepsilon']$, a band Chern number can be defined as
\begin{equation}
\label{band-chern}
\begin{aligned}
&C_{(d+D)/2}(P_{\varepsilon,\varepsilon'})=\tilde{K}_{d+D}\int_{T^d\times S^D} d^dk d^Dr\\
&\times\text{Tr}[\epsilon^{\alpha_1\alpha_2\cdots\alpha_{d+D}} P_{\varepsilon,\varepsilon'}\partial_{\alpha_1}P_{\varepsilon,\varepsilon'} \cdots\partial_{\alpha_{d+D}}P_{\varepsilon,\varepsilon'}],
\end{aligned}
\end{equation}
in which $\tilde{K}_{d+D}$ is the same coefficient as given in Eq.(\ref{static2}). Now we can prove a general relation between the band Chern number in Eq.(\ref{band-chern}) and the gap topological invariant in Eq.(\ref{class-A}). It is a straightforward generalization of a relation in Ref.\cite{rudner2013anomalous,Fruchart2016}. Due to the additive property of winding number, we have
\begin{equation}
W(U_{\varepsilon'})-W(U_{\varepsilon})=W(U_{\varepsilon}^{-1}U_{\varepsilon'}),
\end{equation}
in which $U_{\varepsilon}^{-1}U_{\varepsilon'}$ can be simplified to
\bea
U_{\varepsilon}^{-1}U_{\varepsilon'}&=&\exp(-iH_\varepsilon^{\text{eff}}t)U^{-1}U \exp(iH_{\varepsilon'}^{\text{eff}}t)\nn\\
&=&\exp(i\omega t P_{\varepsilon,\varepsilon'})\nn\\
&=&P_{\varepsilon,\varepsilon'}(\text{e}^{i\omega t}-1)+1.
\eea
Since it takes the same form as Eq.(\ref{time evo}), we can follow the calculations in Appendix \ref{statich} and obtain that
\begin{equation}
 W(U_{\varepsilon}^{-1}U_{\varepsilon'})=C_{(d+D)/2}(P_{\varepsilon,\varepsilon'}),
\end{equation}
from which it follows that
\bea W(U_{\varepsilon'})-W(U_\varepsilon)=C_{(d+D)/2}(P_{\varepsilon,\varepsilon'}). \label{difference} \eea Therefore, the Floquet band Chern numbers $C_{(d+D)/2}(P_{\varepsilon,\varepsilon'})$'s can be obtained from $W(U_\varepsilon)$'s. In contrast, even if one knows all the Chern numbers of Floquet bands, one cannot completely determine the values of $W(U_\varepsilon)$'s. As such, the gap topological invariant $W(U_\varepsilon)$ is more fundamental than the band topological invariant $C_{(d+D)/2}(P_{\varepsilon,\varepsilon'})$. In the case $(d,D)=(2,0)$, concrete models with nonzero $W(U_\varepsilon)$'s but vanishing $C_1(P_{\varepsilon,\varepsilon'})$'s are known, whose edge modes associated with the nonzero $W(U_\varepsilon)$'s are dubbed ``anomalous edge states''\cite{rudner2013anomalous}. These anomalous modes have been experimentally observed in photonic lattices\cite{maczewsky2017observation, Mukherjee2016}.

Concluding this section, we mention that there is yet another band topological invariant, which is defined in terms of the frequency-domain Hamiltonian $\mathcal{H}$ (``Floquet Hamiltonian''). We leave its definition to Appendix \ref{sec:proof}. The effective Hamiltonian $H^{\rm eff}$ contains information of only the full-period time evolution $U(\bk,\br,\tau)$, while the Floquet Hamiltonian $\mathcal{H}$ contains complete information of time evolution. From their definitions, it is not obvious whether the frequency-domain Chern number (see Appendix \ref{sec:proof}) is equal to the effective-Hamiltonian-based Chern number given in Eq.(\ref{band-chern}) or not.   A proof of their being equal is provided in Appendix \ref{sec:proof}.

\subsection{Topological invariants for class  AIII}\label{aiii}

In the presence of chiral symmetry, the periodized time evolution operator satisfies Eq.(\ref{chiral}), which relates $U_\varepsilon$ and $U_{-\varepsilon}$. Only when $\varepsilon=0$ or $\pi$, we can obtain from it symmetry constraint on $U_\varepsilon$ for a fixed $\varepsilon$.

For $\varepsilon=0$, Eq.(\ref{chiral}) implies that\cite{Fruchart2016} \bea S^{-1} U_{\varepsilon=0}(\bk,\br,\frac{\tau}{2}) S =-U_{\varepsilon=0}(\bk,\br,-\frac{\tau}{2}), \eea which, together with the periodicity of evolution operator, \begin{equation} \label{tau/2}
U_{\varepsilon}(\bk,\br,-\frac{\tau}{2})=U_{\varepsilon}(\bk,\br,-\frac{\tau}{2}+\tau) =U_{\varepsilon}(\bk,\br,\frac{\tau}{2}),
\end{equation}
imposes the following symmetry constraint on $U_{\varepsilon=0}(\bk,\br,\frac{\tau}{2})$:
\begin{equation} \label{a}
S^{-1} U_{\varepsilon=0}(\bk,\br,\frac{\tau}{2}) S =-U_{\varepsilon=0}(\bk,\br,\frac{\tau}{2}).
\end{equation}
The $\varepsilon=\pi$ case is slightly more complicated due to the difference in the branch cut involved in $H^{\rm eff}_{\varepsilon=-\pi}$ and $H^{\rm eff}_{\varepsilon=\pi}$, which appears in the definition of $U_{\varepsilon=-\pi}(\bk,\br,\frac{\tau}{2})$ and $U_{\varepsilon=\pi}(\bk,\br,\frac{\tau}{2})$, respectively. In fact, it follows from the relation \bea \ln_{-\varepsilon+2\pi} e^{i\phi} =\ln_{-\varepsilon} e^{i\phi} +2\pi i \eea that \bea U_{\varepsilon=-\pi}(\bk,\br,\frac{\tau}{2})= -U_{\varepsilon=\pi}(\bk,\br,\frac{\tau}{2}). \eea
With this equation as an input, Eq.(\ref{chiral}) leads to
\begin{equation} \label{b}
S^{-1}U_{\varepsilon=\pi}(\bk,\br,\frac{\tau}{2}) S =U_{\varepsilon=\pi}(\bk,\br,\frac{\tau}{2}).
\end{equation}

It is convenient to take the chiral basis, in which
\begin{equation} \label{CSmatrix}
S=
\begin{pmatrix}
I&\\
&-I\\
\end{pmatrix}.
\end{equation}
Now Eq.(\ref{a}) tells us that $U_{\varepsilon=0}$ takes the form of
\begin{equation} \label{block}
U_{\varepsilon=0}(\bk,\br,\frac{\tau}{2})=
\begin{pmatrix}
&U_{\varepsilon=0}^+(\bk,\br)\\
U_{\varepsilon=0}^-(\bk,\br)&\\
\end{pmatrix},
\end{equation} where both $U_{\varepsilon=0}^+$ and $U_{\varepsilon=0}^-$ are unitary matrices. Similarly, Eq.(\ref{b}) implies that
\begin{equation} \label{block-pi}
U_{\varepsilon=\pi}(\bk,\br,\frac{\tau}{2})=
\begin{pmatrix}
U_{\varepsilon=\pi}^+(\bk,\br)&\\
&U_{\varepsilon=\pi}^-(\bk,\br)\\
\end{pmatrix},
\end{equation} Again, both $U_{\varepsilon=\pi}^+$ and $U_{\varepsilon=\pi}^-$ are unitary matrices.
With either $\varepsilon=0$ or $\varepsilon=\pi$, we can define a natural winding number when $\delta=d-D$ is an odd integer (therefore $d+D$ is also an odd integer):
\begin{equation}
\label{AIII-inv}
\begin{aligned}
&W(U^+_\varepsilon(\bk,\br))=K_{d+D}\int_{T^d\times S^D} d^dk d^Dr\\
&\times\text{Tr}\{\epsilon^{\alpha_1\alpha_2\cdots\alpha_{d+D}} [(U_{\varepsilon}^+)^{-1}\partial_{\alpha_{1}}U_{\varepsilon}^+] \cdots[(U_{\varepsilon}^+)^{-1} \partial_{\alpha_{d+D}}U_{\varepsilon}^+]\},
\end{aligned}
\end{equation}
where the coefficient $K_{d+D}$ is the same as given in the previous section (but remember that $d+D$ is an even integer there, while it is an odd integer here):
\begin{eqnarray}
K_{d+D}=\frac{(-1)^{\frac{d+D-1}{2}}(\frac{d+D-1}{2})!}{(d+D)!}\left(\frac{i}{2\pi}\right)^{\frac{d+D+1}{2}}.
\end{eqnarray} The homogeneous (namely, $D=0$) cases of class AIII have been investigated in Ref.\cite{Fruchart2016}, from which the present section benefits considerably. We emphasize that there is no integration over $t$ in the winding number given in Eq.(\ref{AIII-inv}), in contrast to the class A.  Putting together the integer-valued topological invariants $W(U^+_{\varepsilon=0})$ and $W(U^+_{\varepsilon=\pi})$, we get the second line of Table \ref{table}.

It should be mentioned that $W(U^-_\varepsilon(\bk,\br))$ does not generate an additional topological invariant, which can be explained as follows. We start from the winding number of $U(\bk,\br,t)$ at a fixed $t$, which is given by
\begin{equation}
\label{winding}
\begin{aligned}
&W(U_{\varepsilon}(\bk,\br,t))=K_{d+D}\int_{T^d\times S^D} d^dk d^Dr\\
&\times\text{Tr}[\epsilon^{\alpha_1\alpha_2\cdots\alpha_{d+D}} (U_{\varepsilon}^{-1}\partial_{\alpha_{1}}U_{\varepsilon}) \cdots(U_{\varepsilon}^{-1}\partial_{\alpha_{d+D}}U_{\varepsilon})].
\end{aligned}
\end{equation} We emphasize that $t$ here is viewed as a fixed parameter, which is not integrated over, unlike the cases in Sec. \ref{sec:class-A}. Since the time evolution operators $U_{\varepsilon}(\bk,\br,t)$ at different moments (i.e., different $t$'s) can be smoothly connected as $t$ varies, this winding number cannot change as we tune $t$. On the other hand,  when $t=0$, $U_{\varepsilon}(\bk,\br,0)=I$, thus the winding number $W(U_{\varepsilon}(\bk,\br,t))$ vanishes at $t=0$. Therefore, the winding number satisfies \bea W(U_{\varepsilon}(\bk,\br,t))=0 \eea for any value of $t$, including the particular time we are focusing on, $t=\tau/2$. In the chiral basis, the winding number $W(U_{\varepsilon}(\bk,\br,t))$ splits into the sum of two parts, which leads to
\begin{equation}
\label{zero}
\begin{aligned}
0=W(U_{\varepsilon} (\bk,\br,\frac{\tau}{2}))=W(U_{\varepsilon}^+(\bk,\br)) +W(U_{\varepsilon}^-(\bk,\br )),
\end{aligned}
\end{equation} where $\varepsilon=0$ or $\pi$. Therefore, $W(U^-_\varepsilon(\bk,\br))$ is not an independent topological invariant.

We remark that Eq.(\ref{AIII-inv}) can be written in a basis-independent form:
\begin{equation}
\begin{aligned}
&W(U_{\varepsilon}(\bk,\br,\tau/2))=e^{i\varepsilon} K_{d+D}\int_{T^d\times S^D} d^dk d^Dr \text{Tr}\{\epsilon^{\alpha_1\alpha_2\cdots\alpha_{d+D}} \\
&\times[(I-S)/2] [(U_{\varepsilon})^{-1}\partial_{\alpha_{1}}U_{\varepsilon}] \cdots[(U_{\varepsilon})^{-1}\partial_{\alpha_{d+D}}U_{\varepsilon}]\}.
\end{aligned}\label{windingdensity-general} \end{equation} Note that the $(I-S)/2$ factor has been inserted, $I$ being the identity matrix. The factor $e^{i\varepsilon}=\pm 1$ is included so that this definition is consistent with Eq.(\ref{AIII-inv}). The basis-independent Eq.(\ref{windingdensity-general}) is more convenient when non-chiral basis is used (i.e. the chiral matrix $S$ is not diagonal).

Now we prove that this topological invariant reduces to the static topological invariant of Ref.\cite{teo2010} in the time-independent limit. We consider a generic time-independent flat-band Hamiltonian
\bea
H_0(\bk,\br)&=& -E_0P(\bk,\br)+E_0[1-P(\bk,\br)]\nn\\ &=& E_0 Q(\bk,\br),
\eea  where $P$ is the projection operator of the valence bands with energy $-E_0$, and $Q=1-2P$. In the chiral basis, due to the chiral symmetry $\{H_0(\bk,\br),S\}=0$, it takes the off-diagonal form
\bea  Q=
 \begin{pmatrix}
       & q \\
     q^\dag &
\end{pmatrix}.
\eea
In the static limit, we only need to consider the $\varepsilon=0$ gap. Borrowing the calculation in Eq.(\ref{time evo}), we can transform
the periodized time evolution operator into
\begin{equation} \label{}
U_{\varepsilon=0}(\bk,\br,\frac{\tau}{2})=P(\text{e}^{i\omega\tau/2}-1)+1\\
=1-2P=Q,
\end{equation} therefore, $U_{\varepsilon=0}(\bk,\br,\frac{\tau}{2})$ is simply proportional to the static Hamiltonian. More explicitly, we have
\begin{equation} \label{}
\begin{pmatrix}
&U_{\varepsilon=0}^+\\
U_{\varepsilon=0}^-&\\
\end{pmatrix}
=
\begin{pmatrix}
&q\\
q^\dagger&\\
\end{pmatrix}.
\end{equation}
The winding number Eq.(\ref{AIII-inv}) becomes:
\begin{equation}
\label{}
\begin{aligned}
&W(U^+_{\varepsilon=0}(\bk,\br))=K_{d+D}\int_{T^d\times S^D} d^dk d^Dr\\
&\times\text{Tr}[\epsilon^{\alpha_1\alpha_2\cdots\alpha_{d+D}} (q^{-1}\partial_{\alpha_{1}}q)\cdots(q^{-1}\partial_{\alpha_{d+D}}q)],
\end{aligned}
\end{equation}
which is exactly the topological invariant for static topological defects in class AIII\cite{teo2010}.

For the case $(d,D)=(2,1)$, namely, a point defect in a two-dimensional system, we study a concrete lattice model (see Sec. \ref{AIII-point}), for which the topological invariant is numerically evaluated, and the topological defect modes are also numerically calculated. The numbers of topological modes in both the $\varepsilon=0$ and the $\varepsilon=\pi$ quasienergy gaps are exactly determined by the topological invariants.

The correspondence between the values of topological invariant and the chirality of topological modes is somewhat subtle for the chiral class. We will discuss this point in Sec. \ref{AIII-point} in terms of a concrete model.

\section{Topological invariants of real classes}

Now we turn to the eight real classes, which have at least one anti-unitary symmetry, $\Xi$ or $\Theta$, or both. Due to the special role played by the chiral symmetry, we will first study classes D, C, AI, AII, which have no chiral symmetry, and then classes BDI, DIII, CII, CI, which have the chiral symmetry.  The topological invariants take quite different forms for the nonchiral classes and the chiral classes. Although all these topological invariants are winding numbers,  they are defined on different parameter spaces.

\subsection{Topological invariants of the nonchiral classes: D, C, AI, AII }

\subsubsection{The winding number}

For classes D, C, AI, AII without chiral symmetry,  we can define an integer winding number when $\delta=d-D$ is an even integer:
\begin{equation}
\label{nonchiral-winding}
\begin{aligned}
&W(U_\varepsilon(\bk,\br,t))=K_{d+D+1}\int_{T^d\times S^D\times S^1} d^dk d^Dr dt\\
&\times\text{Tr}[\epsilon^{\alpha_1\alpha_2\cdots\alpha_{d+D+1}} (U_{\varepsilon}^{-1}\partial_{\alpha_{1}}U_{\varepsilon}) \cdots(U_{\varepsilon}^{-1}\partial_{\alpha_{d+D+1}}U_{\varepsilon})],
\end{aligned}
\end{equation}
where the coefficient $K_{d+D+1}$ is the same as defined above, namely $K_{d+D+1}=\frac{(-1)^{\frac{d+D}{2}}(\frac{d+D}{2})!}{(d+D+1)!} \left(\frac{i}{2\pi}\right)^{\frac{d+D}{2}+1}$.
The spatial dimensions in which this topological invariant is applicable depend on the symmetry, which will be discussed in Sec. \ref{sec:DC} and Sec. \ref{sec:AIAII} below. It should also be mentioned that Eq.(\ref{nonchiral-winding}) is the expression only for the integer topological invariant\footnote{Recently, we were informed that, for the homogeneous case ($D=0$ case), existence of $\mathbb{Z}$ topological invariant in the presence of an anti-unitary symmetry is independently considered in Ref.\cite{fruchart-phd}.}; the $\mathbb{Z}_2$ topological invariants will be studied in Sec. \ref{sec:WZWDC} and Sec. \ref{sec:WZW-AIAII}.

For bulk systems with translational symmetry, the topological invariant can be simply obtained from Eq.(\ref{nonchiral-winding}) by taking $D=0$, and  accordingly, eliminating $S^D$ and $\br$:
\begin{equation}
\label{nonchiral-winding-bulk}
\begin{aligned}
&W(U_\varepsilon(\bk,t))=K_{d+1}\int_{T^d\times S^1} d^dk  dt\\
&\times\text{Tr}[\epsilon^{\alpha_1\alpha_2\cdots\alpha_{d+1}} (U_{\varepsilon}^{-1}\partial_{\alpha_{1}}U_{\varepsilon}) \cdots(U_{\varepsilon}^{-1}\partial_{\alpha_{d+1}}U_{\varepsilon})].
\end{aligned}
\end{equation} The same is true for other symmetry classes, and we will not mention this $D=0$ case repeatedly.

By similar calculations as Appendix.\ref{statich}, we can find that, for a static Hamiltonian as given by Eq.(\ref{flat}), the winding number reduces to
\begin{equation}
\label{}
\begin{aligned}
&W(U_{\varepsilon=0})=\tilde{K}_{d+D}\int_{T^d\times S^D} d^dk d^Dr\\
&\times\text{Tr}[\epsilon^{\alpha_1\alpha_2\cdots\alpha_{d+D}}P\partial_{\alpha_1}P\cdots\partial_{\alpha_{d+D}}P],
\end{aligned}
\end{equation}
in which $\tilde{K}_{d+D}$ is given in Eq.(\ref{static2}). This is the Chern number $C_{(d+D)/2}$ of the valence bands.

Although the winding number takes the same form as that of class A, the symmetries impose certain constraints on its possible values, which depend on spatial dimensions.  Now we discuss these features.

\subsubsection{Particle-hole symmetry: class D and class C}\label{sec:DC}

Let us recall the effects of symmetries, which we have discussed in Sec. \ref{sec:symmetries}. In particular, the symmetries of the periodized time evolution operator are immediately relevant now.
For the class D and class C, which have the PHS, the periodized time evolution operator with branch cut at $\varepsilon=0$ satisfies
[see Eq.(\ref{particle})]:
\begin{eqnarray}
C^{-1} U_{\varepsilon=0}(\bk,\br,t) C =U^*_{\varepsilon=0}(-\bk,\br,t)\exp(i\frac{2\pi t}{\tau}).\label{particle0}
\end{eqnarray}
Now we would like to obtain its constraints on the topological invariants. By a quite lengthy calculation given in Appendix \ref{symdc}, we obtain the symmetry of the winding number density $w(U_{\varepsilon=0})=K_{d+D+1} \text{Tr}[\epsilon^{\alpha_1\alpha_2\cdots\alpha_{d+D+1}} (U_{\varepsilon=0}^{-1}\partial_{\alpha_{1}}U_{\varepsilon=0}) \cdots (U_{\varepsilon=0}^{-1} \partial_{\alpha_{d+D+1}}U_{\varepsilon=0})]$, which is
\begin{equation}
\begin{aligned}
w(U_{\varepsilon=0})(\bk,\br,t)=w(U_{\varepsilon=0})(-\bk,\br,t)(-1)^{ 1-\delta/2}.
\end{aligned} \label{DC00}
\end{equation}
Therefore, when $\delta=4n$ ($n$ is an integer), we have \bea w(U_{\varepsilon=0})(\bk,\br,t)=-w(U_{\varepsilon=0})(-\bk,\br,t), \eea and the winding number, which is the integral of $w(U_{\varepsilon=0})(\bk,\br,t)$ on $T^d\times S^D\times S^1$, must vanish. This fact indicates the absence of integer topological classification in these dimensions. Only when $\delta=4n+2$ ($n$ is an integer), namely, $\delta=2,6,10,\cdots$, the winding number can be nonzero, indicating the presence of integer classification. This is an example of how topological invariants tell us about topological classifications.

Similarly, for $\varepsilon=\pi$, the PHS implies (see Appendix \ref{symdc})
\begin{equation}
\begin{aligned}
C^{-1} U_{\varepsilon=\pi}(\bk,\br,t) C =U^*_{\varepsilon=\pi}(-\bk,\br,t)\exp(i\frac{4\pi t}{\tau}), \label{particlepi}
\end{aligned}
\end{equation} which is slightly different from Eq.(\ref{particle0}) in that $2\pi t/\tau$ is replaced by $4\pi t/\tau$. It follows from
Eq.(\ref{particlepi}) that
\begin{equation}
\begin{aligned}
w(U_{\varepsilon=\pi})(\bk,\br,t)=w(U_{\varepsilon=\pi})(-\bk,\br,t)(-1)^{1-\delta/2}.
\end{aligned}  \label{DCPI}
\end{equation}
Again, only when $\delta=4n+2$, the winding number can be nonzero.

Before moving on, we would like to emphasize two salient features in Eq.(\ref{DC00}) and Eq.(\ref{DCPI}), which are shared by the topological invariants of other real symmetry classes to be discussed below. First, the symmetry constraints of Eq.(\ref{DC00}) and Eq.(\ref{DCPI}) depend only on $\delta\equiv d-D$, but not on $d$ or $D$ separately. Thus, the symmetry of the time evolution operator automatically leads to the combination $\delta=d-D$, though the winding number is defined on the $(d+D+1)$-dimensional space. Going from $(d,D)$ to $(d+1,D+1)$ or $(d-1,D-1)$ does not change the symmetry constraint. As a result, the classification in Table \ref{table} depends only on $\delta$. In static systems, a similar conclusion is reached via the K-theory\cite{teo2010}. Compared to the situation in static systems, the combination $\delta=d-D$ enters more automatically by Floquet topological invariants. Second, $\delta$ enters as $(-1)^{\delta/2}$, therefore, if the necessary condition for the existence of integer winding number, namely $(-1)^{1-\delta/2}=1$, is satisfied by a given $\delta$, the next $\delta$ satisfying it would be $\delta+4$, which means that the dimensional periodicity of integer winding numbers should be $4$. This periodicity can be appreciated in Table \ref{table}. For class D and class C, integer winding numbers exist when $\delta=2$ (mod $4$) (the difference between $\mathbb{Z}$ and $2\mathbb{Z}$ will be discussed shortly).

\subsubsection{Time-reversal symmetry: class AI and class AII}\label{sec:AIAII}

In the presence of TRS, Eq.(\ref{time}) tells us that
\begin{eqnarray}
T^{-1} U_{\varepsilon}(\bk,\br,t) T =U^*_{\varepsilon}(-\bk,\br,-t).
\end{eqnarray}
Taking advantage of this symmetry, we find that the winding number density satisfies (see Appendix.\ref{syma})
\begin{equation}
\label{windai}
\begin{aligned}
w(U_\varepsilon)(\bk,\br,t)=w(U_\varepsilon)(-\bk,\br,-t)(-1)^{2-\delta/2}.
\end{aligned}
\end{equation} Compared with Eq.(\ref{DC00}), there is an additional $-1$ factor in the right-hand side of Eq.(\ref{windai}), which originates from the fact that TRS reverses $t$ (see Appendix.\ref{syma} for more details).
Therefore, when $\delta=4n+2$ ($n$ is an integer), we have $w(U_\varepsilon)(\bk,\br,t)= -w(U_\varepsilon)(-\bk,\br,-t)$, and the winding number must vanish, indicating the absence of integer classification in these dimensions; only when $\delta=4n$, the winding number can be nonzero, indicating integer classifications.

It is interesting to compare the cases of TRS and PHS. In Eq.(\ref{windai}) and Eq.(\ref{DC00}), we have the $(-1)^{2-\delta/2}$ and the $(-1)^{1-\delta/2}$ factor, respectively. Due to the $1/2$ factor of $\delta$, if one of these two factors is $+1$ for a $\delta$,  the other factor is $+1$ for $\delta\pm 2$. This causes the difference between $4n$ (TRS) and $4n+2$ (PHS).

\subsubsection{Even-integer ($2\mathbb{Z}$) topological invariants}\label{sec:2Z}

In addition to the constraints we have discussed, symmetries also imposes one more constraint, which is that the winding number has to take even-integer values when $\delta-s=4$ (mod $8$), or equivalently, $d-D-s=4$ (mod $8$) (Here, $s=0,1,2,\cdots,7$ labels the eight real symmetry classes, as indicated in the first row of Table \ref{table}). It is not straightforward to see this fact directly from the definition of winding number, nevertheless, we can reach this conclusion from Eq.(\ref{difference}). In fact, it is known\cite{ryu2010,teo2010,Chiu2016rmp} that all the band Chern numbers are even integers when $\delta-s= 4$ (mod $8$) (an intuitive understanding of this fact is to construct minimal Dirac Hamiltonians with given symmetries and spatial dimensions\cite{ryu2010}). In the Floquet systems, the same derivation tells us that all the Floquet band Chern numbers $C_{(d+D)/2}(P_{\varepsilon,\varepsilon'})$ are even integers, where $P_{\varepsilon,\varepsilon'}$ is the Floquet band projection operator previously defined. When $\delta-s = 4$ (mod $8$), Eq.(\ref{difference}) implies that the differences between   $W(U_\varepsilon)$ and $W(U_{\varepsilon'})$ for any pair of $\varepsilon,\varepsilon'$ is always an even integer. Therefore,  all $W$'s must have the same (even/odd) parity. If one of the quasienergy gaps is closed and then reopened as we tune certain Hamiltonian parameters, the change of its $W$ must be an even integer, so that the parity of $W$ remains the same. To ensure that $W=0$ (i.e., the topologically trivial class) can appear somewhere in the phase diagram (which is a natural expectation), we have to take the scenario that the parity of $W$ is always even.

A more rigorous proof of the $2\mathbb{Z}$ topological invariants for $\delta-s=4$ (mod $8$), which does not involve the Floquet band Chern numbers, relies on the representative Dirac Hamiltonians constructed in Appendix \ref{sec:even}. This proof is given in Appendix \ref{sec:even}.

The $2\mathbb{Z}$ topological invariants for $\delta-s=4$ (mod $8$) are related to the absence of $\mathbb{Z}_2$ topological classification for $\delta-s=3$ (mod $8$), while the $\mathbb{Z}$ topological invariants for $\delta-s=0$ (mod $8$) are related to the presence of $\mathbb{Z}_2$ topological classification for $ \delta-s=-1$ or $7$ (mod $8$), which will be studied in more details in due time below.  For the moment, let us simply take the fact that there is no $\mathbb{Z}$ or $\mathbb{Z}_2$ classification for $\delta-s=3$ (mod $8$) (i.e., all phases are topologically trivial). Given this fact, for $\delta-s=4$ (mod $8$), we can smoothly deform the periodized time evolution operator $U_\varepsilon(\bk,\br,t)$ to a new function $\bar{U}_\varepsilon(\bk,\br,t)$ such that $\bar{U}_\varepsilon(k_1=0,k_2,\cdots,k_d,\br,t) =\bar{U}_\varepsilon(k_1=\pi,k_2,\cdots,k_d,\br,t)$, which is always possible because fixing $k_1=0$ and fixing $k_1=\pi$ yield topologically equivalent time evolution operator at $\delta-s=3$ (mod $8$) [Due to the absence of $\mathbb{Z}_2$ and $\mathbb{Z}$ classifications at $\delta-s=3$ (mod $8$), there is only one topological class, namely, the topologically trivial class; therefore, any two time evolution operators can be smoothly deformed to each other]. Now the deformed periodized time evolution operator $\bar{U}_\varepsilon(\bk,\br,t)$ at $\delta-s=4$ (mod $8$) has periodic boundary condition in the half Brillouin zones ( $k_1\in[0,\pi]$ or $k_1\in[-\pi,0]$ ), and the winding number split into the sum of the two winding numbers, one of which is defined on the $k_1\in[0,\pi]$ half, the other defined on the $k_1\in[-\pi,0]$ half. Moreover, these two winding numbers are equal due to the symmetry of winding number density between $\bk$ and $-\bk$ [for instance, see Eq.(\ref{DC00}) with $\delta=4n+2$], therefore, the winding number on the entire Brillouin zone must be an even integer. As such, the absence of $\mathbb{Z}_2$ classification for $\delta-s=3$ (mod $8$) implies the $2\mathbb{Z}$ (instead of $\mathbb{Z}$) topological invariants for $\delta-s=4$ (mod $8$).

\subsubsection{$\mathbb{Z}_2$ topological invariants of Wess-Zumino-Witten form for class D and class C}\label{sec:WZWDC}

As we have discussed, for class D and class C with PHS,
we can define an integer topological invariant when $\delta =4n+2$ (When $\delta=4n$, the same topological invariant would always yield zero, which is not useful). For $\delta=4n+1$ or $4n$, there is no such an integer topological invariant, whereas we can use the construction of Wess-Zumino-Witten term \cite{witten1983,wang2010b,wang2012a,wang2013} to define a $\mathbb{Z}_2$ topological invariant. It is the purpose of this subsection to do so.

When $\delta\equiv d-D=4n+1$, the $(\bk,\br,t)$ parameter space of  $U_{\varepsilon}(\bk,\br,t)$ is even-dimensional, however, a winding number has definition only in an odd-dimensional space, which excludes the possibility of defining a winding number in the $(\bk,\br,t)$ space. Nevertheless, we can extend the parameter space by adding one more momentum-like dimension, so that the winding number can be defined. However, only the parity (even/odd) of the winding number is well defined, which leads to a $\mathbb{Z}_2$ topological invariant in the initial dimensions. It is the purpose of this subsection to explain this construction.

First, we define a relative $\mathbb{Z}_2$ topological invariant, which is constructed as follows. Let us consider two Floquet systems ($a$ and $b$) with PHS, both having a nonzero quasienergy gap at $\varepsilon$ (due to the PHS, we will take $\varepsilon=0$ or $\pi$ below). The time evolution operator is denoted as $U^a (\bk,\br,t)$ and $U^b (\bk,\br,t)$, respectively. We can construct a smooth interpolation $U(\bk,\br,t,\lambda)$ ($\lambda\in[0,\pi]$) between them such that \bea U (\bk,\br,t,0)= U^a (\bk,\br,t),\,  U (\bk,\br,t,\pi)= U^b (\bk,\br,t). \eea  We also require that, like $U^a (\bk,\br,t)$ and $U^b (\bk,\br,t)$, the interpolation $U(\bk,\br,t,\lambda)$ has a nonzero quasienergy gap at $\varepsilon$. The interpolation $U(\bk,\br,t,\lambda)$ induces an interpolation between $U^a_\varepsilon (\bk,\br,t)$ and $U^b_\varepsilon (\bk,\br,t)$, namely $U_\varepsilon(\bk,\br,t,\lambda)= U(\bk,\br,t,\lambda)\exp[iH^{\rm eff}_\varepsilon(\bk,\br,\lambda)t]$, which satisfies $U_\varepsilon(\bk,\br,t,0)= U^a_\varepsilon (\bk,\br,t)$ and $U_\varepsilon(\bk,\br,t,\pi)= U^b_\varepsilon (\bk,\br,t)$. It is always possible to find an interpolation because the trivial homotopy group $\pi_{d+D+1}(U(N))=0$ (note that $d-D=4n+1$; therefore, $d+D+1$ is an even integer here). Hereafter, when we talk about an interpolation $U_\varepsilon(\bk,\br,t,\lambda)$ between $U^a_\varepsilon (\bk,\br,t)$ and $U^b_\varepsilon (\bk,\br,t)$, we always implicitly refer to an interpolation $U(\bk,\br,t,\lambda)$ between $U^a(\bk,\br,t)$ and $U^b(\bk,\br,t)$, and take  $U_\varepsilon(\bk,\br,t,\lambda)$ as the periodized version of this $U(\bk,\br,t,\lambda)$.

When $\lambda\neq 0,\pi$, $U(\bk,\br,t,\lambda)$ does not necessarily have the PHS. To apply the PHS, let us introduce a mirror interpolation in $[-\pi,0]$, which is fully determined by the original interpolation in $[0,\pi]$:   for $\lambda\in[-\pi,0]$, we take $U(\bk,\br,t,\lambda)  = CU^*(-\bk,\br,t,-\lambda)C^{-1}$. Equivalently, it can be written as \bea C^{-1}U(\bk,\br,t,\lambda)C  = U^*(-\bk,\br,t,-\lambda), \eea which takes the same form as Eq.(\ref{particleu}) if we regard $\lambda$ as a momentum-like variable. This equation is consistent with the PHS of $U(\bk,\br,t,\lambda)$ at $\lambda=0$ and $\pi$. Accordingly, the periodized time evolution operator satisfies
\begin{equation}
\label{}
\begin{aligned}
C^{-1} U_{\varepsilon}(\bk,\br,t,\lambda) C =U^*_{-\varepsilon}(-\bk,\br,t,-\lambda)\exp(i\frac{2\pi t}{\tau}),
\end{aligned}
\end{equation}
which takes the same form as Eq.(\ref{particle}), $\lambda$ being a momentum-like variable.

In particular, when $\varepsilon=0$, we have
\begin{eqnarray}
U^*_{\varepsilon=0}(\bk,\br,t,\lambda) =C^{-1} U_{\varepsilon=0}(-\bk,\br,t,-\lambda)C \exp(-i\frac{2\pi t}{\tau}),
\end{eqnarray}
while for $\varepsilon=\pi$, as a result of the relation between $U_{\varepsilon=\pi}$ and $U_{\varepsilon=-\pi}$ (see Appendix \ref{symdc}), we have
\begin{eqnarray}
U^*_{\varepsilon=\pi}(\bk,\br,t,\lambda) =C^{-1} U_{\varepsilon=\pi}(-\bk,\br,t,-\lambda)C \exp(-i\frac{4\pi t}{\tau}),
\end{eqnarray}
which is reminiscent of Eq.(\ref{particlepi}).
With the parameter $\lambda$ included, a winding number can be defined on the $(d+D+2)$-dimensional $(\bk,\br,\lambda,t)$ parameter space:
\begin{equation}
\label{}
\begin{aligned}
&W(U_\varepsilon(\bk,\br,t,\lambda))=K_{d+D+2}\int_{ T^{d+1}\times S^D\times S^1} d^dk d^Dr dtd\lambda\\
&\times\text{Tr}[\epsilon^{\alpha_1\alpha_2\cdots \alpha_{d+D+2}}(U_{\varepsilon}^{-1}\partial_{\alpha_{1}}U_{\varepsilon}) \cdots(U_{\varepsilon}^{-1}\partial_{\alpha_{d+D+2}}U_{\varepsilon})],
\end{aligned}
\end{equation}
where $\varepsilon=0$ or $\pi$. The coefficient
$K_{d+D+2}= \frac{(-1)^{\frac{d+D+1}{2}}(\frac{d+D+1}{2})!}{(d+D+2)!} \left(\frac{i}{2\pi}\right)^{\frac{d+D+1}{2}+1}$.

Given $U^a_{\varepsilon}(\bk,\br,t)$ and $U^b_{\varepsilon}(\bk,\br,t)$, there exist infinitely many ways to interpolate them. For two different interpolations,  $U_{\varepsilon}(\bk,\br,t,\lambda)$ and $U'_{\varepsilon}(\bk,\br,t,\lambda)$, the winding number can be different. Nevertheless, we will show below that their difference is always an even integer, namely,
\begin{eqnarray} \label{mod2}
W(U_\varepsilon(\bk,\br,t,\lambda))-W(U'_\varepsilon(\bk,\br,t,\lambda))=0   \ (\text{mod}\ 2),
\end{eqnarray} therefore, $W(U_\varepsilon)$ (mod $2$) is independent of the interpolation and is well defined.

When $W(U_\varepsilon)=0   \ (\text{mod}\ 2)$, $U^a_\varepsilon(\bk,\br,t)$ and $U^b_\varepsilon(\bk,\br,t)$ are regarded to be in the same topological class; alternatively, when $W(U_\varepsilon) =1 \ (\text{mod}\ 2)$, they belong to different classes. Moreover, if $U^a_\varepsilon(\bk,\br,t )$ and $U^b_\varepsilon(\bk,\br,t )$ can be interpolated by some $U_\varepsilon(\bk,\br,t,\lambda)$ with winding number $W(U_\varepsilon(\bk,\br,t,\lambda))\equiv W_{ab}$ (mod $2$), while $U^b_\varepsilon(\bk,\br,t)$ and $U^c_\varepsilon(\bk,\br,t)$ can be connected by another interpolation whose winding number is $W_{bc}$ (mod $2$), then the combination of these two interpolations yields an interpolation between $U^a_\varepsilon(\bk,\br,t)$ and $U^c_\varepsilon(\bk,\br,t)$, whose winding number is \bea W_{ac}=W_{ab}+W_{bc}\quad (\text{mod}\,2). \eea
As such, our construction yields a $\mathbb{Z}_2$ classification, and $W(U_\varepsilon(\bk,\br,t,\lambda))$ (mod $2$) is the relative $\mathbb{Z}_2$ topological invariant of $U^a_{\varepsilon}(\bk,\br,t)$ and $U^b_{\varepsilon}(\bk,\br,t)$.
We emphasize that the validness of $\mathbb{Z}_2$ classification crucially relies on Eq.(\ref{mod2}), which guarantees that $W(U_\varepsilon) \, (\text{mod}\, 2)$ does not depend on the choice of interpolation. Similar mechanisms of $\mathbb{Z}_2$ topological invariants can be found in the contexts of static gapped Hamiltonian\cite{qi2008} and Green's function\cite{wang2010b}.

\begin{figure}
\includegraphics[width=8.5cm, height=3.9cm]{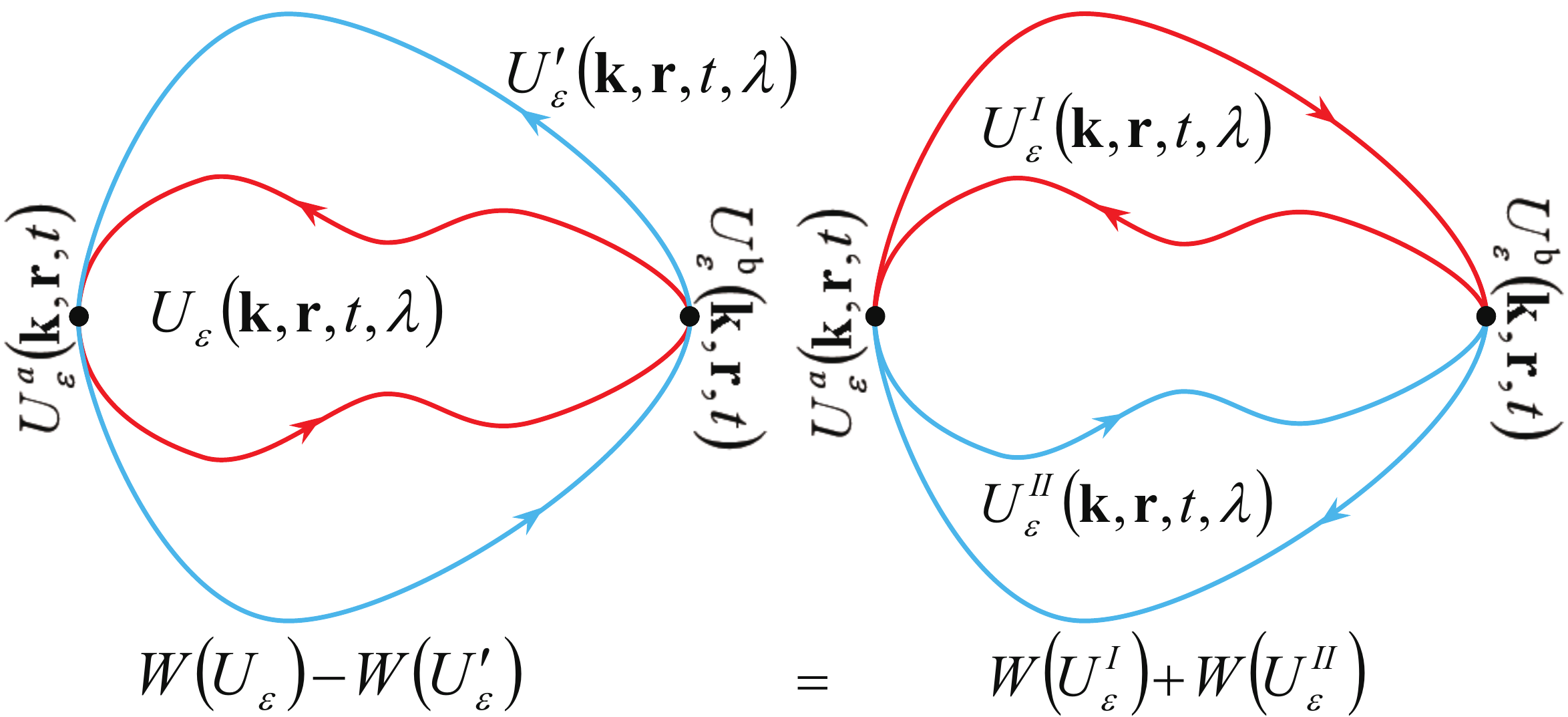}
\caption{ The interpolations $U^{I}(\bk,\br,t,\lambda)$ and $U^{II}(\bk,\br,t,\lambda)$ are constructed from $U(\bk,\br,t,\lambda)$ and $U'(\bk,\br,t,\lambda)$; accordingly, the periodized time evolution operators $U^{I}_\varepsilon(\bk,\br,t,\lambda)$ and $U^{II}_\varepsilon(\bk,\br,t,\lambda)$ are constructed from $U_\varepsilon(\bk,\br,t,\lambda)$ and $U'_\varepsilon(\bk,\br,t,\lambda)$.  }  \label{deformation}
\end{figure}

Now it remains to prove that Eq.(\ref{mod2}) is true.
To this end, let us define two new interpolations, which are reorganizations of $U(\bk,\br,t,\lambda)$ and $U' (\bk,\br,t,\lambda)$:
\begin{equation}
U^{I}(\bk,\br,t,\lambda)=\left\{
\begin{aligned}
&U(\bk,\br,t,\lambda), \,  -\pi<\lambda<0, \\
&U'(\bk,\br,t,-\lambda), \,   0<\lambda<\pi,
\end{aligned}\right.
\end{equation}
and
\begin{equation}
U^{II}(\bk,\br,t,\lambda)=\left\{
\begin{aligned}
&U'(\bk,\br,t,-\lambda), \,  -\pi<\lambda<0,\\
&U(\bk,\br,t,\lambda), \,   0<\lambda<\pi.
\end{aligned}\right.
\end{equation} A pictorial illustration of the construction of $U^{I}(\bk,\br,t,\lambda)$ and $U^{II}(\bk,\br,t,\lambda)$ from $U(\bk,\br,t,\lambda)$ and $U'(\bk,\br,t,\lambda)$ is given in Fig.\ref{deformation}. As a consequence, we have the following two interpolations of the periodized time evolution operators:
\begin{equation}
U_{\varepsilon}^{I}(\bk,\br,t,\lambda)=\left\{
\begin{aligned}
&U_{\varepsilon}(\bk,\br,t,\lambda), \,  -\pi<\lambda<0, \\
&U'_{\varepsilon}(\bk,\br,t,-\lambda), \,   0<\lambda<\pi,
\end{aligned}\right.
\end{equation}
and
\begin{equation}
U_{\varepsilon}^{II}(\bk,\br,t,\lambda)=\left\{
\begin{aligned}
&U'_{\varepsilon}(\bk,\br,t,-\lambda), \,  -\pi<\lambda<0,\\
&U_{\varepsilon}(\bk,\br,t,\lambda), \,   0<\lambda<\pi.
\end{aligned}\right.
\end{equation}
From Fig.\ref{deformation}, it is now quite clear that
\begin{eqnarray}
W(U_\varepsilon)-W(U'_\varepsilon)=W(U_{\varepsilon}^{I})+W(U_{\varepsilon}^{II}).
\end{eqnarray}
From the definition of $U_{\varepsilon}^{I}(\bk,\br,t,\lambda)$ and $U_{\varepsilon}^{II}(\bk,\br,t,\lambda)$, it is also clear that, for $\varepsilon=0$,
\begin{equation}\label{wdc1}
C^{-1} U^I_{\varepsilon=0}(\bk,\br,t,\lambda)C =U^{II*}_{\varepsilon=0}(-\bk,\br,t,-\lambda)\exp(i\frac{2\pi t}{\tau}),
\end{equation}
and, for $\varepsilon=\pi$,
\begin{equation}\label{wdc2}
C^{-1} U^I_{\varepsilon=\pi}(\bk,\br,t,\lambda)C =U^{II*}_{\varepsilon=\pi}(-\bk,\br,t,-\lambda)\exp(i\frac{4\pi t}{\tau}).
\end{equation}
For the $\varepsilon=\pi$ case, the relation between $U_{\varepsilon=\pi}$ and $U_{\varepsilon=-\pi}$ (see Appendix \ref{symdc}) has been used.

By a somewhat lengthy calculation, we have (see Appendix \ref{wzwdc} for details)
\begin{equation}
\label{wzwdc0}
\begin{aligned}
w(U_{\varepsilon=0}^{I})(\bk,\br,t,\lambda)=w(U_{\varepsilon=0}^{II})(-\bk,\br,t,-\lambda)(-1)^{2-(\delta-1)/2}
\end{aligned}
\end{equation}
and
\begin{equation}
\label{wzwdcpi}
\begin{aligned}
w(U_{\varepsilon=\pi}^{I})(\bk,\br,t,\lambda) =w(U_{\varepsilon=\pi}^{II})(-\bk,\br,t,-\lambda)(-1)^{2-(\delta-1)/2}.
\end{aligned}
\end{equation}
Therefore, when $\delta\equiv d-D=4n+1$, the winding numbers for the two interpolations $U_\varepsilon^I$ and $U_\varepsilon^{II}$ are equal:
\begin{equation}
\label{}
\begin{aligned}
W(U^I_\varepsilon)&=\int_{ T^{d+1}\times S^D\times S^1} w(U^{I}_\varepsilon)(\bk,\br,t,\lambda)\\
&=\int_{ T^{d+1}\times S^D\times S^1}  w(U^{II}_\varepsilon)(-\bk,\br,t,-\lambda)(-1)^{2-(\delta-1)/2}\\
&=W(U^{II}_\varepsilon).
\end{aligned}
\end{equation}
It follows that
\begin{eqnarray} \label{}
W(U_\varepsilon)-W(U'_\varepsilon)=W(U_{\varepsilon}^{I})+W(U_{\varepsilon}^{II})=2W(U_{\varepsilon}^{I}),
\end{eqnarray}
which is always an even integer.

So far, $W(U_\varepsilon(\bk,\br,t,\lambda))$ is defined as a relative $\mathbb{Z}_2$ topological invariant between $U^a_\varepsilon(\bk,\br,t)$ and $U^b_\varepsilon(\bk,\br,t)$.
If we choose $U^a_\varepsilon(\bk,\br,t)=U_\varepsilon(\bk,\br,t)$, and $U^b_\varepsilon(\bk,\br,t)$ as a trivial time evolution operator, i.e., $U^b_\varepsilon(\bk,\br,t)$ does not depend on $(\bk,\br)$ (constant function), then we can take  $W(U_\varepsilon(\bk,\br,t,\lambda))$ (mod $2$) as the $\mathbb{Z}_2$ topological invariant of $U_\varepsilon(\bk,\br,t)$. Alternatively, we can define
\begin{equation}
\nu(U_\varepsilon(\bk,\br,t))=(-1)^{W(U_\varepsilon(\bk,\br,t,\lambda))}=\pm 1, \label{DC-Z2} \end{equation} as the $\mathbb{Z}_2$ topological invariant.

Now let us briefly discuss the static limit. Similar to  Appendix.\ref{statich}, for a flat-band Hamiltonian under an infinitesimal driving with a sufficiently large frequency, the static limit of the winding number defined on the extended $(\bk,\br,t,\lambda)$ parameter space can be reduced to
\begin{equation}
\label{chern+1}
\begin{aligned}
&W(U_{\varepsilon=0})=\tilde{K}_{d+D+1}\int_{ T^{d+1}\times S^D } d^dk d^Drd\lambda\\
&\times\text{Tr}[\epsilon^{\alpha_1\alpha_2\cdots\alpha_{d+D+1}}P \partial_{\alpha_1}P\cdots\partial_{\alpha_{d+D+1}}P],
\end{aligned}
\end{equation}
where $\tilde{K}_{d+D+1}$ is given by Eq.(\ref{static2}).
This is exactly the Chern number defined on the $(d+D+1)$-dimensional parameter space. It is indeed the topological invariant of static topological defects\cite{teo2010}.

The Chern character $\text{ch}_{\frac{d+D+1}{2}}\equiv \tilde{K}_{d+D+1}\text{Tr} [\epsilon^{\alpha_1\alpha_2\cdots\alpha_{d+D+1}} P\partial_{\alpha_1}P\cdots\partial_{\alpha_{d+D+1}}P]$ is the exterior derivative of the Chern-Simons form, namely, $\text{ch}_{\frac{d+D+1}{2}}=d\mathcal{Q}_{d+D}$, in which the Chern-Simons form\cite{ryu2010,nakahara2003}
\begin{equation}
\label{}
\begin{aligned}
\mathcal{Q}_{d+D} =\frac{1}{(\frac{d+D-1}{2})!}(\frac{i}{2\pi})^{\frac{d+D+1}{2}}\int_{0}^{1}dt \text{Tr}[\mathcal{A}(td\mathcal{A}+t^2\mathcal{A}^2)^{\frac{d+D-1}{2}}]
\end{aligned}
\end{equation}
is defined in terms of the Berry connection $\mathcal{A}$, whose entries are
$\mathcal{A}^{\alpha \beta}(\bk,\br,\lambda)=\bra{u^{\alpha}(\bk,\br,\lambda)}du^{\beta}(\bk,\br,\lambda)\ra$ (the notation of differential form is used here\cite{nakahara2003}).
Integration over $\lambda$ leads to
\begin{equation}
\label{chernsimons}
\begin{aligned}
&W(U_{\varepsilon=0})=\int_{ T^{d+1}\times S^D}\ d\mathcal{Q}_{d+D}=2\int_{T^d\times S^D}  \  \mathcal{Q}_{d+D}.
\end{aligned}
\end{equation}

For $\delta\equiv d-D=4n$, similar construction of $\mathbb{Z}_2$ topological invariant is still possible.  In these cases, we need two WZW extension parameters $\lambda$ and $\mu$, both in $[-\pi,\pi]$. We define an extension of $U (\bk,\br,t)$ to
$U (\bk,\br,t,\lambda,\mu)$, which satisfies $U (\bk,\br,t,0,0)=U (\bk,\br,t)$. In addition, $U (\bk,\br,t,\pm\pi,\mu)$ and $U (\bk,\br,t,\lambda,\pm\pi)$ are trivial time evolution operators (i.e., they are independent of $\bk$ and $\br$). As an extension of the PHS relation given in Eq.(\ref{particleu}), we require that \begin{equation}
\begin{aligned}
C^{-1} U (\bk,\br,t,\lambda,\mu) C =U^* (-\bk,\br,t,-\lambda,-\mu);
\end{aligned}
\end{equation} accordingly,
\begin{equation}
\label{}
\begin{aligned}
C^{-1} U_{\varepsilon}(\bk,\br,t,\lambda,\mu) C =U^*_{-\varepsilon}(-\bk,\br,t,-\lambda,-\mu)\exp(i\frac{2\pi t}{\tau}).
\end{aligned}
\end{equation}
In particular, for the most relevant cases $\varepsilon=0$ and $\varepsilon=\pi$, we have
\begin{equation}
U^*_{\varepsilon=0}(\bk,\br,t,\lambda,\mu) =C^{-1} U_{\varepsilon=0}(-\bk,\br,t,-\lambda,-\mu)C \exp(-i\frac{2\pi t}{\tau}),
\end{equation}
and
\begin{equation}
U^*_{\varepsilon=\pi}(\bk,\br,t,\lambda,\mu) =C^{-1} U_{\varepsilon=\pi}(-\bk,\br,t,-\lambda,-\mu)C \exp(-i\frac{4\pi t}{\tau}).
\end{equation}

Now we can define a winding number on the $(d+D+3)$-dimensional $(\bk,\br,t,\lambda,\mu)$ parameter space:
\begin{equation}
\label{}
\begin{aligned}
&W(U_\varepsilon(\bk,\br,t,\lambda,\mu))=K_{d+D+3}\int_{ T^{d+2}\times S^D\times S^1 } d^dk d^Dr dtd\lambda d\mu\\
&\times\text{Tr}[\epsilon^{\alpha_1\alpha_2 \cdots\alpha_{d+D+3}}(U_{\varepsilon}^{-1}\partial_{\alpha_{1}}U_{\varepsilon}) \cdots(U_{\varepsilon}^{-1}\partial_{\alpha_{d+D+3}}U_{\varepsilon})],
\end{aligned}
\end{equation}
where the coefficient
$K_{d+D+3} =\frac{(-1)^{\frac{d+D+2}{2}}(\frac{d+D+2}{2})!}{(d+D+3)!} \left(\frac{i}{2\pi}\right)^{\frac{d+D+2}{2}+1}$.

By the same derivation as the case $\delta=4n+1$, we can see that $W(U_{\varepsilon}(\bk,\br,t,\lambda,\mu))$ (mod $2$) defines a $\mathbb{Z}_2$ invariant for $U_{\varepsilon}(\bk,\br,t)$:
\begin{equation}
\nu(U_\varepsilon(\bk,\br,t))=(-1)^{W(U_{\varepsilon}(\bk,\br,t,\lambda,\mu))}=\pm 1.
\end{equation}

In the above construction, we have taken $U_{\varepsilon}(\bk,\br,t,\pm\pi,\mu)$ and $U_{\varepsilon}(\bk,\br,t,\lambda,\pm\pi)$ to be trivial time evolution operators (i.e., they do not depend on $\bk$ and $\br$), so that the $\mathbb{Z}_2$ topological invariant yields the topological class of $U_\varepsilon(\bk,\br,t,0,0)\equiv U_\varepsilon(\bk,\br,t)$. If we do not impose this ``boundary triviality'' requirement, then $U_\varepsilon (\bk,\br,t,\lambda,\mu)$ can be regarded as an interpolation among the time evolution operators of the four Floquet systems ($a,b,c,d$), whose periodized time evolution operators are $U^a_\varepsilon(\bk,\br,t) \equiv U_\varepsilon(\bk,\br,t,0,0)$, $U^b_\varepsilon(\bk,\br,t) \equiv U_\varepsilon(\bk,\br,t,0,\pi)$, $U^c_\varepsilon(\bk,\br,t) \equiv U_\varepsilon(\bk,\br,t,\pi,0)$, and $U^d_\varepsilon(\bk,\br,t) \equiv U_\varepsilon(\bk,\br,t,\pi,\pi)$, respectively. Each of them satisfies the PHS.  The winding number $W(U_\varepsilon (\bk,\br,t,\lambda,\mu))$ (mod $2$) can be defined as the relative $\mathbb{Z}_2$ topological invariant of the four Floquet systems ($a,b,c,d$). For instance, if $W(U_\varepsilon (\bk,\br,t,\lambda,\mu))=1$ (mod $2$), we can conclude that one or three of the four Floquet systems are in the $\mathbb{Z}_2$ nontrivial class.

It is important to note that, $\delta$ taking the values of $4n+1$ or $4n$ does not necessarily guarantee the existence of a $\mathbb{Z}_2$ topological invariant. The $\mathbb{Z}_2$ topological classification cannot be constructed in the above way if $W(U_\varepsilon(\bk,\br,t,\lambda))$ can only take even-integer values, since a ``nontrivial class'' with $W(U_\varepsilon(\bk,\br,t,\lambda))=1$ (mod $2$) can never be obtained. As has been discussed in the previous section, we have the $2\mathbb{Z}$ topological invariants when $\delta-s=4$ (mod $8$), or equivalently, $d-D-s=4$ (mod $8$). For the class D, which is labelled as $s=2$, the winding number takes even integer values when $\delta=6$ (mod $8$), therefore, the $\mathbb{Z}_2$ topological invariants cannot be defined for $\delta=5$ or $\delta=4$. They can only be defined when $\delta=1$ or $\delta=0$, which are the descendants of the integer topological invariants of $\delta=2$. For the class C, which is labelled as $s=6$, we can define $\mathbb{Z}_2$ topological invariants for $\delta=5$ and $\delta=4$, which are descendants of the integer winding number of $\delta=6$. Since the winding number of class C is always an even integer when $\delta=2$, we cannot define $\mathbb{Z}_2$ topological invariants for $\delta=1$ or $\delta=0$.

In the rest part of this section, let us explain the reason why   the above procedures of defining $\mathbb{Z}_2$ topological invariants in $\delta=4n+1$ and $\delta=4n$ dimensions cannot be applied to $\delta\leq 4n-1$. The underlying reason lies in the homotopy groups $\pi_p(U(N))$, which is $\mathbb{Z}$ when $p$ is odd, and $0$ when $p$ is even (we assume that $N\gg p$, namely the \emph{stable} regime of homotopy group).

In the $\delta=4n+1$ case, $U_\varepsilon(\bk,\br,t,\lambda)$ defines a homotopy class in $\pi_{d+D+2}(U(N))$, which can be nontrivial (since $d+D+2$ is an odd integer). As such, two given interpolations $U_\varepsilon(\bk,\br,t,\lambda)$ and $U'_\varepsilon(\bk,\br,t,\lambda)$ in general cannot be smoothly connected; nevertheless, our derivation above, as pictorially illustrated by Fig.\ref{deformation}, shows that in any case, different interpolations yield winding numbers with the same parity (even/odd), even though these interpolations can be in different topological classes. This fact enables the definition of $\mathbb{Z}_2$ topological invariant in the dimension $\delta=4n+1$.

When $\delta=4n$, the parity of the winding number is still unambiguous, and the $\mathbb{Z}_2$ topological invariant is well defined. To arrive at this conclusion, suppose that we have four Floquet systems ($a,b,c,d$), whose periodized time evolution operator is $U^a_\varepsilon(\bk,\br,t)$, $U^b_\varepsilon(\bk,\br,t)$, $U^c_\varepsilon(\bk,\br,t)$, and $U^d_\varepsilon(\bk,\br,t)$, respectively. Let us consider two interpolations, denoted as $U_\varepsilon(\bk,\br,t,\lambda,\mu)$ and $U'_\varepsilon(\bk,\br,t,\lambda,\mu)$, which satisfy  $U_\varepsilon(\bk,\br,t,0/\pi,0/\pi) =U'_\varepsilon(\bk,\br,t,0/\pi,0/\pi) =U^{a/b/c/d}_\varepsilon(\bk,\br,t)$.  At fixed $\mu=0$, it is always possible to smoothly deform one of the interpolations, $U'_\varepsilon(\bk,\br,t,\lambda,0)$, which is a function of $(\bk,\br,t,\lambda)$, to the other interpolation $U_\varepsilon(\bk,\br,t,\lambda,0)$, thanks to the trivialness of $\pi_{d+D+2}(U(N))$ (note that $d+D+2$ is an even integer when $\delta=4n$). Similarly, at fixed $\mu=\pi$, $U'_\varepsilon(\bk,\br,t,\lambda,\pi)$ can be smoothly deformed to $U_\varepsilon(\bk,\br,t,\lambda,\pi)$. Rephrased more precisely, we can smoothly deform $U'_\varepsilon(\bk,\br,t,\lambda,\mu)$ to another function $\bar{U}'_\varepsilon(\bk,\br,t,\lambda,\mu)$, so that it satisfies $\bar{U}'_\varepsilon(\bk,\br,t,\lambda,0) =U_\varepsilon(\bk,\br,t,\lambda,0)$, and $\bar{U}'_\varepsilon(\bk,\br,t,\lambda,\pi) =U_\varepsilon(\bk,\br,t,\lambda,\pi)$. Because a smooth deformation cannot change the value of a topological invariant, the winding number $W(\bar{U}'_\varepsilon)$ is equal to $W(U'_\varepsilon)$.
With $U_\varepsilon(\bk,\br,t,\lambda,0)$ and $U_\varepsilon(\bk,\br,t,\lambda,\pi)$ playing the roles of $U^a_\varepsilon$ and $U^b_\varepsilon$ in Fig.\ref{deformation}, we can take the same construction of Fig.\ref{deformation} (i.e., defining two new interpolations $U^{I}_\varepsilon$ and $U^{II}_\varepsilon$) to show that $W(\bar{U}'_\varepsilon)$ and $W(U_\varepsilon)$ have the same parity. Thus, the parity of the winding number does not depend on the specific choice of interpolation, and it can be defined as the relative $\mathbb{Z}_2$ topological invariant of the four time evolution operators, $U^a_\varepsilon(\bk,\br,t)$, $U^b_\varepsilon(\bk,\br,t)$, $U^c_\varepsilon(\bk,\br,t)$, and $U^d_\varepsilon(\bk,\br,t)$.

The same construction would not work if we move down to $\delta=4n-1$, because different interpolations can yield winding numbers with opposite parity, which is a consequence of the nontrivial homotopy groups of $U(N)$. To see this, suppose that we have two matrix-valued functions $U_\varepsilon(\bk,\br,t,\lambda,\mu,\nu)$ and $U'_\varepsilon(\bk,\br,t,\lambda,\mu,\nu)$, which interpolate the eight time evolution operators $U_\varepsilon(\bk,\br,t,0/\pi,0/\pi,0/\pi) =U'_\varepsilon(\bk,\br,t,0/\pi,0/\pi,0/\pi)$. We would like to deform $U'_\varepsilon(\bk,\br,t,\lambda,\mu,\nu)$ to a new function $\bar{U}'_\varepsilon(\bk,\br,t,\lambda,\mu,\nu)$, such that we have $\bar{U}'_\varepsilon(\bk,\br,t,\lambda,\mu,0/\pi) =U_\varepsilon(\bk,\br,t,\lambda,\mu,0/\pi)$, which, if possible, would enable the application of the construction in Fig.\ref{deformation}. Simply put, we would like to deform $U'_\varepsilon(\bk,\br,t,\lambda,\mu,0/\pi)$ to $U_\varepsilon(\bk,\br,t,\lambda,\mu,0/\pi)$. As a prerequisite, we have to deform $U'_\varepsilon(\bk,\br,t,\lambda,0/\pi,0/\pi)$ to $U_\varepsilon(\bk,\br,t,\lambda,0/\pi,0/\pi)$. However, there are topological obstructions in doing this, namely the nontrivial homotopy group $\pi_{d+D+2}(U(N))=\mathbb{Z}$ (note that there are $d+D+2$ parameters, $\bk,\br,t,\lambda$, and that $d-D=4n-1$ is an odd integer). Therefore, it is not always possible to deform $U'_\varepsilon(\bk,\br,t,\lambda,0/\pi,0/\pi)$ to $U_\varepsilon(\bk,\br,t,\lambda,0/\pi,0/\pi)$. Consequently, in general we cannot deform $U'_\varepsilon(\bk,\br,t,\lambda,\mu,0/\pi)$ to $U_\varepsilon(\bk,\br,t,\lambda,\mu,0/\pi)$. As such, the construction of Fig.\ref{deformation} cannot be applied to $\delta=4n-1$. Similar phenomena also occur in all other cases of $\mathbb{Z}_2$ topological invariants studied in this paper: One-parameter and two-parameter interpolations are able to produce $\mathbb{Z}_2$ topological invariants, while three-parameter interpolations cannot.

Finally, let us mention that the special case of $(d,D)=(2,1)$ ($\delta=1$) in class D is particularly interesting. A point defect such as a vortex in a two-dimensional superconductor falls into this class. The topological Majorana modes have potential applications in topological quantum computations. Based on concrete models, we will study this case further in Sec. \ref{D-point}.

\subsubsection{$\mathbb{Z}_2$ topological invariants of Wess-Zumino-Witten form for class AI and class AII}\label{sec:WZW-AIAII}

For class AI and AII with time reversal symmetry,
we have integer topological invariants (winding numbers) when $\delta\equiv d-D=4n$.
When $\delta=4n-1$, the integer topological invariant cannot be defined, nevertheless, we can still define a $\mathbb{Z}_2$ topological invariant. The method is parallel to the previous section of WZW term for class D and class C. Suppose that we have two Floquet systems with TRS, whose time-evolution operator is denoted as $U^a (\bk,\br,t)$ and $U^b (\bk,\br,t)$, respectively. Given a nonzero quasienergy gap at $\varepsilon$, we can define the periodized time evolution operators $U^a_{\varepsilon}(\bk,\br,t)$ and $U^b_{\varepsilon}(\bk,\br,t)$. An interpolation $U(\bk,\br,t,\lambda)$ ($\lambda\in[-\pi,\pi]$) between $U^a (\bk,\br,t)$ and $U^b (\bk,\br,t)$ can be constructed such that $U (\bk,\br,t,0)=U^a (\bk,\br,t)$, and $U (\bk,\br,t,\pi)=U (\bk,\br,t,-\pi) = U^b (\bk,\br,t)$. This interpolation is required to have the TRS: $U^* (\bk,\br,t,\lambda) =T^{-1} U (-\bk,\br,-t,-\lambda)T$, which can be achieved by first finding an interpolation for $\lambda\in[0,\pi]$, and take the mirror interpolation $U (\bk,\br,t,\lambda) =[T^{-1} U (-\bk,\br,-t,-\lambda)T]^*$ for $\lambda\in[-\pi,0]$.

Apparently, $U_{\varepsilon}(\bk,\br,t,\lambda)$ is an interpolation between $U^a_{\varepsilon}(\bk,\br,t)$ and $U^b_{\varepsilon}(\bk,\br,t)$.
In terms of the periodized time evolution operator, we have the symmetry
\begin{eqnarray}
U^*_{\varepsilon}(\bk,\br,t,\lambda) =T^{-1} U_{\varepsilon}(-\bk,\br,-t,-\lambda)T, \label{TRS-interpolation}
\end{eqnarray} which is consistent with  Eq.(\ref{time}).
Now we can define a winding number on the $(d+D+2)$-dimensional $(\bk,\br,t,\lambda)$ parameter space:
\begin{equation}
\label{}
\begin{aligned}
&W(U_\varepsilon (\bk,\br,t,\lambda))=K_{d+D+2}\int_{ T^{d+1}\times S^D\times S^1} d^dk d^Dr dtd\lambda\\
&\times\text{Tr}[\epsilon^{\alpha_1\alpha_2\cdots\alpha_{d+D+2}}(U_{\varepsilon}^{-1} \partial_{\alpha_{1}}U_{\varepsilon})\cdots(U_{\varepsilon}^{-1}\partial_{\alpha_{d+D+2}}U_{\varepsilon})],
\end{aligned}
\end{equation}
where the coefficient
$K_{d+D+2} =\frac{(-1)^{\frac{d+D+1}{2}} (\frac{d+D+1}{2})!}{(d+D+2)!}\left(\frac{i}{2\pi}\right)^{\frac{d+D+1}{2}+1}$.
The value of winding number may depend on the interpolation we choose. To define a meaningful $\mathbb{Z}_2$ topological invariant, we have to show that any two different interpolations, denoted as $U_{\varepsilon}(\bk,\br,t,\lambda)$ and $U'_{\varepsilon}(\bk,\br,t,\lambda)$, yield winding numbers with the same parity (even/odd); in other words,
\begin{eqnarray} \label{}
W(U_\varepsilon)-W(U'_\varepsilon)=0   \  \text{mod}\ 2.
\end{eqnarray}
To prove this, let us take the same Fig.\ref{deformation} of the previous section as a guide, and define two new interpolations,
\begin{equation} \label{}
U_{\varepsilon}^{I}(\bk,\br,t,\lambda) =\left\{
\begin{aligned}
&U_{\varepsilon}(\bk,\br,t,\lambda), \, -\pi<\lambda<0 \\
&U'_{\varepsilon}(\bk,\br,t,-\lambda), \,   0<\lambda<\pi,
\end{aligned}\right.
\end{equation}
and
\begin{equation} \label{}
U_{\varepsilon}^{II}(\bk,\br,t,\lambda) =\left\{
\begin{aligned}
&U'_{\varepsilon}(\bk,\br,t,-\lambda), \,  -\pi<\lambda<0\\
&U_{\varepsilon}(\bk,\br,t,\lambda), \,   0<\lambda<\pi.
\end{aligned}\right.
\end{equation}  From the definition, it is apparent that \begin{equation} \label{}
W(U_\varepsilon)-W(U'_\varepsilon)=W(U_{\varepsilon}^{I})+W(U_{\varepsilon}^{II}),
\end{equation} which can be readily seen from Fig.\ref{deformation}.

According to Eq.(\ref{TRS-interpolation}), the two interpolations $U_{\varepsilon}^{I}(\bk,\br,t,\lambda)$ and $U_{\varepsilon}^{II}(\bk,\br,t,\lambda)$ satisfy the symmetry relation
\begin{equation}
\label{wa}
T^{-1} U^I_\varepsilon(\bk,\br,t,\lambda)T =U^{II*}_\varepsilon(-\bk,\br,-t,-\lambda),
\end{equation}
from which it follows that (see Appendix.\ref{wzwaa})
\begin{equation}
\label{wzwa}
\begin{aligned}
w(U^I_\varepsilon)(\bk,\br,t,\lambda)=w(U^{II}_\varepsilon)(-\bk,\br,-t,-\lambda)(-1)^{3-(\delta-1)/2}.
\end{aligned}
\end{equation}
When $\delta=4n-1$, we have
\begin{equation}
\label{}
\begin{aligned}
W(U^I_\varepsilon)&=\int_{ T^{d+1}\times S^D\times S^1} w(U^{I}_\varepsilon)(\bk,\br,t,\lambda)\\
&=\int_{T^{d+1}\times S^D\times S^1} w(U^{II}_\varepsilon)(-\bk,\br,-t,-\lambda)(-1)^{3-(\delta-1)/2}\\
&=W(U^{II}_\varepsilon),
\end{aligned}
\end{equation}
thus
\begin{equation}
W(U_\varepsilon)-W(U'_\varepsilon)=W(U_{\varepsilon}^{I})+W(U_{\varepsilon}^{II})=2W(U_{\varepsilon}^{I}), \label{TRS-parity}
\end{equation} which is always an even integer.

When $W(U_\varepsilon (\bk,\br,t,\lambda))$ is an even integer, $U^a_\varepsilon (\bk,\br,t)$ and $U^b_\varepsilon (\bk,\br,t)$ are regarded as belonging to the same $\mathbb{Z}_2$ topological class. This definition is unambiguous due to Eq.(\ref{TRS-parity}). If $U^b_\varepsilon (\bk,\br,t)$ is taken to be a trivial evolution operator, then
$W(U_\varepsilon (\bk,\br,t,\lambda))$ defines a $\mathbb{Z}_2$ topological invariant for $U^a_\varepsilon(\bk,\br,t)\equiv U_\varepsilon(\bk,\br,t)$:
\begin{equation} \label{}
\nu( U_\varepsilon(\bk,\br,t) )=(-1)^{W(U_\varepsilon (\bk,\br,t,\lambda))}=\pm 1.
\end{equation}

The static limit of this WZW term is similar to that studied in the previous section. If we consider a static flat-band Hamiltonian, the winding number in the $(\bk,\br,t,\lambda)$ parameter space can be reduced to
\begin{equation}
\label{chern+1}
\begin{aligned}
&W(U_{\varepsilon=0})=\tilde{K}_{d+D+1}\int_{ T^{d+1}\times S^D } d^dk d^Drd\lambda\\
&\times\text{Tr}[\epsilon^{\alpha_1\alpha_2\cdots\alpha_{d+D+1}}P \partial_{\alpha_1}P\cdots\partial_{\alpha_{d+D+1}}P],
\end{aligned}
\end{equation}
where $\tilde{K}_{d+D+1}$ is given by Eq.(\ref{static2}), and $P$ is the valence band projection operator.
This is exactly the Chern number defined on the $(\bk,\br,t,\lambda)$ parameter space, which is the topological invariant of static topological defects\cite{teo2010}. Analogous to the previous section, this Chern number can be written as a Chern-Simons form after integration over $\lambda$:
\begin{equation}
\label{}
\begin{aligned}
&W(U_{\varepsilon=0}) =2\int_{T^d\times S^D}  \  \mathcal{Q}_{d+D}.
\end{aligned}
\end{equation}

The case $\delta\equiv d-D=4n-2$ of class AI and class AII is parallel to that of $\delta=4n$ case of class D and class C, which has been studied in the previous section. Given a time evolution operator $U_{\varepsilon}(\bk,\br,t)$, we can construct an extension
$U_{\varepsilon}(\bk,\br,t,\lambda,\mu)$, which satisfies $U_{\varepsilon}(\bk,\br,t,0,0)=U_{\varepsilon}(\bk,\br,t)$. In addition, $U_{\varepsilon}(\bk,\br,t,\pm\pi,\mu)$ and $U_{\varepsilon}(\bk,\br,t,\lambda,\pm\pi)$ are required to be trivial evolution operators. The WZW extension has to be consistent with the TRS, namely,
\begin{eqnarray}
U^*_{\varepsilon}(\bk,\br,t,\lambda,\mu) =T^{-1} U_{\varepsilon}(-\bk,\br,-t,-\lambda,-\mu)T.
\end{eqnarray}
With the inclusion of two new parameters $\lambda$ and $\mu$, we can define a winding number in the $(d+D+3)$-dimensional parameter space:
\begin{equation}
\label{}
\begin{aligned}
&W(U_\varepsilon (\bk,\br,t,\lambda,\mu))=K_{d+D+3}\int_{ T^{d+2}\times S^D\times S^1 } d^dk d^Dr dtd\lambda d\mu\\
&\times\text{Tr}[\epsilon^{\alpha_1\alpha_2\cdots \alpha_{d+D+3}}(U_{\varepsilon}^{-1}\partial_{\alpha_{1}}U_{\varepsilon}) \cdots(U_{\varepsilon}^{-1}\partial_{\alpha_{d+D+3}}U_{\varepsilon})],
\end{aligned}
\end{equation}
where the coefficient
$K_{d+D+3}= \frac{(-1)^{\frac{d+D+2}{2}}(\frac{d+D+2}{2})!}{(d+D+3)!} \left(\frac{i}{2\pi}\right)^{\frac{d+D+2}{2}+1}$.

Similar to that discussed in the previous section, this winding number is well defined mod $2$, thus it is a $\mathbb{Z}_2$ topological invariant for $U_\varepsilon (\bk,\br,t)$. Alternatively, we can write the $\mathbb{Z}_2$ topological invariant as
\begin{equation} \label{}
\nu( U_\varepsilon(\bk,\br,t) )=(-1)^{W(U_\varepsilon (\bk,\br,t,\lambda,\mu))}=\pm 1.
\end{equation}
Similar to Eq.(\ref{difference}), we can show that $W(U_{\varepsilon'}  (\bk,\br,t,\lambda,\mu))- W(U_\varepsilon (\bk,\br,t,\lambda,\mu))=C_{(d+D)/2}(P_{\varepsilon,\varepsilon'}(\bk,\br,\lambda,\mu))$ (mod $2$).

In the above construction, we have taken $U_{\varepsilon}(\bk,\br,t,\pm\pi,\mu)$ and $U_{\varepsilon}(\bk,\br,t,\lambda,\pm\pi)$ to be trivial time evolution operators, so that the $\mathbb{Z}_2$ topological invariant yields the topological class of $U_\varepsilon(\bk,\br,t,0,0)\equiv U_\varepsilon(\bk,\br,t)$. If this requirement is removed, then $U_\varepsilon (\bk,\br,t,\lambda,\mu)$ is an interpolation among the time evolution operators of the four (trivial or nontrivial) Floquet systems ($a,b,c,d$), whose periodized time evolution operators are $U^a_\varepsilon(\bk,\br,t) \equiv U_\varepsilon(\bk,\br,t,0,0)$, $U^b_\varepsilon(\bk,\br,t) \equiv U_\varepsilon(\bk,\br,t,0,\pi)$, $U^c_\varepsilon(\bk,\br,t) \equiv U_\varepsilon(\bk,\br,t,\pi,0)$, and $U^d_\varepsilon(\bk,\br,t) \equiv U_\varepsilon(\bk,\br,t,\pi,\pi)$, respectively. Each of them satisfies the TRS. The winding number $W(U_\varepsilon (\bk,\br,t,\lambda,\mu))$ (mod $2$) is then defined as the relative $\mathbb{Z}_2$ topological invariant of the four Floquet systems ($a,b,c,d$).

In the special case of $d=2, D=0$, a different topological invariant\cite{Carpentier2015,carpentier2015construction} for the class AII has been proposed by Carpentier, \emph{et al} from a quite different approach. According to the dimensional reduction scheme\cite{qi2008,wang2010a}, the second Chern number of the Floquet bands in $[\varepsilon,\varepsilon']$, namely $C_{2}(P_{\varepsilon,\varepsilon'}(k_1,k_2,\lambda,\mu))$ (mod $2$), is equal to the Kane-Mele $\mathbb{Z}_2$ topological invariant\cite{kane2005b} of these Floquet bands. Therefore, the difference $W(U_{\varepsilon'}  (\bk,\br,t,\lambda,\mu))- W(U_\varepsilon (\bk,\br,t,\lambda,\mu))$ (mod $2$) is just the Kane-Mele $\mathbb{Z}_2$ topological invariant of the Floquet bands in $[\varepsilon,\varepsilon']$. The same key property is shared by Carpentier \emph{et al}'s topological invariant\cite{Carpentier2015,carpentier2015construction}, therefore, we infer that this topological invariant is equal to ours (This is a statement only for $d=2$ and $D=0$ because, although our unified formulation is directly applicable in higher spatial dimensions, it is unclear how to generalize the topological invariant of Ref.\cite{Carpentier2015,carpentier2015construction} to higher dimensions from their approach).

For the case $(d,D)=(3,1)$, namely, a line defect in a three-dimensional system, we will study a concrete lattice model in class AII (see Sec. \ref{AII-line}), which hosts Floquet helical modes.

\subsection{Topological invariants of the chiral classes: BDI, DIII, CII, CI}\label{sec:chiral}

In class BDI, class DIII, class CII, and class CI, both TRS and PHS are present, and their product $\Xi\Theta$ is a CS. In fact, the PHS in Eq.(\ref{particleu}) and the TRS in Eq.(\ref{timeu}) leads to $U(\bk,\br,t)=S^{-1}U(\bk,\br,-t)S$ with $S=TC^{-1}$, which is a CS.
It will be convenient to take the chiral basis, in which the CS matrix $S$ is diagonal [see Eq.(\ref{CSmatrix})].  According to Eq.(\ref{a}) and Eq.(\ref{b}) in Sec.\ref{aiii}, the evolution operators at $\varepsilon=0$ and $\varepsilon=\pi$ take the following forms
\begin{equation}
U_{\varepsilon=0}(\bk,\br,\frac{\tau}{2})=
\begin{pmatrix}
&U_{\varepsilon=0}^+(\bk,\br)\\
U_{\varepsilon=0}^-(\bk,\br)&\\
\end{pmatrix},
\end{equation}
and
\begin{equation}
U_{\varepsilon=\pi}(\bk,\br,\frac{\tau}{2})=
\begin{pmatrix}
U_{\varepsilon=\pi}^+(\bk,\br)&\\
&U_{\varepsilon=\pi}^-(\bk,\br)\\
\end{pmatrix}.
\end{equation}
And the inverse matrices are
\begin{equation}
U^{-1}_{\varepsilon=0}(\bk,\br,\frac{\tau}{2})=
\begin{pmatrix}
&(U_{\varepsilon=0}^-)^{-1}(\bk,\br)\\
(U_{\varepsilon=0}^+)^{-1}(\bk,\br)&\\
\end{pmatrix},
\end{equation}
and
\begin{equation} \label{}
U^{-1}_{\varepsilon=\pi}(\bk,\br,\frac{\tau}{2})=
\begin{pmatrix}
(U_{\varepsilon=\pi}^+)^{-1}(\bk,\br)&\\
&(U_{\varepsilon=\pi}^-)^{-1}(\bk,\br)\\
\end{pmatrix}.
\end{equation}
We can define an integer winding number in the same way as we did in Sec.\ref{aiii} for class AIII:
\begin{equation}
\label{windingdensity}
\begin{aligned}
&W(U_{\varepsilon}^+(\bk,\br))= K_{d+D}\int_{T^d\times S^D} d^dk d^Dr\\
&\times\text{Tr}\{\epsilon^{\alpha_1\alpha_2\cdots\alpha_{d+D}} [(U_{\varepsilon}^+)^{-1}\partial_{\alpha_{1}}U_{\varepsilon}^+] \cdots[(U_{\varepsilon}^+)^{-1}\partial_{\alpha_{d+D}}U_{\varepsilon}^+]\},
\end{aligned}
\end{equation}
in which $\varepsilon=0$ or $\pi$.
The coefficient reads
$K_{d+D} =\frac{(-1)^{\frac{d+D-1}{2}} (\frac{d+D-1}{2})!}{(d+D)!} \left(\frac{i}{2\pi}\right)^{\frac{d+D+1}{2}}$.

The integrand $w(U_{\varepsilon}^+)(\bk,\br)= K_{d+D} \text{Tr} \{\epsilon^{\alpha_1\alpha_2\cdots\alpha_{d+D}}  [(U_{\varepsilon}^+)^{-1}\partial_{\alpha_{1}} U_{\varepsilon}^+] \cdots[(U_{\varepsilon}^+)^{-1} \partial_{\alpha_{d+D}}U_{\varepsilon}^+]\}$ is referred to as the winding number density.  We also mention that the winding number in Eq.(\ref{windingdensity}) can be written in a basis-independent form like Eq.(\ref{windingdensity-general}). The constraints imposed by the symmetries on Eq.(\ref{windingdensity}) will be discussed in the following subsections. We also mention in advance that the forms of $\mathbb{Z}_2$ topological invariants will be studied in Sec. \ref{wzwdiiici} and Sec. \ref{sec:WZWBDICII}.

Apparently, $\delta\equiv d-D$ has to be an odd integer, otherwise the winding number in Eq.(\ref{windingdensity}) is automatically zero. Compared to the winding numbers of nonchiral classes [see Eq.(\ref{nonchiral-winding})], Eq.(\ref{windingdensity}) does not contain an integration over $t$. In fact, the time $t$ has been fixed as $t=\tau/2$ in the definition of winding number of the chiral classes.

Eq.(\ref{windingdensity}) is the general formula of the integer (not including $\mathbb{Z}_2$) topological invariants for the chiral classes (i.e., classes with a chiral symmetry). Nevertheless, each class of BDI, DIII, CII, and CI has its own symmetry constraints; as a result, the possible values of winding numbers for each symmetry class and for each spatial dimension are different, leading to different topological classifications. Now we study these features.

\subsubsection{Symmetry constraints on the winding number of class CI}\label{sec:CI}

For the class CI, the time reversal operator ($\Theta=T\mathcal{K}$) and particle-hole operator ($\Xi=C\mathcal{K}$) satisfy $\Theta^2=1$ ($T^*T=1$) and $\Xi^2=-1$ ($C^*C=-1$). Let choose the matrices $T$ and $C$ as
\begin{eqnarray} \label{}
T=\tau_x
\end{eqnarray} and
\begin{eqnarray} \label{}
C=\tau_y.
\end{eqnarray}
The CS matrix $S$ is proportional to $TC^{-1}$, which can be taken as $\tau_z$.
For $\varepsilon=0$,   Eq.(\ref{particle0}) tells us that $C^{-1} U_{\varepsilon=0}(\bk,\br,\frac{\tau}{2})C =-U_{\varepsilon=0}^*(-\bk,\br,\frac{\tau}{2})$, which immediately implies that, under the basis choice $C=\tau_y$,
\begin{eqnarray} \label{ci1}
U_{\varepsilon=0}^+(\bk,\br )=U_{\varepsilon=0}^{-*}(-\bk,\br )
\end{eqnarray} and
\begin{eqnarray} \label{ci2}
(U_{\varepsilon=0}^+(\bk,\br ))^{-1}=(U_{\varepsilon=0}^{-*}(-\bk,\br ))^{-1}.
\end{eqnarray}
We can obtain the same relations if we start from Eq.(\ref{time}) to obtain $T^{-1} U_{\varepsilon=0}(\bk,\br,\frac{\tau}{2})T =U^*_{\varepsilon=0}(-\bk,\br,\frac{\tau}{2})$  [recall that $U_{\varepsilon}(\bk,\br,-\frac{\tau}{2})=U_{\varepsilon}(\bk,\br,\frac{\tau}{2})$].

The topological invariant is given by Eq.(\ref{windingdensity}), however, the symmetries impose constraints on the possible values it can take.
Using Eq.(\ref{ci1}) and Eq.(\ref{zero}), we can prove that the winding number satisfies
\begin{equation}
\label{windci}
\begin{aligned}
W(U^+_{\varepsilon=0} (\bk,\br ))=W(U^+_{\varepsilon=0} (\bk,\br ))(-1)^{1-(\delta-1)/2}.
\end{aligned}
\end{equation}
The details of calculation are given in Appendix \ref{chiralci}. We note that this is an identity about the winding number, instead of the winding number density. The symmetry constraint of Eq.(\ref{windci}) depends only on $\delta$, but not on $d$ or $D$ separately. Furthermore, this dependence on $\delta$ has a dimensional periodicity of $4$, because $(-1)^{\delta/2}$ returns to the same value under $\delta\rw\delta+4$. The same features have been noted in Section \ref{sec:DC} for the nonchiral classes. They are actually common features for all the eight real classes.

Since $\delta\equiv d-D$ has to be an odd integer, it can be $\delta=4n+1$ or $\delta=4n+3$ ($n$ is an integer). For the case $\delta=4n+1$, Eq.(\ref{windci}) implies that the winding number satisfies $W(U^+_{\varepsilon=0}(\bk,\br))=-W(U^+_{\varepsilon=0}(\bk,\br))$, thus the winding number vanishes.
Only when $\delta\equiv d-D=4n+3$, the winding number can be nonzero.

For $\varepsilon=\pi$, it follows from Eq.(\ref{particlepi}) that $C^{-1}U_{\varepsilon=\pi}(\bk,\br,\frac{\tau}{2})C =U^*_{\varepsilon=\pi}(-\bk,\br,\frac{\tau}{2})$, therefore we have
\begin{eqnarray} \label{}
U_{\varepsilon=\pi}^+(\bk,\br )=U_{\varepsilon=\pi}^{-*}(-\bk,\br )
\end{eqnarray} and
\begin{eqnarray} \label{}
(U_{\varepsilon=\pi}^+(\bk,\br ))^{-1}=(U_{\varepsilon=\pi}^{-*}(-\bk,\br ))^{-1}.
\end{eqnarray}
Starting from the definition of topological invariant in Eq.(\ref{windingdensity}), and following the calculations in Appendix \ref{chiralci}, we arrive at similar conclusion as the case of $\varepsilon=0$:
\begin{equation}
\label{}
\begin{aligned}
W(U^+_{\varepsilon=\pi}(\bk,\br ))=W(U^+_{\varepsilon=\pi}(\bk,\br)) (-1)^{1-(\delta-1)/2},
\end{aligned}
\end{equation}
therefore, we conclude that when $\delta=4n+1$, the winding number is automatically zero;  when $\delta=4n+3$, the winding number can be nonzero. This is reflected in the last line of table \ref{table} (Note that when $\delta=8n+3$, the winding number has to satisfy an even stronger constraint: it must be an even integer. This will be discussed shortly).

\subsubsection{Symmetry constraints on the winding number of class DIII}\label{sec:DIII}

For class DIII, the time reversal operation and the particle-hole operation satisfy $\Theta^2=-1$ ($T^*T=-1$) and $\Xi^2=1$ ($C^*C=1$). Let us choose the time-reversal-symmetry matrix and particle-hole-symmetry matrix as
\begin{eqnarray} \label{}
T=\tau_y
\end{eqnarray} and
\begin{eqnarray} \label{}
C=\tau_x.
\end{eqnarray}
As a result, the CS matrix $S=\tau_z$.
For $\varepsilon=0$,  Eq.(\ref{time}) and Eq.(\ref{tau/2}) imply that $T^{-1} U_{\varepsilon=0}(\bk,\br,\frac{\tau}{2})T =U^*_{\varepsilon=0}(-\bk,\br,\frac{\tau}{2})$, therefore, we have
\begin{eqnarray} \label{diii1}
U_{\varepsilon=0}^+(\bk,\br )=-U_{\varepsilon=0}^{-*}(-\bk,\br )
\end{eqnarray} and
\begin{eqnarray} \label{diii2}
(U_{\varepsilon=0}^+(\bk,\br ))^{-1}=-(U_{\varepsilon=0}^{-*}(-\bk,\br ))^{-1}.
\end{eqnarray}

According to the calculations given in Appendix \ref{chiraldiii}, we have
\begin{equation}
\label{winddiii}
\begin{aligned}
W(U^+_{\varepsilon=0} (\bk,\br))=W(U^+_{\varepsilon=0}(\bk,\br))(-1)^{1-(\delta-1)/2}.
\end{aligned}
\end{equation}
It follows that when $\delta\equiv d-D=4n+1$ ($n$ is an integer), the winding number is automatically zero. Only when $\delta\equiv d-D=4n+3$, the winding number can be nonzero.

For $\varepsilon=\pi$, we can still use Eq.(\ref{time}) and Eq.(\ref{tau/2}) to get $T^{-1} U_{\varepsilon=\pi}(\bk,\br,\frac{\tau}{2})T =U^*_{\varepsilon=\pi}(-\bk,\br,\frac{\tau}{2})$, therefore we have
\begin{eqnarray} \label{DIII-pi}
U_{\varepsilon=\pi}^+(\bk,\br)=U_{\varepsilon=\pi}^{-*}(-\bk,\br)
\end{eqnarray} and
\begin{eqnarray}
(U_{\varepsilon=\pi}^+(\bk,\br))^{-1}=(U_{\varepsilon=\pi}^{-*}(-\bk,\br))^{-1}.
\end{eqnarray} Notice that there is an additional minus sign in Eq.(\ref{diii1}) compared to Eq.(\ref{DIII-pi}), which is due to the fact that $U_{\varepsilon=0}(\bk,\br,\tau/2)$ is block off-diagonal, while $U_{\varepsilon=\pi}(\bk,\br,\tau/2)$ is block diagonal.

Starting from the formula of winding number in Eq.(\ref{windingdensity}), we can show that
\begin{equation}
\begin{aligned}
W(U^+_{\varepsilon=\pi}(\bk,\br))=W(U^+_{\varepsilon=\pi}(\bk,\br)) (-1)^{1-(\delta-1)/2}.
\end{aligned}
\end{equation} The details of calculations are given in Appendix \ref{chiraldiii}.
It follows that when $\delta=4n+1$, the winding number is automatically zero, indicating the absence of integer classification in these dimensions. When $\delta=d-D=4n+3$, the winding number can be nonzero, indicating integer classification.

\subsubsection{Symmetry constraints on the winding number of class BDI}\label{sec:BDI}

For class BDI, the time-reversal-symmetry operation and the particle-hole-symmetry operation satisfy $\Theta^2=1$ ($T^*T=1$) and $\Xi^2=1$($C^*C=1$), thus we can choose the time-reversal-symmetry matrix and particle-hole-symmetry matrix as
\begin{eqnarray} \label{}
T=\tau_0,
\end{eqnarray} which is simply the identity matrix, and
\begin{eqnarray} \label{}
C=\tau_z.
\end{eqnarray}
Thus, the CS matrix $S=\tau_z$ is diagonal (i.e., the chiral basis).
For $\varepsilon=0$, Eq.(\ref{time}) and Eq.(\ref{tau/2}) imply that $T^{-1} U_{\varepsilon=0}(\bk,\br,\frac{\tau}{2})T =U^*_{\varepsilon=0}(-\bk,\br,\frac{\tau}{2})$, from which it follows that
\begin{eqnarray} \label{bdi1}
U_{\varepsilon=0}^+(\bk,\br )=U_{\varepsilon=0}^{+*}(-\bk,\br )
\end{eqnarray} and
\begin{eqnarray} \label{bdi2}
(U_{\varepsilon=0}^+(\bk,\br ))^{-1}=(U_{\varepsilon=0}^{+*}(-\bk,\br))^{-1}.
\end{eqnarray}
Therefore, the winding number density satisfies
\begin{equation}
\label{windbdi}
\begin{aligned}
w(U^+_{\varepsilon=0})(\bk,\br )=w(U^+_{\varepsilon=0})(-\bk,\br ) (-1)^{-(\delta-1)/2}.
\end{aligned}
\end{equation}
The calculation is given in Appendix \ref{chiralbdi}. After integration over $\bk$ and $\br$, the winding number satisfies the same relation, $W(U^+_{\varepsilon=0}(\bk,\br ))=W(U^+_{\varepsilon=0}(\bk,\br ) ) (-1)^{-(\delta-1)/2}$.  We can see that, when $\delta\equiv d-D=4n+3$, the winding number is automatically zero; when $\delta\equiv d-D=4n+1$, the winding number can be nonzero.

For $\varepsilon=\pi$,  Eq.(\ref{time}) and Eq.(\ref{tau/2}) again tell us that $T^{-1} U_{\varepsilon=\pi}(\bk,\br,\frac{\tau}{2})T =U^*_{\varepsilon=\pi}(-\bk,\br,\frac{\tau}{2})$, therefore, we have
\begin{eqnarray} \label{}
U_{\varepsilon=\pi}^+(\bk,\br )=U_{\varepsilon=\pi}^{+*}(-\bk,\br ),
\end{eqnarray} and
\begin{eqnarray} \label{}
(U_{\varepsilon=\pi}^+(\bk,\br ))^{-1}=(U_{\varepsilon=\pi}^{+*}(-\bk,\br ))^{-1}.
\end{eqnarray}
According to the calculations provided in Appendix \ref{chiralbdi}, we can see that, similar to the case of $\varepsilon=0$, the winding number density satisfies
\begin{equation}
\label{}
\begin{aligned}
w(U^+_{\varepsilon=\pi}(\bk,\br ))=w(U^+_{\varepsilon=\pi}(\bk,\br ))(-1)^{-(\delta-1)/2},
\end{aligned}
\end{equation}
which implies that when $\delta\equiv d-D=4n+3$, the winding number is zero; only when $\delta\equiv d-D=4n+1$, the winding number can be nonzero.

\subsubsection{Symmetry constraints on the winding number of class CII}\label{sec:CII}

For class CII, the time-reversal operation and the particle-hole operation satisfy $\Theta^2=-1$ ($T^*T=-1$) and $\Xi^2=-1$ ($C^*C=-1$), thus we can take the time-reversal-symmetry matrix and the particle-hole-symmetry matrix as
\begin{eqnarray}
T=\tau_0\otimes \sigma_y,
\end{eqnarray} and
\begin{eqnarray}
C=\tau_z\otimes \sigma_y.
\end{eqnarray}
The CS matrix is $S=\tau_z$.

For $\varepsilon=0$,  Eq.(\ref{time}) and Eq.(\ref{tau/2}) lead to $T^{-1} U_{\varepsilon=0}(\bk,\br,\frac{\tau}{2})T =U^*_{\varepsilon=0}(-\bk,\br,\frac{\tau}{2})$, from which it follows that
\begin{eqnarray} \label{cii1}
U_{\varepsilon=0}^+(\bk,\br )=\sigma_yU_{\varepsilon=0}^{+*}(-\bk,\br )\sigma_y,
\end{eqnarray} and
\begin{eqnarray} \label{cii2}
(U_{\varepsilon=0}^+(\bk,\br ))^{-1}=\sigma_y (U_{\varepsilon=0}^{+*}(-\bk,\br ))^{-1}\sigma_y.
\end{eqnarray}
Starting from these equations of symmetry, we can show that the winding number density satisfies
\begin{equation}
\label{windcii}
\begin{aligned}
w(U^+_{\varepsilon=0})(\bk,\br ) =w(U^+_{\varepsilon=0})(-\bk,\br ) (-1)^{-(\delta-1)/2},
\end{aligned}
\end{equation}
whose derivation is given in Appendix.\ref{chiralcii}. After integration over $\bk$ and $\br$, the winding number satisfies the same relation, $W(U^+_{\varepsilon=0}(\bk,\br ))=W(U^+_{\varepsilon=0}(\bk,\br )) (-1)^{-(\delta-1)/2}$.
It follows that when $\delta\equiv d-D=4n+1$, the winding number can be nonzero; when $\delta\equiv d-D=4n+3$, the winding number is automatically zero, indicating the absence of integer topological classifications in these dimensions.

For $\varepsilon=\pi$,  Eq.(\ref{time}) and Eq.(\ref{tau/2}) lead to $T^{-1} U_{\varepsilon=\pi}(\bk,\br,\frac{\tau}{2})T =U^*_{\varepsilon=\pi}(-\bk,\br,\frac{\tau}{2})$, which implies
\begin{eqnarray} \label{}
U_{\varepsilon=\pi}^+(\bk,\br )=\sigma_yU_{\varepsilon=\pi}^{+*}(-\bk,\br )\sigma_y,
\end{eqnarray} and
\begin{eqnarray} \label{}
(U_{\varepsilon=\pi}^+(\bk,\br ))^{-1}=\sigma_y (U_{\varepsilon=\pi}^{+*}(-\bk,\br ))^{-1}\sigma_y.
\end{eqnarray}
According to the calculations provided in Appendix \ref{chiralcii},  we can see that
\begin{equation}
\label{windcii-pi}
\begin{aligned}
w(U^+_{\varepsilon=\pi})(\bk,\br )= w(U^+_{\varepsilon=\pi})(-\bk,\br ) (-1)^{-(\delta-1)/2}.
\end{aligned}
\end{equation}
It follows that the winding number can be nonzero only when
$\delta\equiv d-D=4n+1$, indicating integer topological classifications in these spatial dimensions. A concrete lattice model of a point defect in a two-dimensional system, namely, $(d, D)=(2,1)$ and $\delta=1$, will be put forward in Sec. \ref{CII-point}, for which both the topological invariant and the topological modes are numerically found. The topological invariant accurately counts the number of topological modes.

\subsubsection{Even-integer ($2\mathbb{Z}$) topological invariants}

In Sec. \ref{sec:2Z}, we have shown that the winding numbers of nonchiral classes have to be even integers when $\delta-s= 4$ (mod $8$), or equivalently, $d-D-s= 4$ (mod $8$).
In this section, we will show that the same conclusion is also true for the chiral classes, CI, DIII, BDI, and CII, namely, when $d-D-s= 4$ (mod $8$), the winding numbers of chiral classes must take even integer values, though the definitions of winding numbers take  different forms compared to the nonchiral classes (A prominent difference is that the winding numbers of nonchiral classes contain an integration over $t$, while the chiral classes do not).

The argument leading to this result for the nonchiral classes, which have been outlined in Section \ref{sec:2Z}, can be transferred to the present section for the chiral classes, if a counterpart of Eq.(\ref{difference}) can be obtained.

Let us fix $\varepsilon=0$ and $\varepsilon'=\pi$ for this section. Let $P_{\varepsilon=0,\varepsilon'=\pi}$ be the projection operator for the Floquet bands in $[0,\pi]$  (Accordingly, $1-P_{\varepsilon=0,\varepsilon'=\pi}$ is the projection operator of the Floquet bands in $[-\pi,0]$). It is readily found that
\begin{equation}
\begin{aligned}
U_{\varepsilon=0}^{-1}(\bk,\br,\frac{\tau}{2})U_{\varepsilon'=\pi} (\bk,\br,\frac{\tau}{2})
&=\exp[i(H_{\varepsilon'=\pi}^{\text{eff}}-H_{\varepsilon=0}^{\text{eff}})\frac{\tau}{2}]\\
&=\exp(i\pi P_{\varepsilon=0,\varepsilon'=\pi})\\
&=1-2P_{\varepsilon=0,\varepsilon'=\pi}\\
&\equiv Q_{\varepsilon=0,\varepsilon'=\pi}.
\end{aligned}
\end{equation} The projection operator satisfies $S^{-1} P_{\varepsilon=0,\varepsilon'=\pi}S =1-P_{\varepsilon=0,\varepsilon'=\pi}$, which can be readily verified by Eq.(\ref{chiralh}) or by taking $t=\tau$ in Eq.(\ref{chiralu}) (and remember that $U(\bk,\br,-\tau)=U^{-1}(\bk,\br,\tau)$). Equivalently, we have $S^{-1} Q_{\varepsilon=0,\varepsilon'=\pi}S = -Q_{\varepsilon=0,\varepsilon'=\pi}$. In the chiral basis, the chiral matrix is $S=\tau_z$, therefore, $Q_{0,\pi}$ must take the form of
\begin{equation}
Q_{0,\pi}=
\begin{pmatrix}
&q_{0,\pi}\\
q^{\dagger}_{0,\pi}&\\
\end{pmatrix}.  \label{Q}
\end{equation}
On the other hand,
under the chiral basis, Eq.(\ref{block}) and Eq.(\ref{block-pi}) tell us that $U_{\varepsilon=0}^{-1}U_{\varepsilon'=\pi}$ can be written as
\begin{equation} \label{}
U_{\varepsilon=0}^{-1}(\bk,\br,\frac{\tau}{2})U_{\varepsilon'=\pi}(\bk,\br,\frac{\tau}{2})=
\begin{pmatrix}
&(U_{\varepsilon=0}^-)^{-1}U_{\varepsilon'=\pi}^{-}\\
(U_{\varepsilon=0}^+)^{-1}U_{\varepsilon'=\pi} ^{+}&\\
\end{pmatrix},
\end{equation} which means that $q^{\dagger}_{\varepsilon=0,\varepsilon'=\pi} =(U_{\varepsilon=0}^+)^{-1}U_{\varepsilon'=\pi} ^{+}$.
Due to the additive property of the winding number, we have
\begin{equation}
W(U^+_{\varepsilon'=\pi})-W(U^+_{\varepsilon=0}) =W((U_{\varepsilon=0}^{+})^{-1}U^+_{\varepsilon'=\pi}) =W(q^{\dagger}_{\varepsilon=0,\varepsilon'=\pi}).  \label{difference-chiral}
\end{equation} Eq.(\ref{difference-chiral}) is the chiral-class counterpart of Eq.(\ref{difference}), which we have used to argue that the winding numbers take even integer values for the nonchiral classes when $d-D-s= 4$ (mod $8$).

From the knowledge of static Hamiltonian\cite{ryu2010,teo2010,Chiu2016rmp,schnyder2008}, we know that $W(q^{\dagger}_{\varepsilon=0,\varepsilon'=\pi})$ must be an even integer when $d-D-s=4$ (mod $8$).
The rest of the argument will be the same as Section \ref{sec:2Z}, which we do not need to repeat here. The conclusion is that, when $d-D-s= 4$ (mod $8$), the winding numbers must be even integers for the chiral classes.

A more rigorous proof of the fact that the topological invariants take even-integer values for $d-D-s=4$ (mod $8$) is given in Appendix \ref{sec:even}.

\subsubsection{$\mathbb{Z}_2$ topological invariants of Wess-Zumino-Witten form for the symmetry classes DIII and CI}\label{wzwdiiici}

For the symmetry classes DIII and CI, there are integer topological invariants for $\delta\equiv d-D=4n+3$ ($n$ is an integer). Following similar approach as the nonchiral classes, we can define $\mathbb{Z}_2$ topological invariants for $\delta=4n+2$ and $\delta=4n+1$. Just like the nonchiral classes (see the discussion in Section \ref{sec:WZWDC}), these definitions will be meaningless if the integer winding number taken as the starting point is always even-integer valued. To define $\mathbb{Z}_2$ topological invariants for $\delta=4n+2$ and $\delta=4n+1$, we must have $(4n+3)-s\neq 4$ (mod $8$) (see the previous section on even-integer topological invariants).
Therefore, for the class DIII with $s=3$, $n$ must be an even integer;  for the class CI with $s=7$, $n$ must be an odd integer.

Let us work on the $\delta=4n+2$ case first. We will follow the same scheme as in Section \ref{sec:WZWDC}, namely, we would like to establish a topological equivalence/nonequivalence relation between the time evolution operators of two systems, denoted as $U^a (\bk,\br,t)$ and $U^b (\bk,\br,t)$, based on the parity (even/odd) of winding number defined on a higher-dimensional parameter space. We first find an interpolation $U (\bk,\br,t,\lambda)$ ($\lambda\in[-\pi,\pi]$) between $U^a (\bk,\br,t)$ and $U^b (\bk,\br,t)$, such that $U (\bk,\br,t,0)=U^a (\bk,\br,t)$, $U (\bk,\br,t,\pi)=U (\bk,\br,t,-\pi)=U^b (\bk,\br,t)$. The interpolation is required to satisfy the PHS constraint $C^{-1}U(\bk,\br,t,\lambda)C =U^*(-\bk,\br,t,-\lambda)$, or expressed in terms of the periodized time evolution operator,
\begin{equation}
\begin{aligned}
C^{-1} U_{\varepsilon}(\bk,\br,t,\lambda) C =U^*_{-\varepsilon}(-\bk,\br,t,-\lambda)\exp(i\frac{2\pi t}{\tau}).
\end{aligned} \label{WZW-PHS}
\end{equation}
This equation is consistent with the particle-hole symmetry at $\lambda=0$ and $\pi$ [Eq.(\ref{particle})], as it should be. Meanwhile, the interpolation is also required to satisfy the TRS constraint $T^{-1}U(\bk,\br,t,\lambda)T=U^*(-\bk,\br,-t,-\lambda)$, or
\begin{equation}
\begin{aligned}
T^{-1}U_{\varepsilon}(\bk,\br,t,\lambda) T =U^*_{\varepsilon}(-\bk,\br,-t,-\lambda),
\end{aligned} \label{WZW-TRS}
\end{equation} which is consistent with Eq.(\ref{time}).
Given these two constraints, a CS constraint is automatically satisfied because the product of TRS and PHS necessarily gives rise to a CS. In fact, it follows from the PHS and TRS that $S^{-1}U(\bk,\br,t,\lambda)S=U(\bk,\br,-t,\lambda)$, with $S=TC^{-1}$.
An interpolation with the symmetries given in Eq.(\ref{WZW-PHS}) and Eq.(\ref{WZW-TRS}) can be obtained by first finding an interpolation for $\lambda\in[0,\pi]$, and then taking the interpolation for $\lambda\in[-\pi,0]$ as the mirror interpolation of $[0,\pi]$. In other words, for $\lambda\in[-\pi,0]$, we simply take the $T$-mirror $U (\bk,\br,t,\lambda)=[T^{-1}U(-\bk,\br,-t,-\lambda)T]^*$, or the $C$-mirror $U(\bk,\br,t,\lambda)=[C^{-1}U(-\bk,\br,t,-\lambda)C]^*$, as the definition of the interpolation. Thanks to the chiral symmetry with $S=TC^{-1}$, taking the $T$-mirror interpolation and taking the $C$-mirror one lead to the same result.

At the particular time $t=\tau/2$, we have
\begin{equation}
\label{}
\begin{aligned}
T^{-1}U_{\varepsilon}(\bk,\br,\frac{\tau}{2},\lambda) T =U^*_{\varepsilon}(-\bk,\br,-\frac{\tau}{2},-\lambda)=U^*_{\varepsilon}(-\bk,\br,\frac{\tau}{2},-\lambda).
\end{aligned}
\end{equation}
We also have
\begin{eqnarray}
U^*_{\varepsilon=0}(\bk,\br,\frac{\tau}{2},\lambda) =-C^{-1} U_{\varepsilon=0}(-\bk,\br,\frac{\tau}{2},-\lambda)C
\end{eqnarray}
and
\begin{eqnarray}
U^*_{\varepsilon=\pi}(\bk,\br,\frac{\tau}{2},\lambda) =C^{-1} U_{\varepsilon=\pi}(-\bk,\br,\frac{\tau}{2},-\lambda)C,
\end{eqnarray}
which are consistent with Eq.(\ref{particle0}) and Eq.(\ref{particlepi}).

It follows that, under the chiral basis used in Section \ref{sec:CI}, the interpolation used for the class CI satisfies
\begin{equation} \label{CI}
\left\{
\begin{aligned}
U_{\varepsilon=0}^+(\bk,\br,\frac{\tau}{2},\lambda)=U_{\varepsilon=0}^{-*}(-\bk,\br,\frac{\tau}{2},-\lambda),\\
U_{\varepsilon=\pi}^+(\bk,\br,\frac{\tau}{2},\lambda)=U_{\varepsilon=\pi}^{-*}(-\bk,\br,\frac{\tau}{2},-\lambda);
\end{aligned}\right.
\end{equation}
similarly, again in the chiral basis (used in Section \ref{sec:DIII}), the interpolation used for the class DIII satisfies
\begin{equation} \label{DIII}
\left\{
\begin{aligned}
U_{\varepsilon=0}^+(\bk,\br,\frac{\tau}{2},\lambda)=-U_{\varepsilon=0}^{-*}(-\bk,\br,\frac{\tau}{2},-\lambda),\\
U_{\varepsilon=\pi}^+(\bk,\br,\frac{\tau}{2},\lambda)=U_{\varepsilon=\pi}^{-*}(-\bk,\br,\frac{\tau}{2},-\lambda).
\end{aligned}\right.
\end{equation}

Given the interpolation $U_{\varepsilon}(\bk,\br,t,\lambda)$ defined on the $(\bk,\br,t,\lambda)$ parameter space, we can define a winding number
\begin{equation}
\label{}
\begin{aligned}
&W(U^+_\varepsilon(\bk,\br,\frac{\tau}{2},\lambda))=K_{d+D+1}\int_{  T^{d+1}\times S^D } d^dk d^Drd\lambda \\
&\times\text{Tr}[\epsilon^{\alpha_1\alpha_2\cdots\alpha_{d+D+1}}((U_{\varepsilon}^+)^{-1}\partial_{\alpha_{1}}U_{\varepsilon}^+)\cdots((U_{\varepsilon}^+)^{-1}\partial_{\alpha_{d+D+1}}U_{\varepsilon}^+)]
\end{aligned}
\end{equation}
for $\varepsilon=0$ or $\varepsilon=\pi$, where the coefficient $K_{d+D+1}$ reads
$K_{d+D+1}= \frac{(-1)^{\frac{d+D}{2}}(\frac{d+D}{2})!}{(d+D+1)!} \left(\frac{i}{2\pi}\right)^{\frac{d+D}{2}+1}$.

Just like in Section \ref{sec:WZWDC}, we have to show that any two different interpolations lead to the same winding number modulo $2$. Suppose that we have two interpolations $U_{\varepsilon}(\bk,\br,t,\lambda)$ and $U'_{\varepsilon}(\bk,\br,t,\lambda)$. We need to show that, for $\varepsilon=0$ or $\varepsilon=\pi$,  the difference between the two winding numbers satisfies
\begin{eqnarray} \label{}
W(U_\varepsilon^+(\bk,\br,\frac{\tau}{2},\lambda))-W(U'^{+}_\varepsilon(\bk,\br,\frac{\tau}{2},\lambda))=0   \,  (\text{mod}\,2).
\end{eqnarray}

Let us study class DIII first. We will follow the same approach as Section \ref{sec:WZWDC}.
Let us define two new interpolations (Fig.\ref{deformation} in Section \ref{sec:WZWDC} is still a useful pictorial illustration for the present problem):
\begin{equation} \label{}
U_{\varepsilon}^{I}(\bk,\br,t,\lambda)=\left\{
\begin{aligned}
&U_{\varepsilon}(\bk,\br,t,\lambda),\quad  -\pi<\lambda<0, \\
&U'_{\varepsilon}(\bk,\br,t,-\lambda),\quad   0<\lambda<\pi,
\end{aligned}\right.
\end{equation}
and
\begin{equation} \label{}
U_{\varepsilon}^{II}(\bk,\br,t,\lambda)=\left\{
\begin{aligned}
&U'_{\varepsilon}(\bk,\br,t,-\lambda),\quad  -\pi<\lambda<0,\\
&U_{\varepsilon}(\bk,\br,t,\lambda),\quad  0<\lambda<\pi.
\end{aligned}\right.
\end{equation}
As obvious consequences, we have
\begin{equation}
U_{\varepsilon}^{I\pm}(\bk,\br,\frac{\tau}{2},\lambda)=\left\{
\begin{aligned}
&U^{\pm}_{\varepsilon}(\bk,\br,\frac{\tau}{2},\lambda),\quad  -\pi<\lambda<0, \\
&U'^{\pm}_{\varepsilon}(\bk,\br,\frac{\tau}{2},-\lambda),\quad   0<\lambda<\pi,
\end{aligned}\right.
\end{equation}
and
\begin{equation} \label{}
U_{\varepsilon}^{II\pm}(\bk,\br,\frac{\tau}{2},\lambda)=\left\{
\begin{aligned}
&U'^{\pm}_{\varepsilon}(\bk,\br,\frac{\tau}{2},-\lambda),\quad  -\pi<\lambda<0,\\
&U^{\pm}_{\varepsilon}(\bk,\br,\frac{\tau}{2},\lambda),\quad  0<\lambda<\pi.
\end{aligned}\right.
\end{equation}
It can be readily seen that \bea U^{I+}_{\varepsilon=0}(\bk,\br,\frac{\tau}{2},\lambda) = -U^{II-*}_{\varepsilon=0}(-\bk,\br,\frac{\tau}{2},-\lambda), \label{DIII-WZW-0} \eea and
\bea U^{I+}_{\varepsilon=\pi}(\bk,\br,\frac{\tau}{2},\lambda) = U^{II-*}_{\varepsilon=\pi}(-\bk,\br,\frac{\tau}{2},-\lambda). \label{DIII-WZW-pi} \eea

Taking Fig.\ref{deformation} as a pictorial illustration, it is not difficult to see that
\begin{eqnarray} \label{}
W(U_\varepsilon^+)-W(U_\varepsilon'^{+})=W(U^{I+}_\varepsilon)+W(U^{II+}_\varepsilon).
\end{eqnarray}
Now the two terms at the right-hand side, $W(U^{I+}_\varepsilon)$ and $W(U^{II+}_\varepsilon)$, are not independent. In fact, for $\delta=4n+2$,  we have (this calculation resembles that of Appendix \ref{chiraldiii}):
\begin{equation}
\label{DIIIcalculation}
\begin{aligned}
&W(U^{I+}_\varepsilon(\bk,\br,\frac{\tau}{2},\lambda))=\int_{ T^{d+1} \times S^D }d^dk d^Drd\lambda\, w(U^{I+}_\varepsilon) (\bk,\br,\frac{\tau}{2},\lambda) \\
&=\int d^dk d^Drd\lambda\,  w^*(U^{II-}_\varepsilon) (-\bk,\br,\frac{\tau}{2},-\lambda) (-1)^{(d+D)/2+1}(-1)^{d+1}\\
&=W(U^{II-}_\varepsilon ( \bk,\br,\frac{\tau}{2}, \lambda))  (-1)^{2d+2-\delta/2}\\
&=W(U^{II+}_\varepsilon ( \bk,\br,\frac{\tau}{2}, \lambda))  (-1)^{2d+3-\delta/2}\\
&=W(U^{II+}_\varepsilon( \bk,\br,\frac{\tau}{2}, \lambda)),
\end{aligned}
\end{equation} where Eq.(\ref{DIII-WZW-0}) [or Eq.(\ref{DIII-WZW-pi})], Eq.(\ref{zero}), and the reality of winding number density ($w^*=w$), have been used. Needless to mention that $\varepsilon=0$ or $\pi$ in Eq.(\ref{DIIIcalculation}).

Now we can see that
\begin{eqnarray} \label{}
\label{}
\begin{aligned}
W(U_\varepsilon^+)-W(U_\varepsilon^{'+})=W(U^{I+}_\varepsilon)+W(U^{II+}_\varepsilon)=2W(U^{I+}_\varepsilon),
\end{aligned}
\end{eqnarray} which is always an even integer. This fact essentially establishes the $\mathbb{Z}_2$ classification, as has been discussed in \ref{sec:WZWDC}. When $W(U_\varepsilon^+(\bk,\br,\frac{\tau}{2},\lambda))=0$ or $1$ (mod $2$), $U^a_\varepsilon(\bk,\br,t)$ and $U^b_\varepsilon(\bk,\br,t)$ are regarded as in the same or different $\mathbb{Z}_2$ topological class.
In particular, when $U^b_\varepsilon(\bk,\br,t)$ is taken to be a fixed topologically trivial evolution operator and $U^a_\varepsilon(\bk,\br,t)=U_\varepsilon(\bk,\br,t)$, $W(U_\varepsilon^+(\bk,\br,\frac{\tau}{2},\lambda))$ defines a $\mathbb{Z}_2$ topological invariant for $U_\varepsilon(\bk,\br,t)$. It can also be written as  \begin{equation} \label{}
\nu(U^+_\varepsilon (\bk,\br,t))=(-1)^{W(U^+_\varepsilon (\bk,\br,\tau/2,\lambda))}=\pm 1.
\end{equation}
The case of class CI is almost the same as class DIII, which we will not repeat here.

When $\delta=4n+1$, we have similar construction of $\mathbb{Z}_2$ topological invariant. Again, let us study the class DIII for concreteness (the class CI is similar). For $\delta=4n+1$, we need two WZW extension parameters $\lambda$ and $\mu$, both of which take values in $[-\pi,\pi]$. We define an extension of $U_{\varepsilon}(\bk,\br,t)$ to
$U_{\varepsilon}(\bk,\br,t,\lambda,\mu)$, which satisfies $U_{\varepsilon}(\bk,\br,t,0,0)=U_{\varepsilon}(\bk,\br,t)$. In addition, $U_{\varepsilon}(\bk,\br,t,\pm\pi,\mu)$ and $U_{\varepsilon}(\bk,\br,t,\lambda,\pm\pi)$ are trivial time evolution operators. As an extension of the PHS relation Eq.(\ref{particle}) and the TRS relation Eq.(\ref{time}), we require that
\begin{equation}
\label{}
\begin{aligned}
C^{-1} U_{\varepsilon}(\bk,\br,t,\lambda,\mu) C =U^*_{-\varepsilon}(-\bk,\br,t,-\lambda,-\mu)\exp(i\frac{2\pi t}{\tau}),
\end{aligned}
\end{equation}
and that
\begin{equation}
\label{}
\begin{aligned}
T^{-1}U_{\varepsilon}(\bk,\br,t,\lambda,\mu) T =U^*_{\varepsilon}(-\bk,\br,-t,-\lambda,-\mu).
\end{aligned}
\end{equation}
Apply these symmetry constraints to the half-period $t=\tau/2$, we have
\begin{equation}
\label{}
\begin{aligned}
T^{-1}U_{\varepsilon}(\bk,\br,\frac{\tau}{2},\lambda,\mu) T &=U^*_{\varepsilon}(-\bk,\br,-\frac{\tau}{2},-\lambda,-\mu)\\
 &=U^*_{\varepsilon}(-\bk,\br,\frac{\tau}{2},-\lambda,-\mu),
\end{aligned}
\end{equation} which is valid for both $\varepsilon=0$ and $\varepsilon=\pi$. About the PHS, we have
\begin{eqnarray}
U^*_{\varepsilon=0}(\bk,\br,\frac{\tau}{2},\lambda,\mu) =-C^{-1} U_{\varepsilon=0}(-\bk,\br,\frac{\tau}{2},-\lambda,-\mu)C
\end{eqnarray}
for $\varepsilon=0$, and
\begin{eqnarray}
U^*_{\varepsilon=\pi}(\bk,\br,\frac{\tau}{2},\lambda,\mu) =C^{-1} U_{\varepsilon=\pi}(-\bk,\br,\frac{\tau}{2},-\lambda,-\mu)C
\end{eqnarray}
for $\varepsilon=\pi$, which is consistent with Eq.(\ref{particle0}) and Eq.(\ref{particlepi}), respectively.

Now we can define a winding number on the $(d+D+2)$-dimensional $(\bk,\br,\lambda,\mu)$ parameter space:
\begin{equation}
\label{}
\begin{aligned}
&W(U^{+}_\varepsilon (\bk,\br,\frac{\tau}{2},\lambda,\mu))=K_{d+D+2}\int_{ T^{d+2}\times S^D } d^dk d^Drd\lambda d\mu\\
&\times\text{Tr}[\epsilon^{\alpha_1\alpha_2\cdots\alpha_{d+D+2}} ((U_{\varepsilon}^+)^{-1}\partial_{\alpha_{1}}U_{\varepsilon}^+) \cdots((U_{\varepsilon}^+)^{-1}\partial_{\alpha_{d+D+2}}U_{\varepsilon}^+)],
\end{aligned}
\end{equation}
where $\varepsilon=0$ or $\pi$, and the coefficient is
$
K_{d+D+2}=\frac{(-1)^{\frac{d+D+1}{2}}(\frac{d+D+1}{2})!}{(d+D+2)!} \left(\frac{i}{2\pi}\right)^{\frac{d+D+1}{2}+1}$.

By similar derivation as in the case of $\delta=4n+2$, the parity (even/odd) of the winding number depends only on $U_{\varepsilon}(\bk,\br,t)$, therefore, it serves as a $\mathbb{Z}_2$ topological invariant. It can also be written as
\begin{equation} \label{}
\nu(U^+_\varepsilon (\bk,\br,t))=(-1)^{W(U^+_\varepsilon (\bk,\br,\tau/2,\lambda,\mu))}=\pm 1.
\end{equation}

\subsubsection{$\mathbb{Z}_2$ topological invariants of Wess-Zumino-Witten form for the symmetry classes BDI and CII}\label{sec:WZWBDICII}

Now we study the $\mathbb{Z}_2$ topological invariants of class BDI and class CII. When $\delta\equiv d-D=4n+1$, one can define an integer topological invariant, therefore, $\mathbb{Z}_2$ topological invariant can be defined for $\delta= 4n$ and $\delta=4n-1$, provided that the integer topological invariant is $\mathbb{Z}$ (not $2\mathbb{Z}$). For the class BDI, $\mathbb{Z}$ topological invariants occur at $\delta=4n+1$ when $n$ is an even integer; for the class CII, they occur when $n$ is an odd integer. These choices of $n$ will be assumed in the following study.

Let us work on the $\delta=4n$ case first.
Parallel to the argument in Section \ref{wzwdiiici}, we would like to establish a topological equivalence/nonequivalence relation between the time evolution operators of two Floquet systems, denoted as $U^a (\bk,\br,t)$ and $U^b (\bk,\br,t)$, based on the parity (even/odd) of winding number, which is defined not on the $(\bk,\br)$ space, but on a higher-dimensional parameter space, in accordance with the picture of Wess-Zumino-Witten term. We first find an interpolation $U_{\varepsilon}(\bk,\br,t,\lambda)$ ($\lambda\in[-\pi,\pi]$) between $U^a_\varepsilon(\bk,\br,t)$ and $U^b_\varepsilon(\bk,\br,t)$, in other words, $U_{\varepsilon}(\bk,\br,t,0)=U^a_\varepsilon(\bk,\br,t)$, $U_{\varepsilon}(\bk,\br,t,\pi)=U_{\varepsilon}(\bk,\br,t,-\pi)=U^b_\varepsilon(\bk,\br,t)$. The interpolation is required to satisfy the symmetry constraints
\begin{equation}
\label{}
\begin{aligned}
C^{-1} U_{\varepsilon}(\bk,\br,t,\lambda) C =U^*_{-\varepsilon}(-\bk,\br,t,-\lambda)\exp(i\frac{2\pi t}{\tau}),
\end{aligned}
\end{equation}
and
\begin{equation}
\label{}
\begin{aligned}
T^{-1}U_{\varepsilon}(\bk,\br,t,\lambda) T =U^*_{\varepsilon}(-\bk,\br,-t,-\lambda),
\end{aligned}
\end{equation}
whose forms resemble Eq.(\ref{particle}) and Eq.(\ref{time}).
For the half period $t=\tau/2$, we can readily see that
\begin{equation}
\label{}
\begin{aligned}
T^{-1}U_{\varepsilon}(\bk,\br,\frac{\tau}{2},\lambda) T =U^*_{\varepsilon}(-\bk,\br,-\frac{\tau}{2},-\lambda)=U^*_{\varepsilon}(-\bk,\br,\frac{\tau}{2},-\lambda),
\end{aligned}
\end{equation} which is valid for both $\varepsilon=0$ or $\pi$, and that
\begin{eqnarray}
U^*_{\varepsilon=0}(\bk,\br,\frac{\tau}{2},\lambda) =-C^{-1} U_{\varepsilon=0}(-\bk,\br,\frac{\tau}{2},-\lambda)C,
\end{eqnarray}
and also that
\begin{eqnarray}
U^*_{\varepsilon=\pi}(\bk,\br,\frac{\tau}{2},\lambda) =C^{-1} U_{\varepsilon=\pi}(-\bk,\br,\frac{\tau}{2},-\lambda)C,
\end{eqnarray}
which are consistent with Eq.(\ref{particle0}) and Eq.(\ref{particlepi}). As the product of PHS and TRS, a chiral symmetry is also satisfied.
For the class BDI, under the chiral basis (see Section \ref{sec:BDI}) we have
\begin{equation} \label{BDI}
\left\{
\begin{aligned}
U_{\varepsilon=0}^+(\bk,\br,\frac{\tau}{2},\lambda)=U_{\varepsilon=0}^{+*}(-\bk,\br,\frac{\tau}{2},-\lambda),\\
U_{\varepsilon=\pi}^+(\bk,\br,\frac{\tau}{2},\lambda)=U_{\varepsilon=\pi}^{+*}(-\bk,\br,\frac{\tau}{2},-\lambda).
\end{aligned}\right.
\end{equation} Similarly, for the class CII, under the chiral basis (see Section \ref{sec:CII}) we have
\begin{equation} \label{CII}
\left\{
\begin{aligned}
U_{\varepsilon=0}^+(\bk,\br,\frac{\tau}{2},\lambda)=\sigma_yU_{\varepsilon=0}^{+*}(-\bk,\br,\frac{\tau}{2},-\lambda)\sigma_y,\\
U_{\varepsilon=\pi}^+(\bk,\br,\frac{\tau}{2},\lambda)=\sigma_y U_{\varepsilon=\pi}^{+*}(-\bk,\br,\frac{\tau}{2},-\lambda)\sigma_y.
\end{aligned}\right.
\end{equation}  This symmetry constraints will be useful shortly.

We can define a winding number on the $(d+D+1)$-dimensional $(\bk,\br,\lambda)$ parameter space:
\begin{equation}
\begin{aligned}
&W(U^+_\varepsilon(\bk,\br,\frac{\tau}{2},\lambda)) =K_{d+D+1}\int_{ T^{d+1}\times S^D} d^dk d^Drd\lambda \\
&\times\text{Tr}[\epsilon^{\alpha_1\alpha_2\cdots\alpha_{d+D+1}}((U_{\varepsilon}^+)^{-1} \partial_{\alpha_{1}}U_{\varepsilon}^+)\cdots((U_{\varepsilon}^+)^{-1}\partial_{\alpha_{d+D+1}}U_{\varepsilon}^+)],
\end{aligned}
\end{equation}
where the coefficient is
$K_{d+D+1}= \frac{(-1)^{\frac{d+D}{2}}(\frac{d+D}{2})!}{(d+D+1)!}\left(\frac{i}{2\pi}\right)^{\frac{d+D}{2}+1}$.

Parallel to the previous section, before we are able to define a $\mathbb{Z}_2$ topological invariant, we need to prove that
\begin{eqnarray} \label{}
W(U_\varepsilon^+ (\bk,\br,\tau/2,\lambda))-W(U_\varepsilon'^{+}(\bk,\br,\tau/2,\lambda))=0\,  (\text{mod}\,2), \quad\quad
\end{eqnarray} for any two pairs of interpolations of $U^a_\varepsilon(\bk,\br,t)$ and $U^b_\varepsilon(\bk,\br,t)$, denoted as $U_\varepsilon^{+}(\bk,\br,\tau/2,\lambda)$ and $U_\varepsilon'^{+}(\bk,\br,\tau/2,\lambda)$.

Let us focus on the class CII first. We define two new interpolations from $U_\varepsilon (\bk,\br,\tau/2,\lambda)$ and $U_\varepsilon' (\bk,\br,\tau/2,\lambda)$, which are their reorganizations (we may still take Fig.\ref{deformation} as a pictorial illustration):
\begin{equation} \label{}
U_{\varepsilon}^{I}(\bk,\br,t,\lambda)=\left\{
\begin{aligned}
&U_{\varepsilon}(\bk,\br,t,\lambda),\quad  -\pi<\lambda<0, \\
&U'_{\varepsilon}(\bk,\br,t,-\lambda),\quad   0<\lambda<\pi,
\end{aligned}\right.
\end{equation}
and
\begin{equation} \label{}
U_{\varepsilon}^{II}(\bk,\br,t,\lambda)=\left\{
\begin{aligned}
&U'_{\varepsilon}(\bk,\br,t,-\lambda),\quad  -\pi<\lambda<0,\\
&U_{\varepsilon}(\bk,\br,t,\lambda),\quad  0<\lambda<\pi.
\end{aligned}\right.
\end{equation}
It follows as apparent results that
\begin{equation} \label{}
U_{\varepsilon}^{I+}(\bk,\br,\frac{\tau}{2},\lambda)=\left\{
\begin{aligned}
&U^{+}_{\varepsilon}(\bk,\br,\frac{\tau}{2},\lambda), \quad  -\pi<\lambda<0, \\
&U'^{+}_{\varepsilon}(\bk,\br,\frac{\tau}{2},-\lambda),\quad   0<\lambda<\pi.
\end{aligned}\right.
\end{equation}
and
\begin{equation} \label{}
U_{\varepsilon}^{II+}(\bk,\br,\frac{\tau}{2},\lambda) =\left\{
\begin{aligned}
&U'^{+}_{\varepsilon}(\bk,\br,\frac{\tau}{2},-\lambda),\quad -\pi<\lambda<0, \\
&U^{+}_{\varepsilon}(\bk,\br,\frac{\tau}{2},\lambda),\quad   0<\lambda<\pi.
\end{aligned}\right.
\end{equation}
Given these inputs, it is clear that \bea U^{I+}_\varepsilon(\bk,\br,\frac{\tau}{2},\lambda)= \sigma_y U^{II+*}_\varepsilon(-\bk,\br,\frac{\tau}{2},-\lambda)\sigma_y. \label{CII-WZW-I-II} \eea

As has been illustrated by Fig.\ref{deformation}, which is also useful in the present problem, one can readily see that
\begin{eqnarray} \label{}
W(U_\varepsilon^+)-W(U_\varepsilon'^{+})=W(U^{I+}_\varepsilon)+W(U^{II+}_\varepsilon).
\end{eqnarray}
When $\delta=4n$, we have (the calculation in Appendix \ref{chiralcii} is a useful reference here):
\begin{equation}
\label{}
\begin{aligned}
&W(U^{I+}_\varepsilon (\bk,\br,\frac{\tau}{2},\lambda) )=\int_{ T^{d+1}\times S^D} d^dkd^Drd\lambda\, w(U^{I+}_\varepsilon)(\bk,\br,\frac{\tau}{2},\lambda)\\
&=\int  d^dkd^Drd\lambda\, w^*(\sigma_y U^{II+}_\varepsilon\sigma_y)(-\bk,\br,\frac{\tau}{2},-\lambda)(-1)^{(d+D)/2+1}(-1)^{d+1}\\
&=\int  d^dkd^Drd\lambda\, w^*(U^{II+}_\varepsilon)(-\bk,\br,\frac{\tau}{2},-\lambda)(-1)^{2d+2-\delta/2}\\
&=W(U^{II+}_\varepsilon (\bk,\br,\frac{\tau}{2},\lambda))(-1)^{2d+2-\delta/2}\\
&=W(U^{II+}_\varepsilon ( \bk,\br,\frac{\tau}{2},\lambda)),
\end{aligned}
\end{equation}
in which $\varepsilon=0$ or $\pi$. In this calculation,
Eq.(\ref{CII-WZW-I-II}) and the reality of winding number have been used. Therefore, we have
\begin{eqnarray}
\label{}
\begin{aligned}
W(U_\varepsilon^+)-W(U_\varepsilon'^{+})=W(U^{I+}_\varepsilon)+W(U^{II+}_\varepsilon) =2W(U^{I+}_\varepsilon),
\end{aligned}
\end{eqnarray} which is exactly what we need to formulate the $\mathbb{Z}_2$ topological invariants.
Now $W(U^+_\varepsilon(\bk,\br,\tau/2,\lambda))$ (mod $2$) can be taken as a $\mathbb{Z}_2$ topological invariant to determine the relative triviality/nontriviality of $U^a_\varepsilon(\bk,\br,t)$ and $U^b_\varepsilon(\bk,\br,t)$. When  $W(U^+_\varepsilon(\bk,\br,\tau/2,\lambda))=0$ or $1$ (mod $2$), $U^a_\varepsilon(\bk,\br,t)$ and $U^b_\varepsilon(\bk,\br,t)$ are relatively trivial or nontrivial. If one of them (say $U^b_\varepsilon(\bk,\br,t)$) is fixed as a trivial time evolution operator, then $W(U^+_\varepsilon(\bk,\br,\tau/2,\lambda))$ (mod $2$) is a $\mathbb{Z}_2$ topological invariant for the other one ($U^a_\varepsilon(\bk,\br,t)\equiv U_\varepsilon(\bk,\br,t)$). Written in an equivalent way, it is \begin{equation} \label{}
\nu(U^+_\varepsilon(\bk,\br,t))=(-1)^{W(U^+_\varepsilon(\bk,\br,\tau/2,\lambda))}=\pm 1.
\end{equation}

For class BDI, the formulation is the same as class CII, which we will not repeat here.

Now we move on to $\delta=4n-1$,  the formulation will be parallel to Sec.\ref{wzwdiiici}. We define a two-parameter extension $U_{\varepsilon}(\bk,\br,t,\lambda,\mu)$ of the original time evolution operator $U_{\varepsilon}(\bk,\br,t)$  (with $\lambda,\mu\in[-\pi,\pi]$). It satisfies $U_{\varepsilon}(\bk,\br,t,0,0)=U_{\varepsilon}(\bk,\br,t)$, moreover, $U_{\varepsilon}(\bk,\br,t,\pm\pi,\mu)$ and $U_{\varepsilon}(\bk,\br,t,\lambda,\pm\pi)$ are trivial time evolution operators. As an extension of the PHS relation Eq.(\ref{particle}) and the TRS relation Eq.(\ref{time}), we require that
\begin{equation}
\label{}
\begin{aligned}
C^{-1} U_{\varepsilon}(\bk,\br,t,\lambda,\mu) C =U^*_{-\varepsilon}(-\bk,\br,t,-\lambda,-\mu)\exp(i\frac{2\pi t}{\tau}),
\end{aligned}
\end{equation}
and
\begin{equation}
\label{}
\begin{aligned}
T^{-1}U_{\varepsilon}(\bk,\br,t,\lambda,\mu) T =U^*_{\varepsilon}(-\bk,\br,-t,-\lambda,-\mu).
\end{aligned}
\end{equation}
For the half period $t=\tau/2$, the TRS relation becomes
\begin{equation}
\label{}
\begin{aligned}
 T^{-1}U_{\varepsilon}(\bk,\br,\frac{\tau}{2},\lambda,\mu) T &=U^*_{\varepsilon}(-\bk,\br,-\frac{\tau}{2},-\lambda,-\mu)\\
 &=U^*_{\varepsilon}(-\bk,\br,\frac{\tau}{2},-\lambda,-\mu),
\end{aligned}
\end{equation} for $\varepsilon=0$ or $\pi$.
The PHS becomes, for $\varepsilon=0$,
\begin{eqnarray}
U^*_{\varepsilon=0}(\bk,\br,\frac{\tau}{2},\lambda,\mu) =-C^{-1} U_{\varepsilon=0}(-\bk,\br,\frac{\tau}{2},-\lambda,-\mu)C;
\end{eqnarray}
and for $\varepsilon=\pi$,
\begin{eqnarray}
U^*_{\varepsilon=\pi}(\bk,\br,\frac{\tau}{2},\lambda,\mu) =C^{-1} U_{\varepsilon=\pi}(-\bk,\br,\frac{\tau}{2},-\lambda,-\mu)C.
\end{eqnarray}

We can define a winding number on the $(d+D+2)$-dimensional $(\bk,\br,\lambda,\mu)$ parameter space:
\begin{equation}
\label{}
\begin{aligned}
&W(U_\varepsilon^+ (\bk,\br,\frac{\tau}{2},\lambda,\mu))=K_{d+D+2}\int_{ T^{d+2}\times S^D } d^dk d^Drd\lambda d\mu\\
&\times\text{Tr}[\epsilon^{\alpha_1\alpha_2\cdots\alpha_{d+D+2}} ((U_{\varepsilon}^+)^{-1}\partial_{\alpha_{1}}U_{\varepsilon}^+)\cdots ((U_{\varepsilon}^+)^{-1}\partial_{\alpha_{d+D+2}}U_{\varepsilon}^+)],
\end{aligned}
\end{equation}
where the coefficient is
$K_{d+D+2}= \frac{(-1)^{\frac{d+D+1}{2}}(\frac{d+D+1}{2})!}{(d+D+2)!}\left(\frac{i}{2\pi}\right)^{\frac{d+D+1}{2}+1}$.

Taking advantage of the symmetry constraints discussed above, we can show that the parity (even/odd) of this winding number depends only on $U_{\varepsilon}(\bk,\br,t)$, not on the specific interpolation used. Therefore, it yields a $\mathbb{Z}_2$ invariant for $U_{\varepsilon}(\bk,\br,t)$. The $\mathbb{Z}_2$ topological invariant can be written as
\begin{equation} \label{}
\nu(U^+_\varepsilon(\bk,\br,t))=(-1)^{W(U^+_\varepsilon(\bk,\br,\tau/2,\lambda,\mu))}=\pm 1.
\end{equation}

\section{Frequency-domain numerical algorithm of topological invariants}

The above topological invariants expressed in terms of the time evolution operator can be calculated numerically, and this straightforward approach will be taken to study a few Floquet systems (see Sec. \ref{sec:point}). However, sometimes the computational load becomes high. In this section, we will describe a different numerical algorithm, which is based on the truncated frequency-domain Hamiltonian. This approach circumvent the integration over $t$, though at the price of dealing with matrices of high ranks. We first introduce this scheme for the class A\cite{rudner2013anomalous}, and then discuss the generalization to the cases with symmetries.

In Appendix \ref{sec:proof}, the standard frequency-domain Floquet Hamiltonian is reviewed. It reads
\bea \mathcal{H} =
\begin{pmatrix}
\cdots& & & &\\
&H_0+\omega&H_1&H_2&\\
&H_{-1}&H_0&H_1&\\
&H_{-2}&H_{-1}&H_0-\omega&\\
&&&&\cdots\\
\end{pmatrix}, \eea in which $H_m$'s are the Fourier components of $H(t)$. In practice, $\mathcal{H}$ is truncated to $M^2$ blocks, and the eigenvalues are located approximately in the interval $[-M\omega/2,M\omega/2]$ ($M$ is a large integer). For sufficiently large $M$, the topological defects modes can be faithfully obtained from the truncated $\mathcal{H}$, which takes the mathematical form of a static Hamiltonian. Therefore, according to the bulk-defect correspondence, the truncated $\mathcal{H}(\bk,\br)$ should be able to produce meaningful topological invariants of the defect. For class A, we can add up the Chern numbers of all the bands (of the truncated $\mathcal{H}$) below certain energy $\epsilon_0$, which is denoted as $\mathcal{C}_{(d+D)/2}(\mathcal{P}_{\epsilon<\epsilon_0})$ in the notation of Appendix \ref{sec:proof}, as the topological invariant of the gap $\epsilon_0$. This algorithm has been adopted in Ref.\cite{rudner2013anomalous} for two-dimensional Floquet insulators. It is not a trivial problem to prove by brute force that this ``truncated Chern number'' coincides with the winding number defined in Eq.(\ref{class-A}), nevertheless, there are a few justifications. In particular, if we take two quasienergy gaps $\epsilon_0$ and $\epsilon_1$ ($0\leq \epsilon_0<\epsilon_1<\omega$), then it is apparent that $\mathcal{C}_{(d+D)/2}(\mathcal{P}_{\epsilon<\epsilon_1})- \mathcal{C}_{(d+D)/2}(\mathcal{P}_{\epsilon<\epsilon_0}) =  \mathcal{C}_{(d+D)/2}(\mathcal{P}_{\epsilon_0,\epsilon_1})$, namely, the difference between the two truncated Chern numbers is the Chern number of the bands within $[\epsilon_0,\epsilon_1]$. In Appendix \ref{sec:proof}, it is proved that  $\mathcal{C}_{(d+D)/2}(\mathcal{P}_{\epsilon_0,\epsilon_1})$ is equal to the band Chern number calculated from $H^{\rm eff}$ (see Eq.(\ref{equivalence})), namely $C_{(d+D)/2}(P_{\varepsilon,\varepsilon'})$. Since the winding number satisfies the same relation (see Eq.(\ref{difference})), it is consistent to take $\mathcal{C}_{(d+D)/2}(\mathcal{P}_{\epsilon<\epsilon_0})$ as being equal to the winding number.

\begin{figure}
\includegraphics[width=9.0cm, height=6.0cm]{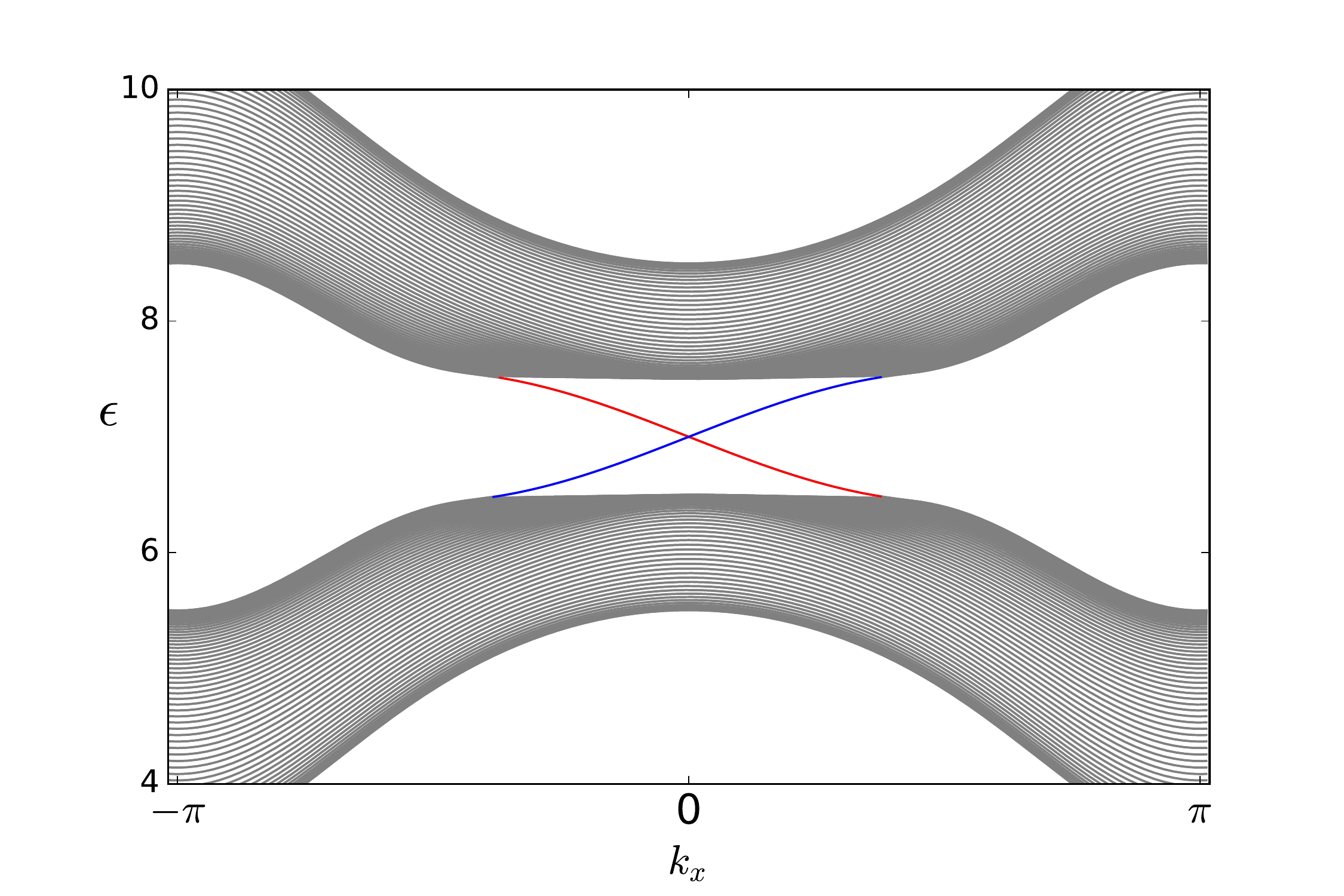}
\caption{ The quasienergy dispersions of a ribbon of Floquet topological insulator along the $x$ direction. The width of ribbon is $60$ unit cells. The blue and red lines (each consists of two almost degenerate lines) are the helical modes along the two edges. The parameters used are  $t=1.0,m_0=-5.5,m_d=4.0,B=0.5,t'=0.2$ and $\omega=14.0$ (see Eq.(\ref{floquet-TI})). }\label{Z2energy}
\end{figure}

For other symmetry classes, we have to study the symmetries of $\mathcal{H}$ before talking about possible $\mathcal{H}$-based topological invariants.  We can readily check that the TRS requires that \bea T^{-1}H_m(\bk,\br) T = H_m^*(-\bk,\br), \eea which leads to \bea \mathcal{T}^{-1}\mathcal{H}(\bk,\br)\mathcal{T}=\mathcal{H}^*(-\bk,\br), \label{T-frequency} \eea where \bea \mathcal{T} =
\begin{pmatrix}
\cdots& & & &\\
& T & & &\\
& & T & &\\
& & & T &\\
&&&&\cdots\\
\end{pmatrix}.  \eea Similarly, we have \bea \mathcal{C}^{-1}\mathcal{H}(\bk,\br)\mathcal{C}=-\mathcal{H}^*(-\bk,\br), \eea and \bea \mathcal{S}^{-1}\mathcal{H}(\bk,\br)\mathcal{S}=-\mathcal{H} (\bk,\br), \eea however, unlike the frequency-domain TRS matrix $\mathcal{T}$, the frequency-domain PHS and CS matrices $\mathcal{C}$ and $\mathcal{S}$ do not take the block-diagonal form, instead, they read \bea  \mathcal{C} =
 \begin{pmatrix}
& & & &\cdots\\
&   & & C &\\
& & C & &\\
& C & & &\\
\cdots&&&&\\
\end{pmatrix},\quad  \mathcal{S} =
 \begin{pmatrix}
& & & &\cdots\\
&   & & S &\\
& & S & &\\
& S & & &\\
\cdots&&&&\\
\end{pmatrix}. \eea These symmetry equations show that the symmetry operations are quite simple on $\mathcal{H}$: They take similar form as in the cases of static Hamiltonian. Thus, given a gap $\epsilon_0$, the familiar static topological invariants can be defined for the truncated ``static Hamiltonian'' $\mathcal{H}(\bk,\br)$, and these topological numbers are effective provided that $M$ is sufficiently large. In practice,
being ``sufficiently large'' means that $M\gg E_0/\omega$, where $E_0$ is a typical energy of the time-dependent Hamiltonian.

In this $\mathcal{H}$-based algorithm, the integration over $t$ involved in the topological invariants of nonchiral classes is circumvented, which is an advantage. In addition, it is not necessary to calculate $H^{\rm eff}$ in this approach. On the other hand, the rank of the truncated $\mathcal{H}$ has to be large, especially for small frequencies, which is a shortcoming.

As an application of this truncated $\mathcal{H}$ algorithm, let us consider the following toy model of a two-dimensional Floquet system in class AII ($d=2,D=0$), namely a Floquet topological insulator\cite{lindner2011floquet}. The Bloch Hamiltonian is given as
\bea H(\bk,t) &=& 2t(\sin k_x \sigma_x s_z + \sin k_y\sigma_y) + 2t'(\sin k_x+\sin k_y)\sigma_x s_x \nn \\ &&
+ [m(t)-2B(\cos k_x +\cos k_y)]\sigma_z ,  \label{floquet-TI} \eea where $m(t)=m_0+ m_d\cos(\omega t)$. This Bloch Hamiltonian is time-reversal-symmetric with $T=is_y$. If we take $m_d=t'=0$, the model is just the Bernevig-Hughes-Zhang model of two-dimensional topological insulators (``quantum spin Hall insulators'') in the HgTe quantum well\cite{bernevig2006c}. With the periodic driving included, the model is essentially a prototype model of Floquet topological insulator\cite{lindner2011floquet}. We have deliberately added the $t'$ term so that the Bloch Hamiltonian cannot be decoupled as two ($s_z=\pm 1$) blocks, otherwise the $\mathbb{Z}_2$ topological invariant is merely the parity of the winding number\cite{rudner2013anomalous} of each block. This model has the inversion symmetry $\sigma_z H(\bk,t)\sigma_z = H(-\bk,t)$. From the time-dependent Bloch Hamiltonian, we can obtain the Floquet Hamiltonian $\mathcal{H}(\bk)$. Due to the symmetry of Eq.(\ref{T-frequency}), we can use the $\mathbb{Z}_2$ Pfaffian topological invariant of Fu and Kane\cite{fu2006}. Taking advantage of the inversion symmetry, the $\mathbb{Z}_2$ topological invariant reduces to the product of parity eigenvalues at the four time-reversal-invariant momenta\cite{fu2007a}. For instance, let us take the parameters to be $t=1.0,m_0=-5.5,m_d=4.0,B=0.5,t'=0.2$ and $\omega=14.0$. The values of $\mathbb{Z}_2$ topological invariant are found to be $\nu_{\epsilon=0}=1$ and $\nu_{\epsilon=\omega/2}=-1$, which are obtained as the product of parity eigenvalues of the eigenvectors of $\mathcal{H}(\bk)$ with eigenvalues below $0$ or $\omega/2$, at the four time-reversal-invariant momenta $(0,0),(0,\pi),(\pi,0),(\pi,\pi)$. The values of topological invariant are consistent with the quasienergy dispersions of a ribbon system, which are shown in Fig.\ref{Z2energy} (only the dispersions near $\omega/2$ are shown; there is no edge mode at $\epsilon=0$).

\section{Line defects}

\begin{figure*}
\subfigure{\includegraphics[width=7.0cm, height=4.5cm]{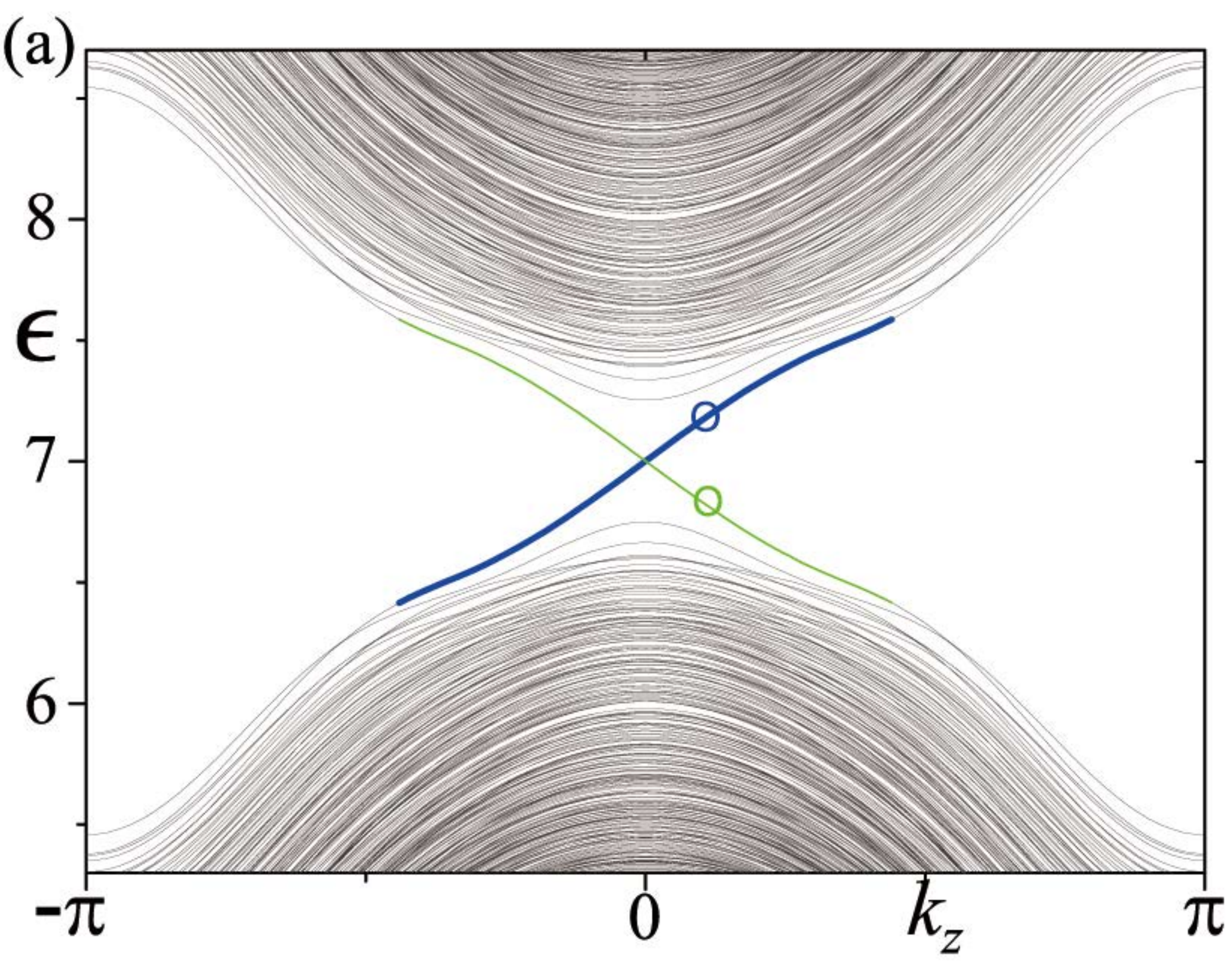}}
\subfigure{\includegraphics[width=4.3cm, height=4.4cm]{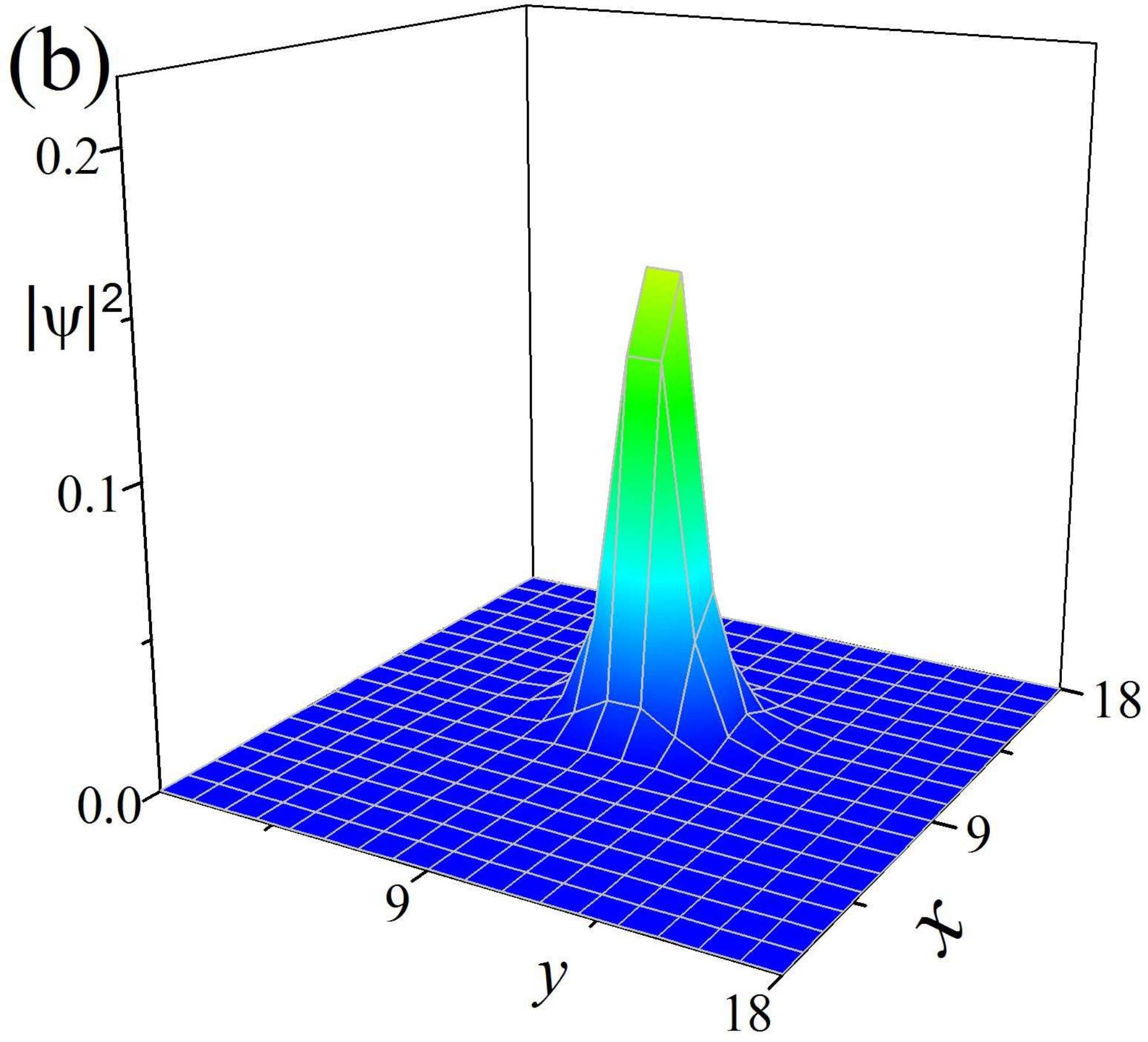}}
\subfigure{\includegraphics[width=4.3cm, height=4.4cm]{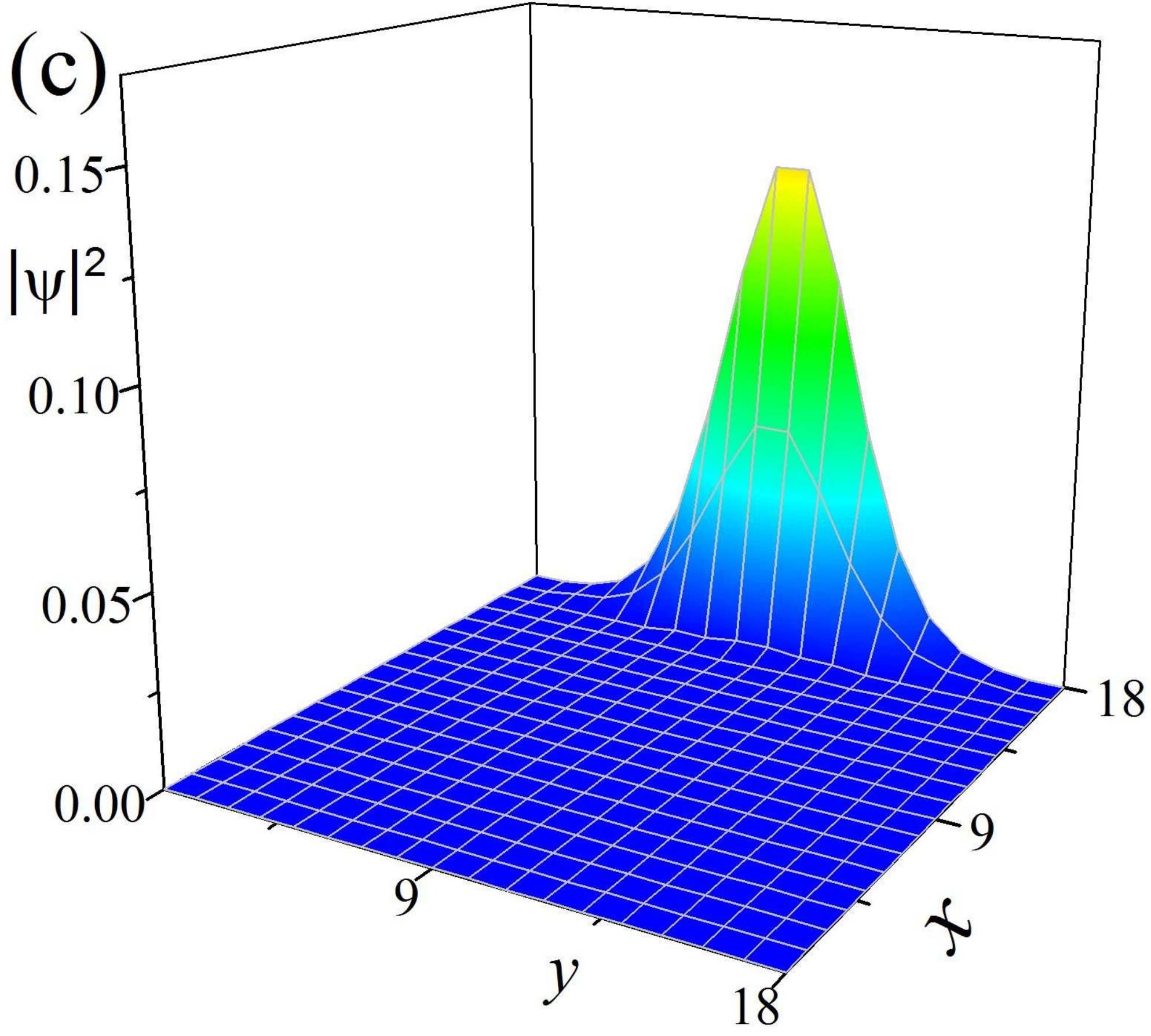}}
\subfigure{\includegraphics[width=7.0cm, height=4.5cm]{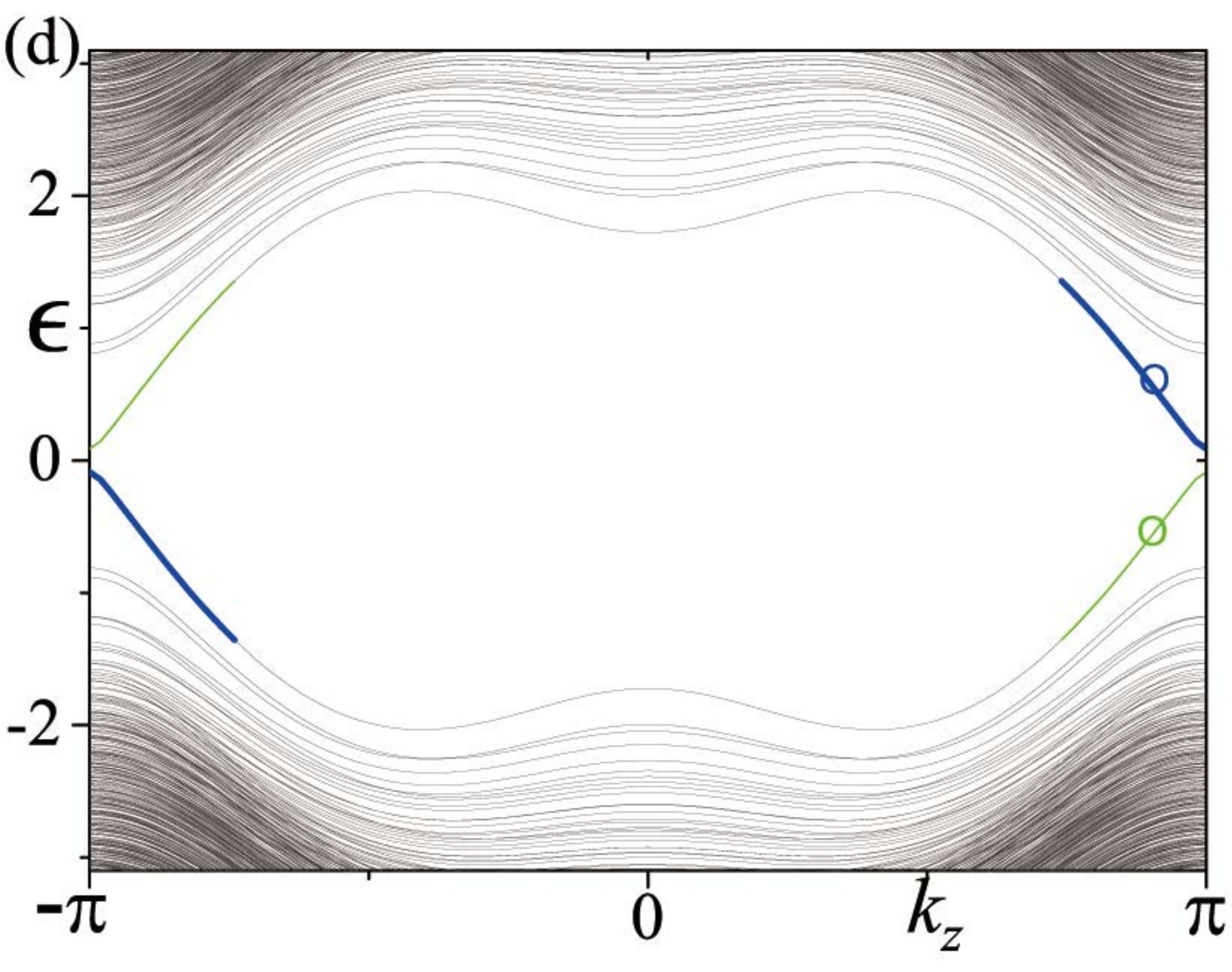}}
\subfigure{\includegraphics[width=4.3cm, height=4.4cm]{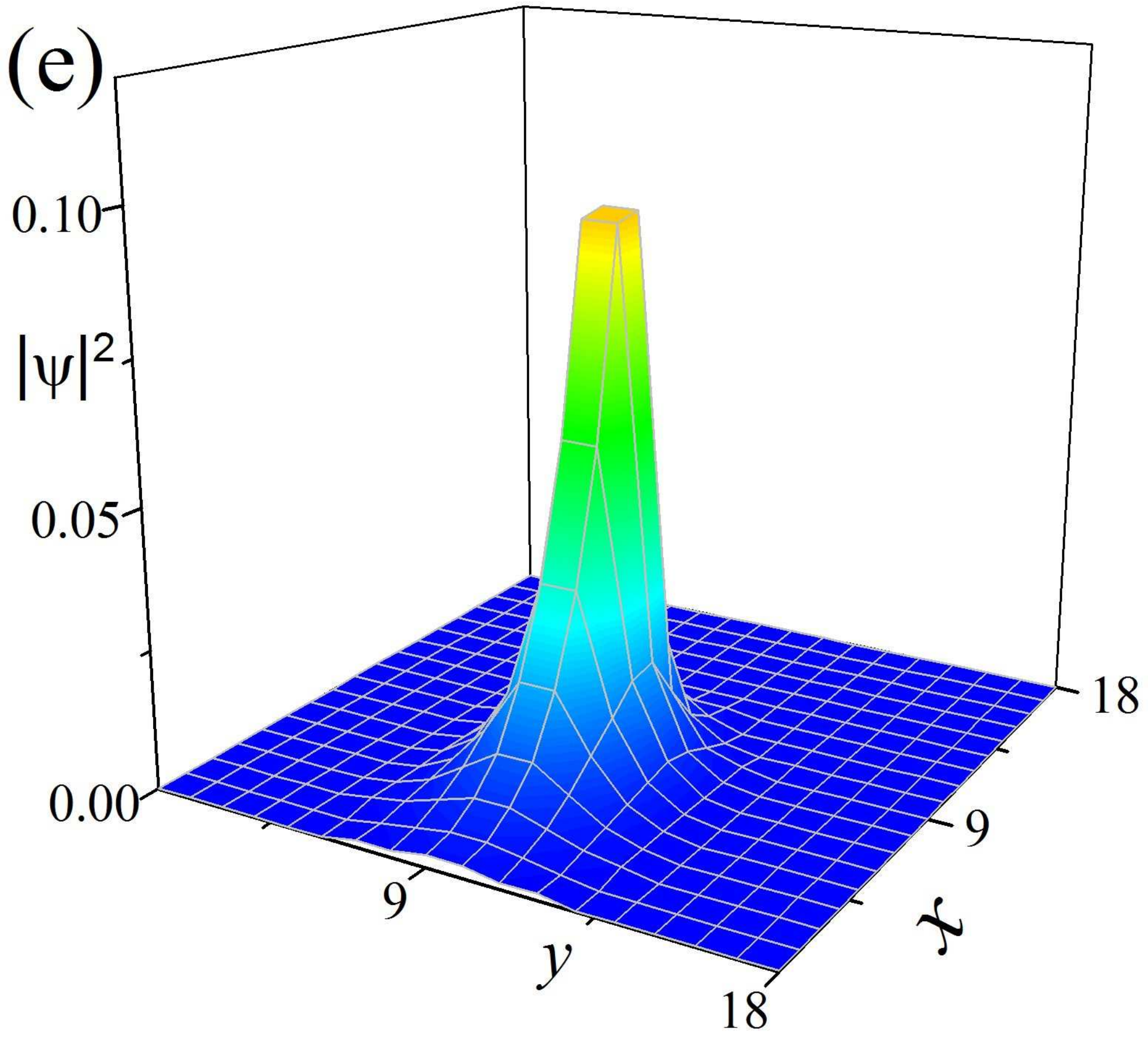}}
\subfigure{\includegraphics[width=4.3cm, height=4.4cm]{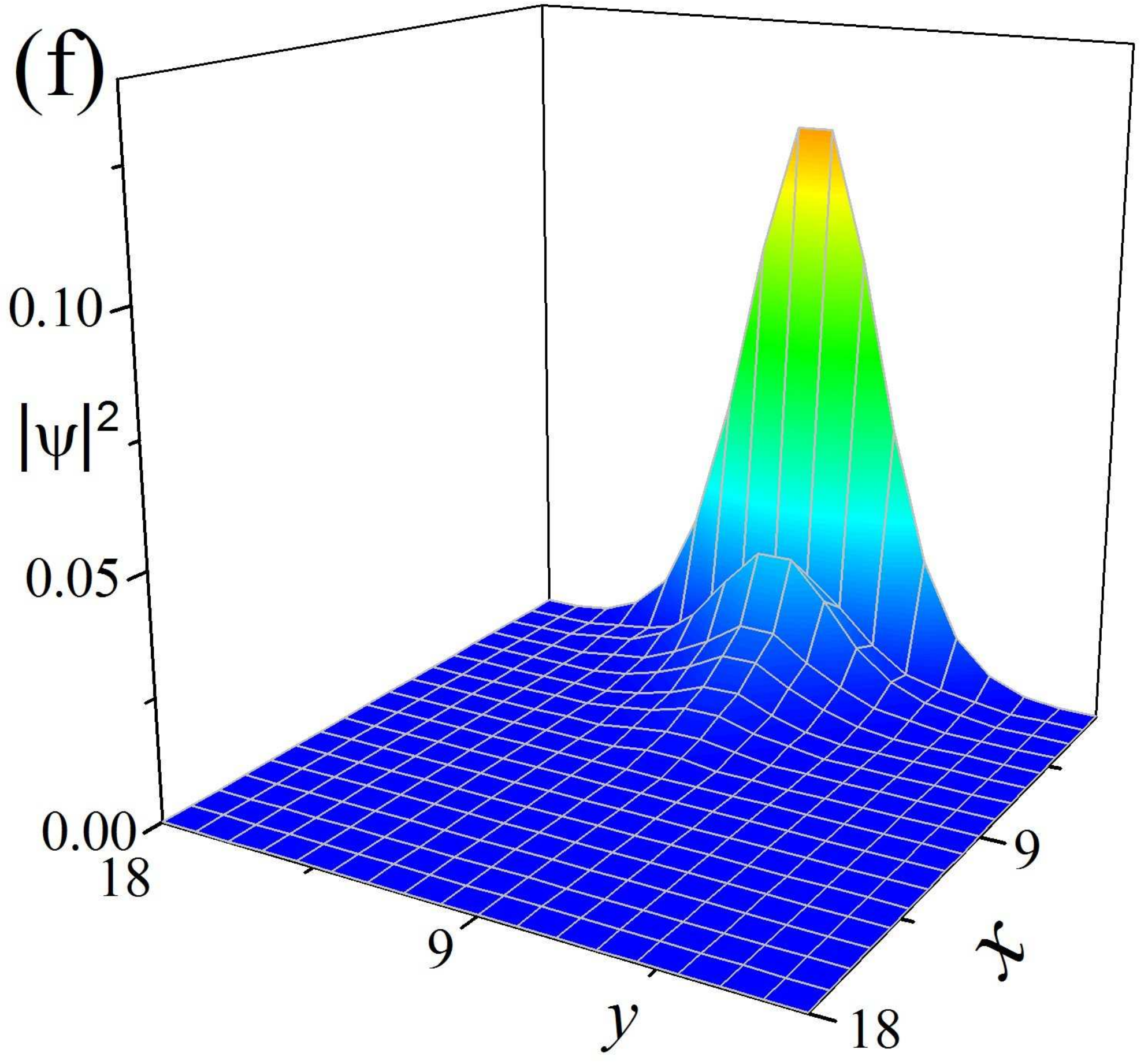}}
\caption{ Quasienergy dispersions and chiral mode profiles [case (i) in the text]. The parameters used here are $t_{1}=1.0$, $t_{2}=0.5$, $m_{0}=-3.5$, $\omega=14.0$, $m_{d}=4.0$, and
$L_{x}\times L_{y}\times L_{z}=18\times 18\times \infty$.
The Floquet Hamiltonian truncation is $M=5$ in the numerical calculations (namely, $5^2=25$ blocks). (a):
Quasienergy dispersion close to $\epsilon=\omega/2$. (b): Chiral mode profiles at the quasienergy marked by the blue circle in (a). (c): Chiral mode profile at the energy marked by the green circle in (a). The momentum for the blue and green circles in (a)
is $k_{z}=0.1\pi$. (d):
Quasienergy dispersion close to $\epsilon=0$. (e): Chiral mode profile at the energy marked by the blue circle in (d). (f):
Chiral mode profile at the energy marked by the green circle in (d).  The momentum for the blue and green circles in (d)
is $k_{z}=0.9\pi$.  }\label{SNTchiral}
\end{figure*}

Experimentally, the most relevant topological defects are low-dimensional ones: line defects and point defects. In this section, we will focus on line defects,
which have $d_{\rm def}=1$, and $\delta=d-D=d_{\rm def}+1=2$. Point defects will be left to the next section. According to our formulations of topological invariants,  line defects in class A, class D, class DIII, class AII, and class C can have topologically protected Floquet modes.  In this section we study a number of important examples. In the numerical calculations of quasienergy spectra, we will use the quasienergy $\epsilon$ instead of the dimensionless quasienergy $\varepsilon\equiv\epsilon\tau$, which has been used extensively above. As has been mentioned, the dimensionless quasienergy $\varepsilon$ has periodicity $2\pi$, while the quasienergy $\epsilon$ has periodicity $\omega$ or $2\pi/\tau$.

\begin{figure}
\subfigure{\includegraphics[width=7.5cm, height=5.5cm]{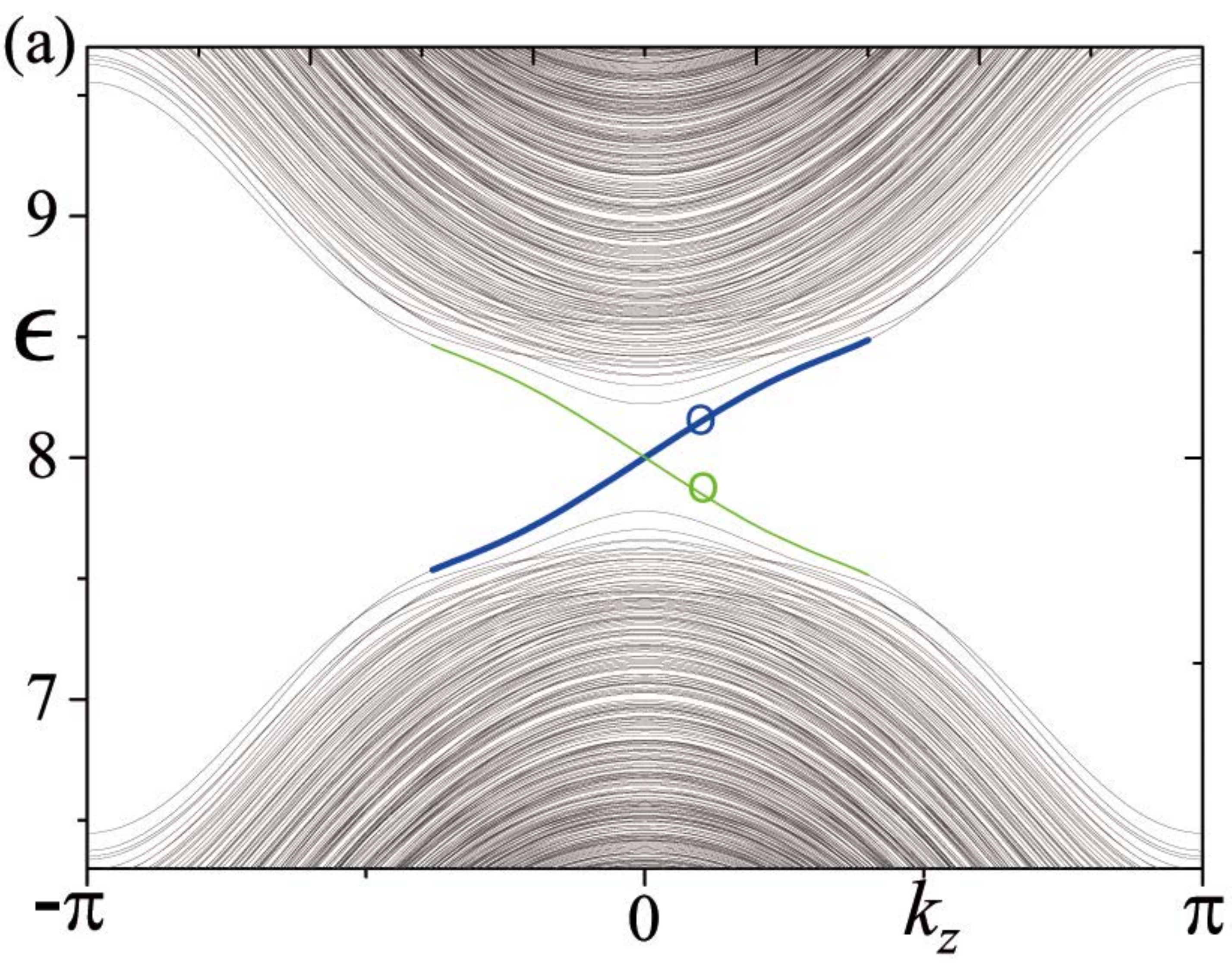}}
\subfigure{\includegraphics[width=3.7cm, height=3.5cm]{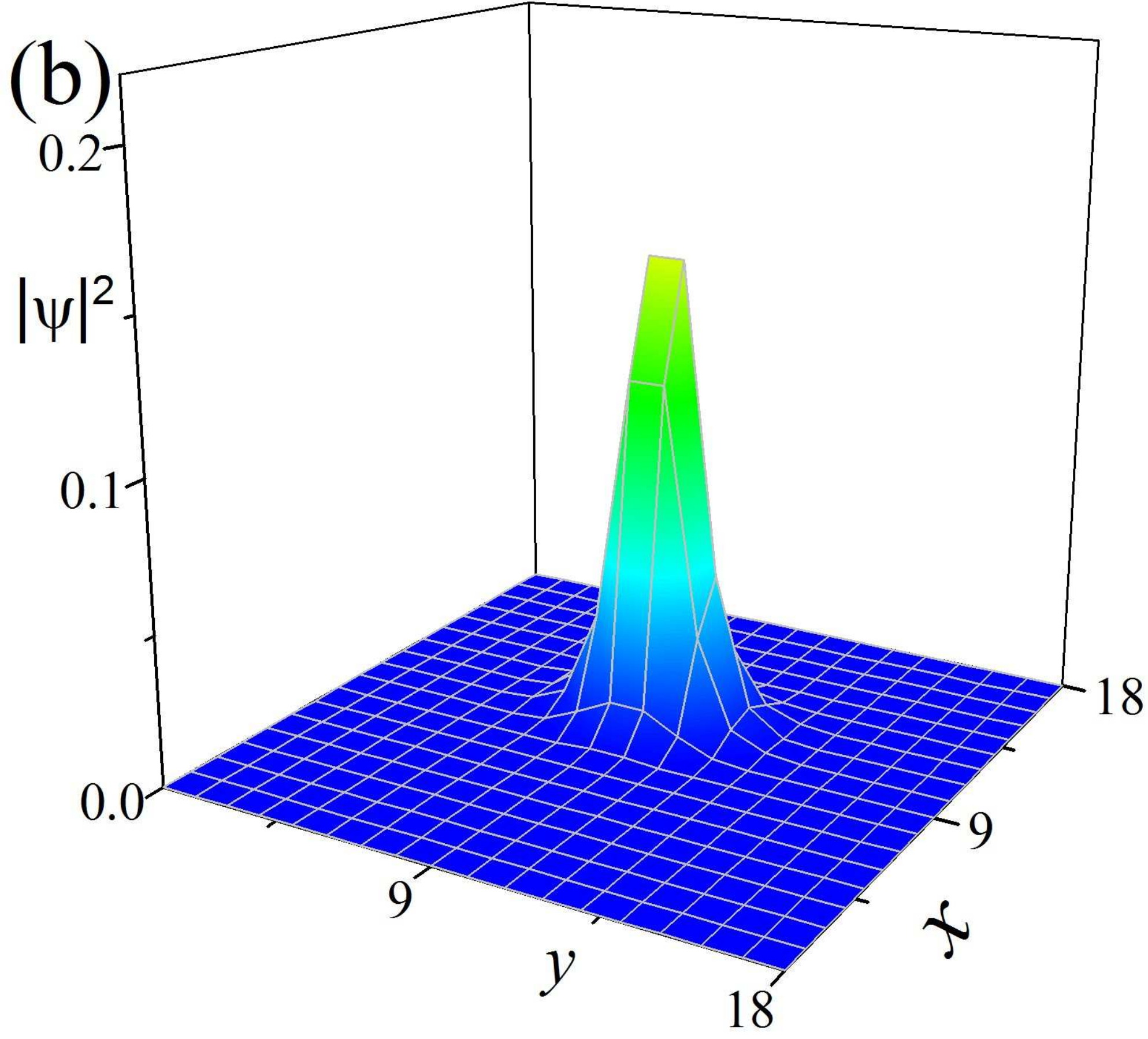}}
\subfigure{\includegraphics[width=3.7cm, height=3.5cm]{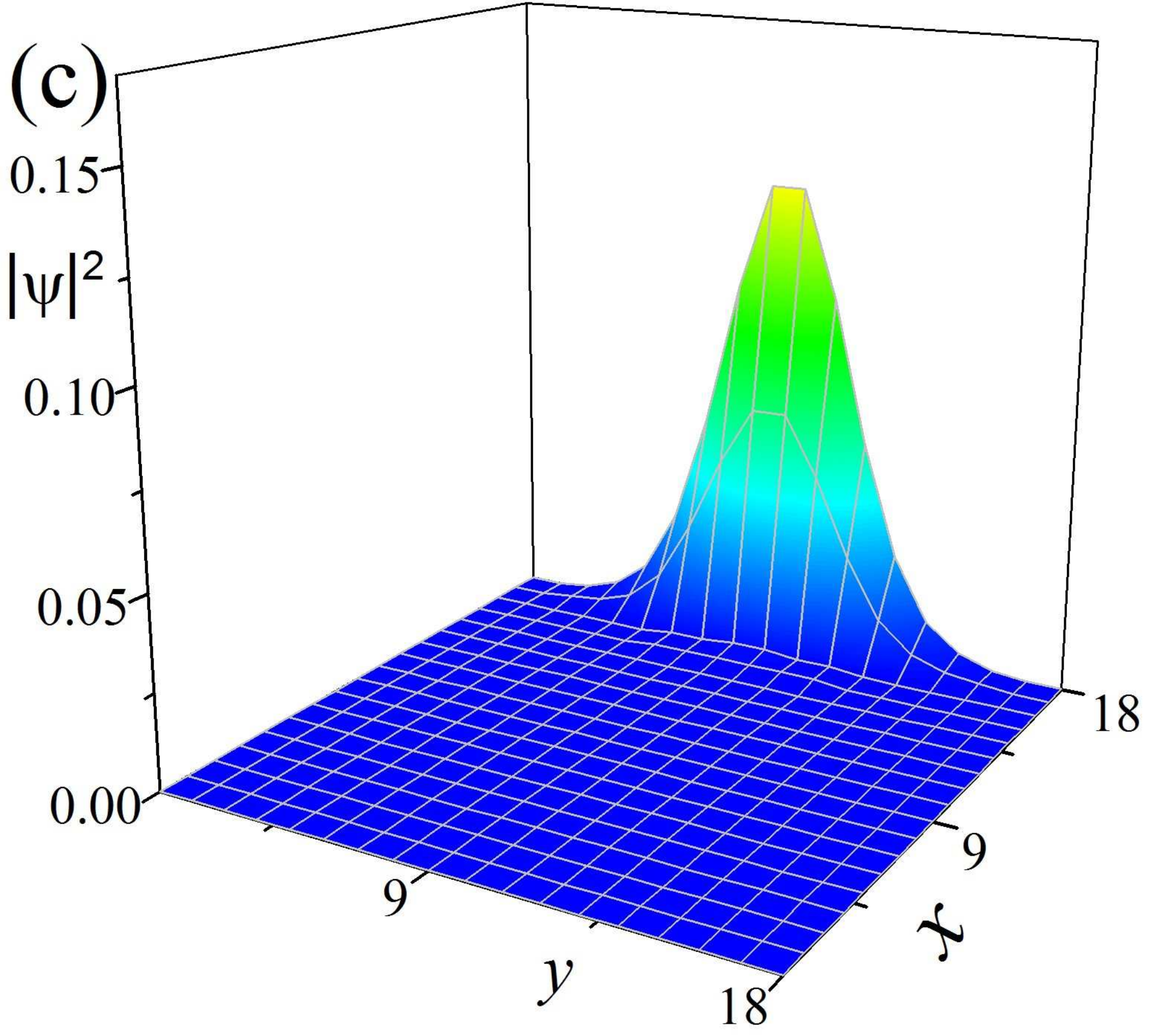}}
\caption{ Quasienergy dispersions and the wavefunction profiles of the Floquet chiral modes [case (ii) in the text].  The parameters used here are $t_{1}=1.0$, $t_{2}=0.5$, $m_{0}=-4.5$, $\omega=16.0$, $m_{d}=4.0$,
$L_{x}\times L_{y}\times L_{z}=18\times 18\times \infty$.
The Floquet Hamiltonian truncation is $M=5$. (a):
Quasienergy dispersions close to $\epsilon=\omega/2$. (b): Wave function profile of the Floquet chiral modes at the energy labeled by the blue circle in (a).  (c):
Wave function profile of the Floquet chiral mode at the energy labeled by the green circle in (a). The momentum for the blue and green circle is
$k_{z}=0.1\pi$.  }\label{STchiral}
\end{figure}

\subsection{Class A: Floquet chiral modes along a line defect}\label{sec:line-class-A}

First, we recall that static line defects in the class A have been studied in several contexts. The number of chiral modes along the line defect is equal to the second Chern number of the occupied bands, defined in the $(k_x,k_y,k_z,\theta)$ parameter space\cite{teo2010,qi2008,fiber}, where $\theta$ is the angular coordinate in the cylindrical coordinate systems (the $z$-axis is taken to be coincident with the line defect). These chiral modes have deep field-theoretical origin in the continuum field theory\cite{callan1985,witten1985superconducting}.
It has been pointed out that the dislocations in the charge density wave in Weyl semimetals carry chiral modes\cite{wang2013a,Bi2015, Roy2015magnetic,Schuster2016cable,You2016}, which brings this conception closer to potential experimental realization. The photonic analog has also been proposed\cite{fiber}.

In this section we study Floquet chiral modes along driven line defects in 3-dimensional space. To be concrete, let us first study a lattice model, which is constructed as follows. Before discussing topological defects,
let us consider a 4-band Bloch Hamiltonian parameterized by $\lambda$ for a homogeneous crystal:
\begin{eqnarray}
H(\bk,t)&=&2t_{1}(\sin k_{x}\sigma_{x}+\sin k_{y}\sigma_{y}+\sin k_{z}\sigma_{z})\tau_{z}+
2t_{1}\sin \lambda\,\sigma_0\tau_{y}\nonumber\\
&+&[m(t)-2t_{2}(\cos k_{x}+\cos k_{y}+\cos k_{z}+\cos \lambda)]\sigma_{0}\tau_{x},\quad\quad \label{class-A-lambda}
\end{eqnarray}
where $\sigma_i$'s and $\tau_i$'s ($i=x,y,z$) are Pauli matrices ($\sigma_0=\tau_0\equiv I$), and \bea m(t)=m_{0}+m_{d}\cos\omega t \eea provides a periodic driving.  The hopping parameters $t_{1,2}$ will be fixed as $t_1=1.0, t_2=0.5$. When $\lambda=0$ or $\pi$, this Bloch Hamiltonian has time-reversal symmetry $T^{-1}H(\bk,t)T =H^*(-\bk,-t)$, with $T=\sigma_y$. In certain parameter regimes, it is a lattice model of 3d Floquet topological insulators. Apparently, $\lambda\neq 0,\pi$ entails time-reversal-symmetry breaking.

Now let us discuss the line defect [Fig.\ref{sketch}d]. Suppose that the line defect is parallel to the $z$ axis, with an $x$-$y$ plane coordinate $(x_0,y_0)$. We can use the cylindrical coordinates, in which the angular coordinate $\theta= \arctan[(y -y_0)/(x -x_{0})]$. As has been explained in Sec. \ref{sec:outline}, in the regions sufficiently far away from the line defect, translational symmetry is restored and one can talk about the Bloch Hamiltonian $H(\bk,\theta,t)$, which is a smooth function of $\theta$. A simplest construction of $H(\bk,\theta,t)$ is to take \bea \lambda = n\theta  \eea in Eq.(\ref{class-A-lambda}), where $n$ is an arbitrary integer. We shall focus on the $n=1$ case for simplicity. Sufficiently far away from the line defect, the Bloch Hamiltonian reads \begin{eqnarray}
H(\bk,\theta, t)&=&2t_{1}(\sin k_{x}\sigma_{x}+\sin k_{y}\sigma_{y}+\sin k_{z}\sigma_{z})\tau_{z}+
2t_{1}\sin\theta\,\sigma_0\tau_{y}\nonumber\\
&+&[m(t)-2t_{2}(\cos k_{x}+\cos k_{y}+\cos k_{z}+\cos \theta)]\sigma_{0}\tau_{x}. \quad\quad \label{line-defect-A}
\end{eqnarray}  If the driving term $m_d\cos\omega t\,\sigma_0\tau_x$ is removed from the Hamiltonian, static line defects with chiral modes\cite{teo2010,Bi2015} can be constructed  in certain regimes of $(t_1,t_2,m_0)$, provided that the second Chern number in the $(k_x,k_y,k_z,\theta)$ parameter space is nonzero. In this work we are more interested in the effects of nonzero $m_d$, which is responsible for the Floquet chiral modes.

The real-space Hamiltonian of this line defect is (being real-space only in the $x,y$ directions, while the good quantum number $k_z$ remains):
\begin{eqnarray}
\hat{H}(k_z,t)&=&\sum_{x,y,k_{z}}\left\{ [-it_{1}(c^{\dag}_{x,y;k_{z}}\sigma_{x}\tau_{z}c_{x+1,y;k_{z}}
+c^{\dag}_{x,y;k_{z}}\sigma_{y}\tau_{z}c_{x,y+1;k_{z}})+h.c.]\right.\nonumber\\
&&+2t_{1}\sin k_{z}c^{\dag}_{x,y;k_{z}}\sigma_{z}\tau_{z}c_{x,y;k_{z}}+2t_{1}\sin\theta_{x,y}
c^{\dag}_{x,y;k_{z}}\sigma_{0}\tau_{y}c_{x,y;k_{z}}\nonumber\\
&&-(t_{2}c^{\dag}_{x,y;k_{z}}\sigma_{0}\tau_{x}c_{x+1,y;k_{z}}
+t_{2}c^{\dag}_{x,y;k_{z}}\sigma_{0}\tau_{x}c_{x,y+1;k_{z}}+h.c.)\nonumber\\
&&\left.+[m(t)-2t_{2}(\cos k_{z}+\cos\theta_{x,y})]c_{x,y;k_{z}}\sigma_{0}\tau_{x}c_{x,y;k_{z}}
\right\}, \label{class-A-real}
\end{eqnarray}
in which $(x,y)$ are integer-valued lattice coordinates (labelling the unit cell), and $c,c^\dag$ stand for particle annihilation/creation operators. As has been defined above, the polar angle $\theta_{x,y}= \arctan[(y -y_0)/(x -x_{0})]$.   For sufficiently large $\sqrt{(x-x_0)^2+(y-y_0)^2}$, the polar angle $\theta_{x,y}$ can be regarded as a constant locally (denoted as $\theta$), and the Fourier transformation of Eq.(\ref{class-A-real}) is just Eq.(\ref{line-defect-A}).  The real-space Hamiltonian will be useful in the numerical calculations of quasienergies and wave functions. We also mention  that modifying this real-space Hamiltonian in the vicinity of line defect core does not change the robust topological properties of the line defect (e.g., the number of chiral modes).

In the numerical calculations of quasienergy spectrum, we use the frequency-domain (repeated zone) formulation, which is now a standard method in Floquet theory  (for example, see Ref.\cite{rudner2013anomalous}). This method is quite simple to practice. With the current model in mind, we briefly introduce this formulation as follows.
Let us start from the time-dependent Schr\"{o}dinger equation,
\begin{eqnarray}
i\partial_{t}|\psi(k_{z},t)\rangle=H(k_{z},t)|\psi(k_{z},t)\rangle,
\end{eqnarray} in which we have kept the $(x,y)$ coordinates implicit. The rank of $H(k_z,t)$ is proportional to the size of the system in the $(x,y)$ plane, i.e., $L_x\times L_y$.
In the presence of time periodicity of $H(k_z,t)$, the Floquet theory
tells us that the time-dependent solutions of the Schr\"{o}dinger equation
can be expressed as \bea |\psi_n(k_{z},t)\rangle
=\exp[-i\epsilon_n(k_{z})t]|\phi_n(k_z,t)\rangle, \eea where $|\phi_n\ra$ satisfies
$|\phi(t+\tau)\rangle=|\phi(t)\rangle$.
The periodicity of  $|\phi_n(t)\rangle$ enables the Fourier expansion: \bea |\phi_n(k_z,t)\rangle
=\sum_{m}e^{im\omega t}|\phi_n^{(m)}(k_z,t)\rangle. \eea As a result,
the time-dependent Schr\"{o}dinger equation is equivalent to
\begin{eqnarray}
\sum_{m'}\mathcal{H}_{mm^{'}}(k_{z})|\phi_n^{(m')}(k_z)\rangle
=\epsilon_n(k_{z})|\phi_n^{(m)}(k_z)\rangle,\label{FH}
\end{eqnarray}
where $\mathcal{H}_{mm^{'}}(k_{z})=m\omega \delta_{mm'}\mathbf{I}+H_{m-m'}(k_{z})$,
in which $H_m(k_z)$'s are the Fourier transformation of $H(k_z,t)$: \bea H_{m}(k_{z})=\frac{1}{\tau}\int_{0}^{\tau}dt H(k_{z},t) e^{-im\omega t}. \label{fourier-component} \eea
More explicitly,  the ``Floquet Hamiltonian'' $\mathcal{H}(k_z)$ appearing in Eq.(\ref{FH}) is a matrix of
infinite rank,
\begin{eqnarray}
\mathcal{H}(k_z)=\left(
                     \begin{array}{ccccccc}
                       \cdots &  &  &  & & & \\
                        & H_0+2\omega & H_1 & H_2 & H_3 & \cdots & \\
                        & H_{-1} & H_{0}+\omega & H_{1} & H_2 & H_3 &  \\
                       & H_{-2} & H_{-1} & H_0 & H_1 &  H_2 & \\
                       & H_{-3} & H_{-2} & H_{-1} & H_0-\omega & H_1 &  \\
                       & \cdots & H_{-3} & H_{-2} & H_{-1} & H_0-2\omega & \\
                      & &  &  &  &  & \cdots \\
                     \end{array}
                   \right).
                   \quad \nn
\end{eqnarray}
In practical calculations, we have to truncate this infinite-rank matrix, keeping $M^2$ blocks ($M$ of them are diagonal blocks). This truncation procedure is valid because the Floquet Hamiltonian takes the form of a ``Wannier-Stark ladder''\cite{emin1987existence}, whose eigenstates are exponentially localized in the $m$ direction\cite{rudner2013anomalous}.

Given the form of $H(k_z,t)$ in our model, we can now solve the spectrum by diagonalizing the Floquet Hamiltonian $\mathcal{H}(k_z)$.  We consider two illustrative cases. In the
case (i), we take $m_0=-3.5$, for which the static defect without the driving term ($m_d=0$) already have chiral modes in the $\epsilon=0$ gap. Let us add a driving with $\omega=14$ and $m_d=4$ (the magnitude of driving is exaggerated for a better illustration, yet the physics is qualitatively the same for smaller $m_d$). The numerical results are shown in Fig.\ref{SNTchiral}. An additional quasienergy gap is generated at $\epsilon=\omega/2$ by the driving. There is a chiral mode localized around the defect in each one of the two gaps, $\epsilon=0$ and $\epsilon=\omega/2$, indicated by the thick blue lines. The thin green lines stand for the back-propagating chiral modes at the system boundary. These boundary modes are localized around certain polar angle $\theta$ instead of extending over the whole boundary, as can be seen from Fig.\ref{SNTchiral}(c) and (f). This is due to the absence of rotational symmetry in this model (similar localization of boundary modes near certain polar angle is also common for static defects\cite{MZM-Hopf}).

In the case (ii), we take $m_0=-4.5$, for which the static defect without driving has no chiral modes at $\epsilon=0$. Under a driving with $\omega=16.0$ and $m_d=4.0$, a chiral mode is generated in the $\epsilon=\omega/2$ gap. The numerical results are shown in Fig.\ref{STchiral}.

We note that in the static heterostructures considered in Ref.\cite{teo2010}, topological insulators are used to yield a nontrivial second Chern number of the line defects (In general, topological insulators are fruitful platforms of topological defects; for example, see Ref.\cite{teo2017topological,lee2007,qi2008b,seradjeh2009,Barkeshli2012}). In our model, even if the static second Chern number is zero, there can still be
Floquet chiral modes (not at $\epsilon=0$, but at $\epsilon=\omega/2$), as illustrated by the case (ii). Topological insulators are not necessary to generate these chiral modes.

Finally, let us discuss the topological invariant of line defect. For line defects of class A in 3d space, the winding number is the $(d,D)=(3,1)$ case of Eq.(\ref{class-A}), which reads
\bea
W(U_\varepsilon)=&& K_5\int_{T^3\times S^1\times S^1} d^3k d\theta dt \text{Tr}[\epsilon^{\alpha_1\alpha_2\cdots\alpha_5}(U_{\varepsilon}^{-1}\partial_{\alpha_{1}}U_{\varepsilon}) \nn\\ && \times(U_{\varepsilon}^{-1}\partial_{\alpha_2}U_{\varepsilon})\cdots (U_{\varepsilon}^{-1}\partial_{\alpha_5}U_{\varepsilon})]. \label{A-line-defect}
\eea
Although this winding number can be calculated numerically in principle, we will follow a different approach. Since the topology is not sensitive to the magnitude of $m_d$, let us take it to be small and treat it as a perturbation. Let us take the message of Eq.(\ref{difference}): $W(U_{\varepsilon=\pi})-W(U_{\varepsilon=0})=C_2(P_{0,\pi})$. The effect of a small $m_d$ is negligible near $\varepsilon=0$, thus $W(U_{\varepsilon=0})$ can be inferred from the static limit with $m_d=0$, which is the second Chern number of a static line defect [see Eq.(\ref{static1})]. For a Dirac-type Hamiltonian $H_0(\bk)=\sum_{\mu=1}^5 d_\mu(\bk)\Gamma_\mu$ with $\{\Gamma_\mu,\Gamma_\nu\}=2\delta_{\mu\nu}$ [In our problem, ${\bf \Gamma}=(\sigma_x\tau_z,\sigma_y\tau_z,\sigma_z\tau_z, \sigma_0\tau_y,\sigma_0\tau_x)$, and $\bd$ can be read from Eq.(\ref{line-defect-A}) with $m_d=0$], the second Chern number can be reduced to the form of\cite{qi2008} \bea W(U_{\varepsilon=0}) = \frac{3}{8\pi^2}\int d\theta d^3k \epsilon^{\mu\nu\rho\sigma\tau} d_\mu\partial_\theta d_\nu\partial_{k_x}d_\rho\partial_{k_y}d_\sigma\partial_{k_z}d_\tau,\quad \label{C2-winding}. \eea  from which we find that, for the case (i) and (ii), $W(U_{\varepsilon=0})=-1$ and $0$, respectively.

Now we proceed to calculate $C_2(P_{0,\pi})$, namely, the Chern number of the Floquet bands with quasienergy $\epsilon\in [0,\omega/2]$ (or dimensionless quasienergy  $\varepsilon\in[0,\pi]$). Thanks to the Eq.(\ref{equivalence}) in Appendix \ref{sec:proof}, we do not need to derive the form of $H^{\rm eff}$; instead, we can calculate $\mathcal{C}_2(\mathcal{P}_{0,\pi})$ using $\mathcal{H}$, which is easier in practice. Near $\varepsilon=\pi$ or $\epsilon=\omega/2$, only four of the Floquet bands of $\mathcal{H}$ are important, which can be well described by a four-band Hamiltonian \bea H_{\rm R}(\bk,\theta) = \bd_{\rm R}\cdot{\bf \Gamma} + \omega/2, \label{RW} \eea where
\bea \bd_{\rm R} = (|\bd|-\omega/2)\hat{\bd} + \tilde{\bd}_\perp, \label{d-R} \eea in which $\hat{\bd}\equiv\bd/|\bd|$ is the unit vector of $\bd= (2t_1\sin k_x, 2t_1\sin k_y, 2t_1 \sin k_z, 2t_1\sin\theta, m_0-2t_2(\sum_i\cos k_i+\cos\theta) )$. The symbol $\tilde{\bd}_\perp\equiv \tilde{\bd}-(\tilde{\bd}\cdot\hat{\bd})\hat{\bd}$ denotes the perpendicular part of the vector $\tilde{\bd}$, which comes from the periodic driving: In the Fourier series $H(\bk,\theta,t)=\sum_m e^{im\omega t}H_m(\bk,\theta)$, the first component is denoted as $H_1(\bk,\theta)=\tilde{\bd}\cdot{\bf \Gamma}$, which defines the vector $\tilde{\bd}$. In our problem, one can readily find that $\tilde{\bd}=(0,0,0,0,m_d/2)$. This form of $H_{\rm R}$ can be derived by inspecting $\mathcal{H}$ near $\omega/2$ (omitting all bands far away from $\omega/2$), or by the rotating-wave approximation\cite{lindner2011floquet} (A two-band counterpart of Eq.(\ref{d-R}) can be found in Ref.\cite{rudner2013anomalous}, to which the interested readers may refer). Now we calculate $\mathcal{C}_2(\mathcal{P}_{0,\pi})$ numerically by replacing $\bd$ in Eq.(\ref{C2-winding}) by $\bd_{\rm R}$, which yields $\mathcal{C}_2(\mathcal{P}_{0,\pi})=2$ and $\mathcal{C}_2(\mathcal{P}_{0,\pi})=1$ for case (i) and case (ii), respectively. The relation $W(U_{\varepsilon=\pi})=W(U_{\varepsilon=0}) +\mathcal{C}_2(\mathcal{P}_{0,\pi})$ then tells us that $W(U_{\varepsilon=\pi})=-1+2=+1$ and $0+1=+1$ for case (i) and case (ii), respectively, which is consistent with the number of Floquet chiral mode in the $\epsilon=\omega/2$ quasienergy gap.

Our model suggests a way of creating Floquet chiral modes in a driven line defect, which is topologically trivial if the driving is removed. Before concluding this section, it is useful to mention that, even if the static system itself is defect-free, a Floquet line defect can be created solely by the periodic driving\cite{bi2016}. To this end, the driving must have spatial modulations\cite{Katan2013modulated,katan2013generation,bi2016}. A possible platform is a Dirac semimetal under spatially modulated driving, which was recently suggested in Ref.\cite{bi2016}.

\subsection{Class D: Floquet chiral Majorana modes}

Floquet chiral Majorana modes along a line defect in a 3d superconductor can be modelled by the following defect Hamiltonian (the notations follow the previous section):
\bea H_{\rm BdG}(\bk,\theta,t) && =  [\mu(t) -\sum_i\cos k_i-\sin\theta]\,\sigma_0\tau_z \nn \\ && +  \Delta_p\sum_i\sin k_i\,\sigma_i \tau_x + \Delta_s\sin\theta\,\sigma_0 \tau_y, \label{line-defect-D} \eea with \bea \mu(t)=\mu_0+\mu_d\cos\omega t. \eea It satisfies the symmetry \bea \sigma_y\tau_y H^*_{\rm BdG} (\bk,\theta,t) \sigma_y\tau_y = -H_{\rm BdG}(-\bk,\theta,t), \eea with $(\sigma_y\tau_y)^*(\sigma_y\tau_y)=1$, therefore, the Hamiltonian belongs to class D.

The structure of this Hamiltonian is essentially the same as Eq.(\ref{line-defect-A}), therefore, Floquet chiral modes should also be present in the model of Eq.(\ref{line-defect-D}). The new ingredient, compared to the previous section, is the physical interpretation. We may interpret it as a Bogoliubov-de Gennes (BdG) equation, $\tau_z=\pm 1$ being the particle/hole subspace. The $\Delta_p$ and $\Delta_s$ term is the $p$-wave and $s$-wave Cooper pairing, respectively. The $\mu(t)$ term is then interpreted as a time-dependent chemical potential. Thus the Hamiltonian describes a superconductor with spatially modulated pairings and a time-dependent chemical potential. In principle, it may be imitated by superconductor heterostructures containing both $s$-wave superconductors and $p$-wave superconductors, which will be left for future investigations.

\subsection{Class AII: Floquet helical modes}\label{AII-line}

\begin{figure*}
\subfigure{\includegraphics[width=7.5cm, height=4.7cm]{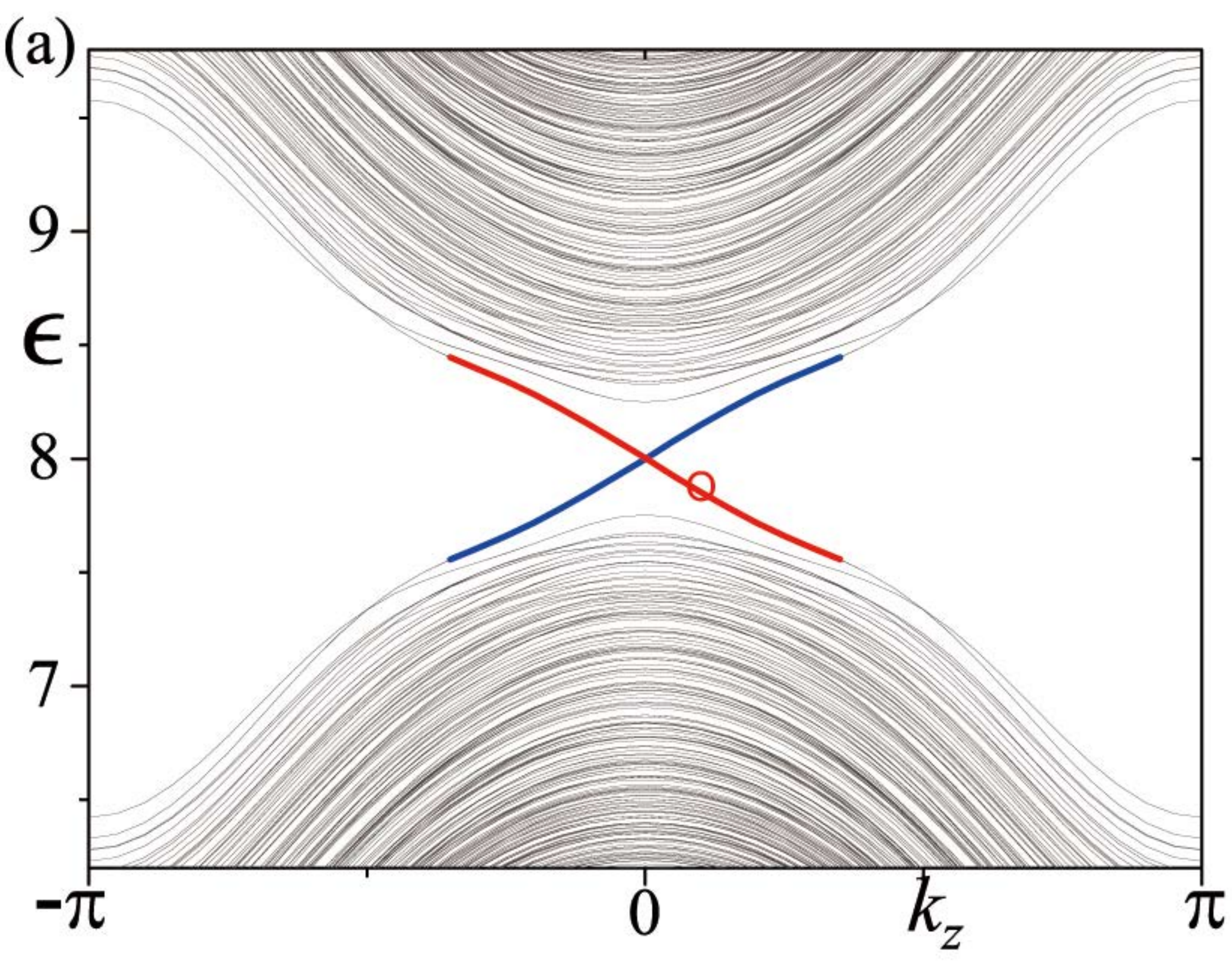}}
\subfigure{\includegraphics[width=4.6cm, height=4.7cm]{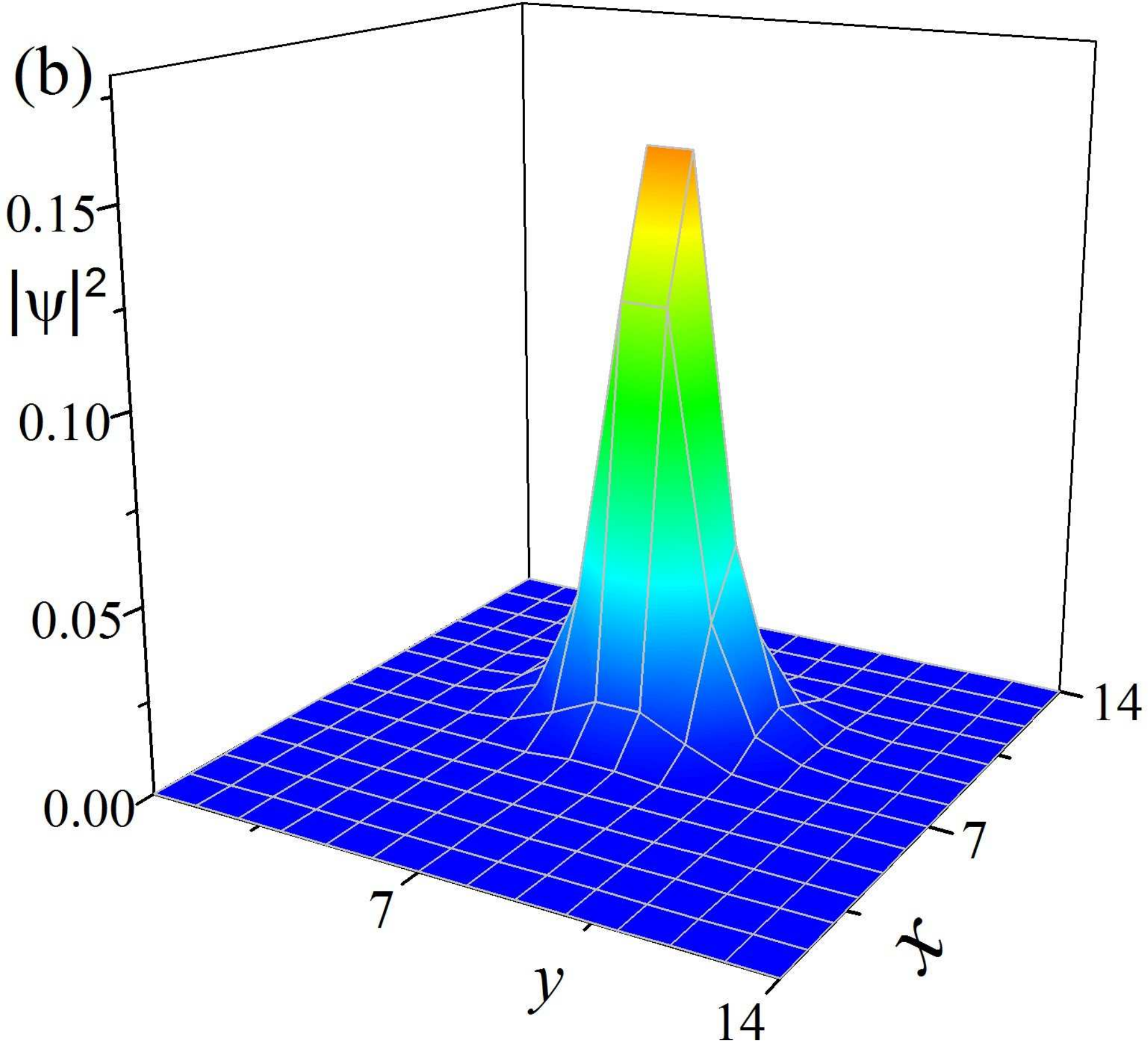}}
\subfigure{\includegraphics[width=4.6cm, height=4.7cm]{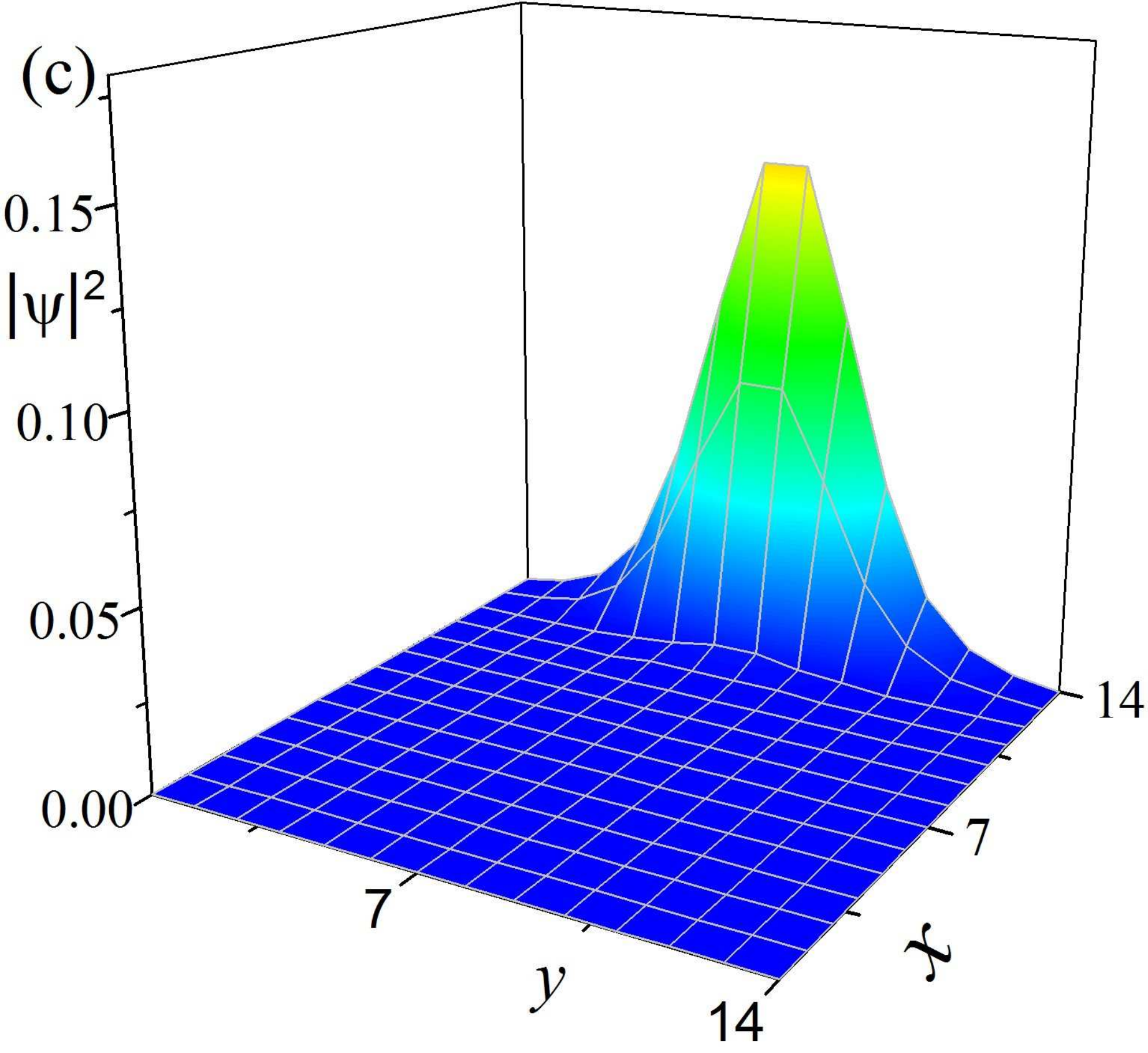}}
\caption{ Helical modes along a line defect in class AII [Eq.(\ref{line-AII})].  Parameters used here are $t_{1}=1.0$, $t_{2}=0.5$, $m_{0}=-4.5$, $\omega=16.0$, $m_{d}=4.0$, and the system size is
$L_{x}\times L_{y}\times L_{z}=14\times 14\times \infty$.
Floquet Hamiltonian truncation is $M=5$. (a):
Quasienergy dispersion near $\epsilon=\omega/2$. Both the blue and red lines are (almost) doubly degenerate.  (b) and
(c): The helical mode profiles at $k_{z}=0.1\pi$ (marked by a red circle in (a)). One of the mode is localized at the sample center, the other is localized at the boundary.  }\label{helical}
\end{figure*}

Helical modes along line defects have been studied in static systems\cite{ran2009one,Slager2014,zhang2015}. It has been pointed out that the screw dislocations in a weak topological insulator carry helical modes\cite{ran2009one}. Helical modes have also been proposed to exist along the lattice dislocations in three-dimensional double-Dirac semimetals with an energy gap generated by symmetry breaking\cite{Wieder2016}. These helical modes belong to static topological defect modes in the AII class\cite{teo2010}.

Here, we study a model of Floquet helical modes along line defects. The Bloch Hamiltonian far from the defect takes the form of
\begin{eqnarray}
H(\bk,\theta,t)&=&2t_{1}(\sin k_{x}\Gamma_{1}+\sin k_{y}\Gamma_{2}+\sin k_{z}\Gamma_{3})+
2t_{1}\sin \theta\Gamma_{4}\nonumber\\
&+& [m(t)-2t_{2}(\cos k_{x}+\cos k_{y}+\cos k_{z}+\cos \theta)]\Gamma_{5},  \quad\quad \label{line-AII}
\end{eqnarray}
where $\Gamma_{1,2,3}=s_{z}\tau_{z}\sigma_{x,y,z}$, $\Gamma_{4}=s_{z}\tau_{x}\sigma_0$,
$\Gamma_{5}=s_{x}\tau_0\sigma_0$, $m(t)=m_{0}+m_{d}\cos\omega t$, and $\theta$ is the polar angle (see Sec. \ref{sec:line-class-A}). This model belongs to class AII because it satisfies \bea T^{-1}H(\bk,\theta,t)T =H^*(-\bk,\theta,-t), \eea with $T=\sigma_y$.
It may be useful to compare this model with Eq.(\ref{line-defect-A}) without the TRS.

The numerical scheme for the defect modes will be similar to Sec. \ref{sec:line-class-A}, which we shall not repeat.
We take the static parameter $m_0=-4.5$, for which the static system has no helical mode at zero energy. With the periodic driving added ($m_d\neq 0$), we find helical modes at $\epsilon=\omega/2$, whose quasienergy dispersion is shown in Fig.\ref{helical}(a). There are in fact two helical modes, one of which is localized at the system center, the other at the boundary. The helical mode profiles at $k_z=0.1\pi$ are shown in Fig.\ref{helical}(b) and Fig.\ref{helical}(c). Their time-reversal partners at $k_z=-0.1\pi$  have the same profiles (thus no need to show repeatedly).

The Floquet helical mode at $\epsilon=\omega/2$ does not require the presence of static helical mode at zero energy, thus it may be realized in the dislocations of trivial insulators, not necessarily of weak topological insulators\cite{ran2009one}.

\begin{figure}
\subfigure{\includegraphics[width=5.5cm, height=4.5cm]{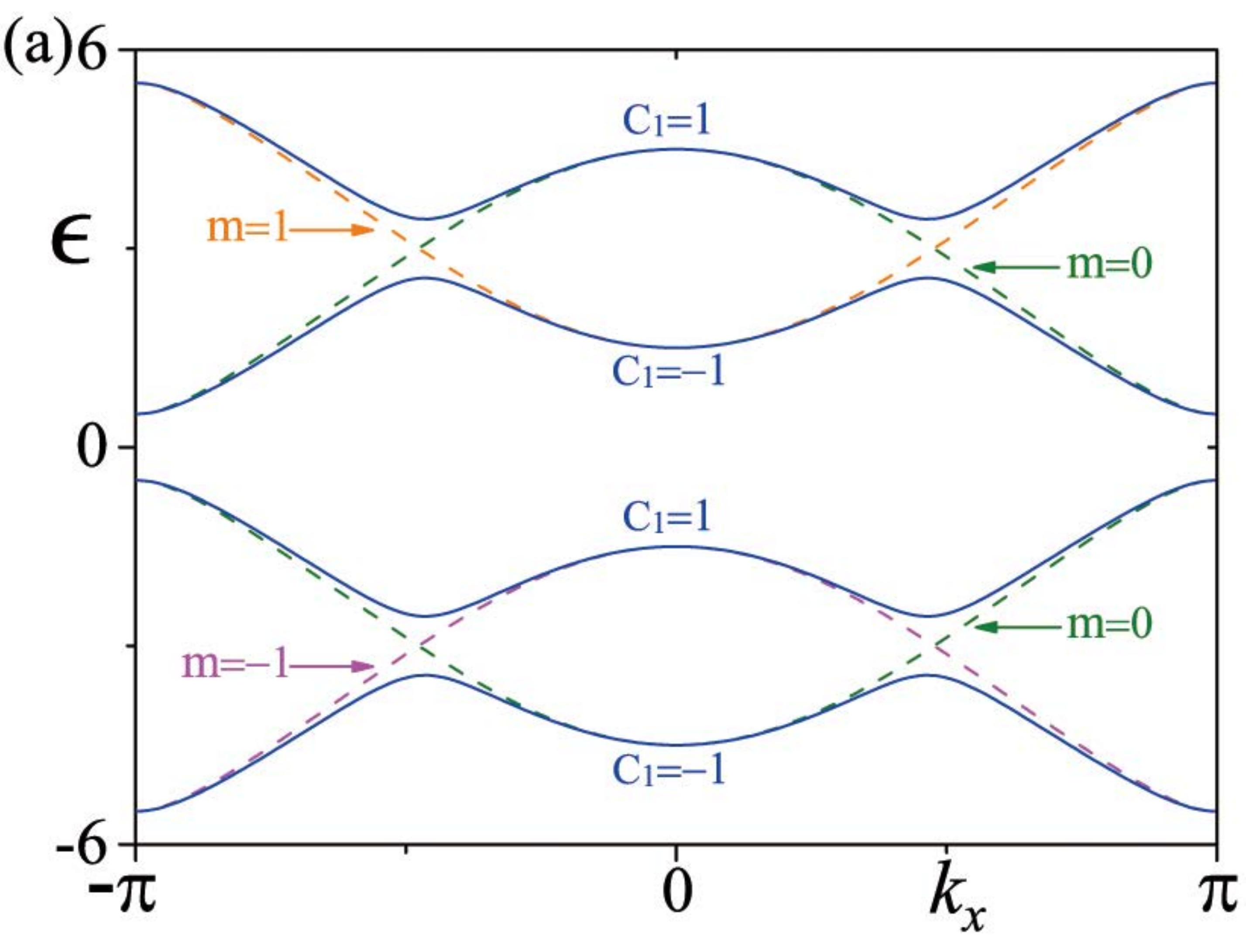}}
\subfigure{\includegraphics[width=5.5cm, height=4.5cm]{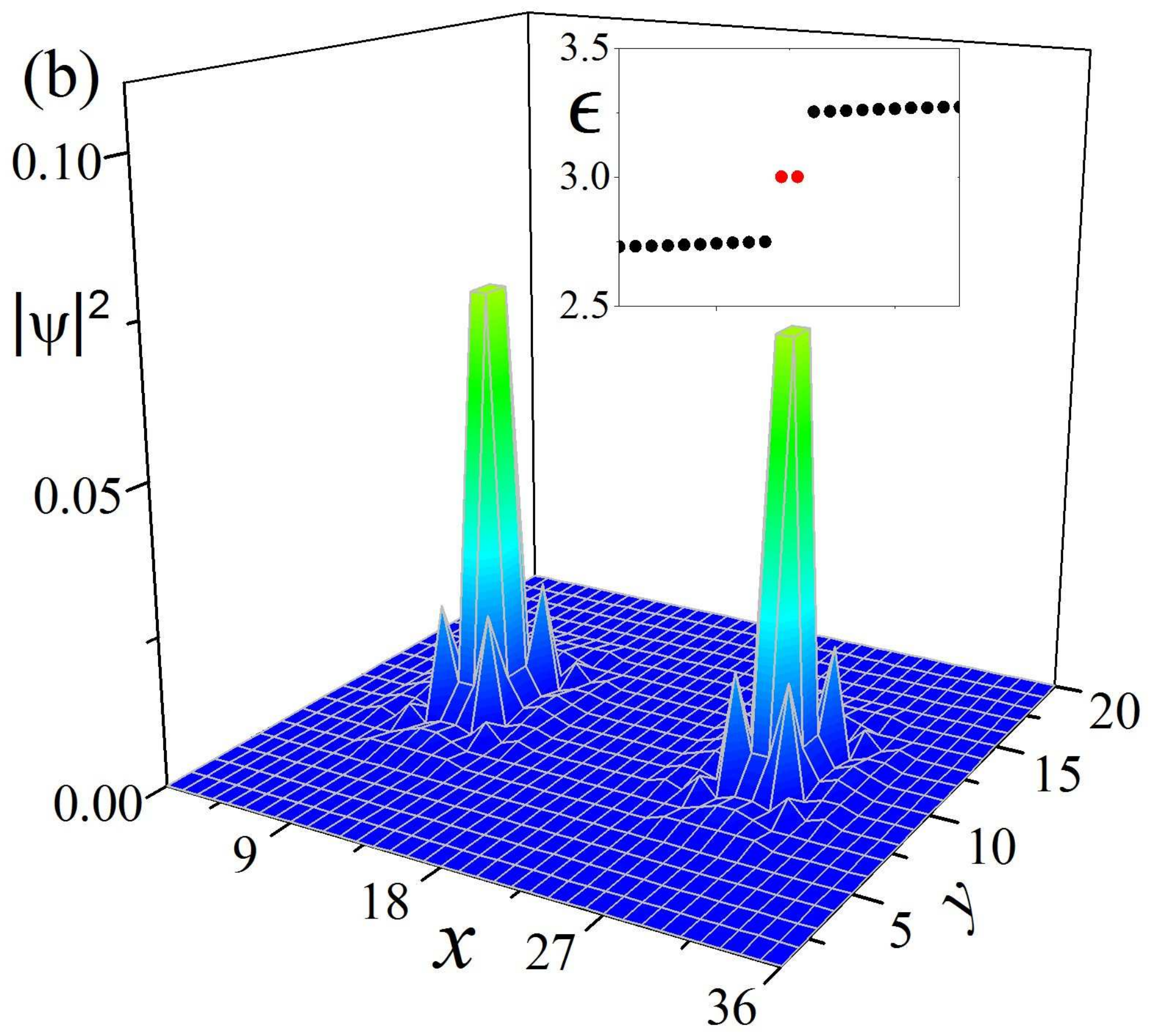}}
\caption{ Bulk Floquet bands and Floquet MPMs for the case (i). Parameters used here are $t=1.0$, $\mu_{0}=-2.5$, $\Delta=1.0$, $\omega=6.0$, and $\mu_{d}=2.0$.
(a): The solid blue lines stand for the bulk Floquet bands plotted along the line $k_{y}=k_{x}$. The $\mu_{d}=0$ bands are shown in dashed lines as a comparison ($m$ stands for the Floquet index) . (b): Two MPMs in the presence of two vortices at $(x,y)=(9.5, 10.5)$ and
$(27.5, 10.5)$, respectively. The system size is
$L_{x}\times L_{y}=36\times20$ with periodic boundary condition.  The inset shows the quasienergy spectra near $\epsilon=\omega/2$. The two quasienergies of MPMs are colored red. Floquet Hamiltonian truncation $M=5$ is taken in the calculation.}\label{STMZM}
\end{figure}

\begin{figure*}
\subfigure{\includegraphics[width=5.5cm, height=4.5cm]{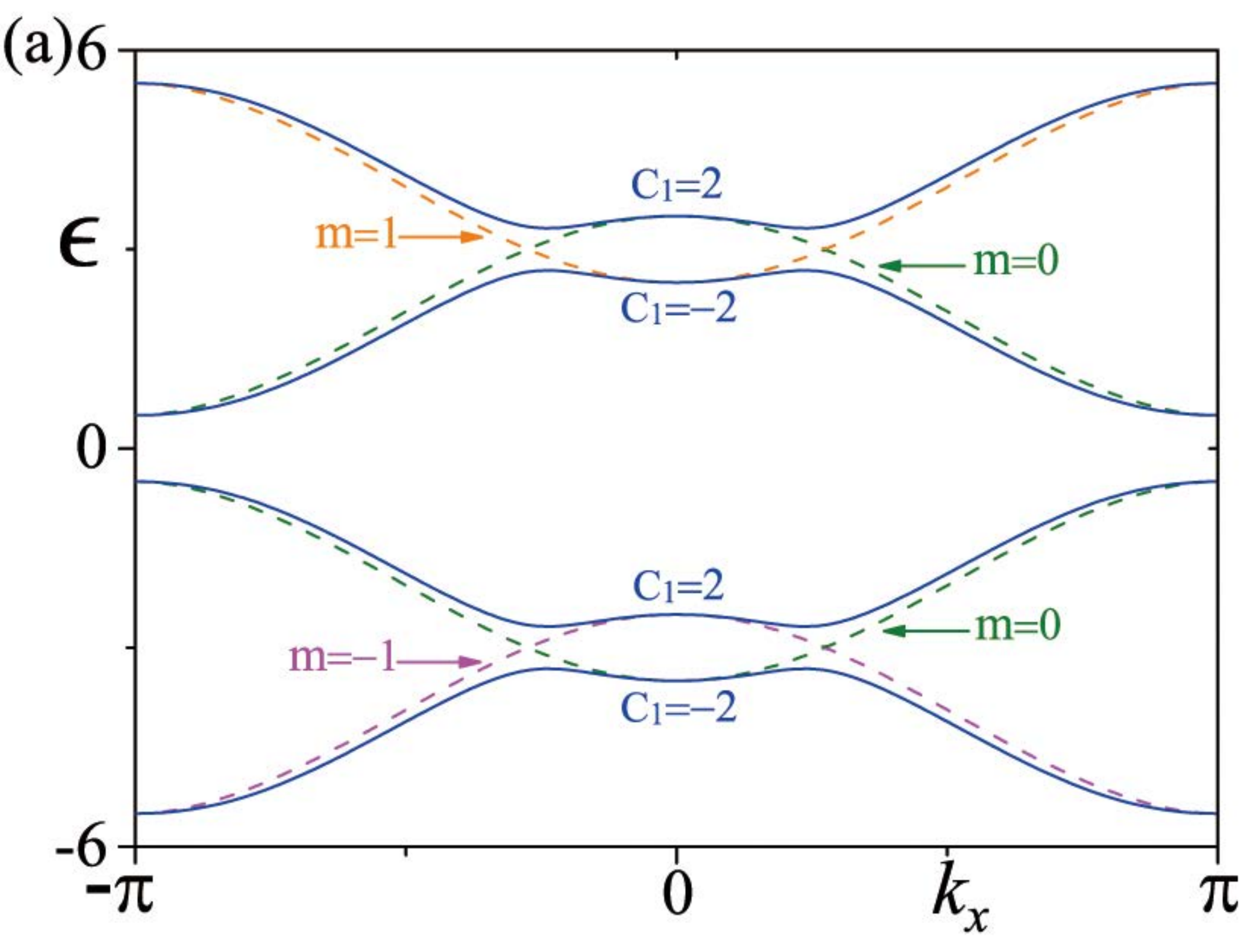}}
\subfigure{\includegraphics[width=5.5cm, height=4.5cm]{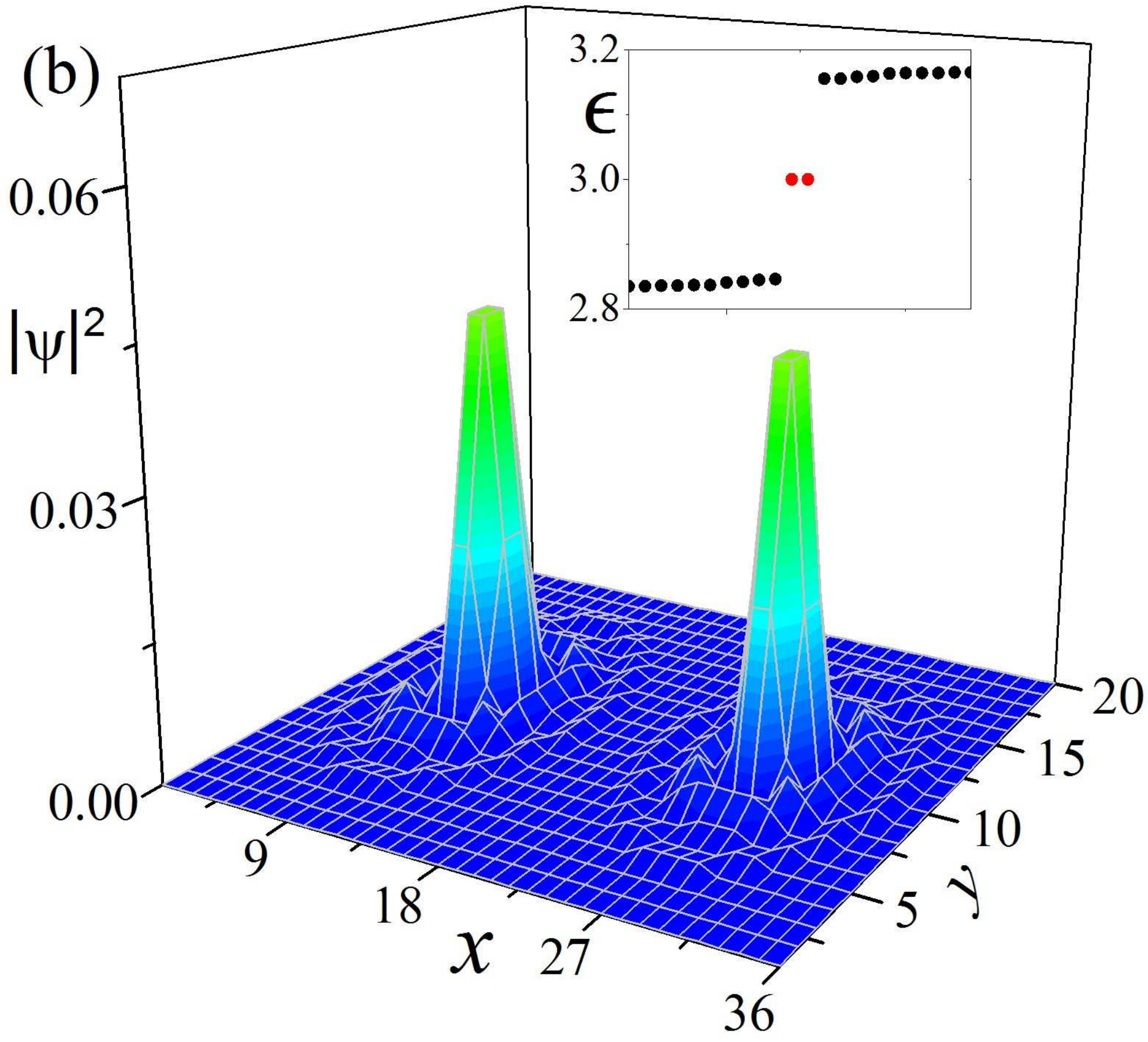}}
\subfigure{\includegraphics[width=5.5cm, height=4.5cm]{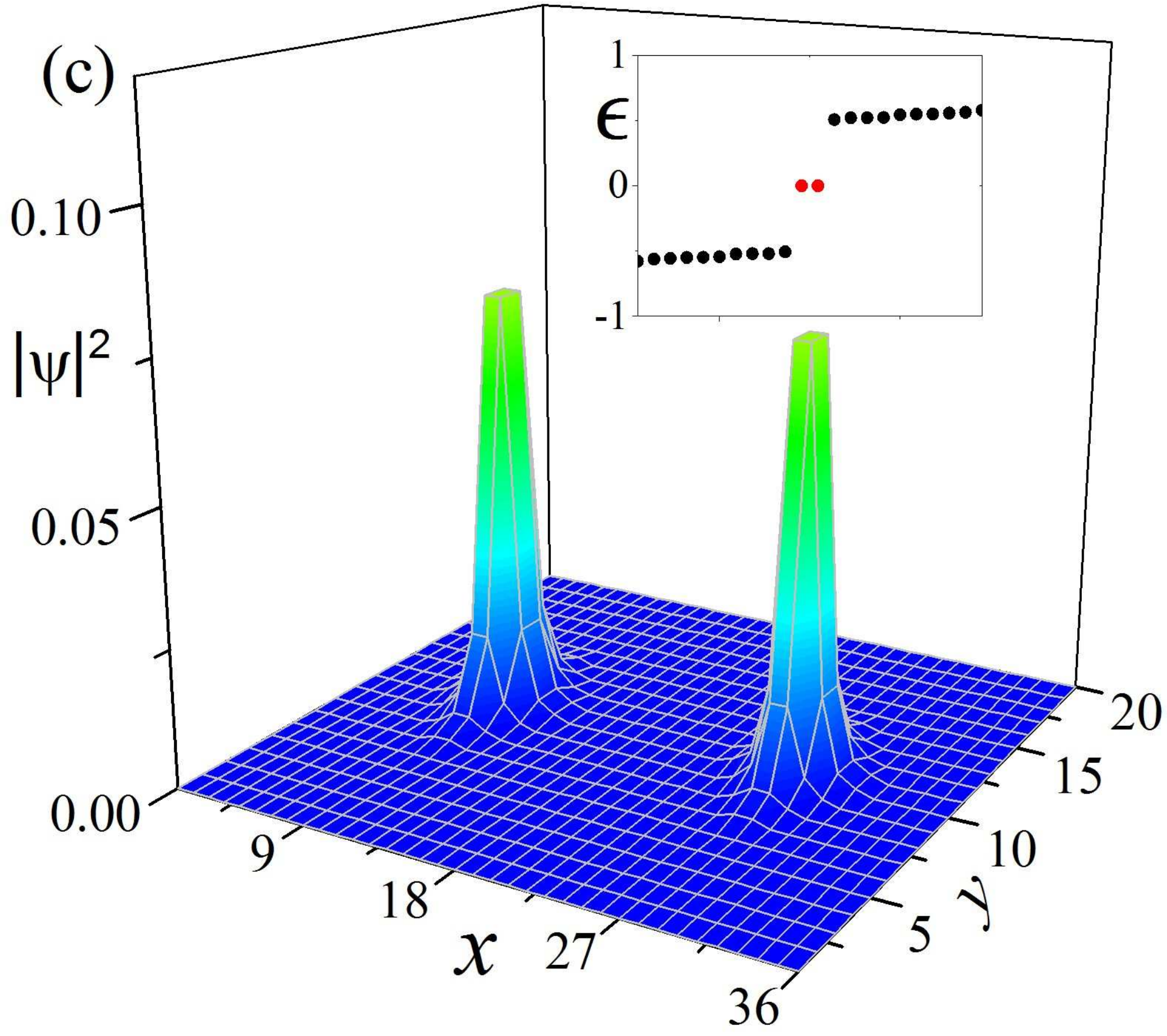}}
\caption{ Bulk Floquet bands and Floquet Majorana modes for the case (ii). Parameters used here are $t=1.0$, $\mu_{0}=-1.5$, $\Delta=1.0$, and $\omega=6.0$.
(a): The solid blue lines stand for the bulk Floquet bands plotted along the line $k_{y}=k_{x}$ (with $\mu_{d}=2.0$). The $\mu_{d}=0$ bands are shown in dashed lines ($m$ stands for the Floquet index). (b)(c): Profiles of the two MPMs (b) and two MZMs (c) in the presence of two vortices at $(x,y)=(9.5, 10.5)$ and
$(27.5, 10.5)$. The system size is
$L_{x}\times L_{y}=36\times20$, with periodic boundary condition. In this calculation, $\mu_d=1.0$, and Floquet Hamiltonian truncation is $M=5$.}\label{SNTMZM}
\end{figure*}

\section{Point defects}\label{sec:point}

Point defects have $d_{\rm def}=0$ and $\delta\equiv d-D=1$. According to Table \ref{table}, topologically nontrivial point defects can exist in classes AIII, BDI, D, DIII, and CII.

\subsection{Class D: Floquet MZMs and MPMs in vortices of topologically trivial superconductors}\label{D-point}

MZMs in static systems have attracted wide attentions in recent years due to their potential applications in topological quantum computations (there are many excellent review articles, for instance, Refs.\cite{alicea2012new,Beenakker2013, leijnse2012introduction,stanescu2013majorana, sarma2015majorana,sato2016majorana,Elliott2015}).

Here, we are concerned with Floquet MZMs, which may also be useful in topological quantum computations\cite{Liu2013Floquet}. We mention that Floquet MZMs at the ends of one-dimensional wires have been investigated before\cite{Jiang2011,Kundu2013,Reynoso2013, Liu2013Floquet,Tong2013Generating,Li2014tunable, Thakurathi2013,Thakurathi2017,Saha2017}. To enable braiding operations, which are crucial in topological quantum computations, two-dimensional systems are more advantageous. In this paper, we study Floquet Majorana modes in the vortex of topologically trivial superconductors under periodic driving.

A simple model of driven homogeneous superconductors is given by the following BdG equation:
\begin{eqnarray}
H(\bk, t)=[t(\cos k_{x}+\cos k_{y}) - \mu(t)]\tau_{z}+\Delta(\sin k_{x}\tau_{x}+\sin k_{y}\tau_{y}),\quad\quad
\end{eqnarray}
where $\Delta$ is a $p$-wave Cooper pairing, and \bea \mu(t)=\mu_{0}+\mu_{d}\cos\omega t \eea stands for a time-dependent fermion energy or chemical potential. In a cold-atom setup, it can be implemented by periodically varying the trap potential of the optical lattice. We will fix $t =1.0$, $\Delta=1.0$, and $\omega=6.0$ below.

Let us consider two representative cases. The case (i) is $\mu_0= -2.5$, for which the Chern numbers are $0$ for the static bands, thus the superconductor is topologically trivial. In fact, the band bottom of $E(\bk)=t(\cos k_x+\cos k_y)$ is $E(\pi,\pi)=-2.0$, and the regime $\mu_0<-2.0$ corresponds to the trivial ``strong-pairing phase''\cite{read2000}. Under the driving of a nonzero $\mu_d$, a quasienergy gap opens at $\epsilon=\omega/2$.  The
bulk Floquet bands are shown in Fig.\ref{STMZM}(a). The Floquet band Chern numbers are also marked in Fig.\ref{STMZM}(a). These Chern numbers are calculated numerically using the Floquet Hamiltonian $\mathcal{H}$, which have been proved equivalent to the Chern numbers calculated from the effective Hamiltonian $H^{\rm eff}$ [see Eq.(\ref{equivalence})].

We are most interested in the quasienergy spectra in the presence of a vortex, for which the Bloch Hamiltonian sufficiently far away from the vortex reads
\begin{eqnarray}
H(\bk,\theta,t)=\left(
         \begin{array}{cc}
          t(\cos k_{x}+\cos k_{y})-\mu(t) & \Delta e^{-i\theta}(\sin k_x - i\sin k_y) \\
           \Delta e^{i\theta}(\sin k_x + i\sin k_y) & -[t(\cos k_{x}+\cos k_{y})-\mu(t)] \\
         \end{array}
       \right),\quad\quad \nn
\end{eqnarray} where $\theta$ is the polar angle viewed from the vortex core. In numerical implementation, we consider a finite size sample with two vortices (more precisely, a vortex and an anti-vortex).  We can do a gauge transformation to eliminate the angle-dependent phase factor $\exp(i\theta)$ in the Cooper pairing. Accordingly, the periodic boundary condition of fermions around the vortex becomes the anti-periodic boundary condition, namely, all the hoppings across the straight line connecting the two vortices are multiplied by a $-1$ factor. We find two localized Majorana modes at $\epsilon=\omega/2$ (equivalently, $\varepsilon=\pi$), as shown in Fig.\ref{STMZM}(b). These Majorana modes are Floquet versions of the MZMs of static systems, which are also protected by the particle-hole symmetry. They are dubbed the Majorana Pi modes (MPMs)\cite{Potter2016classification}. No MZM is found in this case, which is consistent with the static system being a topologically trivial superconductor.

Now we consider the case (ii) with $\mu_0=-1.5$, for which the Chern numbers of the static bands are $\pm 1$.  Without driving, it corresponds to the topologically nontrivial ``weak-pairing phase''\cite{read2000}. The Floquet bands are shown in Fig.\ref{SNTMZM}(a), with their Chern numbers marked. The profiles of the MPMs and MZMs are shown in Fig.\ref{SNTMZM}(b) and Fig.\ref{SNTMZM}(c), respectively. We note that the Floquet band topological invariants are all $\mathbb{Z}_2$ trivial in the case (ii), namely, the band Chern numbers are all even integers. In static systems, an even-integer Chern number implies that a vortex carries no robust MZM. In fact, the static $\mathbb{Z}_2$ topological invariant of point defects of class D is just the product of the Chern number and the vorticity (which is unity here)\cite{teo2010}. In the sense that all Floquet bands are $\mathbb{Z}_2$ trivial, the MPMs and MZMs here are anomalous (in the terminology of Ref.\cite{rudner2013anomalous}).

To summarize, in case (i),  we found MPMs in the vortex of topologically trivial superconductors; in case (ii), we found both MZMs and MPMs, though all the Floquet bands are $\mathbb{Z}_2$ trivial (with even-integer Chern numbers).

A few remarks before concluding this section. First, the Floquet MZMs and MPMs may be detected experimentally by a quantized conductance sum rule\cite{Kundu2013} or heat transfer\cite{Molignini2017}. Second,
possible Floquet MZMs and MPMs located in driven disclinations, whose static counterparts have been studied\cite{Teo2013,benalcazar2014classification}, will also be interesting to study. Third, the potential applications of MPMs in topological quantum computation calls for further investigations.

\subsection{Class AIII: Point defects carrying zero modes and Pi modes }\label{AIII-point}

\begin{figure*}
\subfigure{\includegraphics[width=5.5cm, height=4.5cm]{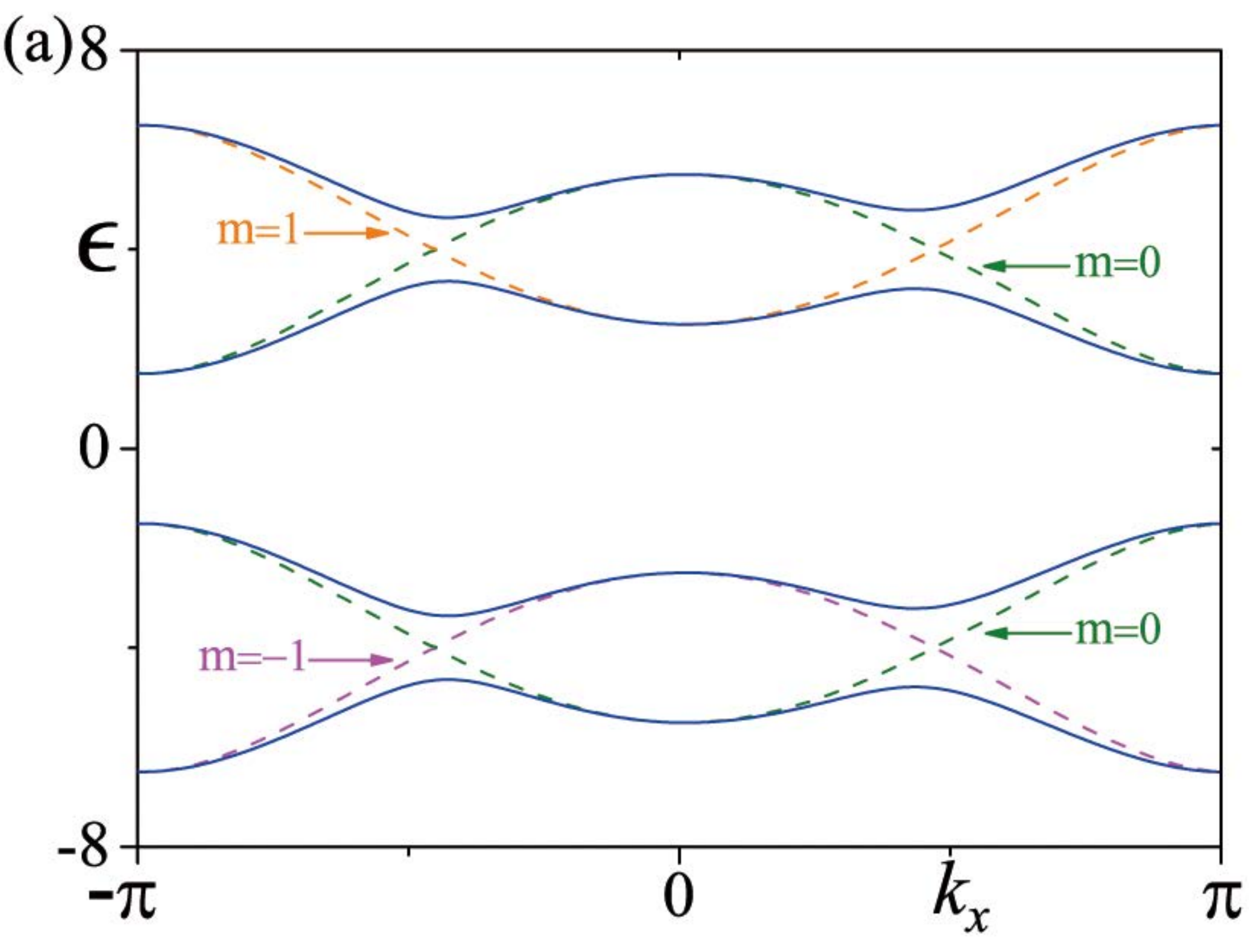}}
\subfigure{\includegraphics[width=5.5cm, height=4.5cm]{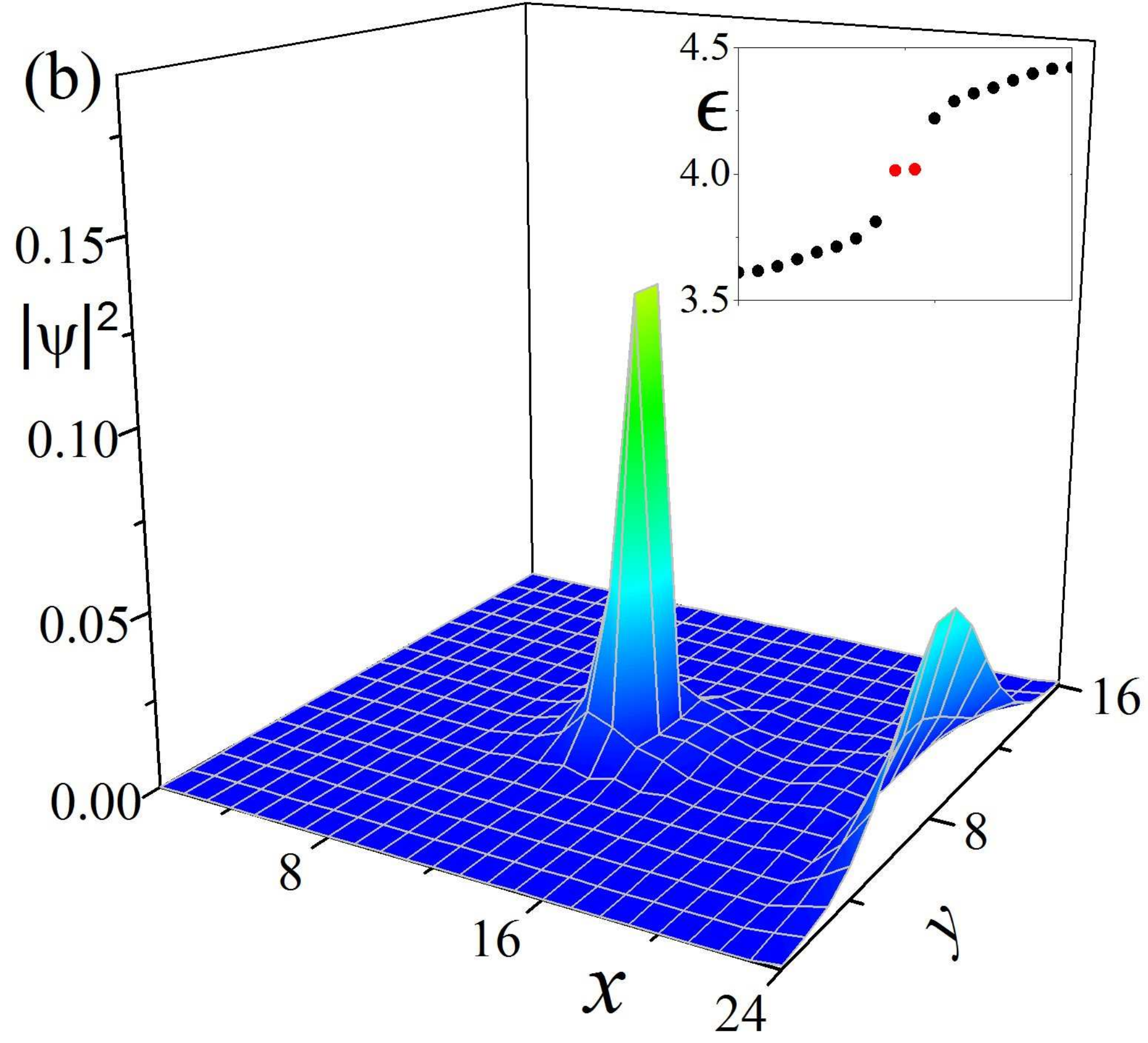}}
\subfigure{\includegraphics[width=5.5cm, height=4.5cm]{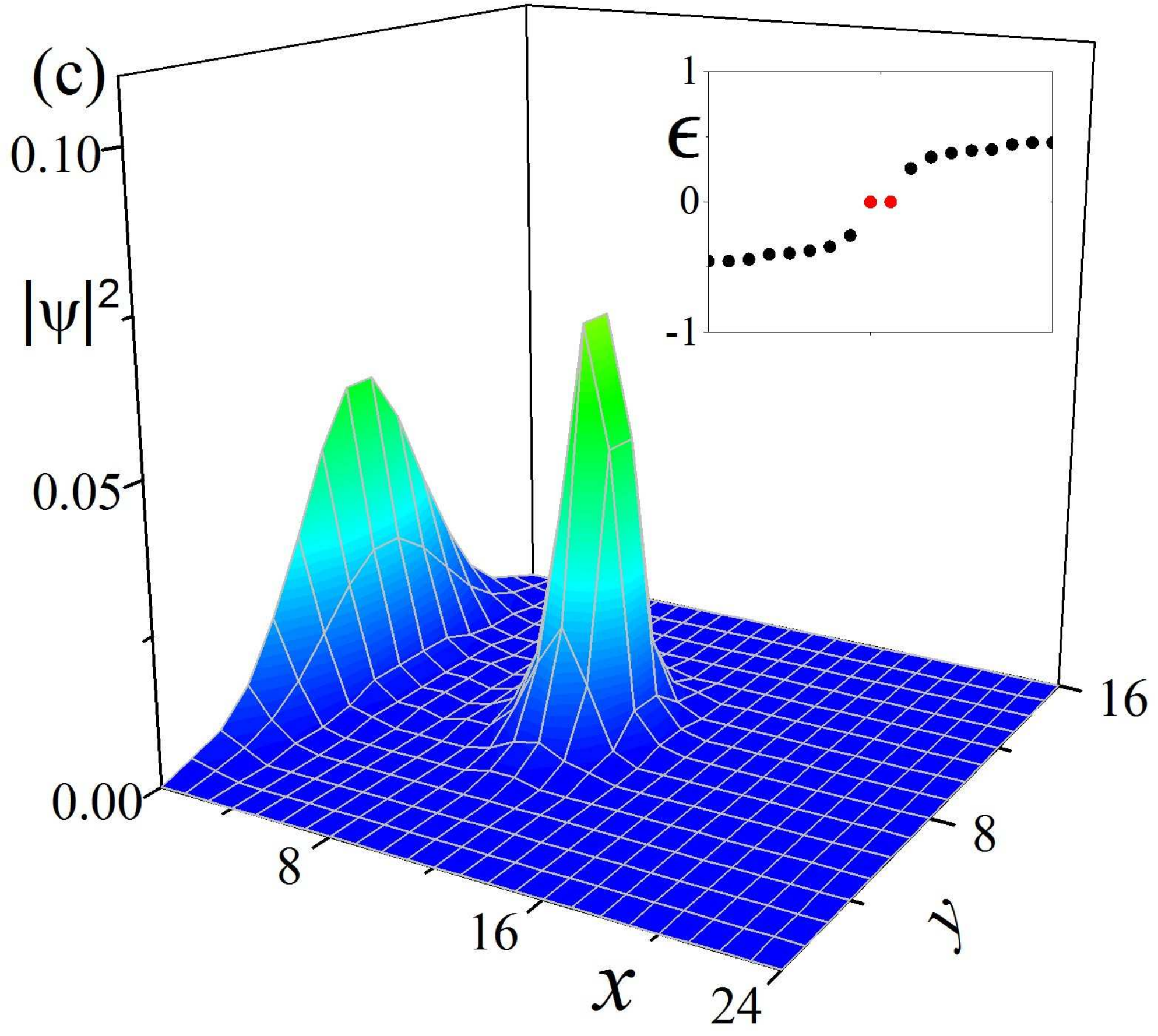}}
\caption{ (a): Bulk Floquet bands plotted along the line $k_{y}=k_{x}$ (with $\theta=0$ fixed).  Parameters used here are $m_{0}=-2.5$, $\omega=8.0$, $\delta=0.2$, and $m_{d}=5.0$.
The Floquet bands of $m_{d}=0$ are shown in dashed lines, with Floquet index $m$ marked.  (b): Floquet Pi modes profiles in a system with size $L_{x}\times L_{y}=24\times16$ (open boundary condition). The point defect center is $(x_{0}, y_{0})=(12.5, 8.5)$. Floquet Hamiltonian truncation is $M=5$. The inset shows quasienergies close to $\omega/2$. (c): Floquet zero modes profiles and quasienergies close to $0$. }\label{SNTchiralinsulator}
\end{figure*}

Let us consider a lattice model of point defect of class AIII in two-dimensional space. Sufficiently far away from the point defect, the translational symmetry is restored and the Bloch Hamiltonian is given as
\begin{eqnarray}
H(\bk,\theta, t)=\left(
         \begin{array}{cc}
           0 & q(\bk,\theta, t) \\
           q^{\dag}(\bk,\theta,t) & 0 \\
         \end{array}
       \right),
\end{eqnarray}
where $\theta$ is the polar angle viewed from the defect center, i.e., $\theta= \arctan[(y -y_0)/(x -x_{0})]$ (Here, $(x_0,y_0)$ is the defect center location), and
\begin{eqnarray}
q(\bk,\theta,t)&=&-i(m(t)-\cos k_{x}-\cos k_{y}-\cos \theta)\sigma_{0}\nonumber\\
&&+(\sin k_{x}+\delta)\sigma_{x}+\sin k_{y}\,\sigma_{y}+\sin\theta\,\sigma_{z},
\end{eqnarray}
with $m(t)=m_{0}+m_{d}\cos \omega t$. This model belongs to class AIII as it satisfies \bea S^{-1}H(\bk,\theta,t)S =-H(\bk,\theta,-t), \eea with
\begin{eqnarray}
S=\left(
         \begin{array}{cc}
          I &   \\
             & -I \\
         \end{array}
       \right).
\end{eqnarray} While preserving the chiral symmetry, the $\delta$ term is included to break other symmetries. Without the $\delta$ term, the Hamiltonian also has the time reversal symmetry. We mention that a mathematically similar Hamiltonian has been studied as a model of three-dimensional Floquet topological insulators in class AIII\cite{Fruchart2016}, though the purpose there was unrelated to topological defects. Other aspects of bulk Floquet systems in class AIII have also been investigated in Ref.\cite{Ho2014,Asboth2014}.

In terms of Dirac matrices, the model can be written as
\begin{eqnarray}
H(\bk,\theta)&=&(\sin k_{x}+\delta)\Gamma_{1}+\sin k_{y}\Gamma_{2}+\sin\theta\Gamma_{3}\nonumber\\
&&+(m(t)-\cos k_{x}-\cos k_{y}-\cos \theta)\Gamma_{4},\label{point-AIII}
\end{eqnarray}
where $\Gamma_{1,2,3}=\sigma_{x,y,z}\tau_{x}$, $\Gamma_{4}=\tau_{y}$. After a Fourier transformation to real space, we have
\begin{eqnarray}
\hat{H}&=&\sum_{x,y}\left\{c^{\dag}_{x,y}[\delta\,\Gamma_{1}+\sin\theta_{x,y}\Gamma_{3}+(m(t)-\cos\theta_{x,y})\Gamma_{4}]c_{x,y}\right.\nonumber\\
&&-(\frac{i}{2}c^{\dag}_{x,y}\Gamma_{1} c_{x+1,y}+\frac{i}{2}c^{\dag}_{x,y}\Gamma_{2} c_{x,y+1}+h.c.)\nonumber\\
&&\left.-(\frac{1}{2}c^{\dag}_{x,y}\Gamma_{4} c_{x+1,y}+\frac{1}{2}c^{\dag}_{x,y}\Gamma_{4} c_{x,y+1}+h.c.)\right\},
\end{eqnarray} where $(x,y)$ are integer-valued real space coordinates. In fact, this real-space Hamiltonian is just one of the many realizations of the point defect described by Eq.(\ref{point-AIII}), since we only require that the Bloch Hamiltonian approaches Eq.(\ref{point-AIII}) far away from the defect, therefore, modifying the real-space Hamiltonian in the vicinity of defect does not change the topological classification.

For a set of parameters, we plot the bulk Floquet bands in
Fig.\ref{SNTchiralinsulator}(a). Floquet Pi modes and Floquet zero modes are found in the presence of a point defect, as shown in Fig.\ref{SNTchiralinsulator}(b) and (c). Closer inspection of the mode wave functions shows that both the zero mode and the Pi mode have sublattice index (or ``chirality'') $S=-1$.  For both the zero mode and Pi mode at the defect center, there is a partner mode at the system boundary, whose chirality is $S=+1$.

\begin{figure}
\includegraphics[width=9cm, height=6.0cm]{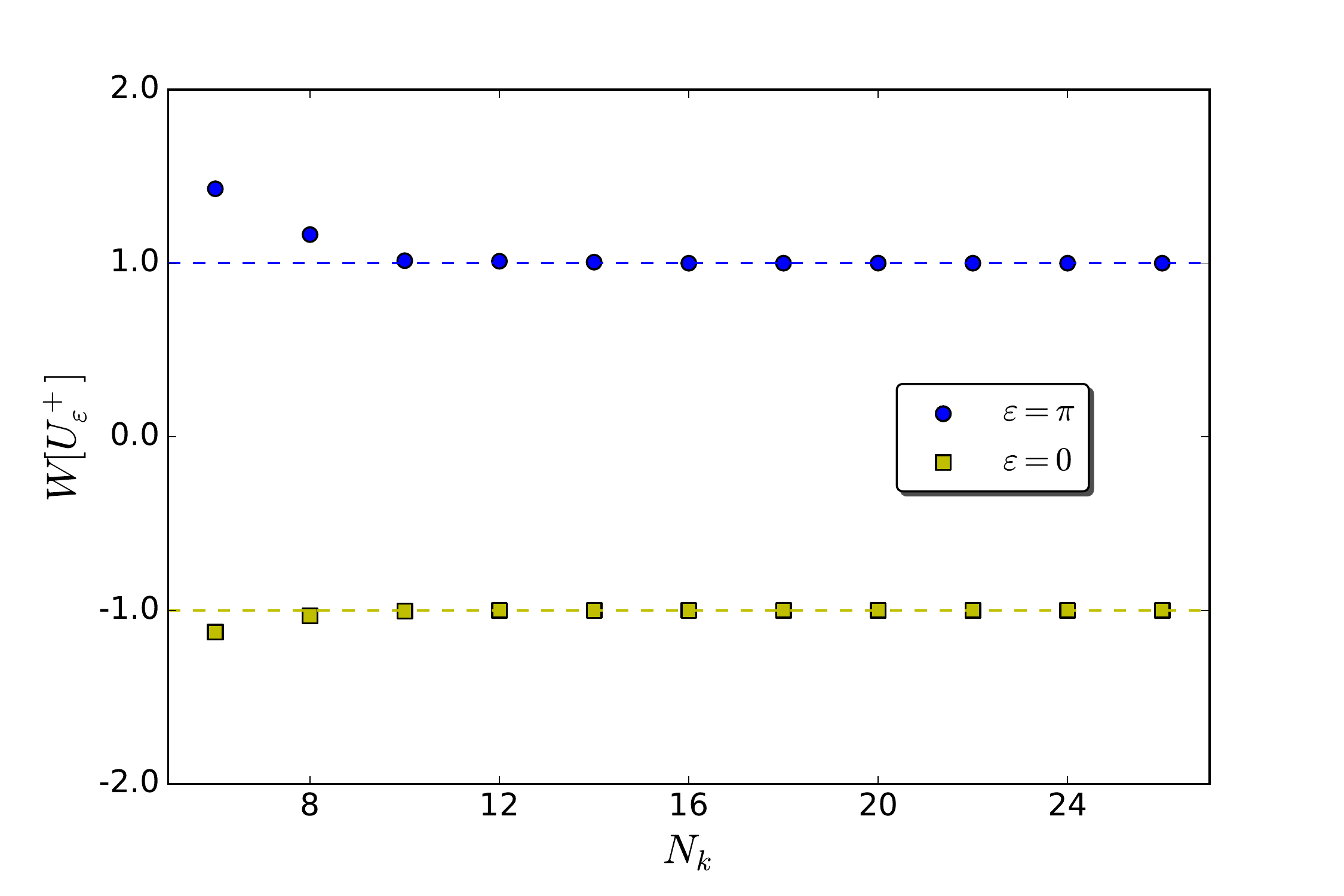}
\caption{ Numerical evaluation of the topological invariant of the point defect in class AIII. Parameters used here are the same as in Fig.\ref{SNTchiralinsulator}. We use $N_k^3$ grid points in the integral domain of $(\bk,\theta)$.   The numerical topological invariant converges rapidly to integers as the grid becomes finer.  For $N_k=26$, we get $W(U^+_{\varepsilon=\pi})= 1.000024$ and $W(U^+_{\varepsilon=0})=-1.000026$.  The grid point number in the $t$ direction is fixed as 800 in calculating $H^\text{eff}_\varepsilon$, which is needed in the definition of $U_\varepsilon^+$. }  \label{topo-inv-aiii}
\end{figure}

Now let us calculate the topological invariant of this point defect.  The topological invariant of class AIII has been given in Eq.(\ref{AIII-inv}), in which we should take $d=2$ and $D=1$ here.
We have numerically calculated this topological invariant, obtaining $W(U^+_{\varepsilon=\pi})=1$ and $W(U^+_{\varepsilon=0})=-1$ (see Fig.\ref{topo-inv-aiii}). These topological numbers are indeed consistent with the presence of one Floquet Pi mode and one zero mode in the defect.

There is nevertheless an important yet subtle point about the sign, which we now explain. Let $n_0$ and $n_\pi$ be the net number of modes at $\epsilon=0$ and $\epsilon=\omega/2$, respectively, namely, $n_{0/\pi}$ is the number of zero (or Pi) modes with $S=+1$ minus that of modes with $S=-1$. By analog with class A in two dimensions, for which the band Chern number measures the difference between the numbers of chiral edge modes above and below the bands, one may tempted to guess that $n_\pi-n_0=W(q_{0,\pi}^\dag)$, where $W(q_{0,\pi}^\dag)$ stands for the winding number of the Floquet bands with $\varepsilon\in[0,\pi]$ (or equivalently, $\epsilon\in[0,\omega/2]$) (for the notation of $q^\dag_{0,\pi}$, see Eq.(\ref{Q})), however, this expectation is incorrect. In fact, the same analog would also suggest that $n_0-n_{-\pi}=W(q_{-\pi,0}^\dag)$, but it is not difficult to check the relation $q_{-\pi,0}^\dag=-q_{0,\pi}^\dag$, and consequently, $W(q_{0,\pi}^\dag)=W(q_{-\pi,0}^\dag)$. Now the relation $n_\pi-n_0=W(q_{0,\pi}^\dag)$ is inconsistent with $n_0-n_{-\pi}=W(q_{-\pi,0}^\dag)$ because $n_{\pi}=n_{-\pi}$. We find that the correct relation should be \bea  -n_\pi -n_0 = W(q_{0,\pi}^\dag), \eea which is consistent with $-n_0 -n_{-\pi}= W(q_{-\pi,0}^\dag)$. This relation has been corroborated by our numerical results: $n_0=n_\pi=-1$ and $W(q_{0,\pi}^\dag)=W(q_{-\pi,0}^\dag)=2$. We have also taken other Hamiltonian parameters so that the values of $n_0,n_\pi$ and $W(q_{0,\pi}^\dag)$ are different, yet the relation remains valid. Because we also have \bea W(U^+_{\varepsilon=\pi})- W(U^+_{\varepsilon=0}) =  W(q_{0,\pi}^\dag), \label{AIII-difference} \eea it is natural to postulate that \bea W(U^+_{\varepsilon=\pi})=-n_\pi, \quad W(U^+_{\varepsilon=0})=n_0. \label{chiral-number} \eea Note the minus sign in the first equation. These relations have indeed been verified in our numerical results for various Hamiltonian parameters. Alternatively, one may redefine the topological invariant by adding a minus sign to $W(U^+_{\varepsilon=\pi})$ so that $W(U^+_{\varepsilon=\pi})=n_\pi$, however, this would modify the desirable relation in Eq.(\ref{AIII-difference}) to the less illuminating one: $-W(U^+_{\varepsilon=\pi})- W(U^+_{\varepsilon=0}) =  W(q_{0,\pi}^\dag)$. Therefore, we do not take this alternative definition.

From the derivation given above, we can see that Eq.(\ref{chiral-number}) should be a general relation for all chiral classes (AIII, BDI, DIII, CII, CI) in all spatial dimensions. We have indeed verified this in a few models (To control the length of this paper, we will not discuss them here).

\subsection{Class CII: Pairs of Pi modes in a point defect}\label{CII-point}

For a point defect in class CII ($\delta=1$), the topological invariant is always an even integer (see Sec. \ref{sec:chiral}), therefore, we expect that there are even numbers of zero modes and Pi modes in the defect.

We put forward a model of point defect as follows. Sufficiently far away from the defect, where the translational symmetry is restored, the Bloch Hamiltonian reads
\begin{eqnarray}
H(\bk,\theta,t)&=&2t_{1}\mu_{z}\tau_{x}(\sin k_{x}\sigma_{x}+\sin k_{y}\sigma_{y})+2t_{1}\sin\theta\mu_{x}\tau_{y}\sigma_{z}\nonumber\\
&&+[m(t)-2t_{2}(\cos k_{x}+\cos k_{y}+\cos\theta)]\mu_{x}\tau_{x}, \label{point-CII}
\end{eqnarray} where $\theta$ is the polar angle viewed from the defect core. The reason of considering $8\times 8$ Dirac matrices can be understood by Appendix \ref{sec:even}, to which the interested readers may refer. The Hamiltonian satisfies the following defining symmetries:
\bea T^{-1}H(\bk,\theta,t)T  &=& H^*(-\bk,\theta,-t), \nn \\ C^{-1}H(\bk,\theta,t)C  &=& -H^*(-\bk,\theta,t), \quad \eea where $T=\tau_0\sigma_y$ and $C=\tau_z\sigma_y$. Since $T^*T=C^*C=-I$, this model belongs to the class CII.

A real space form of the above Hamiltonian is
\begin{eqnarray}
\hat{H}&=&\sum_{x,y}\left\{c^{\dag}_{x,y}[2t_{1} \sin\theta_{x,y}\mu_{x}\tau_{y}\sigma_{z}+(m(t)-2t_{2}\cos\theta_{x,y})\mu_{x}\tau_{x}]c_{x,y}\right.\nonumber\\
&&-(it_{1}c^{\dag}_{x,y}\mu_{z}\tau_{x}\sigma_{x}c_{x+1,y} +it_{1}c^{\dag}_{x,y}\mu_{z}\tau_{x}\sigma_{y}c_{x,y+1}+h.c.)\nonumber\\
&&\left.-(t_{2}c^{\dag}_{x,y}\mu_{x}\tau_{x}c_{x+1,y}+t_{2}c^{\dag}_{x,y}\mu_{x}\tau_{x} c_{x,y+1}+h.c.)\right\}. \label{point-CII-real}
\end{eqnarray} At sufficiently large distance from the defect center, $\theta_{x,y}$ can be taken as a constant locally, and the Fourier transformation of Eq.(\ref{point-CII-real}) is just Eq.(\ref{point-CII}).

\begin{figure}
\includegraphics[width=7cm, height=5cm]{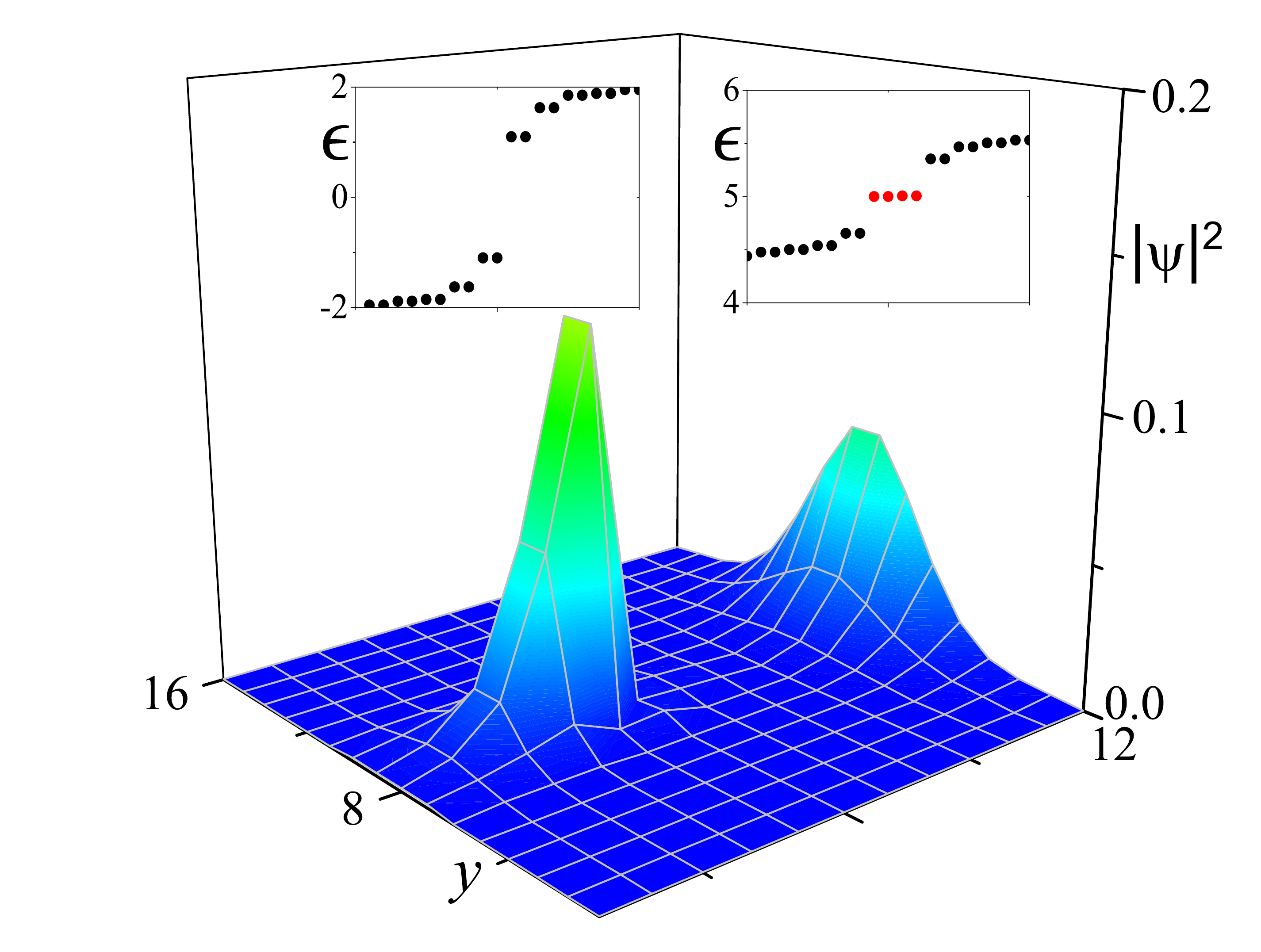}
\caption{ Profiles of localized Pi modes in a point defect in class CII. The parameters used here are $t_{1}=1.0$, $t_{2}=0.5$, $m_{0}=-3.5$, $m_{d}=4.0$, and $\omega=10.0$. The system size is $L_{x}\times L_{y}=12\times16$, with a point defect at $(x_{0},y_{0})=(4.5,8.5)$. The inset shows the quasienergies close to $0$ and $\omega/2$. Two of the four Pi modes are shown (the other two have the same profiles). }\label{TRS8}
\end{figure}

We have calculated the quasienergy spectra of a finite size system with open boundary conditions, and find four Pi modes (Fig.\ref{TRS8}),  two of which are localized at the defect center, while the other two are localized at the system boundary. No zero mode is found for the parameters we choose here.

\begin{figure}
\includegraphics[width=9cm, height=6.0cm]{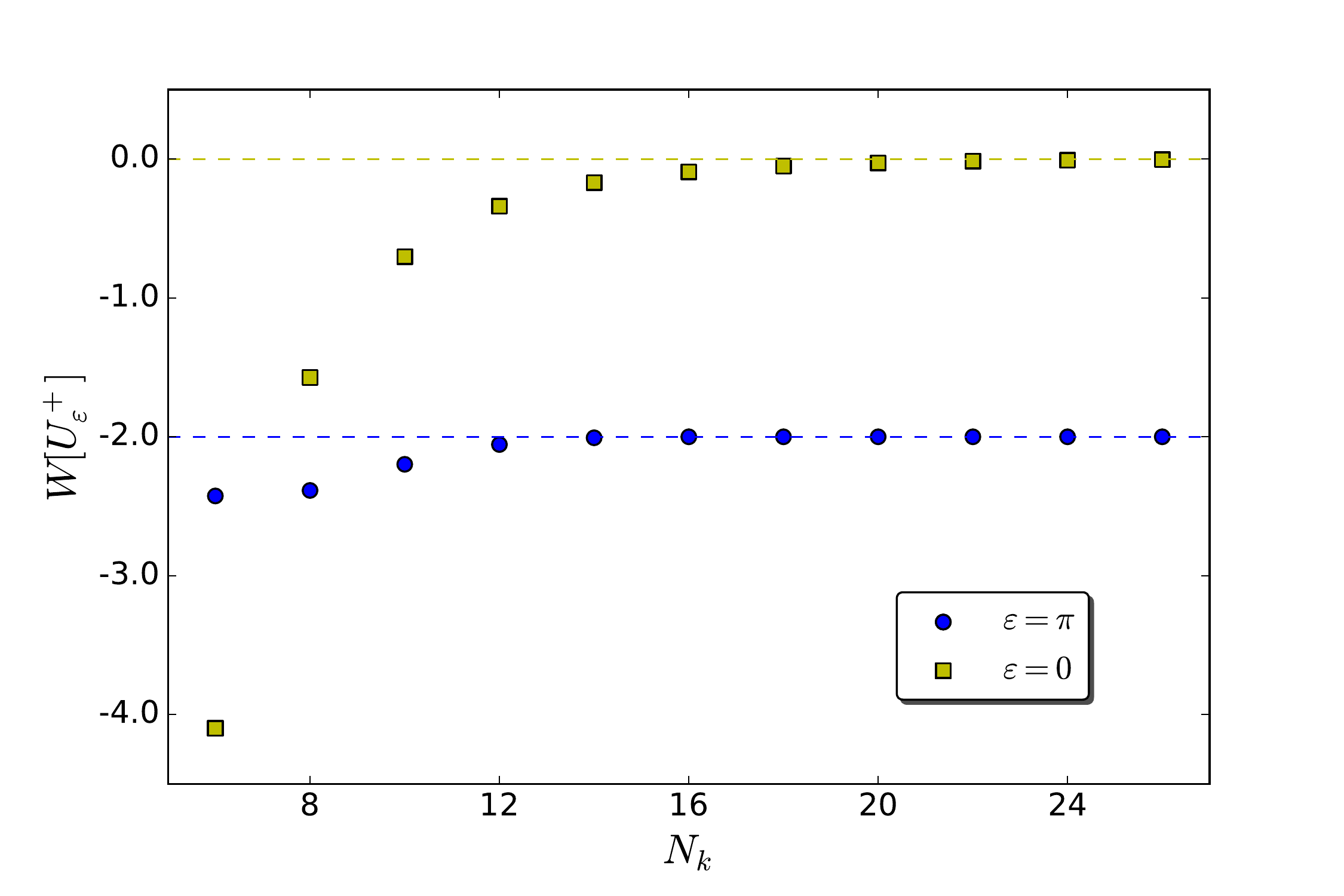}
\caption{ Numerical evaluation of the topological invariant of the point defect in class CII (Eq.(\ref{winding-d2D1})). Parameters used here are the same as those used in Fig.\ref{TRS8}. We use $N_k^3$ grid points in the integral domain of $(\bk,\theta)$.  The numerical topological invariant converges rapidly to integers as the number of grid points increases.  For $N_k=26$, we get $W(U^+_{\varepsilon=\pi})= -2.000062$ and $W(U^+_{\varepsilon=0})=-0.0049$.  The grid point number in the $t$ direction is fixed as 800 in calculating $H^\text{eff}_\varepsilon$, which is needed in the definition of $U_\varepsilon^+$. }  \label{topo_cii}
\end{figure}

The topological invariant for this point defect is the $(d,D)=(2,1)$ case of the general formula of winding number given in Eq.(\ref{windingdensity}), whose explicit form is
\begin{equation}
\begin{aligned}
&W(U_{\varepsilon}^+(\bk,\theta))=K_{3}\int_{T^2\times S^1} d^2k d\theta\, \text{Tr}\{\epsilon^{\alpha_1\alpha_2\alpha_{3}} [(U_{\varepsilon}^+)^{-1}\partial_{\alpha_{1}}U_{\varepsilon}^+] \\
&\times [(U_{\varepsilon}^+)^{-1}\partial_{\alpha_{2}}U_{\varepsilon}^+] [(U_{\varepsilon}^+)^{-1}\partial_{\alpha_{3}}U_{\varepsilon}^+]\}.\label{winding-d2D1}
\end{aligned}
\end{equation}
We have numerically calculated this topological invariant, which yields $W(U^+_{\varepsilon=\pi})=-2$ and $W(U^+_{\varepsilon=0})=0$ to high accuracy, as illustrated in Fig.\ref{topo_cii}. Apparently, the values of topological invariant are consistent with the numbers of zero modes and Pi modes found numerically.

\section{Conclusions}

In conclusion, we have formulated topological invariants of Floquet systems in all spatial dimensions based on the cooperation of topology and symmetries of the time evolution operators. All these topological invariants take the forms of (the usual or the Wess-Zumino-Witten) winding numbers, though the relevant parameter spaces depend on the symmetries and spatial dimensions. The simple combination $\delta=d-D$, the periodicity, the $\mathbb{Z}_2$ topological invariants, and many other results, are obtained in an explicit and natural way. This unified framework of Floquet topological invariants will be useful in future investigations of Floquet systems. In addition, we have raised and clarified several notable issues about topological invariants (such as the equivalence between different approaches to Floquet band topological invariants).

In some sense, the effects of symmetries are more transparent in the Floquet topological invariants than in static topological invariants such as Chern numbers, which are generally expressed in terms of the eigenvectors of the Hamiltonian, instead of the Hamiltonian itself. The way Hamiltonian enters the Floquet topological invariants is more straightforward.

Our formulations of topological invariants are applicable to both homogeneous systems and topological defects. Based on these topological invariants, we have developed a general theory of Floquet topological defects in the tenfold way. This part can be regarded as a Floquet generalization of Ref.\cite{teo2010}, which is a comprehensive study of topological defects in static systems.

Taking lattice models as tools, we have studied a variety of possible approaches of realizing low-dimensional Floquet topological defects, which are  experimentally most relevant. In particular, we show that Majorana Pi modes (the Floquet version of MZMs) can be realized in vortices of topologically trivial superconductors under periodic driving, which suggests an interesting platform for the MZMs and MPMs.  Let us also mention that the higher-dimensional
topological invariants are not merely of theoretical interests; they are useful to the physics of quasicrystals\cite{Kraus2012,Kraus2013,Verbin2013,Prodan2015} and synthetic dimensions\cite{Celi2014,Mancini2015,stuhl2015visualizing, Zeng2015charge,Yan2015}, which can be experimentally studied in low dimensions.

Finally, we should mention that the many-body effects in Floquet systems have been under active investigations\cite{Lazarides2015,Ponte2015,DAlessio2014,Potter2016classification, Po2016chiral,Khemani2016,Bukov2015prethermal,Grushin2014,bordia2017periodically, fidkowski2017interacting,Harper2017,Roy2017,Moessner2017}. Although we have not taken into account the interaction effects, we hope that this systematic symmetry-based study of topological Floquet bands may provide some useful pieces of groundwork for investigation of topology in many-body Floquet systems. In many-body systems with a high driving frequency, the heating time is exponentially long
\cite{Abanin2015,abanin2017effective,Mori2016, kuwahara2016floquet,Else2017prethermal}, possibly allowing many-body generalizations of our topological invariants to this regime. In some other notable regimes of many-body Floquet systems with a modest driving frequency, such as the universal chiral quasi-steady states\cite{Lindner2017Universal}, the interband scattering can be exponentially suppressed, creating a long time window in which only a single or a few Floquet bands are populated. In such regimes, our topological invariants can be directly related to the physical responses. Another interesting future direction is to generalize the present formulation to disordered Floquet systems, which is interesting in its own right, and also valuable in light of the important roles played by disorders in creating Floquet many-body localized states immune to heating up to infinite temperature\cite{Lazarides2015,Ponte2015,ponte2015periodically, Abanin2016,Khemani2016,nandkishore2015many}.

\emph{Acknowledgements.--} We would like to thank Xiao-Liang Qi for helpful discussions.
This work is supported by NSFC under Grant No. 11674189. Z.Y. is supported
in part by China Postdoctoral Science Foundation (No. 2016M590082).

\appendix

\section{Calculations related to symmetries of Floquet systems}

\subsection{Symmetries of the time evolution operator}\label{symmu}

Taking advantages of the particle-hole symmetry of Eq.(\ref{particlem}), and the expansion of time evolution operator as given in Eq.(\ref{divide}), we can derive (taking $t>0$ for concreteness)
\begin{equation} \label{}
\label{}
\begin{aligned}
&C^{-1} U(\bk,\br,t)C \\&=[1-i\Delta tC^{-1}H(\bk,\br,t)C ][1-i\Delta tC^{-1}H(\bk,\br,t-\Delta t)C ]\cdots \\&\cdots[1-i\Delta tC^{-1}H(\bk,\br,\Delta t)C ]\\
&=[1+i\Delta tH^*(-\bk,\br,t)][1+i\Delta tH^*(-\bk,\br,t-\Delta t)]\cdots\\&\cdots[1+i\Delta tH^*(-\bk,\br,\Delta t)]\\
&=U^*(-\bk,\br,t),
\end{aligned}
\end{equation}
which is Eq.(\ref{particleu}) in the main text.

Taking advantage of the time reversal symmetry in Eq.(\ref{timem}), and the expansion of time evolution operator in Eq.(\ref{divide}), we can derive (again taking $t>0$ for concreteness):
\begin{equation} \label{}
\label{}
\begin{aligned}
&T^{-1} U(\bk,\br,t)T \\&=[1-i\Delta tT^{-1}H(\bk,\br,t)T ][1-i\Delta tT^{-1}H(\bk,\br,t-\Delta t)T ]\cdots\\&\cdots[1-i\Delta tT^{-1}H(\bk,\br,\Delta t)T ]\\
&=[1-i\Delta tH^*(-\bk,\br,-t)][1-i\Delta tH^*(-\bk,\br,-t+\Delta t)]\cdots\\&\cdots[1-i\Delta tH^*(-\bk,\br,-\Delta t)]\\
&=\{[1+i\Delta tH^*(-\bk,\br,-\Delta t)]\cdots\\&\cdots[1+i\Delta tH^*(-\bk,\br,-t+\Delta t)][1+i\Delta tH^*(-\bk,\br,-t)]\}^{-1}\\
&=U^{*-1}(-\bk,\br;0,-t)\\
&=U^*(-\bk,\br,-t),
\end{aligned}
\end{equation}
which is Eq.(\ref{timeu}) in the main text. In this calculation, we have used the notation of $U(\bk,\br; t_a,t_b)$, which has been defined in the main text (see Sec. \ref{sec:outline}).

Using the chiral symmetry of Eq.(\ref{chiralm}), and the expansion of time evolution operator as given in Eq.(\ref{divide}), we can derive the constraint of the chiral symmetry on the time evolution operator:
\begin{equation} \label{}
\label{}
\begin{aligned}
&S^{-1} U(\bk,\br,t)S \\&=[1-i\Delta tS^{-1}H(\bk,\br,t)S ][1-i\Delta tS^{-1}H(\bk,\br,t-\Delta t)S ]\cdots\\&\cdots[1-i\Delta t S^{-1}H(\bk,\br,\Delta t)S ]\\
&=[1+i\Delta tH(\bk,\br,-t)][1+i\Delta tH(\bk,\br,-t+\Delta t)]\cdots\\&\cdots[1+i\Delta tH(\bk,\br,-\Delta t)]\\
&=\{[1-i\Delta tH(\bk,\br,-\Delta t)]\cdots\\&\cdots[1-i\Delta tH(\bk,\br,-t+\Delta t)][1-i\Delta tH(\bk,\br,-t)]\}^{-1}\\
&=U^{-1}(\bk,\br;0,-t)\\
&=U(\bk,\br,-t),
\end{aligned}
\end{equation}
which is the Eq.(\ref{chiralu}) in the main text. Again, we have used the notation of $U(\bk,\br; t_a,t_b)$ defined in the main text.

\subsection{Symmetries of the effective Hamiltonian}\label{symmh}

In this section we will derive the symmetry properties of the effective Hamiltonian.

It is apparent that $\exp[-i C^{-1}H^{\rm eff}_\varepsilon(\bk,\br)C \tau]= C^{-1}\exp[-i  H^{\rm eff}_\varepsilon(\bk,\br) \tau]C =C^{-1}U(\bk,\br,\tau)C$, thus we have
\begin{equation}
\label{}
\begin{aligned}
& C^{-1}H^{\rm eff}_\varepsilon(\bk,\br)C  =\\
&=\frac{i}{\tau}\ln_{-\varepsilon}[C^{-1} U(\bk,\br,\tau)C ]\\
&=\frac{i}{\tau}\ln_{-\varepsilon}[( U^* (-\bk,\br,\tau))]\\
&=\frac{i}{\tau}\ln_{-\varepsilon}[( U^\dagger (-\bk,\br,\tau))^T]\\
&=\frac{i}{\tau}\sum\limits_{n}[\ln_{-\varepsilon}(\lambda_n^{-1}(-\bk,\br))\ket{\psi_n (-\bk,\br)}\bra{\psi_n(-\bk,\br)}]^T\\
&=\frac{i}{\tau}\sum\limits_{n}\{[-\ln_{\varepsilon}(\lambda_n(-\bk,\br))-2\pi i]\ket{\psi_n(-\bk,\br)}\bra{\psi_n(-\bk,\br)}\}^T\\
&=-[H^{\text{eff}}_{-\varepsilon}(-\bk,\br)]^T+\frac{2\pi}{\tau}\\
&=-H^{\text{eff}*}_{-\varepsilon}(-\bk,\br)+\frac{2\pi}{\tau},
\end{aligned}
\end{equation}
where the PHS of time evolution operator, given by Eq.(\ref{particleu}), has been used. In rewriting $\ln_{-\varepsilon}(\lambda_n^{-1})$, we have used the mathematical identity
 \begin{equation}
\label{1}
\begin{aligned}
\ln_{-\varepsilon}(e^{-i\phi})=-\ln_{\varepsilon}(e^{i\phi})-2\pi i,
\end{aligned}
\end{equation}
which can be proved as follows. We can always choose $\phi$ to satisfy $-\varepsilon-2\pi<-\phi<-\varepsilon$ (given the value of $e^{-i\phi}$, there is one and only one $\phi$ located in this interval), thus we have $\ln_{-\varepsilon}(e^{-i\phi})=-i\phi$ according to our definition of branch cut (see the main text). Now we also have $\varepsilon<\phi<\varepsilon+2\pi$, and equivalently, $\varepsilon-2\pi<\phi-2\pi<\varepsilon$, therefore, we have $\ln_{\varepsilon}(e^{i\phi})= \ln_{\varepsilon}(e^{i(\phi-2\pi)}) = i(\phi-2\pi)$, from which Eq.(\ref{1}) follows.

Thus, the Eq.(\ref{particleh}) in the main text has been established.

Now let us proceed to proving Eq.(\ref{timeh}). The calculations go as
\begin{equation}
\label{}
\begin{aligned}
&T^{-1} H^{\text{eff}}_{\varepsilon}(\bk,\br)T \\
&=\frac{i}{\tau}\ln_{-\varepsilon}(T^{-1} U(\bk,\br,\tau)T )\\
&=\frac{i}{\tau}\ln_{-\varepsilon}[U^*(-\bk,\br,-\tau)]\\
&=\frac{i}{\tau}\ln_{-\varepsilon}[(U^*(-\bk,\br,\tau))^{-1}]\\
&=\frac{i}{\tau}[\ln_{-\varepsilon}(U(-\bk,\br,\tau))]^T\\
&=\frac{i}{\tau}\sum\limits_{n}[\ln_{-\varepsilon}(\lambda_n(-\bk,\br))\ket{\psi_n(-\bk,\br)}\bra{\psi_n (-\bk,\br)}]^T\\
&=[H^{\text{eff}}_{\varepsilon}(-\bk,\br)]^\text{T}=H^{\text{eff}*}_{\varepsilon}(-\bk,\br),
\end{aligned}
\end{equation} in which Eq.(\ref{timeu}) has been used.

Finally, we would like to prove Eq.(\ref{chiralh}). The calculation is
\begin{equation}
\label{}
\begin{aligned}
&S^{-1} H^{\text{eff}}_{\varepsilon}(\bk,\br)S \\
&=\frac{i}{\tau}\ln_{-\varepsilon}(S^{-1} U(\bk,\br,\tau)S )\\
&=\frac{i}{\tau}\ln_{-\varepsilon}( U(\bk,\br,-\tau))\\
&=\frac{i}{\tau}\ln_{-\varepsilon}( U^{-1}(\bk,\br,\tau))\\
&=\frac{i}{\tau}\sum\limits_{n}\ln_{-\varepsilon}(\lambda_n^{-1}(\bk,\br))\ket{\psi_n(\bk)}\bra{\psi_n(\bk)}\\
&=\frac{i}{\tau}\sum\limits_{n}[-\ln_{\varepsilon}(\lambda_n(\bk,\br))-2\pi i]\ket{\psi_n(\bk,\br)}\bra{\psi_n(\bk,\br)}\\
&=-H^{\text{eff}}_{-\varepsilon}(\bk,\br)+\frac{2\pi}{\tau},
\end{aligned}
\end{equation} in which Eq.(\ref{chiralu}) and
Eq.(\ref{1}) have been used.

\subsection{Symmetries of the periodized time evolution operator }\label{symmuu}

In this section we will derive symmetry properties of the periodized time evolution operators. The derivations are based on the symmetry properties of the time evolution operators and the effective Hamiltonian, which have been studied in the previous two appendices.

Taking advantage of Eq.(\ref{particleu}) and Eq.(\ref{particleh}), we can find that
\begin{equation}
\label{}
\begin{aligned}
&C^{-1} U_{\varepsilon}(\bk,\br,t) C \\
&=C^{-1} U(\bk,\br,t) C \exp[iC^{-1} H^{\text{eff}}_{\varepsilon}(\bk,\br)C t]\\
&=U^*(-\bk,\br,t)\exp[-iH^{\text{eff}*}_{-\varepsilon}(-\bk,\br)t+i\frac{2\pi}{\tau}t]\\
&=U^*_{-\varepsilon}(-\bk,\br,t)\exp(i\frac{2\pi t}{\tau}),
\end{aligned}
\end{equation}
which is Eq.(\ref{particle}) in the main text.

Taking advantage of Eq.(\ref{timeu}) and Eq.(\ref{timeh}), we have, for the time reversal symmetry:
\begin{equation}
\label{}
\begin{aligned}
&T^{-1} U_{\varepsilon}(\bk,\br,t)T \\
&=T^{-1} U(\bk,\br,t) T \exp[iT^{-1} H^{\text{eff}}_{\varepsilon}(\bk,\br)T t]\\
&=U^*(-\bk,\br,-t)\exp[iH^{\text{eff}*}_{\varepsilon}(-\bk,\br)t]\\
&=U_{\varepsilon}^*(-\bk,\br,-t),
\end{aligned}
\end{equation}
which is Eq.(\ref{time}) in the main text.

Finally, taking advantage of Eq.(\ref{chiralu})and Eq.(\ref{chiralh}), we have, for the chiral symmetry:
\begin{equation}
\label{}
\begin{aligned}
&S^{-1} U_{\varepsilon}(\bk,\br,t) S \\
&=S^{-1} U(\bk,\br,t) S \exp(iS^{-1} H^{\text{eff}}_{\varepsilon}(\bk,\br)S t)\\
&=U(\bk,\br,-t)\exp\{i[-H^{\text{eff}}_{-\varepsilon}(\bk,\br)+\frac{2\pi}{\tau}]t\}\\
&=U_{-\varepsilon}(\bk,\br,-t)\exp(i\frac{2\pi t}{\tau}),
\end{aligned}
\end{equation}
which is Eq.(\ref{chiral}) in the main text.

\section{The static limit of the topological invariants of class A}

In this appendix, we will show that the winding number of class A reduces to the Chern number of Teo and Kane in the static limit\cite{teo2010}.

 Since all static Hamiltonian can be smoothly deformed to flat-band ones, we will focus on the flat-band cases. The Hamiltonian takes the general form of Eq.(\ref{flat}), which is reproduced as
\begin{equation}
H_0(\bk,\br)=-E_0P(\bk,\br)+E_0[1-P(\bk,\br)],
\end{equation}
where $P(\bk,\br)$ is the occupied-bands projection operator, which depends on both $\bk$ and $\br$. It satisfies $P^2(\bk,\br)=P(\bk,\br)$. The conduction band projection operator $1-P(\bk,\br)$ must also satisfy $[1-P(\bk,\br)]^2=1-P(\bk,\br)$. The constant $E_0>0$.

When an infinitesimal driving with frequency $\omega$ is added, the system is naturally described by the Floquet theory. The Floquet topological invariants should be consistent with the static topological invariants. We will show that, for sufficiently large $\omega$, the winding number is equal to the Chern number of the static system; for small $\omega$, the winding number is the Chern number multiplied by an integer. This result is consistent with the folding of static energy bands into quasienergy bands, as explained in the main text (see Fig.\ref{fig-static}).

\subsection{Large frequency: The winding number reduces to the Chern number}\label{statich}

For reason to become clear shortly, let us focus on the $\omega>E_0$ (or $2\pi>E_0\tau$) case in this section. It is readily found that
\bea
U(\bk,\br,\tau)
 = e^{-iE_0\tau}[1-P(\bk,\br)]+e^{iE_0\tau}P(\bk,\br),\label{flat-evolution}
\eea
from which it follows that
\begin{equation}
\label{}
\begin{aligned}
&H_{\varepsilon=0}^{\text{eff}}(\bk,\br)=\frac{i}{\tau}\ln_{-\varepsilon=0}U(\bk,\br,\tau)\\ &=\frac{i}{\tau}\ln_{-\varepsilon=0}(e^{-iE_0\tau}) (1-P)+\frac{i}{\tau}\ln_{-\varepsilon=0}(e^{iE_0\tau})P\\
&=\frac{i}{\tau}(-iE_0\tau)(1-P) +\frac{i}{\tau}i(E_0\tau-2\pi)P\\
&=E_0 (1-P) +(-E_0+\frac{2\pi}{\tau} )P,
\end{aligned}
\end{equation} where we have used the fact that $\ln_{-\varepsilon=0} e^{iE_0\tau} = \ln_{-\varepsilon=0} e^{iE_0\tau-2\pi i} = i(E_0\tau-2\pi)$, because $-2\pi<E_0\tau-2\pi<0$ (see the main text for the definition of branch cut).

Therefore, the periodized time evolution operator with branch cut at $-\varepsilon=0$ can be obtained as
\begin{equation}
\label{time evo}
\begin{aligned}
U_{\varepsilon=0}(\bk,\br,t)&=U(\bk,\br,t)\exp[{iH_{\varepsilon=0}^{\text{eff}}(\bk,\br)t}]\\
&=(1-P)\text{e}^{-iE_0t}\text{e}^{iE_0t}+P\text{e}^{iE_0t}\text{e}^{-i(E_0-2\pi/\tau)t}\\
&=(1-P)+ P \text{e}^{i\omega t}\\
&=P(\text{e}^{i\omega t}-1)+1,
\end{aligned}
\end{equation}
in which $\omega\equiv 2\pi/\tau$. Its inverse matrix is
\begin{equation} \label{}
U^{-1}_{\varepsilon=0}(\bk,\br,t)=P(\text{e}^{-i\omega t}-1)+1.
\end{equation} The calculation presented below resembles that of Ref.\cite{rudner2013anomalous,Fruchart2016} (with some minor improvements made). By a straightforward calculation, we have
\begin{equation} \label{}
U^{-1}_{\varepsilon=0}\partial_tU_{\varepsilon=0}=i\omega P,
\end{equation} and
\begin{equation} \label{}
U^{-1}_{\varepsilon=0}\partial_iU_{\varepsilon=0}=u(t)P\partial_i P+v(t)\partial_i P,
\end{equation}
where we have defined the shorthand notations\cite{rudner2013anomalous,Fruchart2016}
\begin{equation} \label{}
u(t)=2[1-\cos(\omega t)],\,v(t)=\text{e}^{i\omega t}-1.
\end{equation}
Inserting them into the definition of winding number, as given by Eq.(\ref{class-A}) in the main text, we have
\begin{equation}
\label{}
\begin{aligned}
&W(U_{\varepsilon=0})=(d+D+1)K_{d+D+1}\int_{T^d\times S^D\times S^1} d^dk d^Dr dt\\
&\times\text{Tr}[\epsilon^{\alpha_1\alpha_2\cdots\alpha_{d+D}}i\omega P(uP\partial_{\alpha_1}P+v\partial_{\alpha_1}P)\cdots(uP\partial_{\alpha_{d+D}}P+v\partial_{\alpha_{d+D}}P)].
\end{aligned}
\end{equation}
To further simplify this expression, we notice the mathematical fact
\begin{equation}
\label{}
\begin{aligned}
P(\partial_i P)P=P\partial_i (PP)-PP\partial_i P=0,
\end{aligned}
\end{equation} from which it follows that the product of two adjacent factors in the expression of winding number can be simplified as
\bea &&(uP\partial_{\alpha_1}P+v\partial_{\alpha_1}P) (uP\partial_{\alpha_2}P+v\partial_{\alpha_2}P) \nn\\
&=& u^2P\partial_{\alpha_1}PP\partial_{\alpha_2}P +uvP\partial_{\alpha_1}P\partial_{\alpha_2}P +uv\partial_{\alpha_1}PP\partial_{\alpha_2}P +v^2\partial_{\alpha_1}P\partial_{\alpha_2}P \nn\\ &=& uvP\partial_{\alpha_1}P\partial_{\alpha_2}P +uv\partial_{\alpha_1}(PP)\partial_{\alpha_2}P -uvP\partial_{\alpha_1}P\partial_{\alpha_2}P +v^2\partial_{\alpha_1}P\partial_{\alpha_2}P \nn\\ &=& uv\partial_{\alpha_1}P\partial_{\alpha_2}P+v^2\partial_{\alpha_1}P\partial_{\alpha_2}P \nn\\ &=&  (u+v)v\partial_{\alpha_1}P\partial_{\alpha_2}P. \eea   The products of any other two adjacent factors, $(uP\partial_{\alpha_3}P+v\partial_{\alpha_3}P) (uP\partial_{\alpha_4}P+v\partial_{\alpha_4}P)$, $\cdots$, can be calculated in the same way, therefore, the winding number simplifies as follows:
 \begin{equation}
\label{c}
\begin{aligned}
&W(U_{\varepsilon=0})=(d+D+1)K_{d+D+1}\int_{T^d\times S^D\times S^1} d^dk d^Dr dt\\
&\times\text{Tr}[\epsilon^{\alpha_1\alpha_2\cdots\alpha_{d+D}}i\omega P(uP\partial_{\alpha_1}P+v\partial_{\alpha_1}P)\cdots(uP\partial_{\alpha_{d+D}}P+v\partial_{\alpha_{d+D}}P)]\\
&=i\omega (d+D+1)K_{d+D+1}\int_{T^d\times S^D\times S^1} d^dk d^Dr dt\\
&\times\text{Tr}\{\epsilon^{\alpha_1\alpha_2\cdots\alpha_{d+D}} [(u+v)v]^{\frac{d+D}{2}}P\partial_{\alpha_1}P\partial_{\alpha_2}P\cdots \partial_{\alpha_{d+D-1}}P\partial_{\alpha_{d+D}}P\}.
\end{aligned}
\end{equation}
The integral of time $t$ can be done as
\begin{equation}
\label{}
\begin{aligned}
&\int_{-\pi/\omega}^{\pi/\omega}dt(u+v)^{\frac{d+D}{2}}v^{\frac{d+D}{2}}\\
&=\frac{1}{\omega}\int_{-\pi}^{\pi}d\phi(1-\text{e}^{-i\phi})^{\frac{d+D}{2}}(\text{e}^{i\phi}-1)^{\frac{d+D}{2}} \,({\rm with}\,\phi\equiv\omega t)\\
&=\frac{2\pi}{\omega}\frac{(D+d)!}{(\frac{d+D}{2})!(\frac{d+D}{2})!}(-1)^{\frac{d+D}{2}}.
\end{aligned}
\end{equation}
Therefore, Eq.(\ref{c}) is simplified to
\begin{equation}
\begin{aligned}
&W(U_{\varepsilon=0})=\tilde{K}_{d+D}\int_{T^d\times S^D} d^dk d^Dr\\
&\times\text{Tr}[\epsilon^{\alpha_1\alpha_2\cdots\alpha_{d+D}}P\partial_{\alpha_1}P\cdots\partial_{\alpha_{d+D}}P],
\end{aligned}
\end{equation}
in which $\tilde{K}_{d+D}$ is:
\begin{equation}
\label{}
\begin{aligned}
\tilde{K}_{d+D}&=i\omega (d+D+1)\frac{2\pi}{\omega}\frac{(D+d)!}{(\frac{d+D}{2})!(\frac{d+D}{2})!}(-1)^{\frac{d+D}{2}}K_{d+D+1}\\
&=-\left(\frac{i}{2\pi}\right)^{\frac{d+D}{2}}\frac{1}{(\frac{d+D}{2})!},
\end{aligned}
\end{equation}
which is exactly the $(d+D)/2$-th Chern number.

\subsection{Small frequency:  The winding number reduces to the Chern number multiplied by an integer}\label{static-small-f}

To have a quasienergy gap at $\epsilon=0$, $E_0/\omega$ cannot be an integer. Let us define the floor function $q=\lfloor E_0/\omega \rfloor$, which denotes the greatest integer below $E_0/\omega$. In other words, $\omega$ satisfies $q\omega<E_0<(q+1)\omega$. In terms of $\tau$, it reads $2\pi q<E_0\tau<2\pi(q+1)$.

The full-period evolution operator is the same as Eq.(\ref{flat-evolution}), while the effective Hamiltonian is now replaced by
\begin{equation}
\label{}
\begin{aligned}
&H_{\varepsilon=0}^{\text{eff}}(\bk,\br)=\frac{i}{\tau}\ln_{-\varepsilon=0}U(\bk,\br,\tau)\\
&=\frac{i}{\tau}\ln_{-\varepsilon=0}(e^{-iE_0\tau}) (1-P)+\frac{i}{\tau}\ln_{-\varepsilon=0}(e^{iE_0\tau})P\\
&=\frac{i}{\tau}i(-E_0\tau+2q\pi)(1-P) +\frac{i}{\tau}i[E_0\tau-2(q+1)\pi]P\\
&=(E_0-\frac{2q\pi}{\tau}) (1-P) +[-E_0+\frac{2(q+1)\pi}{\tau} ]P,
\end{aligned}
\end{equation}
where we have used the fact that $\ln_{-\varepsilon=0} e^{iE_0\tau} = \ln_{-\varepsilon=0} e^{i[E_0\tau-2(q+1)\pi] } = i[E_0\tau-2(q+1)\pi]$, which is a consequence of $-2\pi<E_0\tau-2(q+1)\pi<0$ (see the main text for the definition of branch cut). Similarly, we can see that $\ln_{-\varepsilon=0} e^{-iE_0\tau}=i(-E_0\tau+2q\pi)$.
Therefore, the periodized time evolution operator with branch cut at $-\varepsilon=0$ can be obtained as
\begin{equation}
\label{time evo ad}
\begin{aligned}
U_{\varepsilon=0}(\bk,\br,t)&=U(\bk,\br,t)\exp[{iH_{\varepsilon=0}^{\text{eff}}(\bk,\br)t}]\\
&=(1-P)\text{e}^{-iE_0t}\text{e}^{i(E_0-2q\pi/\tau)t}+P\text{e}^{iE_0t} \text{e}^{i[-E_0+2(q+1)\pi/\tau]t}\\
&=(1-P)\text{e}^{-iq\omega t}+ P \text{e}^{i(q+1)\omega t}.
\end{aligned}
\end{equation}
Its inverse matrix is
\begin{equation} \label{}
U^{-1}_{\varepsilon=0}(\bk,\br,t)=(1-P)\text{e}^{iq\omega t}+ P \text{e}^{-i(q+1)\omega t}.
\end{equation} Straightforwardly, we have
\begin{equation} \label{}
U^{-1}_{\varepsilon=0}\partial_tU_{\varepsilon=0}=-iq\omega+i(2q+1)\omega P,
\end{equation} and
\begin{equation} \label{}
U^{-1}_{\varepsilon=0}\partial_iU_{\varepsilon=0}=u_q(t)P\partial_i P+v_q(t)\partial_i P,
\end{equation}
where we have defined the shorthand notations:
\begin{equation} \label{}
u_q(t)=2\{1-\cos[(2q+1)\omega t]\},\,v_q(t)=\text{e}^{i(2q+1)\omega t}-1.
\end{equation}
Inserting them into Eq.(\ref{class-A}), we have
\begin{equation}
\label{}
\begin{aligned}
&W(U_{\varepsilon=0})=(d+D+1)K_{d+D+1}\int_{T^d\times S^D\times S^1} d^dk d^Dr dt\\
&\times\text{Tr}\{\epsilon^{\alpha_1\alpha_2\cdots\alpha_{d+D}}[-iq\omega+i(2q+1)\omega P]\\
&(u_q P\partial_{\alpha_1}P+v_q\partial_{\alpha_1}P)\cdots(u_qP\partial_{\alpha_{d+D}}P+v_q\partial_{\alpha_{d+D}}P)\}.
\end{aligned}
\end{equation}
Due to the Levi-Civita symbol, the $-iq\omega$ term in ``$[-iq\omega+i(2q+1)\omega P]$'' can be discarded.

The rest part of the calculation is similar to Appendix \ref{statich}, and the final result can be obtained as
\begin{equation}
\begin{aligned}
&W(U_{\varepsilon=0})=(2q+1)\tilde{K}_{d+D}\int_{T^d\times S^D} d^dk d^Dr\\
&\times\text{Tr}[\epsilon^{\alpha_1\alpha_2\cdots\alpha_{d+D}}P\partial_{\alpha_1}P\cdots\partial_{\alpha_{d+D}}P],
\end{aligned}
\end{equation}
which is Eq.(\ref{static-small-freq}) in the main text. This result is consistent with the illustrative Fig.\ref{fig-static}. As can be appreciated from the above calculations, the branch cut of logarithm plays a key role. All the differences among different $\omega$'s come from the location of the branch cut.


\section{Derivations of useful properties of winding number and winding number density}

\subsection{The winding number density is real}\label{realwind}

The winding number density $w(U_\varepsilon)$ is real (see Sec. \ref{sec:class-A} for the definition of $w(U_\varepsilon)$).
Although this should be a well established mathematical fact, we give an explicit derivation here because this calculation can serve as a warm up exercise for later calculations.

To be transparent in the complex conjugation, we extract the ``$i$'' factors in the expression of the winding number density $w(U_\varepsilon)$, writing it as
\begin{equation}
\label{windingden}
\begin{aligned}
&w(U_\varepsilon)=K'_{d+D+1}(i)^{\frac{d+D}{2}+1}\\
&\times\text{Tr}[\epsilon^{\alpha_1\alpha_2\cdots\alpha_{d+D+1}} (U_{\varepsilon}^{-1}\partial_{\alpha_{1}}U_{\varepsilon}) \cdots(U_{\varepsilon}^{-1}\partial_{\alpha_{d+D+1}}U_{\varepsilon})],
\end{aligned}
\end{equation}
where $K'_{d+D+1}$ is a real number:
\begin{eqnarray} \label{windingcoef}
K'_{d+D+1}=\frac{(-1)^{\frac{d+D}{2}}(\frac{d+D}{2})!}{(d+D+1)!}\left(\frac{1}{2\pi}\right)^{\frac{d+D}{2}+1}.
\end{eqnarray} The expression in Eq.(\ref{windingden}) will be useful in keeping track of the signs in the complex conjugation  (for example, see the calculations below and Appendix \ref{symdc}).

The complex conjugate of the winding number density can be calculated as follows
\begin{widetext}
\begin{equation}
\label{}
\begin{aligned}
&w^*(U_\varepsilon)=K'_{d+D+1}(i)^{\frac{d+D}{2}+1}(-1)^{\frac{d+D}{2}+1} \text{Tr}[\epsilon^{\alpha_1\alpha_2\cdots\alpha_{d+D+1}}(U_{\varepsilon}^{T}\partial_{\alpha_{1}}U^*_{\varepsilon})\cdots(U_{\varepsilon}^{T}\partial_{\alpha_{d+D+1}}U^*_{\varepsilon})]\\
&=K'_{d+D+1}(i)^{\frac{d+D}{2}+1}(-1)^{\frac{d+D}{2}+1} \text{Tr}\{\epsilon^{\alpha_1\alpha_2\cdots\alpha_{d+D+1}}[(\partial_{\alpha_{d+D+1}}U^\dagger_{\varepsilon})U_{\varepsilon}]\cdots[(\partial_{\alpha_{1}}U^\dagger_{\varepsilon})U_{\varepsilon}]\}\\
&=K'_{d+D+1}(i)^{\frac{d+D}{2}+1}(-1)^{\frac{d+D}{2}+1} \text{Tr}\{\epsilon^{\alpha_1\alpha_2\cdots\alpha_{d+D+1}}[(\partial_{\alpha_{d+D+1}}U^{-1}_{\varepsilon})U_{\varepsilon}]\cdots[(\partial_{\alpha_{1}}U^{-1}_{\varepsilon})U_{\varepsilon}]\}\\
&=K'_{d+D+1}(i)^{\frac{d+D}{2}+1}(-1)^{\frac{d+D}{2}+1} \text{Tr}[\epsilon^{\alpha_1\alpha_2\cdots\alpha_{d+D+1}}(-U^{-1}_{\varepsilon}\partial_{\alpha_{d+D+1}}U_{\varepsilon})\cdots(-U^{-1}_{\varepsilon}\partial_{\alpha_{1}}U_{\varepsilon})]\\
&=K'_{d+D+1}(i)^{\frac{d+D}{2}+1}(-1)^{\frac{d+D}{2}+1}(-1)^{\frac{(d+D)(d+D+1)}{2}}(-1)^{d+D+1} \text{Tr}[\epsilon^{\alpha_{d+D+1}\alpha_{d+D}\cdots\alpha_1}(U^{-1}_{\varepsilon}\partial_{\alpha_{d+D+1}}U_{\varepsilon})\cdots(U^{-1}_{\varepsilon}\partial_{\alpha_{1}}U_{\varepsilon})]\\
&=K'_{d+D+1}(i)^{\frac{d+D}{2}+1}(-1)^{\frac{(d+D+2)^2}{2} } \text{Tr}[\epsilon^{\alpha_{d+D+1}\alpha_{d+D}\cdots\alpha_1}(U^{-1}_{\varepsilon}\partial_{\alpha_{d+D+1}}U_{\varepsilon})\cdots(U^{-1}_{\varepsilon}\partial_{\alpha_{1}}U_{\varepsilon})]\\
&=K'_{d+D+1}(i)^{\frac{d+D}{2}+1} \text{Tr}[\epsilon^{\alpha_{d+D+1}\alpha_{d+D}\cdots\alpha_1}(U^{-1}_{\varepsilon}\partial_{\alpha_{d+D+1}}U_{\varepsilon})\cdots(U^{-1}_{\varepsilon}\partial_{\alpha_{1}}U_{\varepsilon})]\\
&=w(U_\varepsilon),
\end{aligned}
\end{equation}
\end{widetext}
in which we have used the fact that $(-1)^{(d+D+2)^2/2} =1$ because $d+D$ is an even integer. In fact, the winding number makes sense only when $d+D$ is an even integer, otherwise its expression would yield zero by definition.

\subsection{Symmetry properties of the winding number density of the nonchiral classes}

\subsubsection{Symmetry properties of winding number density of class D and class C}\label{symdc}

In this appendix, we would like to establish Eq.(\ref{DC00}) and Eq.(\ref{DCPI}) in the main text.

Eq.(\ref{particle}) is a symmetry relation between periodized time evolution operators with opposite branch cut, $\varepsilon$ and $-\varepsilon$. At the special point $\varepsilon=0$, the branch cut becomes the same at the left hand side and the right hand side, and the symmetry becomes
\begin{eqnarray}
C^{-1} U_{\varepsilon=0}(\bk,\br,t) C =U^*_{\varepsilon=0}(-\bk,\br,t)\exp(i\frac{2\pi t}{\tau}).
\end{eqnarray}
The winding number density can be written in the form of Eq.(\ref{windingden}):
\begin{equation}
\label{}
\begin{aligned}
&w(U_\varepsilon)=K'_{d+D+1}(i)^{\frac{d+D}{2}+1}\\
&\times\text{Tr}[\epsilon^{\alpha_1\alpha_2\cdots\alpha_{d+D+1}} (U_{\varepsilon}^{-1}\partial_{\alpha_{1}}U_{\varepsilon})\cdots (U_{\varepsilon}^{-1}\partial_{\alpha_{d+D+1}}U_{\varepsilon})],
\end{aligned}
\end{equation}
where $K'_{d+D+1}$ is a real number as defined in Eq.(\ref{windingcoef}). The factors containing ``$i$'' are explicit in this expression.

To establish the PHS relation, given by Eq.(\ref{DC00}), of the winding number density at $\varepsilon=0$, we do the following somewhat lengthy calculations:

\begin{widetext}
\begin{equation}
\label{DC0}
\begin{aligned}
&w(U_{\varepsilon=0})(\bk,\br,t)=K'_{d+D+1}(i)^{\frac{d+D}{2}+1}
\text{Tr}\{\epsilon^{\alpha_1\alpha_2\cdots\alpha_{d+D+1}}[(C^{-1} U_{\varepsilon=0}^{-1}C )\partial_{\alpha_{1}}(C^{-1} U_{\varepsilon=0}C )]\cdots[(C^{-1} U_{\varepsilon=0}^{-1}C )\partial_{\alpha_{d+D+1}}(C^{-1} U_{\varepsilon=0}C )]\}\\
&=K'_{d+D+1}(i)^{\frac{d+D}{2}+1}
\text{Tr}\{\epsilon^{\alpha_1\alpha_2\cdots\alpha_{d+D+1}}[ \exp (-i\frac{2\pi t}{\tau}) U_{\varepsilon=0}^{-1*}(-\bk,\br,t)\partial_{\alpha_{1}} \left(U_{\varepsilon=0}^*(-\bk,\br,t)\exp (i\frac{2\pi t}{\tau} )\right)]\cdots\\&\cdots[ \exp (-i\frac{2\pi t}{\tau} ) U_{\varepsilon=0}^{-1*} (-\bk,\br,t)\partial_{\alpha_{d+D+1}}\left(U_{\varepsilon=0}^*(-\bk,\br,t)\exp (i\frac{2\pi t}{\tau} )\right)]\}\\
&=K'_{d+D+1}(i)^{\frac{d+D}{2}+1} \text{Tr}\{\epsilon^{\alpha_1\alpha_2\cdots\alpha_{d+D+1}} [U_{\varepsilon=0}^{-1*}(-\bk,\br,t)\partial_{\alpha_{1}} U_{\varepsilon=0}^*(-\bk,\br,t)]\cdots[ U_{\varepsilon=0}^{-1*}(-\bk,\br,t)\partial_{\alpha_{d+D+1}} U_{\varepsilon=0}^*(-\bk,\br,t)]\}\\ &+(d+D+1)K'_{d+D+1} (i)^{\frac{d+D}{2}+1} \text{Tr}\{\epsilon^{\alpha_1\alpha_2\cdots\alpha_{d+D}} [\exp (-i\frac{2\pi t}{\tau} )U_{\varepsilon=0}^{-1*}(-\bk,\br,t)\partial_{\alpha_{1}} U_{\varepsilon=0}^*(-\bk,\br,t)]\cdots\\ &\cdots[U_{\varepsilon=0}^{-1*}(-\bk,\br,t) \partial_{\alpha_{d+D-1}}U_{\varepsilon=0}^*(-\bk,\br,t)] [U_{\varepsilon=0}^{-1*}(-\bk,\br,t)\partial_{\alpha_{d+D}} U_{\varepsilon=0}^*(-\bk,\br,t)][\partial_{t}\exp (i\frac{2\pi t}{\tau} )]\},
\end{aligned}
\end{equation}
where we have inserted a number of identity matrix $CC^{-1} $ in the first line. In the last expression, the last term $(d+D+1)K'_{d+D+1} (i)^{\frac{d+D}{2}+1} \text{Tr}\{\epsilon^{\alpha_1\alpha_2\cdots\alpha_{d+D}} \cdots\}$ is actually zero due to the Levi-Civita symbol $\epsilon^{\alpha_1\alpha_2\cdots\alpha_{d+D}}$. Therefore, the winding number becomes
\bea
&&w(U_{\varepsilon=0})(\bk,\br,t)\nn\\ &=& K'_{d+D+1}(i)^{\frac{d+D}{2}+1} \text{Tr}\{\epsilon^{\alpha_1\alpha_2\cdots \alpha_{d+D+1}}    [U_{\varepsilon=0}^{-1*}(-\bk,\br,t)\partial_{\alpha_{1}} U_{\varepsilon=0}^*(-\bk,\br,t)] \cdots [U_{\varepsilon=0}^{-1*}(-\bk,\br,t)\partial_{\alpha_{d+D+1}}U_{\varepsilon=0}^*(-\bk,\br,t)]\}\nn\\
&=& w^*(U_{\varepsilon=0})(-\bk,\br,t)(-1)^{(d+D)/2+1}(-1)^{d}\nn\\ &=& w^*(U_{\varepsilon=0})(-\bk,\br,t)(-1)^{2d+1-\delta/2}\nn \\ &=& w(U_{\varepsilon=0})(-\bk,\br,t)(-1)^{2d+1-\delta/2}\nn \\ &=& w(U_{\varepsilon=0})(-\bk,\br,t)(-1)^{1-\delta/2},
\eea
which is Eq.(\ref{DC00}) in the main article. In this calculation, the factor $(-1)^{(d+D)/2+1}$ and $(-1)^{d}$ comes from the complex conjugation of $(i)^{\frac{d+D}{2}+1}$ and the inversion of $\bk$ (i.e., $\frac{\partial}{\partial k_j} = -\frac{\partial}{\partial (-k_j)}$, for $j=1,2,\cdots,d$), respectively. The simple fact that the complex conjugation of $(i)^{(d+D)/2+1}$ generates a $(-1)^{(d+D)/2+1}$ factor plays an interesting role.
\end{widetext}

Now we turn to Eq.(\ref{particle}) with $\varepsilon=\pi$, which is a symmetry relation between the periodized time evolution operator with branch cut at $\varepsilon=\pi$ and $-\pi$. It is not yet an apparent symmetry relation of $U_{\varepsilon=\pi}$, nevertheless, we can derive such a relation. It follows from Eq.(\ref{particle}) that
\bea
C^{-1} U_{\varepsilon=\pi}(\bk,\br,t) C  =U^*_{\varepsilon=-\pi}(-\bk,\br,t)\exp(i\frac{2\pi t}{\tau}),\label{PHS-pi-minus-pi}
\eea
where the right hand side can be transformed to
\begin{equation}
\label{}
\begin{aligned}
&U^*_{\varepsilon=-\pi}(-\bk,\br,t)\exp(i\frac{2\pi t}{\tau})\\
&=U^*(-\bk,\br,t)\exp(-i H^{\text{eff*}}_{\varepsilon=-\pi}t)\exp(i\frac{2\pi t}{\tau})\\
&=U^*(-\bk,\br,t)\exp\{-i[\frac{i}{\tau}\ln_{-\varepsilon=\pi}(U(-\bk,\br,\tau))]^*t\}\exp(i\frac{2\pi t}{\tau})\\
&=U^*(-\bk,\br,t)\exp\{-i[\frac{i}{\tau}\ln_{-\varepsilon=-\pi}(U(-\bk,\br,\tau))+\frac{i}{\tau}2\pi i]^*t\}\exp(i\frac{2\pi t}{\tau})\\
&=U^*_{\varepsilon=\pi}(-\bk,\br,t)\exp(i\frac{4\pi t}{\tau}),
\end{aligned}
\end{equation} which, after insertion into Eq.(\ref{PHS-pi-minus-pi}), leads to Eq.(\ref{particlepi}) in the main text. In this calculation, we have used the mathematical identity
\begin{equation}
\label{}
\begin{aligned}
\ln_{-\varepsilon+2\pi}(e^{i\phi})=\ln_{-\varepsilon}(e^{i\phi})+2\pi i, \label{2pi-difference}
\end{aligned}
\end{equation} which is readily seen by taking $\phi\in[-\varepsilon-2\pi,-\varepsilon]$, thus we have $\ln_{-\varepsilon}e^{i\phi}=i\phi$; on the other hand, it means that $\phi+2\pi\in [-\varepsilon,-\varepsilon+2\pi]$, therefore, $\ln_{-\varepsilon+2\pi}e^{i\phi}= \ln_{-\varepsilon+2\pi}e^{i(\phi+2\pi)} = i(\phi+2\pi)$, which establishes the desired identity, Eq.(\ref{2pi-difference}).

After these preparations, the Eq.(\ref{DCPI}) in the main text can be derived.
In fact, we have
\begin{equation}
\begin{aligned}
&w(U_{\varepsilon=\pi})(\bk,\br,t)=K'_{d+D+1}(i)^{\frac{d+D}{2}+1}
\text{Tr}\{\epsilon^{\alpha_1\alpha_2\cdots\alpha_{d+D+1}}\\
&\times[(C^{-1} U_{\varepsilon=\pi}^{-1}(\bk,\br,t)C )\partial_{\alpha_{1}}(C^{-1}U_{\varepsilon=\pi}(\bk,\br,t)C )]\cdots\\&\cdots[(C^{-1} U_{\varepsilon=\pi}^{-1}(\bk,\br,t)C )\partial_{\alpha_{d+D+1}}(C^{-1} U_{\varepsilon=\pi}(\bk,\br,t)C )]\}\\
&=K'_{d+D+1}(i)^{\frac{d+D}{2}+1}
\times\text{Tr}\{\epsilon^{\alpha_1\alpha_2\cdots\alpha_{d+D+1}}\\
&\times[\exp (-i\frac{4\pi t}{\tau} ) U_{\varepsilon=\pi}^{-1*}(-\bk,\br,t)\partial_{\alpha_{1}} \left(U_{\varepsilon=\pi}^*(-\bk,\br,t)\exp (i\frac{4\pi t}{\tau} )\right)]\cdots\\&\cdots[ \exp (-i\frac{4\pi t}{\tau} )U_{\varepsilon=\pi}^{-1*}(-\bk,\br,t) \partial_{\alpha_{d+D+1}}\left( U_{\varepsilon=\pi}^*(-\bk,\br,t)\exp (i\frac{4\pi t}{\tau} )\right)]\},
\end{aligned}
\end{equation}
which resembles the situation of Eq.(\ref{DC0}). In fact, it is almost the same except that $2\pi t/\tau$ is replaced by $4\pi t/\tau$. The same calculation below Eq.(\ref{DC0}) leads to
\begin{equation}
\label{}
\begin{aligned}
&w(U_{\varepsilon=\pi})(\bk,\br,t)\\
&=w^*(U_{\varepsilon=\pi})(-\bk,\br,t)(-1)^{(d+D)/2+1}(-1)^{d}\\
&=w^*(U_{\varepsilon=\pi})(-\bk,\br,t)(-1)^{2d+1-\delta/2}\\
&=w(U_{\varepsilon=\pi})(-\bk,\br,t)(-1)^{2d+1-\delta/2}\\
&=w(U_{\varepsilon=\pi})(-\bk,\br,t)(-1)^{1-\delta/2},
\end{aligned}
\end{equation}
which is the Eq.(\ref{DCPI}) in the main text.

\begin{widetext}
\subsubsection{Symmetry properties of winding number density of class AI and class AII}\label{syma}

In this appendix, we would like to establish Eq.(\ref{windai}) in the main text.

Taking advantage of Eq.(\ref{time}), the winding number density in Eq.(\ref{windingden}) can be transformed to
\begin{equation}
\label{}
\begin{aligned}
&w(U_\varepsilon)(\bk,\br,t)=K'_{d+D+1}(i)^{\frac{d+D}{2}+1} \text{Tr}\{\epsilon^{\alpha_1\alpha_2\cdots\alpha_{d+D+1}} T^{-1}U_{\varepsilon}^{-1}(\bk,\br,t)T  \partial_{\alpha_{1}} [T^{-1}U_{\varepsilon}(\bk,\br,t)T ]\cdots  T^{-1}U_{\varepsilon}^{-1}(\bk,\br,t)T \partial_{\alpha_{d+D+1}}[T^{-1}U_{\varepsilon}(\bk,\br,t)T ]\}\\
&=K'_{d+D+1}(i)^{\frac{d+D}{2}+1} \text{Tr}\{\epsilon^{\alpha_1\alpha_2\cdots\alpha_{d+D+1}} U_{\varepsilon}^{-1*}(-\bk,\br,-t)\partial_{\alpha_{1}} U^*_{\varepsilon}(-\bk,\br,-t) \cdots  U_{\varepsilon}^{-1*}(-\bk,\br,-t)\partial_{\alpha_{d+D+1}}U_{\varepsilon}^*(-\bk,\br,-t) \}\\
&=w^*(U_\varepsilon)(-\bk,\br,-t)(-1)^{(d+D)/2+1}(-1)^{d+1}\\
&=w^*(U_\varepsilon)(-\bk,\br,-t)(-1)^{2d+2-\delta/2}\\
&=w(U_\varepsilon)(-\bk,\br,-t)(-1)^{2-\delta/2},
\end{aligned}
\end{equation}
which is Eq.(\ref{windai}) in the main text. Again, the simple mathematical fact $[(i)^{(d+D)/2+1}]^*=(-1)^{(d+D)/2+1}(i)^{(d+D)/2+1}$ plays an interesting role in the calculation.

Compared to the case of PHS (see Eq.(\ref{DC00}) and Eq.(\ref{DCPI})), there is an additional $-1$ factor at the right hand side of  Eq.(\ref{windai}). It is clear in the above calculation that this $-1$ factor originates from the inversion of $t$, which is absent in the PHS case (see the calculation in Appendix \ref{symdc}).

\subsubsection{Symmetry of the Wess-Zumino-Witten terms of class D and class C}\label{wzwdc}

Taking Eq.(\ref{wdc1}) as an input, we have:
\begin{equation}
\label{}
\begin{aligned}
&w(U_{\varepsilon=0}^{I})(\bk,\br,t,\lambda)=K'_{d+D+2}(i)^{\frac{d+D+1}{2}+1}
\text{Tr}\{\epsilon^{\alpha_1\alpha_2\cdots\alpha_{d+D+2}}[ C^{-1} U_{\varepsilon=0}^{I-1}(\bk,\br,t,\lambda)C  \partial_{\alpha_{1}}\left(C^{-1} U^{I}_{\varepsilon=0}(\bk,\br,t,\lambda)C \right)]\cdots\\&\cdots[ C^{-1} U_{\varepsilon=0}^{I-1}(\bk,\br,t,\lambda)C \partial_{\alpha_{d+D+2}}\left(C^{-1} U^{I}_{\varepsilon=0}(\bk,\br,t,\lambda)C \right)]\}\\
&=K'_{d+D+2}(i)^{\frac{d+D+1}{2}+1}
 \text{Tr}\{\epsilon^{\alpha_1\alpha_2\cdots\alpha_{d+D+2}}[ \exp (-i\frac{2\pi t}{\tau} ) U_{\varepsilon=0}^{II-1*}(-\bk,\br,t,-\lambda)\partial_{\alpha_{1}} \left(U_{\varepsilon=0}^{II*}(-\bk,\br,t,-\lambda)\exp (i\frac{2\pi t}{\tau} )\right)]\cdots\\&\cdots[\exp (-i\frac{2\pi t}{\tau} )U_{\varepsilon=0}^{II-1*}(-\bk,\br,t,-\lambda)\partial_{\alpha_{d+D+2}} \left(U_{\varepsilon=0}^{II*}(-\bk,\br,t,-\lambda)\exp (i\frac{2\pi t}{\tau} )\right)]\}\\
&=K'_{d+D+2}(i)^{\frac{d+D+1}{2}+1} \text{Tr}[\epsilon^{\alpha_1\alpha_2\cdots\alpha_{d+D+2}} U_{\varepsilon=0}^{II-1*}(-\bk,\br,t,-\lambda)\partial_{\alpha_{1}} U_{\varepsilon=0}^{II*}(-\bk,\br,t,-\lambda) \cdots  U_{\varepsilon=0}^{II-1*}(-\bk,\br,t,-\lambda)\partial_{\alpha_{d+D+2}} U_{\varepsilon=0}^{II*}(-\bk,\br,t,-\lambda)]\\ &+(d+D+2)K'_{d+D+2}(i)^{\frac{d+D+1}{2}+1} \text{Tr}\{\epsilon^{\alpha_1\alpha_2\cdots\alpha_{d+D+1}} \exp (-i\frac{2\pi t}{\tau}) [U_{\varepsilon=0}^{II-1*}(-\bk,\br,t,-\lambda)\partial_{\alpha_{1}} U_{\varepsilon=0}^{II*}(-\bk,\br,t,-\lambda)] \cdots\\&\cdots[U_{\varepsilon=0}^{II-1*}(-\bk,\br,t,-\lambda) \partial_{\alpha_{d+D+1}}U_{\varepsilon=0}^{II*}(-\bk,\br,t,-\lambda)][\partial_{t}\exp (i\frac{2\pi t}{\tau} )]\}.
\end{aligned}
\end{equation}
\end{widetext}
The last term proportional to $d+D+2$ automatically vanishes due to the Levi-Civita symbol $\epsilon^{\alpha_1\alpha_2\cdots\alpha_{d+D+1}}$ (remember that $d+D+1$ here is an even integer), therefore,
\begin{equation}
\label{}
\begin{aligned}
&w(U_{\varepsilon=0}^{I})(\bk,\br,t,\lambda)=K'_{d+D+2}(i)^{\frac{d+D+1}{2}+1}\\ &\times\text{Tr}[\epsilon^{\alpha_1\alpha_2\cdots\alpha_{d+D+2}} U_{\varepsilon=0}^{II-1*}(-\bk,\br,t,-\lambda)\partial_{\alpha_{1}} U_{\varepsilon=0}^{II*}(-\bk,\br,t,-\lambda) \cdots\\& \cdots  U_{\varepsilon=0}^{II-1*}(-\bk,\br,t,-\lambda)\partial_{\alpha_{d+D+2}}U_{\varepsilon=0}^{II*} (-\bk,\br,t,-\lambda)]\\
&=w^*(U_{\varepsilon=0}^{II})(-\bk,\br,t,-\lambda)(-1)^{(d+D+1)/2+1}(-1)^{d+1}\\
&=w^*(U_{\varepsilon=0}^{II})(-\bk,\br,t,-\lambda)(-1)^{2d+2-(\delta-1)/2}\\
&=w(U_{\varepsilon=0}^{II})(-\bk,\br,t,-\lambda)(-1)^{2d+2-(\delta-1)/2}\\
&=w(U_{\varepsilon=0}^{II})(-\bk,\br,t,-\lambda)(-1)^{2-(\delta-1)/2},
\end{aligned}
\end{equation}
which is Eq.(\ref{wzwdc0}) in the  main text.
For $\varepsilon=\pi$, we can take Eq.(\ref{wdc2}) as input, and do similar calculations to obtain Eq.(\ref{wzwdcpi}) in the main text (except that the $e^{2\pi it/\tau}$ factor involved in the calculation is replaced by $e^{4\pi it/\tau}$).

\subsubsection{Symmetry of the Wess-Zumino-Witten terms of class AI and class AII}\label{wzwaa}

Taking Eq.(\ref{wa}) as an input, we have
\begin{equation}
\label{}
\begin{aligned}
&w(U^I_\varepsilon)(\bk,\br,t,\lambda)= w(T^{-1}U^I_\varepsilon T )(\bk,\br,t,\lambda)\\
&=K'_{d+D+2}(i)^{\frac{d+D+1}{2}+1}\\
&\times\text{Tr}[\epsilon^{\alpha_1\alpha_2\cdots\alpha_{d+D+2}} U_{\varepsilon}^{II-1*}(-\bk,\br,-t,-\lambda)\partial_{\alpha_{1}} U^{II*}_{\varepsilon}(-\bk,\br,-t,-\lambda) \cdots\\&\cdots  U_{\varepsilon}^{II-1*}(-\bk,\br,-t,-\lambda)\partial_{\alpha_{d+D+2}}U^{II*}_{\varepsilon}(-\bk,\br,-t,-\lambda) ]\\
&=w^*(U^{II}_\varepsilon)(-\bk,\br,-t,-\lambda)(-1)^{(d+D+1)/2+1}(-1)^{d+2}\\
&=w^*(U^{II}_\varepsilon)(-\bk,\br,-t,-\lambda)(-1)^{2d+3-(\delta-1)/2}\\
&=w(U^{II}_\varepsilon)(-\bk,\br,-t,-\lambda)(-1)^{3-(\delta-1)/2},
\end{aligned}
\end{equation}
which is the Eq.(\ref{wzwa}) in the main text.

\subsection{Symmetry properties of winding number density or winding number of chiral classes}

\subsubsection{Class CI}\label{chiralci}

In this appendix, we would like to prove Eq.(\ref{windci}) in the main text.

Taking advantage of Eq.(\ref{ci1}) and Eq.(\ref{ci2}), we see that the winding number satisfies
\begin{equation}
\label{}
\begin{aligned}
&W(U^+_{\varepsilon=0}(\bk,\br))=K'_{d+D}(i)^{\frac{d+D+1}{2}}\\
&\times\int_{T^d\times S^D} d^dkd^Dr \text{Tr}\{\epsilon^{\alpha_1\alpha_2\cdots\alpha_{d+D}}[(U_{\varepsilon=0}^+(\bk,\br))^{-1} \partial_{\alpha_{1}}U_{\varepsilon=0}^+(\bk,\br)] \\&\cdots[(U_{\varepsilon=0}^+(\bk,\br))^{-1} \partial_{\alpha_{d+D}}U_{\varepsilon=0}^+(\bk,\br)]\}\\
&=K'_{d+D}(i)^{\frac{d+D+1}{2}}\\
&\times\int_{T^{d}\times S^D}d^dk d^Dr \text{Tr}\{\epsilon^{\alpha_1\alpha_2\cdots\alpha_{d+D}}[(U_{\varepsilon=0}^-(-\bk,\br))^{-1} \partial_{\alpha_{1}}U_{\varepsilon=0}^-(-\bk,\br)]^* \\&\cdots[(U_{\varepsilon=0}^-(-\bk,\br))^{-1} \partial_{\alpha_{d+D}}U_{\varepsilon=0}^-(-\bk,\br)]^*\}\\
&=\int d^dkd^Dr\,w^*(U^-_{\varepsilon=0})(-\bk,\br)(-1)^{(d+D+1)/2}(-1)^d\\
&=W(U^-_{\varepsilon=0}(\bk,\br))(-1)^{(d+D+1)/2}(-1)^{d}\\
&=-W(U^+_{\varepsilon=0}(\bk,\br))(-1)^{2d-(\delta-1)/2}\\
&=W(U^+_{\varepsilon=0}(\bk,\br))(-1)^{1-(\delta-1)/2},
\end{aligned}
\end{equation}
where the real coefficient $K'_{d+D}$
is defined by Eq.(\ref{windingcoef}).
In the above calculation, we have used Eq.(\ref{zero}), namely $W(U^-_{\varepsilon=0}(\bk,\br))=-W(U^+_{\varepsilon=0}(\bk,\br))$.
Thus the Eq.(\ref{windci}) in the main text has been proved. We emphasize that Eq.(\ref{zero}) is an identity of the winding numbers instead of the winding number density, therefore, Eq.(\ref{windci}) is also a symmetry relation of winding numbers, not the winding number densities.

\subsubsection{Class DIII}\label{chiraldiii}

In this section, we would like to prove Eq.(\ref{winddiii}) in the main text.

Taking advantage of Eq.(\ref{diii1}) and Eq.(\ref{diii2}), the winding numbers satisfy
\begin{equation}
\label{}
\begin{aligned}
&W(U^+_{\varepsilon=0}(\bk,\br)) =K'_{d+D}(i)^{\frac{d+D+1}{2}}\\
&\times\int d^dkd^Dr\,\text{Tr}\{\epsilon^{\alpha_1\alpha_2\cdots\alpha_{d+D}}[(U_{\varepsilon=0}^+(\bk,\br))^{-1} \partial_{\alpha_{1}}U_{\varepsilon=0}^+(\bk,\br)] \\&\cdots[(U_{\varepsilon=0}^+(\bk,\br))^{-1} \partial_{\alpha_{d+D}}U_{\varepsilon=0}^+(\bk,\br)]\}\\
&=K'_{d+D}(i)^{\frac{d+D+1}{2}}\\
&\times\int d^dkd^Dr\,\text{Tr}\{\epsilon^{\alpha_1\alpha_2\cdots \alpha_{d+D}}[(-U_{\varepsilon=0}^-(-\bk,\br))^{-1}\partial_{\alpha_{1}} (-U_{\varepsilon=0}^-(-\bk,\br))]^* \\&\cdots[(-U_{\varepsilon=0}^-(-\bk,\br))^{-1} \partial_{\alpha_{d+D}}(-U_{\varepsilon=0}^-(-\bk,\br))]^*\}\\
&=\int d^dkd^Dr\,w^*(U^-_{\varepsilon=0})(-\bk,\br)(-1)^{(d+D+1)/2}(-1)^d\\
&=W(U^-_{\varepsilon=0}(\bk,\br))(-1)^{(d+D+1)/2}(-1)^d\\
&=-W(U^+_{\varepsilon=0}(\bk,\br))(-1)^{2d-(\delta-1)/2}\\
&=W(U^+_{\varepsilon=0}(\bk,\br))(-1)^{1-(\delta-1)/2},
\end{aligned}
\end{equation} which is the Eq.(\ref{winddiii}) in the main text.
Just like the case of class CI studied in Appendix \ref{chiralci}, we have used Eq.(\ref{zero}) in this calculation.  Therefore, the resultant Eq.(\ref{winddiii}) is a statement about the winding numbers instead of winding number densities.

\subsubsection{Class BDI}\label{chiralbdi}

In this appendix, we would like to prove Eq.(\ref{windbdi}) in the main text. Unlike Eq.(\ref{windci}) and Eq.(\ref{winddiii}), which are only statements about winding numbers, Eq.(\ref{windbdi}) is a statement of winding number density.

Taking Eq.(\ref{bdi1}) and Eq.(\ref{bdi2}) as inputs, we can show that the winding number density satisfies
\begin{equation}
\label{}
\begin{aligned}
&w(U^+_{\varepsilon=0})(\bk,\br)=K'_{d+D}(i)^{\frac{d+D+1}{2}}\\
&\times\text{Tr}\{\epsilon^{\alpha_1\alpha_2\cdots\alpha_{d+D}} [(U_{\varepsilon=0}^+(\bk,\br))^{-1}\partial_{\alpha_{1}} U_{\varepsilon=0}^+(\bk,\br)]\\&\cdots [(U_{\varepsilon=0}^+(\bk,\br))^{-1} \partial_{\alpha_{d+D}}U_{\varepsilon=0}^+(\bk,\br)]\}\\
&=K'_{d+D}(i)^{\frac{d+D+1}{2}}\\
&\times\text{Tr}\{\epsilon^{\alpha_1\alpha_2\cdots\alpha_{d+D}} [(U_{\varepsilon=0}^+(-\bk,\br))^{-1}\partial_{\alpha_{1}}(U_{\varepsilon=0}^+(-\bk,\br))]^* \\&\cdots[(U_{\varepsilon=0}^+(-\bk,\br))^{-1}\partial_{\alpha_{d+D}}(U_{\varepsilon=0}^+(-\bk,\br))]^*\}\\
&=w^*(U^+_{\varepsilon=0})(-\bk,\br)(-1)^{(d+D+1)/2}(-1)^d\\
&=w^*(U^+_{\varepsilon=0})(-\bk,\br)(-1)^{2d-(\delta-1)/2}\\
&=w(U^+_{\varepsilon=0})(-\bk,\br)(-1)^{-(\delta-1)/2},
\end{aligned}
\end{equation}
where the real coefficient $K'_{d+D}$
is defined by Eq.(\ref{windingcoef}).

Thus the Eq.(\ref{windbdi}) in the main text has been established.

\subsubsection{Class CII}\label{chiralcii}

In this appendix, we would like to prove Eq.(\ref{windcii}) and Eq.(\ref{windcii-pi}).

Taking Eq.(\ref{cii1}) and Eq.(\ref{cii2}) as inputs, we can show that
\begin{equation}
\label{}
\begin{aligned}
&w(U^+_{\varepsilon=0})(\bk,\br)=K'_{d+D}(i)^{\frac{d+D+1}{2}}\\
&\times\text{Tr}\{\epsilon^{\alpha_1\alpha_2\cdots\alpha_{d+D}}[(U_{\varepsilon=0}^+(\bk,\br))^{-1} \partial_{\alpha_{1}}U_{\varepsilon=0}^+(\bk,\br)] \\&\cdots[(U_{\varepsilon=0}^+(\bk,\br))^{-1} \partial_{\alpha_{d+D}}U_{\varepsilon=0}^+(\bk,\br)]\}\\
&=K'_{d+D}(i)^{\frac{d+D+1}{2}}\\ &\times \text{Tr}\{\epsilon^{\alpha_1\alpha_2\cdots\alpha_{d+D}}
[\sigma_y(U^{+}_{\varepsilon=0}(-\bk,\br))^{-1*} \sigma_y\partial_{\alpha_{1}}\left(\sigma_y(U^{+}_{\varepsilon=0}(-\bk,\br))^*\sigma_y\right)] \\&\cdots [ \sigma_y(U_{\varepsilon=0}^{+}(-\bk,\br))^{-1*}\sigma_y \partial_{\alpha_{d+D}}\left(\sigma_y(U_{\varepsilon=0}^{+} (-\bk,\br))^*\sigma_y\right)]\}\\
&=w^*(U^+_{\varepsilon=0})(-\bk,\br)(-1)^{(d+D+1)/2}(-1)^d\\
&=w(U^+_{\varepsilon=0})(-\bk,\br)(-1)^{2d-(\delta-1)/2}\\
&=w(U^+_{\varepsilon=0})(-\bk,\br)(-1)^{-(\delta-1)/2}.
\end{aligned}
\end{equation}

Thus the Eq.(\ref{windcii}) in the main text has been proved. Eq.(\ref{windcii-pi}) can be proved similarly.

\section{Representative Dirac Hamiltonians and a proof of $2\mathbb{Z}$ topological invariants}\label{sec:even}

In this appendix, we study representative Dirac Hamiltonians for all symmetry classes, which are useful in model constructions of Floquet topological insulators and Floquet topological defects. We also take these representative Dirac Hamiltonians to explain the reason why the topological invariants always take even-integer values when $d-D-s=4$ (mod $8$). Again, the combination $\delta=d-D$ naturally comes out.

To proceed, we define the following Dirac matrices for a pair of non-negative integers $l$ and $m$ ($l+m$ is an even integer):
\begin{equation}\label{gamma}
\begin{aligned}
&\Gamma_{(l+m+1)}^1=\sigma_1\otimes\underbrace{\sigma_3\otimes\cdots\otimes\sigma_3}_{(l+m)/2-1},\\
&\Gamma_{(l+m+1)}^2=\sigma_2\otimes\underbrace{\sigma_3\otimes\cdots\otimes\sigma_3}_{(l+m)/2-1},\\
&\Gamma_{(l+m+1)}^3=\sigma_0\otimes\sigma_1\otimes\underbrace{\sigma_3\otimes\cdots\otimes\sigma_3}_{(l+m)/2-2},\\
&\Gamma_{(l+m+1)}^4=\sigma_0\otimes\sigma_2\otimes\underbrace{\sigma_3\otimes\cdots\otimes\sigma_3}_{(l+m)/2-2},\\
&\cdots\\
&\Gamma_{(l+m+1)}^{l+m-1}=\underbrace{\sigma_0\otimes\cdots\otimes\sigma_0}_{(l+m)/2-1}\otimes\sigma_1,\\
&\Gamma_{(l+m+1)}^{l+m}=\underbrace{\sigma_0\otimes\cdots\otimes\sigma_0}_{(l+m)/2-1}\otimes\sigma_2,\\
&\Gamma_{(l+m+1)}^{l+m+1}=\underbrace{\sigma_3\otimes\cdots\otimes\sigma_3}_{(l+m)/2}.\\
\end{aligned}
\end{equation}
Note that $\Gamma_{(l+m+1)}^{a}$ is real when $a$ is odd, and purely imaginary when $a$ is even. These expressions of Dirac matrices are standard (for instance, see Ref.\cite{ryu2010}), though the notation ``$l+m$'' is specific to the present study of topological defects.  Because the combination $l-m$ will appear frequently, let us define the shorthand notation  \bea \eta=l-m, \eea which is always an even integer.

Let us introduce two important matrices, which are crucial in studying the symmetries of Dirac Hamiltonians. They are given as
\bea
B_{(l+m+1)}^1 &=& \prod_{\text{even}\,a\leq m}\Gamma_{(l+m+1)}^{a} \prod_{m<\text{odd}\,a\leq m+l}\Gamma_{(l+m+1)}^{a},\\
B_{(l+m+1)}^2 &=& \prod_{\text{odd}\,a\leq m}\Gamma_{(l+m+1)}^{a} \prod_{m<\text{even}\,a\leq m+l}\Gamma_{(l+m+1)}^{a}.
\eea
To be more explicit, when $l$ and $m$ are odd integers (recall that $l+m$ is always an even integer), these two matrices read
\begin{equation}
\begin{aligned}
&B_{(l+m+1)}^1\\&=\Gamma_{(l+m+1)}^{2}\Gamma_{(l+m+1)}^{4}\cdots \Gamma_{(l+m+1)}^{m-1}\Gamma_{(l+m+1)}^{m+2}\Gamma_{(l+m+1)}^{m+4}\cdots\Gamma_{(l+m+1)}^{l+m-1},\\
&B_{(l+m+1)}^2\\&=\Gamma_{(l+m+1)}^{1}\Gamma_{(l+m+1)}^{3}\cdots\Gamma_{(l+m+1)}^{m} \Gamma_{(l+m+1)}^{m+1}\Gamma_{(l+m+1)}^{m+3}\cdots\Gamma_{(l+m+1)}^{l+m},\\
\end{aligned}
\end{equation} in which the number of Dirac matrices at the right hand side is $\frac{m+l}{2}-1$ and $\frac{m+l}{2}+1$, respectively;
when $l$ and $m$ are even integers, the $B_{(l+m+1)}^{1,2}$ matrices are
\begin{equation}
\begin{aligned}
&B_{(l+m+1)}^1\\&=\Gamma_{(l+m+1)}^{2}\Gamma_{(l+m+1)}^{4}\cdots \Gamma_{(l+m+1)}^{m}\Gamma_{(l+m+1)}^{m+1}\Gamma_{(l+m+1)}^{m+3}\cdots\Gamma_{(l+m+1)}^{l+m-1},\\
&B_{(l+m+1)}^2\\&=\Gamma_{(l+m+1)}^{1}\Gamma_{(l+m+1)}^{3}\cdots\Gamma_{(l+m+1)}^{m-1}\Gamma_{(l+m+1)}^{m+2} \Gamma_{(l+m+1)}^{m+4}\cdots\Gamma_{(l+m+1)}^{l+m}, \\
\end{aligned}
\end{equation} in which the number of Dirac matrices at the right hand side is $\frac{m+l}{2}$.
These two matrices are generalizations of the ones of Ref.\cite{ryu2010}, which are useful in constructing Dirac Hamiltonians of topological insulators with symmetries. Our generalizations here are useful in model construction of topological defects. The introduction of the two integers $l$ and $m$, instead of a single integer like Ref.\cite{ryu2010}, is to accommodate both momentum-like and space-like coefficients (This will become clear shortly). When $m=0$, the definition of $B_{(l+m+1)}^{1,2}$ reduces to that of Ref.\cite{ryu2010}.

The construction of $B_{(l+m+1)}^{1,2}$ is motivated by the following useful mathematical relations, which will be exploited soon. The first relation is
\begin{equation}\label{b1g}
\begin{aligned}
&(B_{(l+m+1)}^1)^{-1}\Gamma_{(l+m+1)}^{a}B_{(l+m+1)}^1=\\&\left\{
\begin{aligned}
&(-1)^{-\eta/2}\Gamma_{(l+m+1)}^{a*},\,\,  a\leq m\,,\\
&(-1)^{-\eta/2-1}\Gamma_{(l+m+1)}^{a*},\,\, m< a\leq l+m\,.\\
\end{aligned}\right.
\end{aligned}
\end{equation} The additional ``$-1$'' sign for $m< a\leq l+m$ will be crucial shortly. Eq.(\ref{b1g}) can be verified as follows.
For odd-integer $l$ and $m$ we have
\begin{equation}
\begin{aligned}
&(B_{(l+m+1)}^1)^{-1}\Gamma_{(l+m+1)}^{a}B_{(l+m+1)}^1=\\&\left\{
\begin{aligned}
&(-1)^{l-\eta/2-1}\Gamma_{(l+m+1)}^{a*},\,\,  a\leq m\,,\\
&(-1)^{l-\eta/2-2}\Gamma_{(l+m+1)}^{a*},\,\, m< a\leq l+m\, ;\\
\end{aligned}\right.
\end{aligned}
\end{equation}
while for even-integer $l$ and $m$ we have
\begin{equation}
\begin{aligned}
&(B_{(l+m+1)}^1)^{-1}\Gamma_{(l+m+1)}^{a}B_{(l+m+1)}^1=\\&\left\{
\begin{aligned}
&(-1)^{l-\eta/2}\Gamma_{(l+m+1)}^{a*},\,\,  a\leq m\,,\\
&(-1)^{l-\eta/2-1}\Gamma_{(l+m+1)}^{a*},\,\, m< a\leq l+m\,.\\
\end{aligned}\right.
\end{aligned}
\end{equation} Thus, Eq.(\ref{b1g}) is proved.
Similarly, for $B_{(l+m+1)}^2$ we have
\begin{equation}\label{b2g}
\begin{aligned}
&(B_{(l+m+1)}^2)^{-1}\Gamma_{(l+m+1)}^{a}B_{(l+m+1)}^2=\\&\left\{
\begin{aligned}
&(-1)^{-\eta/2+1}\Gamma_{(l+m+1)}^{a*},\,\,  a\leq m\,,\\
&(-1)^{-\eta/2}\Gamma_{(l+m+1)}^{a*},\,\, m< a\leq l+m\,.\\
\end{aligned}\right.
\end{aligned}
\end{equation}
We emphasize that the coefficients at the right hand side, such as $(-1)^{-\eta/2}$, depend only on $\eta=l-m$ (not on $l+m$), which heralds the fact that the topological classifications depend only on $\delta=d-D$ (not on $d+D$).

The matrices $B_{(l+m+1)}^{1,2}$ have the following important properties. For odd-integer $l$ and $m$:
\begin{equation}\label{}
\begin{aligned}
&B_{(l+m+1)}^{1*}B_{(l+m+1)}^1\\&=(-1)^{(m-1)/2}(-1)^{\frac{[(l+m-2)/2-1](l+m-2)/2}{2}}\\
&=(-1)^{(l-\eta-1)/2}(-1)^{\frac{(l-\eta/2-2)(l-\eta/2-1)}{2}}\\
&=(-1)^{-\eta/4}(-1)^{\frac{(l-\eta/2-1)^2}{2}};
\end{aligned}
\end{equation}
similarly, for even-integer $l$ and $m$:
\begin{equation}\label{}
\begin{aligned}
&B_{(l+m+1)}^{1*}B_{(l+m+1)}^1\\&=(-1)^{m/2}(-1)^{\frac{[(l+m-2)/2](l+m)/2}{2}}\\
&=(-1)^{(l-\eta)/2}(-1)^{\frac{(l-\eta/2)(l-\eta/2-1)}{2}}\\
&=(-1)^{-\eta/4}(-1)^{\frac{(l-\eta/2)^2}{2}}.\\
\end{aligned}
\end{equation}
It follows that, irrespective of the parity of $l$ and $m$,
\begin{equation}
B_{(l+m+1)}^{1*}B_{(l+m+1)}^1=\left\{
\begin{aligned}
&(-1)^{-\eta/4},\quad \eta=8n/8n+4,\\
&(-1)^{-(\eta-2)/4},\quad \eta=8n+2/8n+6,\\
\end{aligned}\right.
\end{equation}
where $n$ stands for an integer. Equivalently, we have \begin{equation}
B_{(l+m+1)}^{1*}B_{(l+m+1)}^1=\left\{
\begin{aligned}
& 1,\quad \eta=8n/8n+2,\\
&-1,\quad \eta=8n+4/8n+6,\\
\end{aligned}\right. \label{b1}
\end{equation}
Similarly, for odd-integer $l$ and $m$ we have
\begin{equation}
\begin{aligned}
&B_{(l+m+1)}^{2*}B_{(l+m+1)}^2\\&=(-1)^{\eta/4}(-1)^{\frac{(m+\eta/2+1)^2}{2}},
\end{aligned}
\end{equation}
while for even-integer $l$ and $m$ we have
\begin{equation}
\begin{aligned}
&B_{(l+m+1)}^{2*}B_{(l+m+1)}^2\\&=(-1)^{\eta/4}(-1)^{\frac{(m+\eta/2)^2}{2}}.
\end{aligned}
\end{equation}
Therefore we have, irrespective of the parity of $l$ and $m$,
\begin{equation}
B_{(l+m+1)}^{2*}B_{(l+m+1)}^2=\left\{
\begin{aligned}
&(-1)^{\eta/4},\quad \eta=8n/8n+4,\\
&(-1)^{(\eta+2)/4},\quad \eta=8n+2/8n+6.\\
\end{aligned}\right.
\end{equation} In other words, we have
\begin{equation}
B_{(l+m+1)}^{2*}B_{(l+m+1)}^2=\left\{
\begin{aligned}
& 1,\quad \eta=8n/8n+6,\\
&-1,\quad \eta=8n+2/8n+4,\\
\end{aligned}\right. \label{b2}
\end{equation}
Two more useful matrices are defined as
\begin{equation}\label{tilde}
\begin{aligned}
&\tilde{B}_{(l+m+1)}^1=B_{(l+m+1)}^1\Gamma_{(l+m+1)}^{l+m},\\
&\tilde{B}_{(l+m+1)}^2=B_{(l+m+1)}^2\Gamma_{(l+m+1)}^{l+m-2},\\
\end{aligned}
\end{equation}
which satisfy
\begin{equation}
\tilde{B}_{(l+m+1)}^{1*}\tilde{B}_{(l+m+1)}^1=\left\{
\begin{aligned}
&(-1)^{-(\eta-4)/4},\quad \eta=8n/8n+4,\\
&(-1)^{-(\eta-2)/4},\quad \eta=8n+2/8n+6,\\
\end{aligned}\right.
\end{equation} and
\begin{equation}
\tilde{B}_{(l+m+1)}^{2*}\tilde{B}_{(l+m+1)}^2=\left\{
\begin{aligned}
&(-1)^{\eta/4},\quad \eta=8n/8n+4,\\
&(-1)^{(\eta-2)/4},\quad \eta=8n+2/8n+6.\\
\end{aligned}\right.
\end{equation} Equivalently, we have
\begin{equation}
\tilde{B}_{(l+m+1)}^{1*}\tilde{B}_{(l+m+1)}^1=\left\{
\begin{aligned}
&1,\quad \eta=8n+2/8n+4,\\
&-1,\quad \eta=8n/8n+6,\\
\end{aligned}\right. \label{b1t}
\end{equation} and
\begin{equation}
\tilde{B}_{(l+m+1)}^{2*}\tilde{B}_{(l+m+1)}^2=\left\{
\begin{aligned}
&1,\quad \eta=8n/8n+2,\\
&-1,\quad \eta=8n+4/8n+6.\\
\end{aligned}\right. \label{b2t}
\end{equation}

Following the scheme of dimensional reduction of static topological insulators without topological defects\cite{qi2008,ryu2010}, we construct four generations of representative Dirac Hamiltonians for Floquet topological defects. The Bloch Hamiltonian of the first generation is:
\begin{equation}
\begin{aligned}
(i):\,H_{(l+m+1)}^{l+m}(\bk,\br,t)=\sum_{a=1}^{l+m}d_a(\bk,\br)\Gamma_{(l+m+1)}^a+m(t)\Gamma_{(l+m+1)}^{l+m+1},
\end{aligned}
\end{equation} in which the coefficients $d_a(\bk,\br)$'s satisfy
\begin{equation}
d_a(\bk,\br)=\left\{
\begin{aligned}
&d_a(-\bk,\br),\quad a\leq m,\\
&-d_a(-\bk,\br),\quad m<a\leq l+m, \\
\end{aligned}\right.
\end{equation} in other words, $d_a$'s ($1\leq a\leq m$) are space-like, and $d_a$'s ($m<a\leq l+m$) are momentum-like.
For simplicity, we take $m(t)$ to be an even function of $t$, namely $m(t)=m(-t)$. In this generation, we have $d=l,D=m$, and $\delta=\eta$.  In the treatment of static systems\cite{qi2008,ryu2010}, the momentum-like coefficients are simply taken as $k_i$ ($i=1,2,\cdots,d$), nevertheless, we keep the more general form here. We can also take $m(t)$ as a function of $\bk$ and $\br$, with the property $m(\bk,\br,t)=m(-\bk,\br,t)$. Most of the model Hamiltonians we use in the main text take the Dirac forms. For example, the Dirac Hamiltonians in Eq.(\ref{line-defect-A}) and Eq.(\ref{line-defect-D}) describe Floquet line defects in the three-dimensional space. In these two cases, we have $l=3,m=1$, therefore the rank of Dirac matrices is $2^{(l+m)/2}=4$.

According to Eq.(\ref{b1g}), this Hamiltonian satisfies the symmetry
\begin{equation}
\begin{aligned}
&(B_{(l+m+1)}^1)^{-1}H_{(l+m+1)}^{l+m}(\bk,\br,t)B_{(l+m+1)}^1\\
&=(-1)^{-\delta/2}H_{(l+m+1)}^{l+m*}(-\bk,\br,\pm t).\\
\end{aligned} \label{i-sym}
\end{equation}
The symmetry matrix $B_{(l+m+1)}^1$ satisfies Eq.(\ref{b1}), in which we can replace $\eta$ by $\delta$. We emphasize that replacing $B_{(l+m+1)}^1$ by $B_{(l+m+1)}^2$ at the left hand side of Eq.(\ref{i-sym}) does not produce a simple relation, because the mass term would acquire an opposite sign compared to all other terms at the right hand side.

When $\delta=4n$, the system satisfies the time-reversal symmetry, therefore we can write $B_{(l+m+1)}^1=T$; while when $\delta=4n+2$, the system satisfies the particle-hole symmetry, and we can write $B_{(l+m+1)}^1=C$. When $\delta=8n$, we have $T^*T=1$, therefore the Bloch Hamiltonian belongs to the AI class. When $\delta=8n+4$, we have $T^*T=-1$, therefore the Hamiltonian belong to the AII class. Similarly, when $\delta=8n+2$ we have $C^*C=1$ and the Hamiltonian belongs to class D, while when $\delta=8n+6$, we have $C^*C=-1$ and the Hamiltonian belongs to the class C (see Table \ref{rep}).

\begin{table}[htp!]
\caption{Symmetry classes of representative Dirac Hamiltonians. }
\centering
\begin{tabular*}{7.5cm}{@{\extracolsep{\fill}}   c c| c c c c ccccc }
\hline  \hline
  \multicolumn{1}{c}{Symmetry} &&&  \multicolumn{8}{c}{$\delta=d-D$}       \\

      AZ &    & &     $0$ &     $1$ &     $2$ &     $3$ &     $4$ &     $5$ &     $6$ &     $7$\\
\hline

        AI &      &&     $i$ &      &      &    &     $iii$ &      &     &   \\

         BDI &     &&      &     $ii$ &      &      &      &     $iv$ &      &    \\

         D &     &&     &      &     $i$ &      &      &     &     $iii$ &      \\

         DIII &     & &     &    &     &     $ii$ &      &     &     &    $iv$\\

         AII &      &&     $iii$ &     &     &      &     $i$ &     &     &   \\

        CII &      & &     &     $iv$ &     &    &      &     $ii$ &     &     \\

         C &       & &      &     &     $iii$ &     &     &      &     $i$ &    \\

         CI &       &&     &     &     &     $iv$ &     &     &      &  $ii$ \\
\hline \hline
\end{tabular*} \label{rep}
\end{table}

The second generation is obtained from the first generation by removing one of the momentum-like term $d_{l+m}$:
\begin{equation}
\begin{aligned}
(ii):\,H_{(l+m+1)}^{l+m-1}(\bk,\br,t)=\sum_{a=1}^{l+m-1}d_a(\bk,\br)\Gamma_{(l+m+1)}^a+m(t)\Gamma_{(l+m+1)}^{l+m+1}.
\end{aligned}
\end{equation}
Now we have $d=l-1$ and $D=m$, and $\delta=\eta-1$. According to Eq.(\ref{b1g}) and Eq.(\ref{tilde}), the Hamiltonian has the symmetries
\begin{equation}
\begin{aligned}
&(B_{(l+m+1)}^1)^{-1}H_{(l+m+1)}^{l+m-1}(\bk,\br,t)B_{(l+m+1)}^1\\&=(-1)^{-(\delta+1)/2}H_{(l+m+1)}^{l+m-1*}(-\bk,\br,\pm t),\\
\end{aligned}
\end{equation} and
\begin{equation}
\begin{aligned}
&(\tilde{B}_{(l+m+1)}^1)^{-1}H_{(l+m+1)}^{l+m-1}(\bk,\br,t)\tilde{B}_{(l+m+1)}^1\\&=(-1)^{-(\delta+1)/2-1}H_{(l+m+1)}^{l+m-1*}(-\bk,\br,\pm t).\\
\end{aligned}
\end{equation}
Apparently, the Hamiltonian enjoys a chiral symmetry with $S=\Gamma_{(l+m+1)}^{l+m}$. The symmetry matrices $B_{(l+m+1)}^1$ and $\tilde{B}_{(l+m+1)}^1$ satisfy Eq.(\ref{b1}) and Eq.(\ref{b1t}), which are re-expressed in terms of $\delta$ as
\begin{equation}
B_{(l+m+1)}^{1*}B_{(l+m+1)}^1=\left\{
\begin{aligned}
&1,\quad \delta=8n-1/8n+1,\\
&-1,\quad \delta=8n+3/8n+5,\\
\end{aligned}\right.
\end{equation} and
\begin{equation}
\tilde{B}_{(l+m+1)}^{1*}\tilde{B}_{(l+m+1)}^1=\left\{
\begin{aligned}
&1,\quad \delta=8n+1/8n+3,\\
&-1,\quad \delta=8n-1/8n+5.\\
\end{aligned}\right.
\end{equation}
As an example, when $\delta=8n+1$, the Hamiltonian enjoys both the time-reversal symmetry and particle-hole symmetry with $T^*T=1$ and $C^*C=1$, which indicates that it belongs to the class BDI. Similarly, when $\delta=8n+3$, one can check that the Hamiltonian belongs to the class DIII. The symmetry classes of other values of $\delta$ are listed in Table \ref{rep}.

We remark that the representative Dirac Hamiltonians for $\mathbb{Z}_2$ topological classes can be obtained from the Hamiltonians (i) and (ii) by removing one or two momentum-like terms (thus reducing $\delta$ by $1$ or $2$), while keeping the same symmetries. For example, the representative Dirac Hamiltonian for class AII in the dimension $\delta=3$ can be obtained from the Hamiltonian (i) by removing a momentum-like term, which is simply the Hamiltonian (ii). We have seen that the Hamiltonian (ii) also serves as a representative for the class DIII in dimension $\delta=3$. This is not a problem: The same Dirac Hamiltonian can be taken as the representative of more than one symmetry classes, though the symmetry operation is different for each class. This is also true for static systems. In fact, for the case of $d=3, D=0$, the BdG Hamiltonian of $^3$He-B phase (class DIII) and the model Hamiltonian of the topological insulator Bi$_2$Se$_3$ (class AII) indeed take the same form (up to a basis change)\cite{qi2011}.

Apparently, the Hamiltonians (i) and (ii) can also be taken as the representative Dirac Hamiltonians of class A and class AIII, respectively, though we focus on the eight real classes in this appendix.

The third generation of Bloch Hamiltonian is given as
\begin{equation}
\begin{aligned}
&(iii):\,H_{(l+m+1)}^{l+m-2}(\bk,\br,t)\\&=\sum_{a=1}^{l+m-2}d_a(\bk,\br)\Gamma_{(l+m+1)}^a+im(t)\Gamma_{(l+m+1)}^{l+m+1}\Gamma_{(l+m+1)}^{l+m}\Gamma_{(l+m+1)}^{l+m-1}.
\end{aligned}
\end{equation}
Now we have $d=l-2, D=m$ and $\delta=\eta-2$. We emphasize that the mass term has been chosen as  $im(t)\Gamma_{(l+m+1)}^{l+m+1}\Gamma_{(l+m+1)}^{l+m}\Gamma_{(l+m+1)}^{l+m-1}$. The $m(t)\Gamma_{(l+m+1)}^{l+m+1}$ term is not qualified as a mass term because the resultant Hamiltonian would have two chiral symmetries with chiral matrices $S_1=\Gamma_{(l+m+1)}^{l+m}$ and $S_2=\Gamma_{(l+m+1)}^{l+m-1}$, which is not in the framework of the tenfold way classes.

According to Eq.(\ref{b2g}), the Hamiltonian (iii) satisfies the symmetry
\begin{equation}
\begin{aligned}
&(B_{(l+m+1)}^2)^{-1}H_{(l+m+1)}^{l+m-2}(\bk,\br,t)B_{(l+m+1)}^2\\&=(-1)^{-\delta/2}H_{(l+m+1)}^{l+m-2*}(-\bk,\br,\pm t),\\
\end{aligned}
\end{equation}
in which the symmetry matrix satisfies
\begin{equation}\label{}
B_{(l+m+1)}^{2*}B_{(l+m+1)}^2=\left\{
\begin{aligned}
&1,\quad \delta=8n+4/8n+6,\\
&-1,\quad \delta=8n/8n+2.\\
\end{aligned}\right.
\end{equation}
As an example, when  $\delta=8n$, the Hamiltonian has the time-reversal symmetry with $T^*T=-1$ ($T=B_{(l+m+1)}^2$), which belongs to class AII.

The last generation of Dirac Hamiltonian is
\begin{equation}
\begin{aligned}
&(iv): H_{(l+m+1)}^{l+m-3}(\bk,\br,t)\\&=\sum_{a=1}^{l+m-3}d_a(\bk,\br)\Gamma_{(l+m+1)}^a+im(t)\Gamma_{(l+m+1)}^{l+m+1} \Gamma_{(l+m+1)}^{l+m}\Gamma_{(l+m+1)}^{l+m-1},
\end{aligned}
\end{equation} which has $d=l-3$, $D=m$ and $\delta=\eta-3$.
Taking advantage of Eq.(\ref{b2g}) and Eq.(\ref{tilde}), we can see that this Hamiltonian has the symmetries
\begin{equation}
\begin{aligned}
&(B_{(l+m+1)}^2)^{-1}H_{(l+m+1)}^{l+m-3}(\bk,\br,t)B_{(l+m+1)}^2\\&=(-1)^{-(\delta+1)/2}H_{(l+m+1)}^{l+m-3*}(-\bk,\br,\pm t),\\
\end{aligned}
\end{equation}
and
\begin{equation}
\begin{aligned}
&(\tilde{B}_{(l+m+1)}^2)^{-1}H_{(l+m+1)}^{l+m-3}(\bk,\br,t)\tilde{B}_{(l+m+1)}^2\\&=(-1)^{-(\delta+3)/2}H_{(l+m+1)}^{l+m-3*}(-\bk,\br,\pm t).\\
\end{aligned}
\end{equation}
Apparently, the Hamiltonian also has a chiral symmetry with $S=\Gamma_{(l+m+1)}^{l+m-2}$, which is proportional to $\tilde{B}_{(l+m+1)}^2 (B_{(l+m+1)}^2)^{-1}$.
The symmetry matrices satisfy Eq.(\ref{b2}) and Eq.(\ref{b2t}), which are
\begin{equation}
B_{(l+m+1)}^{2*}B_{(l+m+1)}^2=\left\{
\begin{aligned}
&1,\quad \delta=8n+3/8n+5,\\
&-1,\quad \delta=8n+1/8n+7,\\
\end{aligned}\right.
\end{equation}
and
\begin{equation}\label{}
\tilde{B}_{(l+m+1)}^{2*}\tilde{B}_{(l+m+1)}^2=\left\{
\begin{aligned}
&1,\quad \delta=8n+5/8n+7,\\
&-1,\quad \delta=8n+1/8n+3.\\
\end{aligned}\right.
\end{equation}
As an example, when $\delta=8n+7$, the Hamiltonian has both time-reversal symmetry and particle-hole symmetry with $T^*T=-1$ ($T=B_{(l+m+1)}^2$) and $C^*C=1$ ($C=\tilde{B}_{(l+m+1)}^2$), which belongs to class DIII. As another example, when $\delta=8n+1$, it has time-reversal symmetry and particle-hole symmetry with $T^*T=C^*C=-1$ ($C=B_{(l+m+1)}^2$, $T=\tilde{B}_{(l+m+1)}^2$), which indicates that the Hamiltonian is in class CII. The model Hamiltonian in Eq.(\ref{point-CII}) belongs to this class. In that specific case, we should take $l=5$ and $m=1$ ($d=l-3=2$, $D=m=1$) as the starting point, therefore, the Dirac matrices are $8\times 8$ ones.

Let us reemphasize that all the symmetry properties depend only on $\delta=d-D$, not on $d+D$, which is consistent with the approach from topological invariants discussed in the main text.

Finally, let us explain the $2\mathbb{Z}$ topological invariants from the perspective of Dirac Hamiltonian. We will show that the representative Hamiltonians (iii) and (iv) have $2\mathbb{Z}$ winding numbers. Let us focus on (iii) first.
It is readily seen that $H_{(l+m+1)}^{l+m-2}(\bk,\br,t)$ satisfies $[H_{(l+m+1)}^{l+m-2}(\bk,\br,t),\Gamma_{(l+m+1)}^{i}\Gamma_{(l+m+1)}^{j}]=0$, in which $(i,j)=\{(l+m-1,l+m),(l+m,l+m+1),(l+m+1,l+m-1)\}$, which implies that the Hamiltonian can be made block-diagonal. In our basis of Dirac matrices in Eq.(\ref{gamma}), the Hamiltonian reads
\begin{equation}
\begin{aligned}
H_{(l+m+1)}^{l+m-2}(\bk,\br,t)=\sum_{a=1}^{l+m-2}d_a(\bk,\br)\Gamma_{(l+m-1)}^a\otimes\sigma_3+m(t)\Gamma_{(l+m-1)}^{l+m-1}\otimes\sigma_0.
\end{aligned}
\end{equation}
It can also be rewritten as
\begin{equation}
\label{}
\begin{aligned}
H_{(l+m+1)}^{l+m-2}(\bk,\br,t)=
\begin{pmatrix}
H_{(l+m-1)}^{l+m-2}(\bk,\br,t)&  \\
&-\bar{H}_{(l+m-1)}^{l+m-2}(\bk,\br,t)\\
\end{pmatrix},
\end{aligned}
\end{equation}
in which $\bar{H}_{(l+m-1)}^{l+m-2}(\bk,\br,t)$ stands for the expression obtained from $H_{(l+m-1)}^{l+m-2}(\bk,\br,t)$ by replacing $m(t)$ by $-m(t)$:
\begin{equation}
\begin{aligned}
\bar{H}_{(l+m-1)}^{l+m-2}(\bk,\br,t)=\sum_{a=1}^{l+m-2}d_a(\bk,\br)\Gamma_{(l+m-1)}^a-m(t)\Gamma_{(l+m-1)}^{l+m-1}.
\end{aligned}
\end{equation}
Let us introduce the matrix
\begin{equation}
\label{}
\begin{aligned}
D_{(l+m+1)}=
\begin{pmatrix}
I&  \\
&\Gamma_{(l+m-1)}^{l+m-1}\\
\end{pmatrix},
\end{aligned}
\end{equation} so that
\begin{equation}
\label{}
\begin{aligned}
&D_{(l+m+1)}^{-1}H_{(l+m+1)}^{l+m-2}(\bk,\br,t)D_{(l+m+1)}=\\&
\begin{pmatrix}
H_{(l+m-1)}^{l+m-2}(\bk,\br,t)&  \\
&H_{(l+m-1)}^{l+m-2}(\bk,\br,t)\\
\end{pmatrix}.
\end{aligned}
\end{equation}

Now let us consider the time evolution operator. In our basis, it takes a block-diagonal form:
\begin{widetext}
\begin{equation}
\begin{aligned}
&U_{(l+m+1)}^{l+m-2}(\bk,\br,t)=[1-i\Delta tH_{(l+m+1)}^{l+m-2}(\bk,\br,t-\Delta t)] [1-i\Delta tH_{(l+m+1)}^{l+m-2}(\bk,\br,t-2\Delta t)]\cdots[1-i\Delta tH_{(l+m+1)}^{l+m-2}(\bk,\br,0)]\\&=
\begin{pmatrix}
\begin{aligned}&[1-i\Delta tH_{(l+m-1)}^{l+m-2}(\bk,\br,t-\Delta t)][1-i\Delta tH_{(l+m-1)}^{l+m-2}(\bk,\br,t-2\Delta t)]\\&\cdots[1-i\Delta tH_{(l+m-1)}^{l+m-2}(\bk,\br,0)]\end{aligned}&  \\
&\begin{aligned}&[1+i\Delta t\bar{H}_{(l+m-1)}^{l+m-2}(\bk,\br,t-\Delta t)][1+i\Delta t\bar{H}_{(l+m-1)}^{l+m-2}(\bk,\br,t-2\Delta t)]\\&\cdots[1+i\Delta t\bar{H}_{(l+m-1)}^{l+m-2}(\bk,\br,0)]\end{aligned}\\
\end{pmatrix},
\end{aligned}
\end{equation} \end{widetext}
which satisfies
\bea
 && D_{(l+m+1)}^{-1}U_{(l+m+1)}^{l+m-2}(\bk,\br,t)D_{(l+m+1)} \nn \\ &=&
\begin{pmatrix}
U_{(l+m-1)}^{l+m-2}(\bk,\br,t)&  \\
&U_{(l+m-1)}^{l+m-2}(\bk,\br,t)\\
\end{pmatrix}.
\eea
In this way, we can further show that the periodized time evolution operator (with branch cut at $\varepsilon$) satisfies
\bea
&& D_{(l+m+1)}^{-1}U_{\varepsilon,(l+m+1)}^{l+m-2}(\bk,\br,t)D_{(l+m+1)} \nn \\ &=&
\begin{pmatrix}
U_{\varepsilon,(l+m-1)}^{l+m-2}(\bk,\br,t)&  \\
&U_{\varepsilon,(l+m-1)}^{l+m-2}(\bk,\br,t)\\
\end{pmatrix}. \eea
Therefore, the winding number becomes the sum of the contributions from two identical blocks:  \bea W[U_{\varepsilon(l+m+1)}^{l+m-2}(\bk,\br,t)]=2W[U_{\varepsilon(l+m-1)}^{l+m-2}(\bk,\br,t)], \eea which is always an even integer.
Similar analysis is applicable to the $(iv)$ generation. Therefore, the winding numbers always take even-integer values for (iii) and (iv) in Table \ref{rep}.

Under the assumption that any Hamiltonian can be smoothly deformed to a Dirac representative in the same topological class (this assumption is quite natural because we have found a Dirac representative for each tenfold-way class), we can see that all the (iii)'s and (iv)'s in Table \ref{rep} have $2\mathbb{Z}$ topological invariants.

\section{Equivalence of the effective-Hamiltonian-based band topological invariants and the frequency-domain band topological invariants}\label{sec:proof}

As we have mentioned in the main text, there are two natural Chern numbers of Floquet bands. The first one is defined in terms of the effective Hamiltonian $H^{\rm eff}$, while the second one is defined in terms of the Floquet Hamiltonian $\mathcal{H}$ in the frequency-domain formulation. Let us recall their definitions and study their relation.

The first band Chern number has been defined in the main text [Eq.(\ref{band-chern})]. In line with the frequency-domain formulation to be discussed shortly, we will use the quasienergy $\epsilon$ instead of the dimensionles quasienergy $\varepsilon=\epsilon\tau$ in this appendix. Accordingly, we reproduce Eq.(\ref{band-chern}) here using the notation of $\epsilon$.  For the Floquet bands with quasienergy in $[\epsilon,\epsilon']$ ($0\leq\epsilon<\epsilon'<\omega$), the band Chern number reads
\begin{equation}
\label{band-chern-appendix}
\begin{aligned}
&C_{(d+D)/2}(P_{\epsilon,\epsilon'})=\tilde{K}_{d+D}\int_{T^d\times S^D} d^dk d^Dr\\
&\times\text{Tr}[\epsilon^{\alpha_1\alpha_2\cdots\alpha_{d+D}} P_{\epsilon,\epsilon'}\partial_{\alpha_1}P_{\epsilon,\epsilon'} \cdots\partial_{\alpha_{d+D}}P_{\epsilon,\epsilon'}],
\end{aligned}
\end{equation} in which the Floquet band projection operator  $P_{\epsilon,\epsilon'}=\sum_{\epsilon<\epsilon_n<\epsilon'} \ket{\psi_n(\bk,\br)}\bra{\psi_n(\bk,\br)}$, where $\ket{\psi_n(\bk,\br)}$'s are the eigenvectors of $U(\bk,\br,\tau)$, or equivalently, eigenvectors of the effective Hamiltonian $H^{\rm eff}(\bk,\br)$, which is given by the logarithm of $U(\bk,\br,\tau)$. The numerical coefficient $\tilde{K}_{d+D}$ has been defined in Eq.(\ref{static2}). The simple property of Eq.(\ref{difference}) indicates that this Chern number is a very natural band topological invariant, because it is exactly the difference between the winding numbers defined at $\epsilon'$ and $\epsilon$. Since $P_{\epsilon,\epsilon'}$ is determined by the full-period time evolution operator $U(\bk,\br,\tau)$, or by the effective Hamiltonian $H^{\rm eff}(\bk,\br)$, the Chern number in Eq.(\ref{band-chern-appendix}) may be called the ``effective-Hamiltonian-based band Chern number''.

Nevertheless, there is yet another very natural band Chern number, which is based on the frequency-domain formulation. Since the frequency-domain formulation is widely used in numerical calculations of Floquet systems, this Chern number is also a valuable one (Indeed, it has been used in the main text of this paper).  To define it, let us start from the time-dependent Schr\"odinger equation
\begin{equation}
\label{}
i\partial_t\ket{\psi_n(\bk,\br,t)}=H(\bk,\br,t)\ket{\psi_n(\bk,\br,t)},
\end{equation}
and take the standard Fourier transformation to the frequency domain,
\begin{equation}
\label{time-dep-solution}
\ket{\psi_n(\bk,\br,t)}=\exp[-i\epsilon_n(\bk,\br)t]\sum_m \exp(im\omega t) \ket{\phi_n^{(m)}(\bk,\br)},
\end{equation} where $\epsilon_n(\bk,\br)$ is the quasienergy, and $\ket{\phi_n^{(m)}(\bk,\br)}$'s are $N$-component column vectors ($N$ is the number of static bands if the driving is removed).
In this frequency domain, the Schr\"odinger equation becomes
\begin{eqnarray}
\sum_{m'}\mathcal{H}_{mm^{'}}(\bk,\br)|\phi_n^{(m')}(\bk,\br)\rangle
=\epsilon_n(\bk,\br)|\phi_n^{(m)}(\bk,\br)\rangle,\label{} \label{freq-domain}
\end{eqnarray}
where \bea \mathcal{H}_{mm^{'}}(\bk,\br)=m\omega \delta_{mm'}\mathbf{I}+H_{m-m'}(\bk,\br), \eea
in which $H_{m-m'}(\bk,\br)$'s are the Fourier components of $H(\bk,\br,t)$, namely, \bea H_{m}(\bk,\br)=\frac{1}{\tau}\int_{0}^{\tau}dt H(\bk,\br,t) \exp(-im\omega t),  \eea and $H_{m-m'}$ is obtained by the replacement $m\rw m-m'$ in this expression. The matrix $\mathcal{H}$ is referred to as the ``Floquet Hamiltonian'', whose rank is infinite. In practice, we may take a truncation, keeping $M^2$ Floquet blocks with a sufficiently large $M$ (i.e., the Floquet index $m$ is restricted to $m\in [-M/2,M/2]$). To be more explicit, Eq.(\ref{freq-domain}) reads
\begin{equation}
\begin{pmatrix}
\cdots& & & &\\
&H_0+\omega&H_1&H_2&\\
&H_{-1}&H_0&H_1&\\
&H_{-2}&H_{-1}&H_0-\omega&\\
&&&&\cdots\\
\end{pmatrix}
\begin{pmatrix}
\cdots \\
\phi_n^{(1)}\\
\phi_n^{(0)}\\
\phi_n^{(-1)}\\
\cdots \\
\end{pmatrix}
=\epsilon_n
\begin{pmatrix}
\cdots \\
\phi_n^{(1)}\\
\phi_n^{(0)}\\
\phi_n^{(-1)}\\
\cdots \\
\end{pmatrix}.
\end{equation} The column eigenvector $(\cdots,\phi_n^{(1)},\phi_n^{(0)},\phi_n^{(-1)},\cdots)^T$ here will be denoted as $\ket{\Phi_n}$ for brevity. As long as $\omega$ is nonzero, the wavefunction profile of $\ket{\Phi_n}$ is localized in the $m$ space (mathematically, the problem resembles the Wannier-Stark ladder, for which the localization in the $m$ direction has been studied before\cite{emin1987existence}).

A notable property of this eigenvalue problem is the periodicity in the Floquet $m$ space. It can be readily checked that
\begin{equation}
\begin{pmatrix}
\cdots& & & &\\
&H_0+\omega&H_1&H_2&\\
&H_{-1}&H_0&H_1&\\
&H_{-2}&H_{-1}&H_0-\omega&\\
&&&&\cdots\\
\end{pmatrix}
\begin{pmatrix}
\cdots \\
\phi_n^{(1-m)}\\
\phi_n^{(-m)}\\
\phi_n^{(-1-m)}\\
\cdots \\
\end{pmatrix}
=(\epsilon_n+m\omega)
\begin{pmatrix}
\cdots \\
\phi_n^{(1-m)}\\
\phi_n^{(-m)}\\
\phi_n^{(-1-m)}\\
\cdots \\
\end{pmatrix}, \label{shifting}
\end{equation}
in other words,  shifting of the eigenvectors in the $m$ space are also eigenvectors, with eigenvalues increased or decreased by multiples of $\omega$.

Having explained the above background knowledge, let us define the frequency-domain Chern number. It is defined in the same way as the usual Chern number of static systems, taking $\mathcal{H}(\bk,\br)$ as the ``static Hamiltonian''. For the Floquet bands with quasienergy in $[\epsilon,\epsilon']$, the Chern number can be defined in terms of the projection operator \bea \mathcal{P}_{\epsilon,\epsilon'}(\bk,\br)=\sum_{\epsilon<\epsilon_n<\epsilon'} \ket{\Phi_n(\bk,\br)}\bra{\Phi_n(\bk,\br)}, \eea where $\ket{\Phi_n(\bk,\br)}$ is the eigenvector of $\mathcal{H}$ with eigenvalue $\epsilon_n(\bk,\br)$, as mentioned above. Here, $\ket{\Phi_n(\bk,\br)}$'s form an orthonormal basis, namely, $\bra{\Phi_n(\bk,\br)}\Phi_{n'}(\bk,\br)\ra =\sum_m \bra{\phi_n^{(m)}(\bk,\br)}\phi_{n'}^{(m)}(\bk,\br)\ra =\delta_{nn'}$.  The frequency-domain band
Chern number is defined as  \begin{equation}
\label{band-chern-2}
\begin{aligned}
&\mathcal{C}_{(d+D)/2}(\mathcal{P}_{\epsilon,\epsilon'})=\tilde{K}_{d+D}\int_{T^d\times S^D} d^dk d^Dr\\
&\times\text{Tr}[\epsilon^{\alpha_1\alpha_2\cdots\alpha_{d+D}} \mathcal{P}_{\epsilon,\epsilon'}\partial_{\alpha_1}\mathcal{P}_{\epsilon,\epsilon'} \cdots\partial_{\alpha_{d+D}}\mathcal{P}_{\epsilon,\epsilon'}],
\end{aligned}
\end{equation} where the coefficient $\tilde{K}_{d+D}$ is the same as in Eq.(\ref{band-chern-appendix}).

What is the relation between this frequency-domain Chern number in Eq.(\ref{band-chern-2}) and the effective-Hamiltonian-based Chern  number in Eq.(\ref{band-chern-appendix})? They look quite different: Eq.(\ref{band-chern-appendix}) uses the $N$-component vectors $\ket{\psi_n(\bk,\br)}$'s, while Eq.(\ref{band-chern-2}) uses the $MN$-component vectors $\ket{\Phi_n(\bk,\br)}$'s.

Let us first have a closer inspection. In the definition of $\mathcal{C}_{(d+D)/2}(\mathcal{P}_{\epsilon,\epsilon'})$ in Eq.(\ref{band-chern-2}), the derivative always look like $\bra{\phi_n^{(m)}}\partial_{\alpha_i}\phi_{n'}^{(m)}\ra$ or $\bra{\partial_{\alpha_i}\phi_n^{(m)}}\phi_{n'}^{(m)}\ra$. There is no mixing of $m$ and $m'$ for $m\neq m'$. This fact can be seen from the more explicit expression of the projection operator
\begin{equation}
\label{}
\begin{aligned}
&\mathcal{P}_{\epsilon,\epsilon'}=\sum_{\epsilon<\epsilon_n<\epsilon'}
\begin{pmatrix}
\cdots \\
\ket{\phi_n^{(1)}}\\
\ket{\phi_n^{(0)}}\\
\ket{\phi_n^{(-1)}}\\
\cdots \\
\end{pmatrix}
\begin{pmatrix}
\cdots, &
\bra{\phi_n^{(1)}},&
\bra{\phi_n^{(0)}},&
\bra{\phi_n^{(-1)}},&
\cdots
\end{pmatrix}\\
&=\sum_{\epsilon<\epsilon_n<\epsilon'}
\begin{pmatrix}
\cdots& & && \\
&\ket{\phi_n^{(1)}}\bra{\phi_n^{(1)}}&\ket{\phi_n^{(1)}}\bra{\phi_n^{(0)}}&\ket{\phi_n^{(1)}}\bra{\phi_n^{(-1)}}&\\
&\ket{\phi_n^{(0)}}\bra{\phi_n^{(1)}}&\ket{\phi_n^{(0)}}\bra{\phi_n^{(0)}}&\ket{\phi_n^{(0)}}\bra{\phi_n^{(-1)}}&\\
&\ket{\phi_n^{(-1)}}\bra{\phi_n^{(1)}}&\ket{\phi_n^{(-1)}}\bra{\phi_n^{(0)}}&\ket{\phi_n^{(-1)}}\bra{\phi_n^{(-1)}}&\\
&&&&\cdots \\
\end{pmatrix}.
\end{aligned}
\end{equation}

On the other hand, in Eq.(\ref{band-chern-appendix}), we used the effective-Hamiltonian-based projection operator $P_{\epsilon,\epsilon'}=\sum_{\epsilon<\epsilon_n<\epsilon'} \ket{\psi_n(\bk,\br)}\bra{\psi_n(\bk,\br)}$. It is not difficult to see that $\ket{\psi_n(\bk,\br)}$'s here are simply $\ket{\psi_n(\bk,\br,t=0)}$'s, where $\ket{\psi_n(\bk,\br,t)}$'s are the solutions to the time-dependent Schr\"odinger equation given by Eq.(\ref{time-dep-solution}), which justifies using similar notation ``$\psi_n$'' for the two vectors, $\ket{\psi_n(\bk,\br)}$ and $\ket{\psi_n(\bk,\br,t)}$. In fact, we have $U(\bk,\br,\tau)\ket{\psi_n(\bk,\br,t=0)}= \exp[-i\epsilon_n(\bk,\br)\tau) \ket{\psi_n(\bk,\br,t=0)}$.

It follows from Eq.(\ref{time-dep-solution}) that \bea \ket{\psi_n(\bk,\br)} = \ket{\psi_n(\bk,\br,t=0)}= \sum_m \ket{\phi_n^{(m)}(\bk,\br)}. \label{psi-expansion} \eea
Now we can see that the $N$-component vector $\ket{\psi_n(\bk,\br)}$, which is used in Eq.(\ref{band-chern-appendix}), comes from the $m$-summation of the $MN$ components of $\ket{\Phi_n(\bk,\br)}$. Given the orthonormal condition of $\ket{\Phi_n(\bk,\br)}$, namely, $\bra{\Phi_n(\bk,\br)}\Phi_{n'}(\bk,\br)\ra =\sum_m \bra{\phi_n^{(m)}(\bk,\br)}\phi_{n'}^{(m)}(\bk,\br)\ra =\delta_{nn'}$, we can show that $\ket{\psi_n(\bk,\br)}$'s are also orthonormal. In fact, $\bra{\psi_n(\bk,\br)}\psi_{n'}(\bk,\br)\ra =\sum_m\sum_{m'} \bra{\phi_n^{(m)}(\bk,\br)}\phi_{n'}^{(m')}(\bk,\br)\ra= \sum_{l} (\sum_m \bra{\phi_n^{(m)}(\bk,\br)}\phi_{n'}^{(m+l)}(\bk,\br)\ra)$, in which the $l\neq 0$ terms all vanish due to the fact that the eigenvectors of the Hermitian matrix $\mathcal{H}$ with different eigenvalues are orthogonal [let us also recall Eq.(\ref{shifting})]. Therefore, we have $\bra{\psi_n(\bk,\br)}\psi_{n'}(\bk,\br)\ra= \bra{\Phi_n(\bk,\br)}\Phi_{n'}(\bk,\br)\ra$.

Given Eq.(\ref{psi-expansion}), the effective-Hamiltonian-based projection operator reads \bea P_{\epsilon,\epsilon'} = \sum_{\epsilon<\epsilon_n<\epsilon'}\sum_{m_1,m_2} \ket{\phi_n^{(m_1)}(\bk,\br)}\bra{\phi_n^{(m_2)}(\bk,\br)}, \eea
therefore, the expression of Eq.(\ref{band-chern-appendix}) in terms of $\ket{\phi_n^{(m)}(\bk,\br)}$ involves $\bra{\phi_n^{(m)}}\partial_{\alpha_i}\phi_{n'}^{(m')}\ra$ with $m\neq m'$, which is unlike the case of Eq.(\ref{band-chern-2}), whose expression only involves $\bra{\phi_n^{(m)}}\partial_{\alpha_i}\phi_{n'}^{(m)}\ra$.

If one of the $m$ components of $\ket{\Phi_n(\bk,\br)}$, say $\ket{\phi_n^{(m_0)}(\bk,\br)}$, dominates over all other components, then $\ket{\psi_n(\bk,\br)}= \sum_m \ket{\phi_n^{(m)}(\bk,\br)}\approx \ket{\phi_n^{(m_0)}(\bk,\br)}$, and we can expect that Eq.(\ref{band-chern-2}) and Eq.(\ref{band-chern-appendix}) yield the same integer. However, in the most interesting cases, the profile of $\ket{\Phi_n}$ can be quite extended in the $m$ direction, with several $m$ components having comparable weights, then there seems to be no straightforward relation between the Chern number calculated from the $MN$-component vectors $\ket{\Phi_n(\bk,\br)}$'s and that calculated from the $N$-component vecotr $\ket{\psi_n(\bk,\br)}$'s. At this stage, one may wonder whether Eq.(\ref{band-chern-2}) and Eq.(\ref{band-chern-appendix}) are equal or not. Although the bulk-defect correspondence suggests an affirmative answer (both topological invariants are expected to count the topological defect modes), a straightforward proof is desirable.

In this appendix, we are able to prove the general result:  \bea C_{(d+D)/2}(P_{\epsilon,\epsilon'})= \mathcal{C}_{(d+D)/2}(\mathcal{P}_{\epsilon,\epsilon'}),\label{equivalence} \eea namely, the Chern numbers in Eq.(\ref{band-chern-appendix}) and Eq.(\ref{band-chern-2}) are equivalent. This is the main result of this appendix.

To prove Eq.(\ref{equivalence}), we observe that the Floquet bands of $\mathcal{H}$ have periodicity of $\omega$, as manifested in Eq.(\ref{shifting}). The eigenvectors with eigenvalues in $[\epsilon+m\omega,\epsilon'+m\omega]$ are the same as in $[\epsilon,\epsilon']$, except for a shifting. The Chern number for $\mathcal{P}_{\epsilon+m\omega,\epsilon'+m\omega}$ must be the same as that of $\mathcal{P}_{\epsilon,\epsilon'}$: \bea \mathcal{C}_{(d+D)/2}(\mathcal{P}_{\epsilon+m\omega,\epsilon'+m\omega}) = \mathcal{C}_{(d+D)/2}(\mathcal{P}_{\epsilon,\epsilon'}), \eea which, combined with the summation property of Chern number, tells us that
\bea \mathcal{C}_{(d+D)/2}(\sum_m \mathcal{P}_{\epsilon+m\omega,\epsilon'+m\omega}) &=&  \sum_m\mathcal{C}_{(d+D)/2} (\mathcal{P}_{\epsilon+m\omega,\epsilon'+m\omega}) \nn \\ &=& M\,\mathcal{C}_{(d+D)/2}(\mathcal{P}_{\epsilon,\epsilon'}). \label{sum-chern} \eea Strictly speaking, the Chern numbers of the bands near the truncation, namely, $m\approx \pm M/2$, may be different from $\mathcal{C}_{(d+D)/2}(\mathcal{P}_{\epsilon,\epsilon'})$ due to the truncation error, therefore, the ``$=$'' in Eq.(\ref{sum-chern}) is accurate only to the order of $M$, which nevertheless suffices our purpose as we take sufficiently large $M$.  The strategy of proving Eq.(\ref{equivalence}) is to calculate the left hand side of Eq.(\ref{sum-chern}), $\mathcal{C}_{(d+D)/2}(\sum_m \mathcal{P}_{\epsilon+m\omega,\epsilon'+m\omega})$, and then divide it by $M$, hoping that the result can be related to Eq.(\ref{band-chern-appendix}).

To be explicit, the projection operator of the bands in  $[\epsilon+m\omega,\epsilon'+m\omega]$ reads
\begin{widetext}
\bea
\mathcal{P}_{\epsilon+m\omega,\epsilon'+m\omega}&=&\sum_{\epsilon<\epsilon_n<\epsilon'}
\begin{pmatrix}
\cdots \\
\ket{\phi_n^{(1-m)}}\\
\ket{\phi_n^{(-m)}}\\
\ket{\phi_n^{(-1-m)}}\\
\cdots \\
\end{pmatrix}
\begin{pmatrix}
\cdots, &
\bra{\phi_n^{(1-m)}},&
\bra{\phi_n^{(-m)}},&
\bra{\phi_n^{(-1-m)}},&
\cdots
\end{pmatrix} \nn \\
&=&\sum_{\epsilon<\epsilon_n<\epsilon'}
\begin{pmatrix}
\cdots& & && \\
&\ket{\phi_n^{(1-m)}}\bra{\phi_n^{(1-m)}}&\ket{\phi_n^{(1-m)}}\bra{\phi_n^{(-m)}}&\ket{\phi_n^{(1-m)}}\bra{\phi_n^{(-1-m)}}&\\
&\ket{\phi_n^{(-m)}}\bra{\phi_n^{(1-m)}}&\ket{\phi_n^{(-m)}}\bra{\phi_n^{(-m)}}&\ket{\phi_n^{(-m)}}\bra{\phi_n^{(-1-m)}}&\\
&\ket{\phi_n^{(-1-m)}}\bra{\phi_n^{(1-m)}}&\ket{\phi_n^{(-1-m)}}\bra{\phi_n^{(-m)}}&\ket{\phi_n^{(-1-m)}}\bra{\phi_n^{(-1-m)}}&\\
&&&&\cdots \\
\end{pmatrix},
\eea
therefore, the sum is
\begin{equation}
\begin{aligned}
\sum_m\mathcal{P}_{\epsilon+m\omega,\epsilon'+m\omega}
=\sum_{\epsilon<\epsilon_n<\epsilon'}
\begin{pmatrix}
\cdots& & && \\
&\sum_m\ket{\phi_n^{(1-m)}}\bra{\phi_n^{(1-m)}}&\sum_m\ket{\phi_n^{(1-m)}}\bra{\phi_n^{(-m)}}&\sum_m\ket{\phi_n^{(1-m)}}\bra{\phi_n^{(-1-m)}}&\\
&\sum_m\ket{\phi_n^{(-m)}}\bra{\phi_n^{(1-m)}}&\sum_m\ket{\phi_n^{(-m)}}\bra{\phi_n^{(-m)}}&\sum_m\ket{\phi_n^{(-m)}}\bra{\phi_n^{(-1-m)}}&\\
&\sum_m\ket{\phi_n^{(-1-m)}}\bra{\phi_n^{(1-m)}}&\sum_m\ket{\phi_n^{(-1-m)}}\bra{\phi_n^{(-m)}}&\sum_m\ket{\phi_n^{(-1-m)}}\bra{\phi_n^{(-1-m)}}&\\
&&&&\cdots \\
\end{pmatrix}.
\end{aligned}
\end{equation}
To simplify the expression, let us introduce the shorthand notation:
\begin{equation}
\begin{aligned}
P_{(m)}&= \sum_{\epsilon<\epsilon_n<\epsilon'}\sum_{m_1}\sum_{m_2} \ket{\phi_{n}^{(m_1)}}\bra{\phi_{n}^{(m_2)}} \delta_{m_1-m_2-m}\\ &=\sum_{\epsilon<\epsilon_n<\epsilon'}\sum_{m'}\ket{\phi_{n}^{(m'+m)}}\bra{\phi_{n}^{(m')}},
\end{aligned} \label{(m)}
\end{equation} where the subscript ``$(m)$'' here indicates that the Floquet index (or the sum of indices) of the ket-vectors minus that of the bra-vectors is $m$, and the quasienergies $\epsilon,\epsilon'$ are implicit. The same notation will be used below. Apparently, we have \bea P_{\epsilon,\epsilon'}=\sum_m P_{(m)}. \eea
With the shorthand notations, the projection operator $\sum_m\mathcal{P}_{\epsilon+m\omega,\epsilon'+m\omega}$ reads
\begin{equation}
\label{}
\begin{aligned}
\sum_m\mathcal{P}_{\epsilon+m\omega,\epsilon'+m\omega} =
\begin{pmatrix}
\cdots& & && \\
&P_{(0)}&P_{(1)}&P_{(2)}&\\
&P_{(-1)}&P_{(0)}&P_{(1)}&\\
&P_{(-2)}&P_{(-1)}&P_{(0)}&\\
&&&&\cdots \\
\end{pmatrix}.
\end{aligned}
\end{equation}
Note that all the diagonal blocks are the same, which is an advantage of summation over $m$. Similarly, we have
\begin{equation}
\begin{aligned}
\partial_{\alpha_1}(\sum_m\mathcal{P}_{\epsilon+m\omega,\epsilon'+m\omega})=
\begin{pmatrix}
\cdots& & && \\
&(\partial_{\alpha_1}P)_{(0)}&(\partial_{\alpha_1}P)_{(1)}&(\partial_{\alpha_1}P)_{(2)}&\\
&(\partial_{\alpha_1}P)_{(-1)}&(\partial_{\alpha_1}P)_{(0)}&(\partial_{\alpha_1}P)_{(1)}&\\
&(\partial_{\alpha_1}P)_{(-2)}&(\partial_{\alpha_1}P)_{(-1)}&(\partial_{\alpha_1}P)_{(0)}&\\
&&&&\cdots \\
\end{pmatrix},
\end{aligned}
\end{equation}
in which
\begin{equation}
(\partial_{\alpha_1}P)_{(m)}=\sum_{\epsilon<\epsilon_n<\epsilon'}\sum_{m_3}\sum_{m_4}\partial_{\alpha_1} (\ket{\phi_{n}^{(m_3)}}\bra{\phi_{n}^{(m_4)}})\delta_{m_3-m_4-m}.
\end{equation}
One can readily check that \bea \sum_{m'} P_{(m')} (\partial_{\alpha_1} P)_{(m-m')}  = (P\partial_{\alpha_1} P)_{(m)}, \eea
from which it follows that
\begin{equation}
\begin{aligned}
&(\sum_m\mathcal{P}_{\epsilon+m\omega,\epsilon'+m\omega})\partial_{\alpha_1}
(\sum_m\mathcal{P}_{\epsilon+m\omega,\epsilon'+m\omega}) =
\begin{pmatrix}
\cdots& & && \\
&(P\partial_{\alpha_1}P)_{(0)}&(P\partial_{\alpha_1}P)_{(1)}&(P\partial_{\alpha_1}P)_{(2)}&\\
&(P\partial_{\alpha_1}P)_{(-1)}&(P\partial_{\alpha_1}P)_{(0)}&(P\partial_{\alpha_1}P)_{(1)}&\\
&(P\partial_{\alpha_1}P)_{(-2)}&(P\partial_{\alpha_1}P)_{(-1)}&(P\partial_{\alpha_1}P)_{(0)}&\\
&&&&\cdots \\
\end{pmatrix},
\end{aligned}
\end{equation}
in which
\begin{equation}
\begin{aligned}
(P\partial_{\alpha_1}P)_{(m)} =\sum_{n_1,n_2} \sum_{m_1,m_2,m_3,m_4} (\ket{\phi_{n_1}^{(m_1)}}\bra{\phi_{n_1}^{(m_2)}})\partial_{\alpha_1} (\ket{\phi_{n_2}^{(m_3)}}\bra{\phi_{n_2}^{(m_4)}})\delta_{m_1-m_2+m_3-m_4-m},
\end{aligned}
\end{equation} namely, the sum of the indices of ket-vectors minus that of the bra-vectors is $m$. The summation of $n_i$ is done within $\epsilon<\epsilon_{n_i}<\epsilon'$.

In the same way, we have
\bea
&& \left(\sum_m\mathcal{P}_{\epsilon+m\omega,\epsilon'+m\omega}\right) \partial_{\alpha_1}\left(\sum_m\mathcal{P}_{\epsilon+m\omega,\epsilon'+m\omega}\right) \cdots\partial_{\alpha_{d+D}}\left(\sum_m\mathcal{P}_{\epsilon+m\omega,\epsilon'+m\omega}\right)\nn \\&=&
\begin{pmatrix}
\cdots& & && \\
&(P\partial_{\alpha_1}P\cdots\partial_{\alpha_{d+D}}P)_{(0)}&(P\partial_{\alpha_1}P\cdots\partial_{\alpha_{d+D}}P)_{(1)}&(P\partial_{\alpha_1}P\cdots\partial_{\alpha_{d+D}}P)_{(2)}&\\
&(P\partial_{\alpha_1}P\cdots\partial_{\alpha_{d+D}}P)_{(-1)}&(P\partial_{\alpha_1}P\cdots\partial_{\alpha_{d+D}}P)_{(0)}&(P\partial_{\alpha_1}P\cdots\partial_{\alpha_{d+D}}P)_{(1)}&\\
&(P\partial_{\alpha_1}P\cdots\partial_{\alpha_{d+D}}P)_{(-2)}&(P\partial_{\alpha_1}P\cdots\partial_{\alpha_{d+D}}P)_{(-1)}&(P\partial_{\alpha_1}P\cdots\partial_{\alpha_{d+D}}P)_{(0)}&\\
&&&&\cdots \\
\end{pmatrix}.
\label{many-product}
\eea
\end{widetext}

Taking Eq.(\ref{many-product}) as an input, the frequency-domain Chern number of $\sum_m\mathcal{P}_{\epsilon+m\omega,\epsilon'+m\omega}$, according to the definition in Eq.(\ref{band-chern-2}), is given by
\bea
&&\mathcal{C}_{(d+D)/2}(\sum_m\mathcal{P}_{\epsilon+m\omega,\epsilon'+m\omega}) = \nn \\ &=&  \tilde{K}_{d+D}\int_{T^d\times S^D} d^dk d^Dr \text{Tr} [\epsilon^{\alpha_1\alpha_2\cdots\alpha_{d+D}} \left(\sum_m\mathcal{P}_{\epsilon+m\omega,\epsilon'+m\omega}\right) \nn\\ && \times \partial_{\alpha_1}\left(\sum_m\mathcal{P}_{\epsilon+m\omega,\epsilon'+m\omega}\right) \cdots\partial_{\alpha_{d+D}}\left(\sum_m\mathcal{P}_{\epsilon+m\omega,\epsilon'+m\omega}\right)]  \nn \\ &=& M \tilde{K}_{d+D}\int_{T^d\times S^D} d^dk d^Dr \text{Tr}[\epsilon^{\alpha_1\alpha_2\cdots\alpha_{d+D}} (P\partial_{\alpha_1}P\cdots\partial_{\alpha_{d+D}}P)_{(0)}]. \nn
\eea Note that all the off-diagonal blocks in Eq.(\ref{many-product}) do not contribute to the trace, therefore, only $(P\partial_{\alpha_1}P\cdots\partial_{\alpha_{d+D}}P)_{(0)}$ remains in the last line.

Due to Eq.(\ref{sum-chern}), the band Chern number of $P_{\epsilon,\epsilon'}$ is
\begin{equation}
\begin{aligned}
&\mathcal{C}_{(d+D)/2}( \mathcal{P}_{\epsilon,\epsilon'})   \\   &= \frac{1}{M}\mathcal{C}_{(d+D)/2}(\sum_m\mathcal{P}_{\epsilon+m\omega,\epsilon'+m\omega})     \\ &=\tilde{K}_{d+D}\int_{T^d\times S^D} d^dk d^Dr \text{Tr}[\epsilon^{\alpha_1\alpha_2\cdots\alpha_{d+D}} (P\partial_{\alpha_1}P\cdots\partial_{\alpha_{d+D}}P)_{(0)}].
\end{aligned} \label{bridge}
\end{equation}
Compared to the original frequency-domain Chern number in Eq.(\ref{band-chern-2}), the above expression looks much closer to the effective-Hamiltonian-based band Chern number in Eq.(\ref{band-chern-appendix}), yet it is not exactly the same.

The proof of Eq.(\ref{equivalence}) will be completed if we can also transform Eq.(\ref{band-chern-appendix}) to Eq.(\ref{bridge}). This is indeed the case.  To this end, let us define a time-dependent projection operator \bea P_{\epsilon,\epsilon'}(\bk,\br,t) =\sum_{\epsilon<\epsilon_n<\epsilon'} \ket{\psi_n(\bk,\br,t)}\bra{\psi_n(\bk,\br,t)}. \eea Apparently, $P_{\epsilon,\epsilon'}(\bk,\br,t=0)$ is simply the original projection operator $P_{\epsilon,\epsilon'}(\bk,\br)$. More explicitly, we have
\bea P_{\epsilon,\epsilon'}(\bk,\br,t) =\sum_{\epsilon<\epsilon_n<\epsilon'}\sum_{m_1,m_2} e^{i(m_1-m_2)\omega t} \ket{\phi_n^{(m_1)}(\bk,\br)}\bra{\phi_n^{(m_2)}(\bk,\br)}.\nn \eea
In terms of our shorthand notation of ``$(m)$'' in Eq.(\ref{(m)}), $P_{\epsilon,\epsilon'}(\bk,\br,t)$ reads \bea P_{\epsilon,\epsilon'}(\bk,\br,t) =\sum_m e^{im\omega t} P_{(m)}.  \eea

Since $\ket{\psi_n(\bk,\br,t)}$'s are smooth functions of $t$, the Chern number defined in terms of $P_{\epsilon,\epsilon'}(t)$ cannot change as a function of $t$, i.e., \bea C_{(d+D)/2}(P_{\epsilon,\epsilon'}(t)) = C_{(d+D)/2}(P_{\epsilon,\epsilon'}(t=0))\equiv C_{(d+D)/2}(P_{\epsilon,\epsilon'}), \nn \eea
therefore, $C_{(d+D)/2}(P_{\epsilon,\epsilon'})$ can be calculated as the time average of $C_{(d+D)/2}(P_{\epsilon,\epsilon'}(t))$:
\begin{equation}
\begin{aligned}
&C_{(d+D)/2}(P_{\epsilon,\epsilon'})= \frac{1}{\tau}\int_{0}^{\tau}dt\, C_{(d+D)/2}(P_{\epsilon,\epsilon'}(t))\\&=\frac{1}{\tau}\tilde{K}_{d+D}\int_{0}^{\tau}dt\int_{T^d\times S^D} d^dk d^Dr\\
&\times\text{Tr}[\epsilon^{\alpha_1\alpha_2\cdots \alpha_{d+D}} P_{\epsilon,\epsilon'}(t)  \partial_{\alpha_1} P_{\epsilon,\epsilon'}(t) \cdots\partial_{\alpha_{d+D}} P_{\epsilon,\epsilon'}(t)].
\end{aligned} \label{chern-fourier}
\end{equation}
It can be readily found that
\begin{equation}
\label{}
\begin{aligned}
&P_{\epsilon,\epsilon'}(t)  \partial_{\alpha_1} P_{\epsilon,\epsilon'}(t) \cdots\partial_{\alpha_{d+D}} P_{\epsilon,\epsilon'}(t)\\
&=\sum\limits_{m}(P\partial_{\alpha_1}P\cdots\partial_{\alpha_{d+D}}P)_{(m)}\text{e}^{im\omega t},
\end{aligned}
\end{equation} where the shorthand notation of the subscript ``$(m)$'' is used in the same way as defined above.
Inserting it into Eq.(\ref{chern-fourier}), we have
\begin{equation}
\begin{aligned}
&C_{(d+D)/2}(P_{\epsilon,\epsilon'}) =\frac{1}{\tau}\tilde{K}_{d+D}\int_{0}^{\tau}dt\int_{T^d\times S^D} d^dk d^Dr\\
&\times\text{Tr}[\epsilon^{\alpha_1\alpha_2\cdots \alpha_{d+D}} \sum_m (P\partial_{\alpha_1}P\cdots\partial_{\alpha_{d+D}}P)_{(m)} e^{im\omega t} ].
\end{aligned}
\end{equation}
The integration over $t$ can be straightforwardly done, which keeps only the $m=0$ Fourier component:
\begin{equation}
\begin{aligned}
&C_{(d+D)/2}(P_{\epsilon,\epsilon'}) = \tilde{K}_{d+D} \int_{T^d\times S^D} d^dk d^Dr\\
&\times\text{Tr}[\epsilon^{\alpha_1\alpha_2\cdots \alpha_{d+D}} (P\partial_{\alpha_1}P\cdots\partial_{\alpha_{d+D}}P)_{(0)}].
\end{aligned}
\end{equation}
Since we have already transformed the frequency-domain band Chern number in Eq.(\ref{band-chern-2}) to the same formula [see Eq.(\ref{bridge})], we have proved Eq.(\ref{equivalence}), which states that the effective-Hamiltonian-based Chern number ($H^{\rm eff}$-based Chern number) is equal to the frequency-domain Chern number ($\mathcal{H}$-based Chern number).


\bibliography{dirac}

\end{document}